%% file: paper.tex
\documentclass[tighten,times,twocolumn]{aastex63}

\input{preamble}
\shorttitle{Detection of Cosmological \tcm Emission with CHIME}

\begin{document}

\title{Detection of Cosmological \tcm Emission with the Canadian Hydrogen Intensity Mapping Experiment}

\input{authorlist}

\begin{abstract}
We present a detection of \tcm emission from large-scale structure (LSS) between redshift 0.78 and 1.43 made with the Canadian Hydrogen Intensity Mapping Experiment (CHIME).  Radio observations acquired over 102 nights are used to construct maps which are foreground filtered and stacked on the angular and spectral locations of luminous red galaxies (LRG), emission line galaxies (ELG), and quasars (QSO) from the eBOSS clustering catalogs.  We find decisive evidence for a detection when stacking on all three tracers of LSS, with the logarithm of the Bayes Factor equal to 18.9 (LRG), 10.8 (ELG), and 56.3 (QSO).  An alternative frequentist interpretation, based on the likelihood-ratio test, yields a detection significance of $7.1\sigma$ (LRG), $5.7\sigma$ (ELG), and $11.1\sigma$ (QSO).  These are the first \tcm intensity mapping measurements made with an interferometer.  We constrain the effective clustering amplitude of neutral hydrogen (HI), defined as $\mathcal{A}_{\sHI}\equiv10^{3}\,\Omega_{\sHI}\left(b_{\sHI}+\langle\,f\mu^{2}\rangle\right)$, where $\Omega_{\sHI}$ is the cosmic abundance of HI, $b_{\sHI}$ is the linear bias of HI, and $\langle\,f\mu^{2}\rangle=0.552$ encodes the effect of redshift-space distortions at linear order.  We find $\mathcal{A}_{\sHI}=1.51^{+3.60}_{-0.97}$ for LRGs $(z=0.84)$, $\mathcal{A}_{\sHI}=6.76^{+9.04}_{-3.79}$ for ELGs $(z=0.96)$, and $\mathcal{A}_{\sHI}=1.68^{+1.10}_{-0.67}$ for QSOs $(z=1.20)$, with constraints limited by modeling uncertainties at nonlinear scales.  We are also sensitive to bias in the spectroscopic redshifts of each tracer, and find a non-zero bias $\Delta\,v=\SI{-66}{}\pm\SI{20}{\kilo\meter\per\second}$ for the QSOs.  We split the QSO catalog into three redshift bins and have a decisive detection in each, with the upper bin at $z=1.30$ producing the highest redshift \tcm intensity mapping measurement thus far.
\end{abstract}


\input{sections/introduction}

\input{sections/data}

\input{sections/processing}

\input{sections/stacking}

\input{sections/simulations}

\input{sections/results}

\input{sections/validation}

\input{sections/cosmological}

\input{sections/conclusions}

\input{sections/acknowledgment}

\appendix

\input{appendix/delay_spectrum}

\input{appendix/beam_calibration}

\input{appendix/lognormal_stacking}

\input{appendix/template_calculation}

\bibliography{paper}{}
\bibliographystyle{aasjournal}

\end{document}

%% file: preamble.tex
\graphicspath{{./}{figures/}}

\input{mathdef}

\usepackage{hyperref}
\usepackage{xspace}
\usepackage{graphicx}
\usepackage{booktabs}
\usepackage[enable]{easy-todo}
\usepackage{comment}

\usepackage{savesym}
\savesymbol{tablenum}
\usepackage{siunitx}
\restoresymbol{SIX}{tablenum}
\DeclareSIUnit\Msolar{M_\odot}
\DeclareSIUnit\degr{deg}
\DeclareSIUnit\parsec{pc}
\DeclareSIUnit\dBm{dBm}
\DeclareSIUnit\jansky{Jy}
\DeclareSIUnit\beam{beam}
\DeclareSIUnit\h{h}

\DeclareMathOperator*{\var}{Var}

\definecolor{ucol}{rgb}{.6,0,0}
\definecolor{webgreen}{rgb}{0,.5,0}
\definecolor{halfgray}{gray}{0.55}

\usepackage[capitalize]{cleveref}
\newcommand{\secref}[1]{Section~\ref{#1}}

\newcommand{\ha}{\ensuremath{\textsc{ha}}} 
\newcommand{\synth}{\ensuremath{b_{\rm synth}}} 
\newcommand{\kB}{\ensuremath{k_{\rm B}}}
\newcommand{\taucut}{\ensuremath{\tau_{\rm cut}}}
\newcommand{\sinza}{\ensuremath{\sin{\left(\theta - \Lambda\right)}}}
\newcommand{\sinzap}{\ensuremath{\sin{\left(\theta_{\rm alias}^{\prime}(\nu) - \Lambda\right)}}}
\newcommand{\HI}{\ensuremath{{\rm HI}}}
\newcommand{\sHI}{\ensuremath{{\scriptscriptstyle {\rm HI}}}}
\newcommand{\sg}{{\rm g}}

\newcommand{\Tb}{\ensuremath{T_{\rm b}}}
\newcommand{\Tbar}{\ensuremath{\bar{T}_{\rm b}}}

\newcommand{\OmegaHI}{\ensuremath{\Omega_\sHI}}
\newcommand{\nbarHI}{\ensuremath{\bar{n}_\sHI}}
\newcommand{\nHI}{\ensuremath{n_\sHI}}
\newcommand{\deltaHI}{\ensuremath{\delta_\sHI}}
\newcommand{\deltag}{\ensuremath{\delta_\sg}}
\newcommand{\deltaG}{\ensuremath{\delta_{\rm G}}}
\newcommand{\deltam}{\ensuremath{\delta_{\rm m}}}
\newcommand{\tcm}{21$\,$cm\xspace}  
\newcommand{\nuHI}{\ensuremath{\nu_{\scriptscriptstyle 21}}}

\newcommand{\bHI}{\ensuremath{b_\sHI}}
\newcommand{\bg}{\ensuremath{b_\sg}}

\newcommand{\zeff}{\ensuremath{z_\text{eff}}}
\newcommand{\AHI}{{\ensuremath{\mathcal{A}_\sHI}}}

\newcommand{\MHIg}{\ensuremath{\left\langle M_\sHI \right\rangle_\sg}}
\newcommand{\Mten}{\ensuremath{M_{10}}}
\newcommand{\fid}{\ensuremath{\text{fid}}}
\newcommand{\alphaNL}{\ensuremath{\alpha_{\rm NL}}}
\newcommand{\PL}{\ensuremath{P_{\rm L}}}
\newcommand{\PNL}{\ensuremath{P_{\rm NL}}}
\newcommand{\alphaFoG}[1]{\ensuremath{\alpha_{\rm FoG,#1}}}
\newcommand{\DFoG}[1]{\ensuremath{D_{#1}^{\rm FoG}}}
\newcommand{\Dnu}{\ensuremath{\Delta\nu_0}}
\newcommand{\fmu}{\langle f \mu^2\rangle}
\newcommand{\alphaOmega}{\alpha_\Omega}
\newcommand{\alphah}{\alpha_\sHI}
\newcommand{\alphag}{\alpha_\sg}
\newcommand{\vf}{\vec{f}}
\newcommand{\vz}{\vec{z}}
\newcommand{\mD}{\mat{D}}

%% file: mathdef.tex
\usepackage{amsmath}
\usepackage{bm}
\usepackage{mathtools}

\renewcommand{\la}{\ensuremath{\left\langle}}  
\newcommand{\ra}{\ensuremath{\right\rangle}}

\newcommand{\lv}{\ensuremath{\left\lvert}}   
\newcommand{\rv}{\ensuremath{\right\rvert}}

\newcommand{\ls}{\ensuremath{\left[}}        
\newcommand{\rs}{\ensuremath{\right]}}

\newcommand{\lp}{\ensuremath{\left(}}        
\newcommand{\rp}{\ensuremath{\right)}}

\def\beq{\begin{equation}}
\def\eeq{\end{equation}}

\renewcommand{\vec}[1]{{\boldsymbol{\mathbf{#1}}}}
\newcommand{\mat}[1]{{\boldsymbol{\mathbf{#1}}}}

\newcommand{\hconj}{\ensuremath{\dagger}}

\newcommand{\vd}{\vec{d}}
\newcommand{\vk}{\vec{k}}
\newcommand{\vn}{\vec{n}}

\renewcommand{\vr}{\vec{r}}
\newcommand{\vs}{\vec{s}}

\newcommand{\vv}{\vec{v}}
\newcommand{\vw}{\vec{w}}
\newcommand{\vx}{\vec{x}}

\newcommand{\vmu}{\vec{\mu}}
\newcommand{\vtheta}{\vec{\theta}}

\newcommand{\mA}{\mat{A}}

\newcommand{\mC}{\mat{C}}

\newcommand{\mF}{\mat{F}}

\newcommand{\mI}{\mat{I}}

\newcommand{\mM}{\mat{M}}
\newcommand{\mN}{\mat{N}}

\newcommand{\mS}{\mat{S}}

\newcommand{\mU}{\mat{U}}
\newcommand{\mV}{\mat{V}}
\newcommand{\mW}{\mat{W}}
\newcommand{\mX}{\mat{X}}

\newcommand{\mSigma}{\mat{\Sigma}}

\DeclareMathOperator{\sinc}{sinc}

\newcommand{\appropto}{\mathrel{\vcenter{
  \offinterlineskip\halign{\hfil$##$\cr
    \propto\cr\noalign{\kern2pt}\sim\cr\noalign{\kern-2pt}}}}}

\newcommand{\calG}{\mathcal{G}}

\newcommand{\calL}{\mathcal{L}}
\newcommand{\calM}{\mathcal{M}}
\newcommand{\calP}{\mathcal{P}}

\newcommand{\calZ}{\mathcal{Z}}

\newcommand\numberthis{\addtocounter{equation}{1}\tag{\theequation}}

%% file: authorlist.tex


\newcommand{\UBC}{Department of Physics and Astronomy, University of British Columbia, Vancouver, BC, Canada}
\newcommand{\MITP} {Department of Physics, Massachusetts Institute of Technology, Cambridge, MA, USA}
\newcommand{\MITK} {MIT Kavli Institute for Astrophysics and Space Research, Massachusetts Institute of Technology, Cambridge, MA, USA}
\newcommand{\TRU}{Department of Physical Sciences, Thompson Rivers University, Kamloops, BC, Canada}
\newcommand{\PI}{Perimeter Institute for Theoretical Physics, Waterloo, ON, Canada}
\newcommand{\DRAO}{Dominion Radio Astrophysical Observatory, Herzberg Astronomy \& Astrophysics Research Centre, National Research Council Canada, Penticton, BC, Canada}
\newcommand{\UBCO}{Department of Computer Science, Math, Physics, and Statistics, University of British Columbia-Okanagan, Kelowna, BC, Canada}
\newcommand{\McGill}{Department of Physics, McGill University, Montreal, QC, Canada}
\newcommand{\UofTastro}{David A.\ Dunlap Department of Astronomy \& Astrophysics, University of Toronto, Toronto, ON, Canada}
\newcommand{\UofTphys}{Department of Physics, University of Toronto, Toronto, ON, Canada}
\newcommand{\WVU} {Department of Computer Science and Electrical Engineering, West Virginia University, Morgantown WV, USA}
\newcommand{\WVUA} {Department of Physics and Astronomy, West Virginia University, Morgantown, WV, USA}
\newcommand{\WVUGWAC} {Center for Gravitational Waves and Cosmology, West Virginia University, Morgantown, WV, USA}
\newcommand{\Yale}{Department of Physics, Yale University, New Haven, CT, USA}
\newcommand{\YaleA}{Department of Astronomy, Yale University, New Haven, CT, USA}
\newcommand{\Dunlap}{Dunlap Institute for Astronomy and Astrophysics, University of Toronto, Toronto, ON, Canada}
\newcommand{\RRI}{Raman Research Institute, Sadashivanagar,   Bengaluru, India}
\newcommand{\ASIAA}{Institute of Astronomy and Astrophysics, Academia Sinica, Taipei, Taiwan}
\newcommand{\CITA}{Canadian Institute for Theoretical Astrophysics, Toronto, ON, Canada}
\newcommand{\CIFAR}{Canadian Institute for Advanced Research,  Toronto, ON, Canada}
\newcommand{\WVUphysastro} {Department of Physics and Astronomy, West Virginia University, Morgantown, WV, USA}

\shortauthors{CHIME Collaboration}
\collaboration{100}{The CHIME Collaboration:}
%

%
%
%
\author[0000-0001-6523-9029]{Mandana Amiri}
\affiliation{\UBC}
\author[0000-0003-3772-2798]{Kevin Bandura}
\affiliation{\WVU}
\affiliation{\WVUGWAC}
\author[0000-0003-0173-6274]{Tianyue Chen}
\affiliation{\MITK}
\author[0000-0001-8123-7322]{Meiling Deng}
\affiliation{\DRAO}
\affiliation{\PI}
\affiliation{\UBC}
\author[0000-0001-7166-6422]{Matt Dobbs}
\affiliation{\McGill}
\author[0000-0002-6899-1176]{Mateus Fandino}
\affiliation{\UBC}
\affiliation{\TRU}
\author[0000-0002-0190-2271]{Simon Foreman}
\affiliation{\PI}
\affiliation{\DRAO}
\author[0000-0002-1760-0868]{Mark Halpern}
\affiliation{\UBC}
\author[0000-0001-7301-5666]{Alex S. Hill}
\affiliation{\UBCO}
\affiliation{\DRAO}
\author[0000-0002-4241-8320]{Gary Hinshaw}
\affiliation{\UBC}
\author[0000-0003-4887-8114]{Carolin H\"ofer}
\affiliation{\UBC}
\author[0000-0002-3354-3859]{Joseph Kania}
\affiliation{\WVUphysastro}
\author[0000-0003-1455-2546]{T.L. Landecker}
\affiliation{\DRAO}
\author[0000-0001-8064-6116]{Joshua MacEachern}
\affiliation{\UBC}
\author[0000-0002-4279-6946]{Kiyoshi Masui}
\affiliation{\MITK}
\affiliation{\MITP}
\author[0000-0002-0772-9326]{Juan Mena-Parra}
\affiliation{\MITK}
\author[0000-0001-8292-0051]{Nikola Milutinovic}
\affiliation{\UBC}
\author[0000-0002-2626-5985]{Arash Mirhosseini}
\affiliation{\UBC}
\author[0000-0002-7333-5552]{Laura Newburgh}
\affiliation{\Yale}
\author[0000-0002-2465-8937]{Anna Ordog}
\affiliation{\UBCO}
\affiliation{\DRAO}
\author[0000-0003-2155-9578]{Ue-Li Pen}
\affiliation{\CITA}
\affiliation{\ASIAA}
\affiliation{\UofTastro}
\affiliation{\PI}
\author[0000-0002-9516-3245]{Tristan Pinsonneault-Marotte}
\affiliation{\UBC}
\author[0000-0002-5283-933X]{Ava Polzin}
\affiliation{\YaleA}
\author[0000-0001-6967-7253]{Alex Reda}
\affiliation{\Yale}
\author[0000-0003-3463-7918]{Andre Renard}
\affiliation{\Dunlap}
\author[0000-0002-4543-4588]{J. Richard Shaw}
\affiliation{\UBC}
\author[0000-0003-2631-6217]{Seth R. Siegel}
\affiliation{\McGill}
\author[0000-0001-7755-902X]{Saurabh Singh}
\affiliation{\McGill}
\affiliation{\RRI}
\author[0000-0003-4535-9378]{Keith Vanderlinde}
\affiliation{\UofTastro}
\affiliation{\Dunlap}
\author[0000-0002-1491-3738]{Haochen Wang}
\affiliation{\MITK}
\affiliation{\MITP}
\author[0000-0002-6669-3159]{Donald V. Wiebe}
\affiliation{\UBC}
\author[0000-0001-7314-9496]{{\rm and} Dallas Wulf}
\affiliation{\McGill}

\correspondingauthor{J. Richard Shaw,  Seth Siegel}
\email{richard@phas.ubc.ca \quad  seth.siegel@mcgill.ca}



%% file: sections/introduction.tex

\section{Introduction}

Measurements of the large-scale clustering of matter have great potential to improve our understanding of both the early and late universe, probing phenomena ranging from cosmic inflation to dark energy to galaxy evolution. This large-scale structure can be mapped in a variety of ways, including tabulating the locations of luminous objects, using gravitational lensing to relate the distorted appearance of
galaxy shapes to mass along the line of sight, identifying the absorption of Lyman-alpha photons in the spectra of distant quasars, and isolating so-called secondary anisotropies in maps of the cosmic microwave background (CMB).

Another approach to mapping large-scale structure, {\em \tcm intensity mapping}, uses the hyperfine ``spin-flip" transition in neutral hydrogen (hereafter HI), which has rest wavelength \SI{21.106}{\centi\meter} (rest frequency \SI{1420.406}{\mega\hertz}). The probability of this transition occurring spontaneously
in a given hydrogen atom is extremely low, but this is balanced by the large cosmic abundance of HI in such a way that extragalactic \tcm emission (and/or absorption) is measurable in aggregate. 
The lack of comparably strong spectral lines at frequencies below $\SI{1420}{\mega\hertz}$ and the optical thinness of the hyperfine transition together imply that we can, if foregrounds can be removed, directly observe a redshift of the \tcm line. This can then be related to a distance from the observer.
Thus, maps of the radio sky at different frequencies contain information about the distribution of HI at different cosmic times, and the spectral and angular fluctuations of these maps can provide us with a three-dimensional picture of this distribution \citep{battye2004,chang2008,wyithe2008,peterson2009}. This idea extends beyond the \tcm line, and intensity mapping is now being pursued across a wide range of atomic and molecular transitions \citep{kovetz2019}.

At $z\lesssim 6$, after cosmic reionization has completed, the vast majority of HI is concentrated in the surroundings of galaxies, where it is shielded from ionizing radiation \citep{villaescusa-navarro2018}. Thus, a post-reionization \tcm intensity mapping survey is effectively a coarse-grained galaxy survey, in which galaxies are detected in bulk via their HI content.

\tcm brightness temperature fluctuations are therefore highly correlated with galaxy catalogs from other surveys, 
and this fact has enabled the first detections of large-scale structure using \tcm intensity mapping. After the initial detection by \cite{pen2009}, which combined existing spectral intensity data from the HIPASS survey with the 6dF galaxy survey, subsequent analyses have used dedicated observations by the Green Bank and Parkes radio telescopes, in concert with galaxy catalogs from the DEEP2, WiggleZ, 2dF, and eBOSS surveys \citep{chang2010,masui2013,anderson2018,tramonte2020,li2021-parkesxwigglez,wolz2021}, to detect 
cross-correlations with signal to noise ratios between $4$ and $13$. Several of these studies have placed constraints on the product $\OmegaHI \bHI r$, where $\OmegaHI$ is the mean HI density as a fraction of the present-day critical density, $\bHI$ is the linear bias of HI with respect to matter, and $r$ is a cross-correlation parameter that absorbs uncertainties in the modelling.

In principle, much more powerful measurements of large-scale structure are possible with custom-built telescopes that are optimized for \tcm observations. This, alongside several other science targets, motivated the design and construction of the Canadian Hydrogen Intensity Mapping Experiment (CHIME)\footnote{\url{http://chime-experiment.ca/}}. CHIME is a transit radio interferometer composed of four \SI{20}{\meter} $\times$ \SI{100}{\meter} cylindrical reflectors, each instrumented with 256 dual-polarized feeds observing at \SIrange[range-units = single, range-phrase=$-$]{400}{800}{\mega\hertz}. Signals from each feed are processed by an FX correlator and stored for offline cosmological analysis. These signals are also fed to separate backends devoted to studying fast radio bursts \citep{FRB2018} and pulsars \citep{CHIMEPulsar:2021}. 
\cite{overview-paper} provides an overview of the key features and operational status of the telescope.

In this paper, we report the first detection of large-scale structure with 21cm intensity mapping data from CHIME\footnote{Large-scale structure has previously been detected by cross-correlating CHIME's first catalog of fast radio bursts with photometric galaxy catalogs \citep{rafiei-ravandi2021}.}, in cross-correlation with galaxies and quasars measured by the extended Baryon Oscillation Spectroscopic Survey (eBOSS; \citealt{dawson2016}). We make use of a
stacking approach, which averages sky maps constructed from CHIME observations at the locations of each eBOSS object.
The data processing involved in this approach is more straightforward than other cross-correlation methods (e.g.\ a cross-power spectrum), and involves intermediate data products (such as sky maps) that can be interpreted in terms of features of the telescope and analysis pipeline. These interpretations are vital for examining the performance of our analysis methods, several of which have been custom-designed for CHIME.

Using 102 nights
of CHIME data, we have achieved significant detections of cross-correlations with eBOSS catalogs of luminous red galaxies (LRGs), emission-line galaxies (ELGs), and quasars (QSOs). We quantify this significance within a Bayesian framework, finding Bayes factors $\mathcal{Z}_1 / \mathcal{Z}_0$ (comparing our signal model with a noise-only model) of $\ln{(\mathcal{Z}_1 / \mathcal{Z}_0)} \approx 18.9$ (LRGs), 10.8 (ELGs), and 56.3 (QSOs), each corresponding to decisive evidence on the Jeffreys scale \citep{Jeffreys1961}; an alternative quantification, using a frequentist likelihood ratio test, yields signal to noise ratios of $7.1$ (LRGs), $5.7$ (ELGs), and $11.1$ (QSOs).

HI stacking analyses have previously been carried out on interferometric data from the Westerbork Synthesis Radio Telescope \citep{rhee2013,hu2019,hu2020}, the Giant Metrewave Radio Telescope \citep{lah2007,kanekar2016,rhee2016,rhee2018,bera2019,chowdhury2020} and the Very Large Array \citep{chen2021}, as well as on
single-antenna data from Parkes \citep{delhaize2013,tramonte2019,tramonte2020} and the Arecibo Legacy Fast ALFA Survey \citep{guo2020}. The primary motivation of many of these studies was to improve our understanding of galaxy evolution by probing the reservoirs of HI that serve as fuel for star formation. At $z \gtrsim 0.2$, the \tcm line is too faint to detect in individual galaxies, but stacking enables a measurement of the average \tcm flux (and therefore the average HI mass) across all objects in a given catalog, and a sufficiently small beam (possessed by the interferometers above) acts to limit the associated confusion noise. Under certain assumptions about the HI mass-luminosity relation, as well as the completeness and luminosity function of the catalog used for stacking, these measurements can also be used to constrain $\OmegaHI(z)$, which controls the overall amplitude of the large-scale \tcm fluctuations that can be used for cosmology (see \citealt{chen2021} for a recent summary of these constraints).

In contrast to the interferometers mentioned above, CHIME is designed to make \tcm observations that are intentionally confusion-dominated, allowing efficient mapping of the large-scale clustering of \tcm sources via the corresponding fluctuations in measured \tcm intensity in broad spatial pixels.
Thus, instead of exclusively probing the HI within individual objects in an external catalog, our stacking measurements are broadly sensitive to the nearby structures that are correlated with each object. To infer the value of $\OmegaHI(z)$, 
we must model gravitational and baryonic clustering in addition to the properties of the catalog objects themselves.

This modelling is most straightforward at the largest spatial scales, but as part of our analysis, we have needed to apply aggressive filtering that has removed the sensitivity of the data to these well-understood scales. 
Nevertheless, after marginalizing over the uncertainty associated with modelling of smaller-scale clustering, we are able to constrain an effective HI clustering amplitude $\AHI$, defined as $\OmegaHI (\bHI + \langle f\mu^2 \rangle)$, where $\bHI$ is a linear bias factor that relates large-scale clustering of HI to the clustering of all matter, and $\langle f \mu^2\rangle$ is an effective quantity involving the linear growth rate, $f$, and the relative contribution of line of sight
and
transverse information  (described in detail in \secref{sec:modelfitting}). With $\langle f \mu^2\rangle = 0.552$, we obtain $\AHI=1.51_{-0.97}^{+3.60}$ (LRGs), $\AHI=6.76_{-3.74}^{+9.04}$ (ELGs), and $\AHI=1.68_{-0.67}^{+1.10}$ (QSOs), to be compared with fiducial model values of $1.13$, $1.21$, and $1.37$ respectively. While this precision is lower than previous
single-antenna measurements, it is significantly more robust in its incorporation of modelling uncertainty: if we were able to fix the values of all small-scale parameters {\em a priori}, the precision on $\AHI$ would improve to between 10 and 20\% for each sample.

\begin{figure*}
   \centering \includegraphics[width=0.98\linewidth,keepaspectratio]{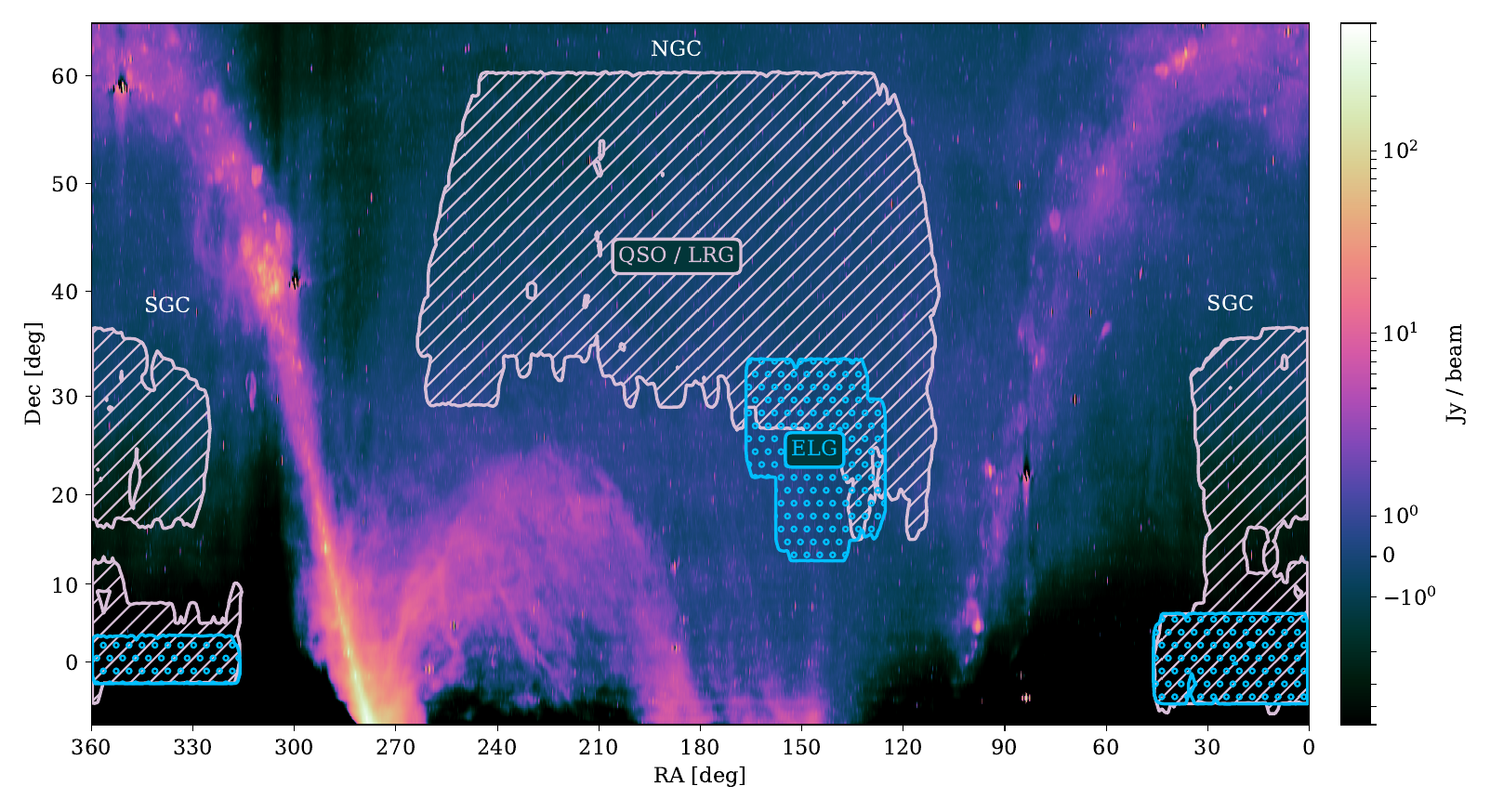}
    \caption{Map of the northern radio sky as measured by CHIME.  Shown is the average spectral flux density over the \SIrange[range-units = single, range-phrase=$-$]{587.5}{800}{\mega\hertz} sub-band.  The hashed regions indicate the spatial footprints of the eBOSS catalogs.  The LRG and QSO catalog share a common footprint indicated by the light-pink, forward-slash hash marks.  The footprint of the ELG catalog is indicated by the blue, circular hash marks.  The eBOSS catalogs are spread across two fields: the North Galactic Cap (NGC) and South Galactic Cap (SGC).  We only present results for the NGC field in this work.  The color scale is linear between \SI{-1}{} and \SI{1}{\jansky\per\beam} and logarithmic otherwise.  The map contains negative values because the autocorrelation data have been excluded.  The zero point is defined by setting the median value of a quiet part of the map with RA between 135 and 150 deg equal to zero for each declination and frequency prior to averaging over the sub-band.}
    \label{fig:footprint}
\end{figure*}

\begin{figure}
   \centering \includegraphics[width=0.98\linewidth,keepaspectratio]{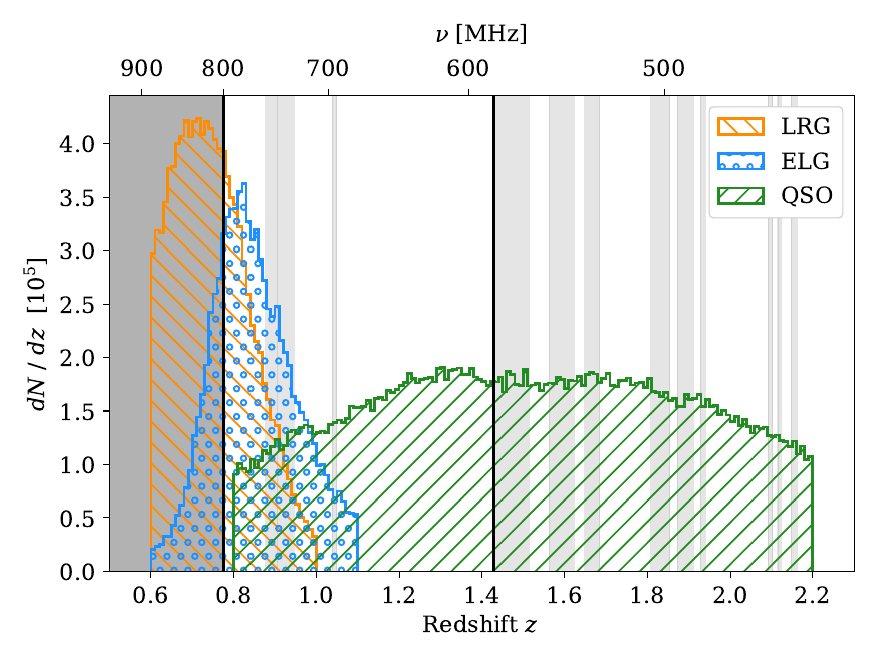}
    \caption{The redshift distribution of the LRG, ELG, and QSO catalogs for the North Galactic Cap (NGC) field.
The y-axis indicates the number of sources per unit redshift.  The upper x-axis indicates the frequency of \tcm emission from a source at the redshift indicated by the lower x-axis.  The dark gray band denotes the range of frequencies that are outside the CHIME band (\SIrange[range-units = single, range-phrase=$-$]{400}{800}{\mega\hertz}).  The light gray bands denote ranges of frequencies that are inaccessible to CHIME because they are contaminated by a persistent source of RFI.  The black solid lines mark the edges of the \SIrange[range-units = single, range-phrase=$-$]{587.5}{800}{\mega\hertz} sub-band that will be used in this analysis.}
    \label{fig:zdist}
\end{figure}

This paper (which includes descriptions of several analysis methods that have not previously appeared in the literature) is organized as follows:
\begin{itemize}
\item In \secref{sec:data}, we describe the CHIME and eBOSS data we use, visualizing the sky coverage in \cref{fig:footprint} and redshift coverage in \cref{fig:zdist}.
\item In \secref{sec:processing} and \secref{sec:stacking_pipeline}, we describe how CHIME data are processed into
stacks
at the locations of eBOSS catalog objects, including our procedures for real-time processing (\secref{sec:real_time}), applying additional corrections to individual days of data (\secref{sec:daily}), averaging over days (\secref{sec:sidereal_avg}), map making (\secref{sec:mapmaking}), beam calibration (\secref{sec:beams}), foreground filtering (\secref{sec:filter}), masking (\secref{sec:map_mask}), stacking (\secref{sec:stacking}), and covariance estimation (\secref{sec:covariance}).
\item In \secref{sec:modelling_simulations}, we discuss the cosmological scales our analysis probes (\secref{sec:scales}), our model for the stacking signal (\secref{sec:skymaps_signal}), our simulation framework (\secref{sec:simulations}), and our simulation-based approach to model fitting (\secref{sec:template}).
\item We begin \secref{sec:results} by presenting our stacking measurements and discussing several null tests in \secref{sec:stackmeasurements}. The main results are shown in Figs.~\ref{fig:stack2d} and~\ref{fig:stack1d}. We then introduce our model fitting procedure (\secref{sec:modelfitting}), visualize the constraints on the parameters of our model and discuss degeneracies (\secref{sec:constraints}; see \cref{fig:qso_reduced_params,fig:elg_reduced_params,fig:lrg_reduced_params}), and quantify the significance of the detected signal (\secref{sec:significance}; see \cref{tab:significance}).
\item In \secref{sec:validation}, we present the results of several validation tests that were performed on the data, related to consistency between the two instrumental polarizations, consistency between jackknives in observing time, beam calibration accuracy, and linearity of the stacking procedure.
\item In \secref{sec:interpretation}, we discuss several aspects of the interpretation of these results: confirmation of a systematic bias in the reported QSO redshifts (\secref{sec:quasar_redshift_errors}), uncertainties on our constraints on the HI clustering amplitude $\AHI$ (\secref{sec:AHI_errors}; see \cref{tab:constraints}), comparisons of the corresponding $\OmegaHI$ constraints with previous results from the literature (\secref{sec:OmegaHIresults}; see \cref{fig:omegaHI_comparison}), and prospects for constraining the mean HI mass of objects in external catalogs (\secref{sec:M10results}).
\item In \secref{sec:conclusions}, we state our conclusions and discuss the prospects for future \tcm measurements by CHIME.
\end{itemize}

We also include four appendices, detailing our Gibbs-sampling--based approach to delay spectrum estimation (Appendix~\ref{sec:delay_spectrum}), the construction of our main beam model using catalogs of point source fluxes (Appendix~\ref{sec:beam_calibration_ptsrc}), the justification for stacking simulated Gaussian \tcm maps on log-normal mock catalogs (Appendix~\ref{sec:lognormal_stacking}), and our construction of simulation-based signal templates used for model fitting (Appendix~\ref{app:template_calculation}).

%% file: sections/data.tex

\section{Data}
\label{sec:data}

\subsection{CHIME}
\label{sec:CHIMEdata}

Our analysis uses the CHIME \texttt{stack} dataset acquired between January 1 and November 5, 2019.  The \texttt{stack} dataset is described in \cite{overview-paper} and consists of the $N_{\rm feed}^{2}$ visibilities (with $N_{\rm feed} = 2048$) after they have been integrated to $\Delta t = \SI{9.9405}{\second}$ cadence, calibrated for complex gain variations, and compressed by averaging subsets of redundant baselines.  We selected 102 nights from this period to include in the analysis, using criteria that will be described in \secref{sec:bad_days}.   After masking intervals of poor data quality, these 102 nights contain \SI{521}{\hour} of total integration time on the relevant eBOSS field.

CHIME is sensitive to radio frequencies from \SIrange[range-units = single]{400}{800}{\mega\hertz}, which corresponds to \tcm emission from redshifts \SIrange{2.55}{0.78}{}.  However, frequencies from \SIrange[range-units = single]{400}{500}{\mega\hertz} suffer from frequent narrow-band, transient radio frequency interference (RFI).  In addition, approximately \SI{60}{\percent} of frequencies between \SIrange[range-units = single, range-phrase={~and~}]{488}{584}{\mega\hertz} are corrupted by persistent RFI from locally broadcast TV channels.  Hence, for this initial analysis we have restricted our attention to the CHIME data acquired in the relatively clean portion of the band between \SIrange[range-units = single, range-phrase={~and~}]{587.5}{800}{\mega\hertz},
corresponding to \tcm emission from redshifts \SIrange{1.42}{0.78}{}.  The spectral resolution of the \texttt{stack} dataset is $\Delta \nu = \SI{0.390625}{\mega\hertz}$, resulting in 544 frequency channels within this range.  We anticipate that the real-time, RFI-excision algorithm that was deployed on the CHIME correlator in mid-October 2019 and recent improvements to the offline RFI excision algorithm will enable the inclusion of the lower half of the CHIME band in future analyses.

\subsection{eBOSS catalogs}

eBOSS \citep{dawson2016}, the cosmological survey within SDSS-IV \citep{blanton2017}, was conducted over 4.5 years using spectrographs previously used for the BOSS survey \citep{smee2013}, mounted on the Sloan Telescope \citep{gunn2006} at the Apache Point Observatory. eBOSS produced four distinct samples of objects, each of which has been used to measure large-scale clustering and place constraints on a variety of cosmological parameters (see \citealt{alam2021} for a summary of these results). In this work, we cross-correlate three of these samples, from SDSS Data Release 16 \citep{ahumada2020}, with CHIME measurements.

The eBOSS emission line galaxy (ELG) sample \citep{raichoor2021} selected targets using imaging from the Dark Energy Camera Legacy Survey \citep{dey2019}, making special use of emission in the [OII] doublet at ($\lambda 3727$, $\lambda3729$~\si{\angstrom}) to obtain efficient and accurate redshift estimates. This resulted in a catalog of \SI{173736}{} unique objects over $0.6<z<1.1$, spread across two fields: the $550\,{\rm deg}^2$ North Galactic Cap (NGC) and the $620\,{\rm deg}^2$ South Galactic Cap (SGC). We show both fields, superimposed on a representative CHIME sky map, in \cref{fig:footprint}.

The luminous red galaxy (LRG) sample \citep{ross2020} is composed of objects from optical imaging taken during previous phases of SDSS \citep{albareti2017}, along with infrared data from the Wide Field Infrared Survey Explorer satellite \citep{lang2016}. Selection criteria were designed to target galaxies with $z>0.6$, with the resulting final catalog containing \SI{174816}{} objects over $0.6<z<1.0$, distributed between a $2566\,{\rm deg}^2$ NGC field and a $1676\,{\rm deg}^2$ SGC field (shown in \cref{fig:footprint}).

The quasar (QSO) sample \citep{ross2020,lyke2020} is composed of objects observed during previous phases of SDSS and new objects selected from the same imaging data as the LRGs. The QSO catalog used for clustering (as opposed to the QSOs used for Lyman-$\alpha$ forest studies) contains \SI{343708}{} objects over $0.8<z<2.2$, covering the same two fields as the LRGs.

\Cref{fig:zdist} shows the redshift distribution of the LRG, ELG, and QSO samples for the NGC field, along with vertical bands indicating redshift ranges that are outside of the CHIME band (dark gray) or excluded due to persistent RFI (light gray).

The stack of the SGC catalog on the CHIME data is a factor of \SIrange{3}{3.5}{} times noisier than the stack on the NGC catalog for the same tracer of large-scale structure.  There are several reasons for this.  First, in the case of the LRG and QSO catalogs there are \SI{50}{\percent} fewer sources in the SGC field compared to the NGC field.  Second, we have less integration time on the SGC field because the range of right ascension (RA) occupied by the SGC field transits at CHIME at night in the summer time, whereas the NGC field transits at night in the winter time when the nights are longer.  Finally, the SGC field is at a lower declination where the CHIME primary beam response is reduced and where we are forced to use a more aggressive delay filter because of aliasing of foregrounds.  For these reasons, we have only a modest detection ($\sim 4 \sigma$) of \tcm emission when stacking on the QSOs in the SGC field, and do not have a detection for the ELGs and LRGs in the SGC field.  In what follows, we present the results for the NGC field only.  We note, however, that our measurements in the SGC field are consistent with the amplitude of the \tcm signal inferred from the catalogs in the NGC field, given the increased noise.

Each object in the eBOSS clustering catalogs includes weight values that account for imaging systematics, close pairs (which can be affected by spectroscopic fiber collisions), and the probability of a catastrophic redshift failure.  We found that incorporating these weights into our analysis had a negligible impact on our result.  Therefore we do not employ the eBOSS weights in what follows.  The weight given to each object is determined entirely by the sensitivity of the CHIME data at that object's angular and spectral location.

\subsection{Effective Redshift of Tracer Cross Correlations}

\begin{deluxetable*}{ccccccc}[htb]
    \tablecaption{The redshift distribution of each tracer used in the cross-correlation analysis.\label{tab:zeff}}
    \tablecolumns{7}
    \tablehead{
        \colhead{Tracer} &
        \colhead{Frequency Range} &
        \multicolumn{2}{c}{Source Number} &
        \colhead{Effective Redshift} &
        \multicolumn{2}{c}{Redshift Range} \\
        \colhead{} &
        \colhead{(\si{\mega\hertz})} &
        \colhead{Total} &
        \colhead{Non-zero weight} &
        \colhead{$\zeff$} &
        \colhead{$z_\mathrm{min}$--$z_\mathrm{max}$} &
        \colhead{$z_{0.16}$--$z_{0.84}$}
    }
    \startdata
         LRG    & 585--800 & 39706 & 21615 & 0.84 & 0.78--1.00 & 0.81--0.87 \\
         ELG    & 585--800 & 63381 & 31181 & 0.96 & 0.78--1.10 & 0.83--1.03 \\
         QSO    & 585--800 & 94706 & 48046 & 1.20 & 0.80--1.43 & 1.00--1.36 \\
         \hline
         QSOb0  & 700--800 & 26908 & 11960 & 0.97 & 0.80--1.03 & 0.85--1.01 \\
         QSOb1  & 650--700 & 23760 & 12311 & 1.12 & 1.03--1.19 & 1.07--1.16 \\
         QSOb2  & 585--650 & 44038 & 23775 & 1.30 & 1.19--1.43 & 1.23--1.39 \\
         \hline
         QSOb00 & 745--800 & 11095 & 5299  & 0.84 & 0.80--0.91 & 0.82--0.87 \\
         QSOb01 & 700--745 & 15813 & 6661  & 0.99 & 0.91--1.03 & 0.96--1.01 \\
    \enddata
    \tablecomments{Frequency range gives the band that the analysis is limited to, whereas the redshift range gives the spread of source redshifts within that band. The $z_{0.16}$--$z_{0.84}$ span gives the 16 and 84\% weighted percentiles giving an effective range within `$1 \sigma$' of the effective redshift (the weighted median of the source redshifts). For later analysis we further split the QSO catalog into sub-bands, denoted by the QSObX and QSObXY tracers.}
\end{deluxetable*}

The effective redshift, $\zeff$, of each catalog, when cross correlated with CHIME data, is a combination of the redshift distribution of the sources in the catalog, the RFI mask used for the CHIME analysis, and the sensitivity of the CHIME data outside the masked regions. To determine $\zeff$, we first take each catalog, and for every source within it, we extract the inverse variance weights for that source in the processed CHIME data (these weights and how they are propagated through our pipeline will be described in \secref{sec:processing}).  We then use these to construct a weighted median of the redshifts of the catalog. Similarly, to define an effective range of each catalog we take the 16\% and 84\% weighted percentiles of the redshift distribution (i.e., the 68\% equal-tailed interval), which gives a region within `$1\sigma$' of the effective redshift. This differs substantially from the minimum--maximum redshift range where the source number density drops at the edges of the redshift distribution, most notably for the low redshift end of the quasar distribution and the high redshift end of the LRG distribution. These are all summarised in \cref{tab:zeff}.

Some sources have zero weight due to RFI masking and outlier cuts. In \cref{tab:zeff} we give the total number of sources within the frequency band being analysed, and an effective source number, defined as the number of sources lying within a voxel with \emph{non-zero} weight. Depending on the frequency range this is typically $\sim 50 \%$ of the total source number.

In \cref{tab:zeff} we also list five additional catalogs which are subdivisions of the QSO catalog, the largest and broadest redshift sample. The three catalogs QSOb0, QSOb1 and QSOb2 divide the redshift span into three roughly equal parts from lowest to highest redshift, the two catalogs QSOb00 and QSOb01 further divide the lowest redshift catalog into two more catalogs. These additional catalogs will be used in later analysis of the data.

\subsection{Coordinate Systems}

CHIME is a transit instrument, and as such we are acutely sensitive to the precession of the Earth's polar axis. Historically, the celestial coordinate system has been anchored to the vernal equinox, which makes the coordinate system sensitive to both an unavoidable precession of the Earth's polar axis, and an artificial shift in the zero point of the Right Ascension coordinate (B1950 and J2000 coordinates are realisations of this anchored at their respective epochs).

The new system outlined in \citet{IERS} and \citet{USNO179} fixes some of these problems. The fundamental position of sources is given in International Celestial Reference System (ICRS) coordinates, which are fixed and unchanging coordinates that are essentially aligned with J2000 coordinates. Position as seen by an observer on Earth can be given in Celestial Intermediate Reference System (CIRS) coordinates, a frame in which the polar axis shifts with the Earth's precession and the RA origin is \emph{minimally rotated}. Unlike previous equinox based coordinates, CIRS coordinates only contain the minimal shift required to keep the polar alignment. As such they are much more suited to use in CHIME: over a 5 year period a typical equinox RA position shifts by \SI{4.3}{\arcmin}, or around a quarter of a CHIME pixel, whereas a CIRS position changes only by \SI{1.5}{\arcmin}, about a tenth of a pixel. This means that we are able to trivially align and average data products such as maps over much longer periods.

In this new system Greenwich Apparent Sidereal Time is replaced by Earth Rotation Angle. Instead of local sidereal time we use \emph{local Earth-rotation angle}, which is equivalent to the current CIRS RA of the local meridian.

Throughout this paper the celestial coordinates we use will be CIRS coordinates at the average epoch of the data being analyzed, and any maps presented will be in those coordinates. In the absence of better terminology, we will use sidereal day to refer to the interval between transitions of the Earth-rotation angle through zero.

%% file: sections/processing.tex

\section{CHIME Data Processing Pipeline}
\label{sec:processing}

\begin{figure*}[t]
   \centering \includegraphics[width=1\linewidth,keepaspectratio, trim = 50 25 50 0]{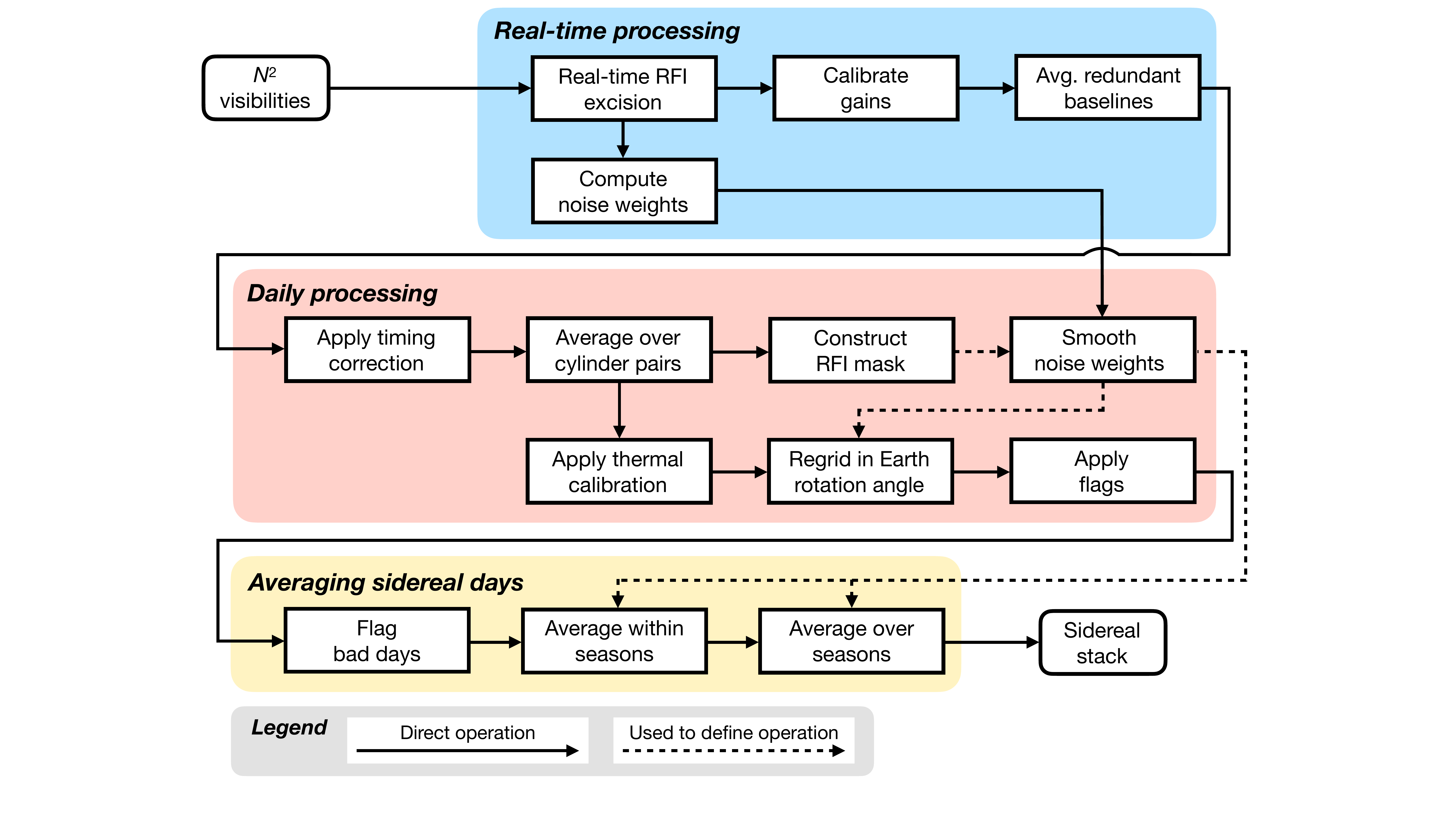}
    \caption{
    A schematic representation of the data processing pipeline, starting from visibilities and culminating in the generation of the calibrated average over \SI{102}{} nights that we refer to as a ``sidereal stack." The real-time pipeline (\secref{sec:real_time}) performs spectral-kurtosis-based RFI excision (for a subset of the timespan used in this work; see \secref{sec:realtimeRFI}), gain calibration, and averaging over redundant baselines within each cylinder pair. It also computes fast-cadence noise estimates that are used as weights in several later steps. The daily processing pipeline (\secref{sec:daily}) applies a correction for clock drift between different ADCs in the CHIME F-engine, further averages redundant baselines over cylinder pairs, applies an ambient-temperature-dependent gain correction factor, regrids the time axis of each day onto a common grid in local Earth rotation angle, and applies data quality flags. This pipeline also constructs a time-frequency mask that targets longer-timescale RFI, and incorporates this mask in a smoothing operation applied to the noise weights. Finally, we average over sidereal days (\secref{sec:sidereal_avg}), first manually identifying and excluding bad days of data before forming several seasonal averages and then averaging these seasons together.
    }
    \label{fig:siderealstack_flowchart}
\end{figure*}

The CHIME data processing pipeline can be divided into two parts, the real-time and offline pipeline.  The real-time pipeline runs on the CHIME correlator and supporting computing infrastructure.  It operates on the digitized voltages measured by the 2048 antenna feeds and outputs calibrated visibilities at 1024 frequency channels spanning the \SIrange[range-units = single]{400}{800}{\mega\hertz} band.  These are integrated to roughly $\SI{10}{\second}$ cadence and further compressed by averaging over a subset of the redundant baselines.  The offline pipeline runs on Compute Canada's Cedar cluster.  It operates on an archived copy of the visibility data and applies additional RFI masking and calibration, averages over all redundant baselines, interpolates onto a fixed grid in local Earth rotation angle, flags bad data, and averages over sidereal days. These real-time and offline operations, which produce the data product we refer to as a ``sidereal stack," are illustrated in \cref{fig:siderealstack_flowchart}.

The CHIME real-time pipeline, offline pipeline and analysis code used in this work is open source and publicly available. It can be found at \url{https://github.com/kotekan/}, \url{https://github.com/radiocosmology/} and \url{https://github.com/chime-experiment/}.

\subsection{Real-time Processing}
\label{sec:real_time}

We refer the reader to \cite{overview-paper} for a description of the CHIME correlator, the real-time pipeline, and the archived data products. Below we highlight several aspects of the real-time processing that are relevant for interpreting what follows.

\subsubsection{Real-time RFI Excision}
\label{sec:realtimeRFI}

\cite{overview-paper} describes an RFI-excision algorithm that runs on the CHIME correlator and is based on the spectral kurtosis statistic calculated at \SI{0.66}{\milli\second} cadence.  This algorithm was deployed for a test period in June 2019 and then turned off until mid-October 2019.  Hence, the majority of the data (\SI{82}{} of \SI{102}{} nights) that is used for this analysis did not benefit from fast-cadence RFI excision, and relies entirely on the offline, $\sim\SI{10}{\second}$ cadence excision algorithms that will be described in \secref{sec:rfi}. This mixed dataset is processed consistently in our analysis, but simply has a higher rate of flagging in the offline pipeline for the days where real-time excision was not used.

\subsubsection{Complex Gain Calibration}

The complex gain of each feed is calibrated once per sidereal day by fitting a model to the eigendecomposition of the $N_{\rm feed}^{2}$ visibility matrix during the transit of the brightest radio source that is available at night.  The primary calibration source is Cygnus A because it is the brightest radio point source in the sky between \SIrange[range-units = single, range-phrase = ~and~]{400}{800}{\mega\hertz}.  It is also unresolved by the longest CHIME baselines and has a stable, well-characterized spectral flux density.  Cassiopeia A, Taurus A, and Virgo A are used as alternative calibration sources when Cygnus A is transiting during the day.  If a source other than Cygnus A was used for calibration, then the resulting gains are corrected for differences in the primary beam pattern of each feed at the location of the calibrator relative to the location of Cygnus A.  This ``beam ratio'' is characterized by averaging the ratio of the gains from the two point sources over many nights.  Hence, the complex gain calibration effectively normalizes the primary beam response at each frequency to unity on meridian at the declination of Cygnus A.

The complex gains are scaled by the flux density of the calibrator source, such that application of the gains converts the visibility data to units of Jansky/beam.  The flux density of these sources was measured with the Karl G.~Jansky Very Large Array (VLA) in 2014 and 2016 at frequencies ranging from \SI{220}{\mega\hertz} to \SI{48.1}{\giga\hertz}.  VLA legacy observations from 1998 also exist at \SI{73.8}{\mega\hertz} for all sources but Casseopia A.  These measurements are interpolated to the CHIME band using the polynomial expressions provided in \cite{perley2017} (henceforth P17). The uncertainty on the relative spectral flux density of the calibration sources in the CHIME band is less than \SI{1}{\percent}.  The absolute flux of the P17 scale at these frequencies is determined by measurements of Cygnus~A by \cite{baars1977}, which the authors estimate is accurate at \SIrange[range-units = single, range-phrase = $-$]{3}{5}{\percent}.

\subsubsection{Compression}
\label{sec:compression}

The CHIME feeds are located on a regular grid, and as a result the $N^{2}_{\rm feed}$ visibilities contain many redundant measurements for each baseline.  In order to compress the data, the real-time pipeline performs a weighted average of all redundant baselines formed from feeds on the same pair of cylinders.  Correlator inputs that are malfunctioning or otherwise anomalous are identified and flagged in semi-real-time using 10 different tests based on a variety of data products and housekeeping metrics.  The weight given to a particular baseline is 0 if either of the inputs that form the baseline are currently flagged and 1 otherwise.  This uniform weighting scheme will result in lower sensitivity compared to an inverse variance weighting scheme that accounts for feed-to-feed differences in the noise referred to the sky.  We estimate that the magnitude of this degredation in sensitivity is approximately $\SI{5}{\percent}$.  Note that redundant baselines formed from feeds on different pairs of cylinders are not averaged at this stage.  This baseline collation strategy allows for cylinder-dependent corrections and calibrations to be applied offline.  Below, we refer to the resulting visibility for baseline $\vec{b}$ at frequency $\nu$ and time $t$ as $V_{\rm raw}(\vec{b}, \nu, t)$.

\subsubsection{Weights}
\label{sec:weights}

The real-time pipeline estimates the variance of the visibility for each baseline, frequency channel, and $\sim\SI{10}{\second}$ integration by differencing the even and odd \SI{30}{\milli\second} sub-integrations.  Since the observed foregrounds do not change significantly on \SI{30}{\milli\second} timescales, they cancel for this difference, leaving contributions from RFI and intrinsic radiometric noise.   This ``fast-cadence estimate'' of the variance is propagated through each stage of the real-time and offline pipeline.

In general, there is percent-level agreement between the fast-cadence estimate of the variance and the radiometric estimate calculated from the measured autocorrelation, frequency channel width, and total integration time.  Most cases where the two estimates differ correspond to known periods of bad data quality or have a temporal and spectral extent that is characteristic of transient RFI.  The fast-cadence estimate is used to construct inverse variance weights that are used to average over sidereal days, average over baselines during map making, and average over sources when stacking on external catalogs.  The inverse variance weights are \emph{not} used to average over redundant baselines, instead we use the uniform weighting scheme described in \secref{sec:compression}.

\subsection{Daily Processing}
\label{sec:daily}

Here we describe the daily pipeline that applies additional processing to a copy of the archived visibility data.  This includes correcting for clock drift, averaging over redundant baselines on different cylinder pairs, identifying and masking RFI, correcting common-mode thermal variations in the amplitude of the gain, interpolating the data onto a common grid in local Earth-rotation angle, and finally masking ranges of time with poor data quality.  We briefly describe each of these stages.  The primary data product output by this processing is the visibility for all unique baselines on each local sidereal day as a function of local Earth-rotation angle at $\Delta \nu = \SI{0.390625}{\mega\hertz}$ spectral resolution.

\subsubsection{Timing Correction}
\label{sec:timing}

The sampling rate of the analog to digital converters (ADCs) that digitize the signal measured by the CHIME feeds is derived from a \SI{10}{\mega\hertz} clock that originates from a GPS-disciplined, oven-controlled crystal oscillator and is distributed to the circuit boards that house the ADCs through a hierarchical network consisting of coaxial cables, power splitters, and amplifiers.  Thermal susceptibility of this distribution network results in copies of the clock drifting with respect to one another on timescales set by the different refrigeration cycles of the water chillers used to control the temperature of the electronics.  The magnitude of this effect is particularly large between copies of the clock provided to ADCs in different receiver huts, which are temperature controlled by independent chillers.

The CHIME ADCs are housed in eight electronics crates.  The thermal drift between the eight copies of the clock that are distributed to the eight crates is measured using a broadband noise source following the procedure described in \cite{overview-paper}.  This yields a proxy for the drift, $\delta \tau^{\rm clock}_{cd}(t)$, between the copies of the clock provided to electronics crate $c$ relative to electronics crate $d$.  The visibility is then corrected as follows
\begin{align}
\nonumber
    V_{\rm cal,1}(\vec{b}, \nu, t) &= \exp \! {\left\{-2 \pi j \nu\, \langle \delta \tau^{\rm clock}_{cd}(t) \rangle_{cd \in \vec{b}}\right\}} \\
    &\quad\times V_{\rm raw}(\vec{b}, \nu, t) \ .
\end{align}
Here $\langle \delta \tau^{\rm clock}_{cd}(t)\rangle_{cd \in \vec{b}}$ is constructed by averaging the estimates of the relative clock drift between the pairs of crates that digitize the pairs of inputs that form every redundant baseline averaged by the real-time pipeline to obtain $V_{\rm raw}(\vec{b}, \nu, t)$.  Applying this correction reduces the standard deviation of the delay noise on timescales less than \SI{20}{\minute} from \SI{4.25}{\pico\second} to \SI{1.5}{\pico\second} on average, as inferred from the phase stability of the signal from bright point sources.  Note that further improvements have been achieved by applying a more complicated, ADC-dependent correction in real-time, but the analysis described in this work uses the simpler, offline correction described above.

\subsubsection{Redundant Baseline Collation}
\label{sec:collate}

The timing correction described in the previous section is the only cylinder-dependent correction that was applied for this analysis.  The next stage of the pipeline averages \emph{all} redundant baselines by performing a weighted average over the redundant baselines measured by different cylinder pairs.  The weighting scheme used is consistent with the scheme used by the real-time pipeline.  Specifically, each cylinder pair is weighted by the number of redundant baselines that were previously averaged by the real-time pipeline.  These weights are constructed from the set of correlator input flags that were used by the real-time pipeline at each time sample.

\subsubsection{RFI Excision}
\label{sec:rfi}

Narrowband RFI will contaminate the high-delay modes that our analysis relies on to avoid the spectrally smooth foregrounds.  Hence, identifying and masking times and frequency channels that are corrupted by RFI is critical to detect the \tcm signal.  The RFI excision occurs in three stages, with each stage generating a single 2D mask in (frequency, time) that is applied to the weight dataset for all baselines before proceeding to the next stage.  The first stage masks any frequency channel that coincides with a known, persistent source of RFI.  There were two sources of persistent RFI in the \SIrange[range-units = single, range-phrase=$-$]{587.5}{800}{\mega\hertz} band: the mobile LTE bands and the local oscillator (LO) used by the Synthesis Telescope at DRAO \citep{landecker2000}. These two sources occupy \SI{14.2}{\percent} of the band.

The second stage creates a mask by identifying variations in the autocorrelation that have a spectral and temporal extent characteristic of RFI.  The average autocorrelation over the 2048 inputs is normalized at each frequency by the median value over the local sidereal day to remove static variations in the bandpass.  The median and median absolute deviation (MAD) are then calculated over a 2D moving window in (frequency, time) of size (\SI{10}{\mega\hertz},  \SI{7}{\minute}).  Any time and frequency where the autocorrelation deviates from the median by more than 5 times the MAD over the window centered on its location is masked.  The window size was calibrated by first identifying RFI events through manual inspection of the autocorrelations acquired on a few typical days, and then searching for a window that maximized the fraction of RFI corrupted data that is masked while minimizing the amount of clean sky data that is masked.

The third stage creates a mask by identifying RFI-like variations in the visibility data from the cross-polar, intra-cylinder baseline with \SI{10}{\meter} separation.  This baseline has a relatively large number of redundant copies, and thus low radiometric noise compared to most other baselines.  RFI events are more easily discriminated from the background radio sky in a cross-polar visibility because the RFI is in general polarized, whereas the radio sky is largely unpolarized at the scales at which the \SI{10}{\meter} baseline is sensitive.  The algorithm for identifying RFI events is similar to the algorithm applied to the autocorrelations.  The median is calculated over a 2D moving window in (frequency, time) of size (\SI{4.3}{\mega\hertz}, \SI{1}{\minute}) and subtracted to remove background radio emission from the sky.  The MAD is then calculated over a 2D window of size (\SI{8.2}{\mega\hertz}, \SI{7.4}{\minute}).  Any time and frequency where the visibility deviates from the median by more than 5 times the MAD over the window centered on its location is masked.  The window sizes were chosen using a procedure similar to that described in the previous paragraph.

One common source of transient RFI arises from the reflection of distant broadcast TV channels off meteor ionisation trails and aircraft.  These appear in known \SI{6}{\mega\hertz} wide bands and last $\sim$\SI{5}{\second}.  A targeted search for these events is performed on the moving median subtracted, cross-polar visibility by identifying time samples where the majority of frequencies within each TV channel are outliers.  The entire TV channel is masked if more than \SI{50}{\percent} of the frequencies within that TV channel exceed 1.8 times the moving MAD.  This results in a false positive rate equal to the \SI{5} MAD cut used in the standard third stage excision, assuming a Gaussian noise model.

\subsubsection{Thermal Calibration}
\label{sec:thermal_cal}

Common-mode variations in the amplitude of the complex receiver gain are corrected using a linear regression model based on measurements of the outside temperature.  Details of the model construction and an evaluation of its performance are provided in \cite{overview-paper}.  To briefly summarize, fractional variations in the amplitude of the complex gain inferred from hundreds of bright point source transits are regressed against the outside temperature as measured by the DRAO weather station at the time of transit.  The resulting thermal susceptibility increases with frequency from \SI{0.07}{\percent\per\kelvin} at \SI{400}{\mega\hertz} to \SI{0.2}{\percent\per\kelvin} at \SI{800}{\mega\hertz}, and varies across inputs at the \SI{0.05}{\percent\per\kelvin} level.  The susceptibility is averaged over the 2048 inputs and the frequency dependence is fit to a quadratic function.  The visibility measured by baseline $\vec{b}$ at frequency $\nu$ and time $t$ is then corrected as follows
\begin{align}
    V_{\rm cal,2}(\vec{b}, \nu, t) & = \left[1 + \alpha(\nu) \left(T(t) - T(t_{*})\right)\right]^2 \ V_{\rm cal,1}(\vec{b}, \nu, t)
\end{align}
where $\alpha(\nu)$ is the quadratic model for the thermal susceptibility, $T(t)$ is the outside temperature at time $t$, and $T(t_{*})$ is the outside temperature at time $t_{*}$ at which the complex gain calibration was derived by the real-time pipeline.  The quantity $t_{*}$ is a step function that changes once per sidereal day to the most recent time of transit of the calibrator source.  This procedure improves the stability from roughly \SI{0.8}{\percent} to \SI{0.5}{\percent} (standard deviation in fractional power units) by correcting the common-mode drift in the amplitude caused by changes in the outside temperature between daily point source calibrations.

\subsubsection{Weight Smoothing}

The radiometric noise is not expected to change appreciably on short timescales.  In order to reduce the uncertainty on our estimate of the variance of the radiometric noise and also make our estimate less sensitive to transient RFI events, a rolling median filter with a \SI{5}{\minute} window is applied to the time axis of the inverse variance weight dataset.  Any time that was masked is ignored when calculating the median and also remains masked after the filtering is applied.

\subsubsection{Sidereal Regridding}
\label{sec:regrid}

The next stage of the daily processing pipeline interpolates the visibilities onto a fixed grid in local Earth-rotation angle $\phi$ that ranges from \SIrange{0}{360}{\degree} with 4096 samples, giving a spacing $d_{\phi} = \SI{5.27}{\arcmin}$.  The interpolation algorithm assumes that the processed visibilities, $V_{\rm cal,2}(\vec{b}, \nu, t)$, are sampled from some regularly-gridded sky visibility, $V_{\rm grid}(\vec{b}, \nu, \phi)$, and corrupted by both noise and RFI, denoted as $n(\vec{b}, \nu, t)$.  Since the sky visibility is band-limited by its maximum fringe rate, and the chosen sample rate is more than twice that, the following relation holds:
\begin{align}
    \label{eq:lanczos_forward}
    V_{\rm cal,2}(\vec{b}, \nu, t_{g}) & = \sum_{h} \ K_{gh} \ V_{\rm grid}(\vec{b}, \nu, \phi_{h}) + n(\vec{b}, \nu, t_{g}) \ ,
\end{align}
where $K_{gh} = \sinc{\left(\frac{\Delta\phi_{gh}}{d_{\phi}}\right)}$ is the interpolation kernel with $\Delta\phi_{gh} = \phi(t_{g}) - \phi_{h}$, and the summation runs over the regular grid in local Earth rotation angle.  The infinite support of the kernel is computationally problematic, so a common approximation involves truncating the kernel by multiplying it with a window function.  We use the Lanczos kernel, which is given by
\begin{equation}
    K_{gh} = \begin{cases}
                     \sinc{\left(\frac{\Delta\phi_{gh}}{d_{\phi}}\right)}  \sinc{\left( \frac{\Delta\phi_{gh}}{a d_{\phi}} \right)} & |\Delta\phi_{gh}| \leq a d_{\phi} \\
                     0 & |\Delta\phi_{gh}| > a d_{\phi}
             \end{cases} \ .
\end{equation}
The parameter $a$ controls the kernel width and couples at most $4 a + 1$ samples of $V_{\rm grid}$.

We use a Wiener filter to invert \cref{eq:lanczos_forward} and solve for the regularly-gridded sky visibility, given the noisy, RFI contaminated data.  Let $\vec{v}_{\rm cal,2}$ denote the vector containing the time-ordered visibility for a given frequency and baseline.  The regularly-gridded visibility $\vec{v}_{\rm grid}$ for that frequency and baseline is estimated as
\begin{align}
    \vec{\hat{v}}_{\rm grid} & = \vec{C} \vec{K}^{T} \vec{N}^{-1} \vec{v}_{\rm cal,2}
    \label{eq:hatvgrid}
\end{align}
where
\begin{align}
    \vec{C} & = \left(\vec{S}^{-1} + \vec{K}^{T} \vec{N}^{-1} \vec{K}\right)^{-1} \ ,
    \label{eq:Cregrid}
\end{align}
$\vec{N}^{-1} = \left(\vec{N}_{\rm noise} + \vec{N}_{\rm RFI}\right)^{-1}$ is the inverse covariance of the noise and RFI, and $\vec{S}^{-1}$ is the inverse covariance of the sky visibility.  The noise covariance is assumed to be diagonal and equal to the fast-cadence estimate of the variance described in \secref{sec:weights}.  The RFI covariance is also assumed to be diagonal, and equal to infinity for times and frequencies that are missing or have been masked by the procedure described in \secref{sec:rfi} and equal to zero otherwise.  Finally the sky covariance is assumed to be diagonal and constant as a function of baseline, frequency, and sidereal angle, such that $\vec{S}^{-1} = s_\text{max}^{-2} \, \vec{I}$ where $\vec{I}$ is the identity matrix and $s_\text{max} = \SI{e4}{\jansky\per\beam}$ is chosen to be around the maximum flux observed on the sky.  Each frequency and baseline is solved independently.  This is made computationally tractable by utilizing the fact that $\vec{C}^{-1}$ is a band matrix, which is a consequence of the compact support of the Lanczos kernel.  Choosing the kernel width, $a$, is a balance between the computational cost of the regridding (which is $O(a^2)$), and the accuracy of the reconstruction. We use $a = 5$ in this work, which has deviations away from the ideal sinc transfer of $\lesssim 10^{-3}$ for the typical range of fringe rates in this analysis.

The covariance of the filtered signal is $\langle\vec{\hat{v}}_{\rm grid} \vec{\hat{v}}_{\rm grid}^{\dagger} \rangle = \vec{C}$ as given by \cref{eq:Cregrid}.  The weight dataset that tracks the inverse variance of the noise present in the visibilities is therefore updated to $w(\vec{b}, \nu, \phi_{h}) = 1 / C_{hh}(\vec{b}, \nu)$.  The interpolation scheme introduces ringing in the visibilities at the edge of any large gap of missing or masked data, with the post-interpolation weights at these edges gradually transitioning to $s_\text{max}^{-2}$.  In order to mask these artifacts we apply a baseline-dependent threshold to the weight dataset, setting it to zero if it is less than \SI{50}{\percent} of the average weight over all frequencies and sidereal angles. As these samples lie at the edge of large periods of missing data, the relative increase in the amount of data flagged is small.

\subsubsection{Daytime, Moon, and Data Flags}
\label{sec:data_flags}

Next we apply a series of flags that exclude certain time ranges from further analysis.  The weight dataset is set to zero for any time sample that meets one or more of the following criteria: (1) the Sun is above the horizon (\SI{52}{\percent} flagged), (2) the Moon is within \SI{5}{\degree} of the meridian (\SI{3}{\percent} flagged), (3) occurred during an interval of poor data quality as indicated by a ``bad data flag'' in our database (\SI{36}{\percent} flagged).

The database that is used for the last item is updated external to the daily pipeline and contains a variety of flag types based on different metrics for data quality.  The following data flag types were employed in this analysis:
\begin{description}
    \item[Rain] Mask any time where the accumulated rainfall during the \SI{30}{\hour} prior was greater than \SI{1}{\milli\meter}.  This condition finds intervals where a large number of feeds are likely to be wet.  Precipitation at the site causes analogue signal corruption in \SIrange{4}{12}{\percent} (inter-quartile range) of the feeds due to water pooling on the focal line \citep{overview-paper}.  (\SI{10}{\percent} flagged)
    \item[Jumps] Mask any time where the autocorrelation for five or more feeds has shown a sudden ($\lesssim \SI{30}{\minute}$), broadband increase of more than \SI{20}{\percent} in the past \SI{30}{\hour}.  This condition too is designed to find intervals where a large number of feeds are likely to be wet. (\SI{29}{\percent} flagged)
    \item[Correlator restart] Mask the interval between a correlator restart and the next daily point source calibration.  The FPGA re-synchronization that occurs during a correlator restart introduces a change in the relative phase between feeds digitized by different ADC chips that is non-negligible with respect to our requirements on phase stability. (\SI{7.1}{\percent} flagged)
    \item[Acquisition restart] Mask the interval between a restart of the data acquisition software and application of the calibration gains. (\SI{2.8}{\percent} flagged)
    \item[Bad calibration] Mask any time where the calibration gains were not updated in the past \SI{24}{\hour}.  Also mask intervals where poor quality gains were applied to the visibility data as determined by several metrics which are generated by the real-time pipeline and monitored by the telescope operator. (\SI{0.4}{\percent} flagged)
\end{description}

CHIME acquired \SI{245}{\day} of integration time during the \SI{309}{\day} period between January 1 and November 6, 2019.  The \SI{64}{\day} of instrument downtime consisted of \SI{55}{\day} of planned hardware maintenance and software upgrades and \SI{9}{\day} of unintended interruptions due to power failures, cooling failures, and other accidental outages.  The flags described above exclude \SI{70}{\percent} of the remaining data from the stacking analysis, with the daytime and rain/jumps flags representing the primary sources of data loss.  After applying these flags, the total integration time is
\SI{1760}{\hour}, of which \SI{834}{\hour} was spent observing the range of right ascension containing the NGC field.  This total is further reduced by the sidereal day flags that will be described in \secref{sec:bad_days}.

\subsection{Averaging Sidereal Days}
\label{sec:sidereal_avg}

\begin{figure*}[htbp]
    \centering
    \includegraphics[width=0.8\textwidth]{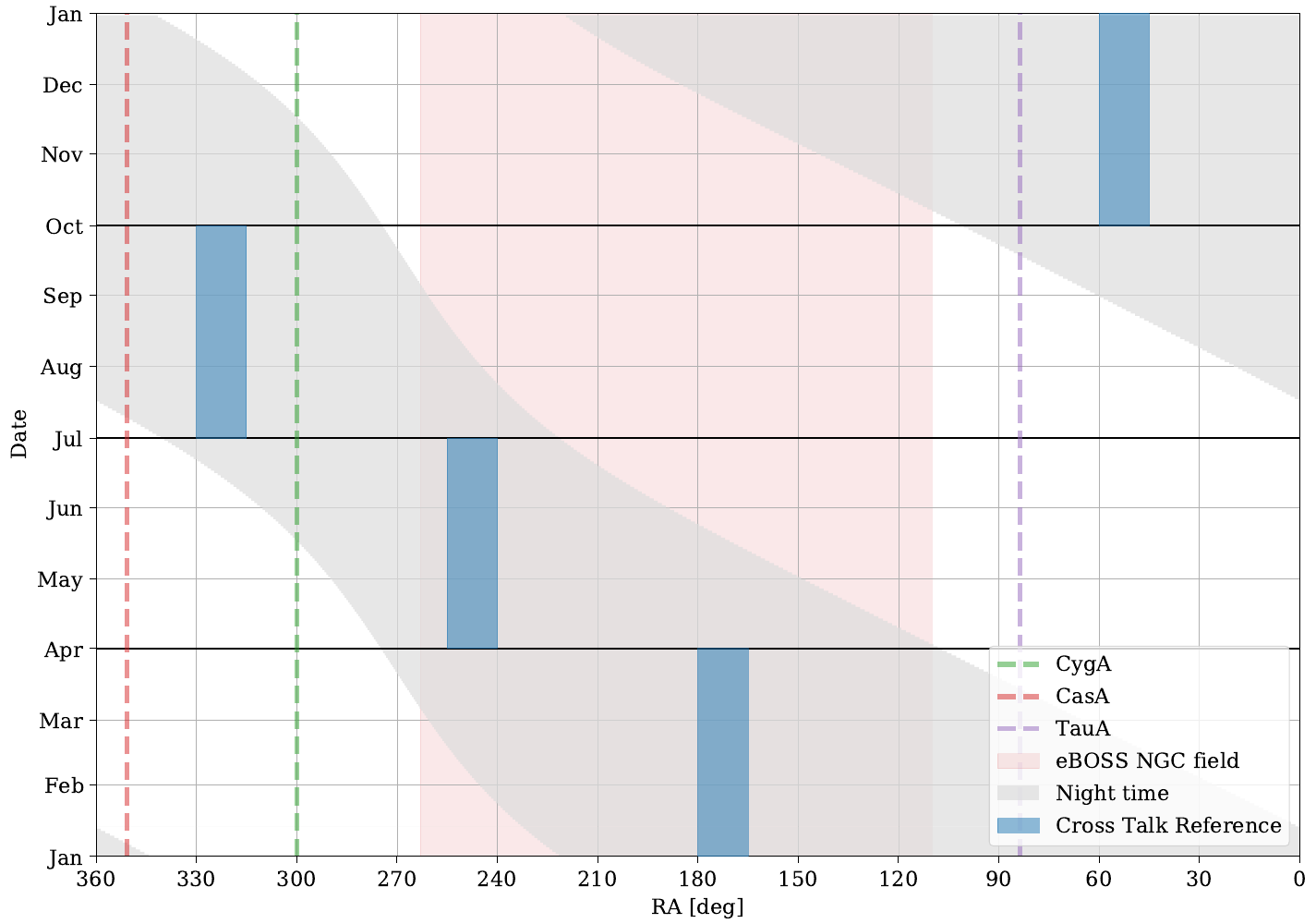}

    \caption{
        In the first stage of combining individual days, we combine sidereal days within each quarter of the year. The figure above shows the right ascension range of nighttime data for each day within a quarter (gray shaded region). The bulk of the sensitivity to the eBOSS NGC field (pink band) comes from the first two quarters of the year, where the local nighttime better overlaps with the NGC RA range. In order to combine the days, we need to consistently reference the mean level of each day to remove crosstalk. For each quarter, we pick a single hour of local Earth rotation angle 
(blue boxes) for which we compute the median and subtract it from each day's sidereal stream. These regions are chosen to be within the nighttime and avoid the transit of bright point sources, in order to minimise the bias from gain variations.}
    \label{fig:stack_design}
\end{figure*}

After the individual days have been flagged and processed to a common grid in local Earth rotation angle, the days are averaged together to produce a high-sensitivity measurement of the sky.

This process is complicated by the presence of \emph{noise crosstalk}, a bias in the zero-level of a non-autocorrelation visibility. Physically the mechanism for this is the leaking of thermal noise generated within the low-noise amplifier
on one signal chain that is broadcast by the antenna and received (directly, or by an indirect path) by another antenna. This common noise term (and the term from the reverse path) gives a bias in the visibility between the two antennas. This crosstalk contribution drops off rapidly with increasing separation between the antennas, and is much smaller (but still non-zero) for antenna pairs on different cylinders where there is no direct line-of-sight.

We observe the crosstalk to be relatively stable in time, varying slowly over the course of one day. In practice this allows the crosstalk removal to be performed by estimating and removing a single time-independent signal from each day for each frequency and baseline.  However, as the crosstalk signal is not known a priori and must be measured from the data, it is degenerate with any constant sky signal within the time period being used to estimate it.

As we use only nighttime data spread over a year, there is no single period in common between all days that we can choose as a reference. To account for this we break the sidereal averaging into two stages: the first operates on data taken from each quarter of the year, and the second combines those into a full stacking of the data.

\subsubsection{Sidereal Day Flags}
\label{sec:bad_days}

Prior to averaging the sidereal streams we make further cuts to the data. Any sidereal day with less than \SI{80}{\percent} of the day remaining after applying the data flags in \secref{sec:data_flags} is rejected, as is any day where less than half the crosstalk reference range is available (see next section).

Finally each day is manually inspected via a standardised set of visualisations:
\begin{itemize}
    \item A delay power spectrum for each baseline generated by averaging over all unmasked time samples (see Appendix~\ref{sec:delay_spectrum}). This presents a holistic summary of all elements of the data, and is particularly powerful for illustrating poor RFI flagging and misbehaving baselines.
    \item A sensitivity plot showing the estimated point source flux sensitivity found by appropriately averaging the fast-cadence estimate of the variance over all baselines at each time and frequency. This is another summary of the whole dataset and is a good diagnostic of RFI excision performance.
    \item A sky map (see \secref{sec:mapmaking})
    at two different frequencies and its difference from a day-averaged map. This is not a complete summary as it does not incorporate information from every frequency, but is very effective at identifying poor calibration.
\end{itemize}
Each day was inspected by at least two people and any day flagged by at least one person was removed from further analysis.

After all these cuts are applied, 102 sidereal days remain for averaging.  After also applying the flags described in \secref{sec:data_flags}, the 102 sidereal days contain \SI{1073}{\hour} of integration time.  Of this, \SI{521}{\hour} was spent observing the range of right ascension containing the NGC field.

\subsubsection{Sidereal Averaging (Seasonal)}
\label{sec:sidereal_avg_seasonal}

The first stage of sidereal averaging combines data from a single quarter of each calendar year and assigns each ``good'' day of data into alternating partitions of the data. By splitting into partitions per quarter we are able to produce two jackknife splits of our data that have approximately the same sensitivity and sidereal coverage; these will be used for consistency tests in \secref{sec:consistent_even_odd}.

For each quarter, we pick a single hour-long range in local Earth-rotation angle that is observed within the nighttime for the entire quarter and avoids the transits of bright point sources. This time range is used to reference the crosstalk signal for the entire quarter. We illustrate these ranges for each quarter, and how the quarters overlap with the eBOSS NGC field, in \cref{fig:stack_design}.

Every day we calculate the median over this time range for each visibility and subtract it from the data for that day.
Assuming that the crosstalk signal is approximately constant across the day, this procedure will remove that day's crosstalk contamination and a small amount of the sky signal, which is the same across all days within the quarter. It is important to use consistent estimates of the crosstalk; therefore, if more than 70\% of the data within this reference range is missing for a frequency on a given day, the entire frequency will be flagged out for the whole day. This differs from the initial selection discussed in \secref{sec:bad_days} as it is determined from the full frequency dependent missing data mask for that day, not just the frequency independent data flags.

After the crosstalk has been removed consistently from all days within the quarter, the days within
each partition are averaged together with an inverse variance weighting.

\subsubsection{Sidereal Averaging (All)}
\label{sec:sidereal_avg_all}

The second stage of sidereal averaging is to combine the data for all quarters and partitions. As the crosstalk removal uses a different sky reference region for each quarter, a simple averaging would introduce discontinuities at the boundaries. To account for this we exploit the 
overlap in local Earth rotation angle of the nighttime data for each quarter with its neighbours to solve for the differences and set a common reference.

To do this we treat our estimate of the regularly-gridded visibility $\vv_{\mathrm{grid},i}$ (where we have dropped the $~\hat{}~$ symbol to simplify notation) for each frequency and baseline within a partition $i$ (out of $p$ total partitions) as being composed of a signal $\vv$ that we are interested in that is constant for all partitions, a noise $\vn_i$ and a residual crosstalk contribution $\vx_i$ that is different for each partition and also incorporates the bias from the per-partition crosstalk referencing. We write this as
\begin{equation}
    \vv_{\mathrm{grid},i} = \vv + \vn_i + \vx_i \; .
\end{equation}
We model the statistics of each component as having zero-mean with covariance matrices $\langle \vv \vv^\hconj \rangle = \mS$, $\langle \vn_i \vn_i^\hconj \rangle = \mN_i$ and $\langle \vx_i \vx_i^\hconj \rangle = \mX$.
As the crosstalk has little time variation,
we model the residuals as a low-rank contribution (with rank $k$), allowing us to factorize the covariance as $\mX = \mU \mU^\hconj$, where $\mU$ is a rectangular matrix. Though the crosstalk referencing means the modes $\vx_i$ may be very different, we assume the crosstalk statistics are the same across all partitions, so $\mX$ does not depend on $i$.
The noise matrix $\mN_i$ is assumed to be diagonal and includes both the noise expected in the data (\secref{sec:weights}) and any masking that has been applied (\secref{sec:rfi} and \ref{sec:data_flags}), encoded in the standard way of setting the inverse-variance to zero for masked samples. Although the averaging over sidereal days has reduced the number of samples that are flagged entirely, ranges of RA observed during the daytime for the entire quarter, and badly RFI contaminated frequencies, will still be masked.

To solve for the signal we start by writing a Wiener estimator for $\vs$ treating both $\vn_i$ and $\vx_i$ as a generalised noise
\begin{equation}
\hat{\vv} = \mC \Bigl[ \sum_i (\mN_i + \mX)^{-1} \vv_{\mathrm{grid},i} \Bigr]
\label{eq:vest1}
\end{equation}
where the covariance matrix $\mC$ is defined by
\begin{equation}
    \label{eq:mC_1}
    \mC^{-1} = \mS^{-1} + \sum_i (\mN_i + \mX)^{-1} \; .
\end{equation}
A naive application of this scheme would require tracking and inverting a large matrix for each frequency, but we can simplify it by repeated application of the Woodbury matrix identity%
\footnote{The Woodbury matrix identity allows us to expand the inverse of a low-rank update to a matrix with known inverse. In its most general form it is written as
\begin{equation}
    \notag
    \lp \mA + \mU \mC \mV \rp^{-1} = \mA^{-1} - \mA^{-1} \mU \lp \mC^{-1} + \mV \mA^{-1}\mU \rp^{-1} \mV \mA^{-1} \; ,
\end{equation}
with $\mA$ and $\mC$ square, but potentially different sizes.
}%
.
First we expand the $\mN_i + \mX$ term allowing us to regroup \cref{eq:mC_1} as
\begin{equation}
    \mC^{-1} = \mC_0^{-1} + \mW \mW^\hconj
    \label{eq:CinvC0W}
\end{equation}
where
\begin{equation}
    \mC_0^{-1} = \mS^{-1} + \sum_i \mN_i^{-1}
\end{equation}
and $\mW$ is a block matrix,
\begin{equation}
    \mW = \lp\begin{array}{c|c|c|c}
        & & &\\
        \mW_0 & \mW_1 & \ldots & \mW_{p - 1} \\
        & & &
    \end{array} \rp
\end{equation}
with one block for each partition, and within each block are $k$ columns for each crosstalk mode and a row for each RA sample. The blocks are
\begin{equation}
   \mW_i = \mN_i^{-1} \mU (\mI_k + \mU^\hconj \mN_i^{-1} \mU)^{-1/2} \; .
\end{equation}
where $\mI_k$ is the identity matrix of size $k$, and each $\mW_i$ can be interpreted as a noise-weighted projection operator onto the crosstalk basis for each partition.

The estimator in \cref{eq:vest1} can be rewritten as
\begin{equation}
    \hat{\vv} = \mC \Bigl[ \sum_i (\mN_i^{-1} - \mW_i \mW_i^\hconj) \vv_{\mathrm{grid},i} \Bigr] \; .
\end{equation}
A second application of the Woodbury identity, this time to \cref{eq:CinvC0W}, allows us to write $\mC$ in a more easily applied form
\begin{equation}
    \label{eq:mC_2}
    \mC = \mC_0 + \mC_0 \mW (\mI_{(k\times p)} - \mW^\hconj \mC_0 \mW)^{-1} \mW^\hconj \mC_0 \; .
\end{equation}
To produce the final estimate for the stacked signal we need to: generate $\mW_i$ for each day and retain it; accumulate $(\mN^{-1}_i - \mW_i \mW_i^\hconj) \vv_{\mathrm{grid},i}$ to generate $\mC^{-1} \hat{\vv}$; and then finally apply the deconvolving matrix $\mC$, which can be done efficiently by evaluating matrix-vector products from right to left in \cref{eq:mC_2} using the accrued $\mW_i$ rather than explicit construction of $\mC$. Conceptually this final step uses the noise weighted overlaps between the different partitions (in the $\mI - \mW^\hconj \mC_0 \mW$ term) to solve for a consistent bias and remove it.

In the implementation within our pipeline we model the crosstalk as a single time-independent constant mode per day (i.e. $k = 1$ and $\mU \propto \mathbf{1}$, where $\mathbf{1}$ is a column vector filled with ones).
We also assume that both $\mX$ and $\mS$ are both much larger than the instrumental noise $\mN$ for unmasked samples. This means that the estimator we use does not depend on $\mS$ at all, nor on the scale of $\mX$ (but it does depend on the form), and, importantly, means that $\mC_0$ is a diagonal matrix. However, this does produce one singular mode, the sidereal average of each visibility, that must be regularized externally. Finally, rather than using the $\mN$ for each baseline, we use an average over all baselines, which ensures that the same linear combinations of partitions are used for all baselines at a given frequency. We use these same linear combinations when updating the baseline-dependent weights in the final stack, although we drop the small correction to the weights that comes from removing the crosstalk, which primarily affects the off-diagonal elements of the noise covariance that we do not track in our analysis, for memory reasons.

As the sidereal-time-independent component
of the sky is entirely degenerate with a constant noise bias, the mean of each sidereal stream is a singular mode. To regularize this degenerate mode we add a constant offset to set the median in time of the full sidereal day to zero.

%% file: sections/stacking.tex

\section{Stacking Pipeline}
\label{sec:stacking_pipeline}

\begin{figure*}[t]
   \centering \includegraphics[width=1\linewidth,keepaspectratio, trim = 0 0 170 0]{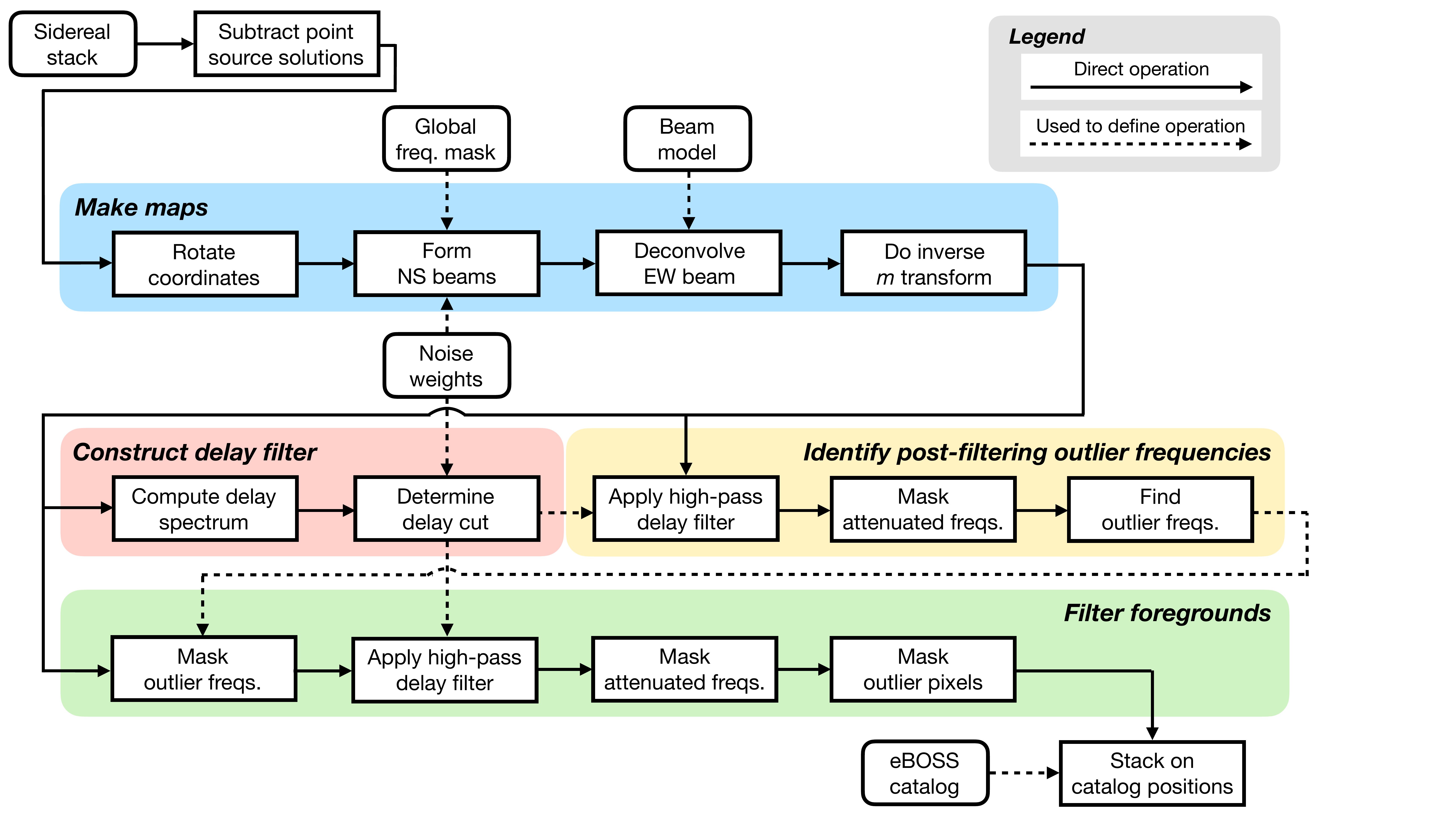}
    \caption{A schematic representation of the stacking pipeline, proceeding from the sidereal stack described in \secref{sec:sidereal_avg}. After subtraction of the four brightest point sources (\secref{sec:point_source_subtraction}), sky maps are formed (\secref{sec:mapmaking}), accounting for a global frequency mask (\secref{sec:freq_mask}) and a model for the 
primary-beam pattern (\secref{sec:beams}). Our foreground filtering scheme is designed to reject components of the data with variance far in excess of the expected thermal noise, and includes a high-pass delay filter (\secref{sec:filter}) and several additional masking operations (\secref{sec:map_mask}), including masking of frequencies that are attenuated by the delay filter. Finally, the filtered maps are stacked at the positions of objects in each eBOSS catalog (\secref{sec:stacking}).}
    \label{fig:stacking_flowchart}
\end{figure*}

We have developed a dedicated pipeline to stack the CHIME data on the angular and spectral locations of the sources in a spectroscopic catalog.  The pipeline takes as input the sidereal stack that is generated by the CHIME data processing pipeline as described in \secref{sec:processing}.  It subtracts the signal from the four brightest point sources and masks corrupted frequency channels.  Next, it constructs a map of the sky at each frequency channel, deconvolving a model for the primary-beam pattern in the process.  It applies a high-pass filter to the frequency axis of each map pixel to remove foregrounds.  It then masks frequency channels and pixels that are outliers.  Finally, it stacks the maps on the angular and spectral locations of the sources in a catalog.  The entire process is visualized in \cref{fig:stacking_flowchart}.  In what follows, we describe each stage of the pipeline.

\subsection{Point Source Subtraction}
\label{sec:point_source_subtraction}

The signal from the four brightest point sources -- Cygnus A, Cassiopeia A, Taurus A, and Virgo A -- is estimated and subtracted from the data.  The following model for the visibility measured by baseline $\vec{b}$ at local Earth rotation angle $\phi$ and frequency $\nu$ is assumed:
\begin{align}
    \label{eq:model_bright_psrc}
    V_{\rm psrc}(\vec{b}, \nu, \phi) = \sum_{s=1}^{4} a_{s}(\nu, \phi) \ e^{j 2 \pi \nu \vec{b} \cdot \vec{\hat{n}}(\theta_{s}, \phi - \phi_{s}) / c} \ ,
\end{align}
where $a_{s}(\nu, \phi)$, $\theta_{s}$, and $\phi_{s}$ denote the primary-beam-modulated amplitude, declination, and right ascension of source $s$, $\vec{\hat{n}}$ is the unit vector pointing towards the source's location, and $c$ is the speed of light.  At every frequency and local Earth rotation angle we estimate the set of source amplitudes $\vec{a}$ using weighted linear regression:
\begin{align}
    \vec{\hat{a}} & = \left(\vec{Z}^{\dagger} \vec{N}^{-1} \vec{Z}\right)^{-1} \vec{Z}^{\dagger} \vec{N}^{-1} \vec{v} \ .
\end{align}
Here \vec{v} is a vector containing the visibilities for a selection of baselines,
\begin{align}
    Z_{is} & = e^{j 2 \pi \nu \vec{b}_{i} \cdot \vec{\hat{n}}(\theta_{s}, \phi - \phi_{s}) / c}
\end{align}
is the geometric phase factor for baseline $i$ and source $s$, and $\vec{N}$ is the noise covariance.  As before, we assume the noise covariance is diagonal and equal to the propagated fast-cadence estimate of the variance (see \secref{sec:weights}).

The amplitude $a_{s}$ is equal to the spectral flux density of source $s$ modulated by the power beam pattern of the instrument at the source's coordinates, and is expected to vary slowly as a function of frequency and hour angle.  To improve the signal to noise, the best-fit amplitude for each source is smoothed in $(\nu, \phi)$ by iteratively applying a 2D moving average window with size (\SI{1.2}{\mega\hertz}, \SI{0.44}{\degree}) and number of iterations (12, 8).  The model for the four brightest sources is then computed using \cref{eq:model_bright_psrc} and subtracted from the data.  Note that only visibilities measured by baselines consisting of feeds on different cylinders are used to solve for the source amplitudes -- because contamination from diffuse Galactic emission and noise crosstalk is significantly reduced for these \emph{inter-cylinder baselines} -- but the resulting model is subtracted from all baselines.

\subsection{Frequency Mask}
\label{sec:freq_mask}

The inverse variance weights are multiplied by a global frequency mask that completely excludes certain frequency channels from the stacking analysis.  The list below gives the conditions under which a frequency channel is masked and the fraction of the \SIrange[range-units = single, range-phrase=$-$]{587.5}{800}{\mega\hertz} band that meets each condition.
\begin{itemize}
    \item Mask any frequency channel that coincides with a known, persistent source of RFI.  There were two sources of persistent RFI in the \SIrange[range-units = single, range-phrase=$-$]{587.5}{800}{\mega\hertz} band: the mobile LTE bands and the local oscillator (LO) used by the Synthesis Telescope at DRAO \citep{landecker2000}.  \textbf{(14.2\% masked)}
    \item Mask any frequency channel where the sidereal stack is missing a subset (or all) of the full sidereal day, which prevents a straightforward application of the $m$-mode transform required for map making.  This could be due to a GPU node that was not operational for a significant portion of 2019, as one example. \textbf{(12.5\% masked)}
    \item Mask any frequency channel where the total integration time over the range of RA coinciding with the NGC field is less than \SI{75}{\percent} of the maximum over frequencies.  Again, most often this is due to a temporarily non-operational GPU node.  \textbf{(5.9\% masked)}
    \item Mask any frequency channel where manual inspection of the foreground-filtered map in an initial iteration of the analysis revealed residuals that are large relative to the expected radiometric noise and corrupt a significant fraction of the NGC field.  This procedure is described in greater detail in \secref{sec:map_mask}.  \textbf{(14.7\% masked)}
\end{itemize}
In total, these four conditions mask \SI{47.2}{\percent} of the \SIrange[range-units = single, range-phrase=$-$]{587.5}{800}{\mega\hertz} band.

\subsection{Map Making}
\label{sec:mapmaking}

The next step in the data processing is to construct a map from the sidereal visibilities.  We use a map-making technique that is tailored to CHIME, or effectively any transit radio interferometer consisting of cylindrical telescopes oriented in the north-south direction with a close-packed array of antennas along the axis of each cylinder.  The technique draws on the work presented in \cite{shaw2014} and \cite{masui2017}, but is distinct and has not been described elsewhere, so we go into considerable detail in this section.  Note that these types of maps are referred to as \emph{deconvolved ringmaps} in \cite{overview-paper}.

\subsubsection{Baseline Configuration}

To good approximation, the CHIME baselines $\vec{b_{c}}$ are located on a 2D grid that lies in the plane tangent to the earth's surface at (latitude, longitude) $\equiv (\Lambda, \ \Phi) = (\SI{49.320709}{\degree}, \ \SI{-119.623677}{\degree})$.  The 2D grid is given by
\begin{align}
    \vec{b_{c}} &= x d_{x} \vec{\hat{x}_{c}} + y d_{y} \vec{\hat{y}_{c}}
\end{align}
where $\vec{\hat{x}_{c}}$ is the unit vector that is orthogonal to the cylinder, $\vec{\hat{y}_{c}}$ is the unit vector parallel to the cylinder, $d_{x} = \SI{22.0}{\meter}$ is the (center-to-center) cylinder spacing, $d_{y} = \SI{0.3048}{\meter}$ is the spacing of the feeds along the focal line, and the grid indices are denoted by $x \in [-3, 3]$ and $y \in [-255, 255]$.

The sidereal visibilities are arranged onto this 2D grid.  Let $V_{xy}^{pq}(\nu, \phi)$ denote the visibility measured at frequency $\nu$ and local Earth-rotation angle $\phi$ by the baseline at the $(x, y)$ grid position.  The variables $p,q \in \{X, Y\}$ refer to the polarisations of the two antennas that form the baseline, with the dipole of the $X$ and $Y$ polarisations oriented in the $\vec{\hat{x}_{c}}$ and $\vec{\hat{y}_{c}}$ directions, respectively.  The analysis presented in this work will only use the co-polar baselines, $XX$ and $YY$, so that $p = q$, and we drop the redundant index in the notation going forward.  Note that it is assumed that the visibilities have conjugate symmetry about the origin, specifically
\begin{align}
    V_{-x,-y}^{p} \equiv \left(V_{xy}^{p}\right)^{*} \ .
\end{align}

The CHIME cylinders were aligned with the north-south direction by design.  However, we have empirically determined that the cylinders are rotated by $\psi = \SI{-0.071}{\degree}$ with respect to true astronomical north using observations of a large number of bright point sources \citep{overview-paper}. Let $\vec{b} = \vec{R}(\psi) \vec{b_{c}}$ denote the baselines in a coordinate system where $\vec{\hat{x}}$ is aligned with the east-west direction, $\vec{\hat{y}}$ is aligned with the north-south direction, and
\begin{align}
    \vec{R}(\psi) & = \left[\begin{matrix}
                        \cos{\psi} & -\sin{\psi} \\
                        \sin{\psi} & \cos{\psi} \\
                    \end{matrix} \right]
\end{align}
is the rotation matrix that transforms between the cylinder-based coordinate system and the north-south based coordinate system.

The measured visibility is the true sky visibility $\mathcal{V}$ corrupted by noise $n$,
\begin{align}
    \label{eq:vis_meas_eq}
    V_{xy}^{p}(\nu, \phi) = \mathcal{V}_{xy}^{p}(\nu, \phi) + n_{xy}^{p}(\nu, \phi) \ .
\end{align}
The sky visibility $\mathcal{V}$ is the integral of the spectral flux density, $S$, of the sky multiplied by the primary-beam pattern, $A$, of the two feeds and a geometric phase factor set by the baseline between the feeds:
\begin{align}
    \label{eq:sky_vis}
    \mathcal{V}_{xy}^{p}(\nu, \phi) = \int & |A^{p}(\nu, \theta', \phi - \phi')|^2 \ e^{j 2 \pi \nu \vec{b} \cdot \vec{\hat{n}}(\theta', \phi - \phi') / c} \nonumber \\
    & S(\nu, \theta', \phi')  \cos{\theta'} d\theta' d\phi' \ .
\end{align}
Here $c$ is the speed of light and $\vec{\hat{n}}(\theta', \phi - \phi')$ is the unit vector pointing towards declination $\theta'$ and hour angle $\ha \equiv \phi - \phi'$ and is given by
\begin{align}
    \label{eq:sky_unit_vector}
    \vec{\hat{n}}(\theta', \phi - \phi') = & -\cos{\theta'} \sin{(\phi - \phi')} \ \vec{\hat{x}} + \nonumber \\
        & (\cos{\Lambda} \sin{\theta'} - \sin{\Lambda} \cos{\theta'} \cos{(\phi - \phi')}) \ \vec{\hat{y}} + \nonumber \\
        & (\sin{\Lambda} \sin{\theta'} + \cos{\Lambda} \cos{\theta'} \cos{(\phi - \phi')}) \ \vec{\hat{z}} \ ,
\end{align}
with $\Lambda$ denoting the latitude of the telescope.  Note that \cref{eq:sky_vis} assumes that the primary beam pattern is the same for all feeds of a given polarisation.  It also assumes that there are no residual complex gain variations.

\subsubsection{North-South Beamforming}

The CHIME power beam, $|A|^{2}$, is reasonably compact in the hour-angle direction.  The FWHM of the main lobe is $\lesssim \SI{2.1}{\degree} (\SI{2.5}{\degree})$ for the $Y$ ($X$) polarisation in the \SIrange[range-units = single, range-phrase=$-$]{587.5}{800}{\mega\hertz} band and the sidelobes are $\lesssim \SI{1}{\percent}$ \citep{overview-paper}.  If we restrict the integral in \cref{eq:sky_vis} to the range of hour angles covering the main lobe of the primary beam, then we can expand the geometric phase to first order in the small angles $\ha$ and $\psi$ to obtain
\begin{align}
    \label{eq:phase_linear}
    \vec{b} \cdot \vec{\hat{n}}(\theta', \phi - \phi') = \ & x d_{x} \left[ \sin{\left(\theta' - \Lambda\right)} \psi - \cos{\theta'} \ (\phi - \phi') \right] + \nonumber \\
    &  y d_{y} \sin{\left(\theta' - \Lambda\right)} \ .
\end{align}
The geometric phase due to the $y$ component of the baseline is given by the second term in \cref{eq:phase_linear}.  Since this term depends only on declination and does not depend on hour angle, we can form a linear combination of all visibilities with the same $x$ so that only signal from a specific declination, $\theta$, adds coherently:
\begin{align}
    \label{eq:beamform_ns}
    V_{x}^{p}&(\nu, \phi, \theta) = \nonumber \\
    & \sum_{y} W_{xy}^{p}(\nu, \phi) \ V_{xy}^{p}(\nu, \phi) \ e^{-j 2 \pi \nu y d_{y} \sin{(\theta - \Lambda)} / c} \,\, ,
\end{align}
where
\begin{align}
    \label{eq:beamform_ns_weights}
    W_{xy}^{p}(\nu, \phi) & = w_{xy}^{p}(\nu, \phi) / \sum_{y'} w_{xy'}^{p}(\nu, \phi)
\end{align}
denotes the relative weights, which are normalized to preserve point-source flux.  This beamforming operation is repeated for a grid of pointings that span from horizon to horizon and are equally spaced in $\sin{(\theta - \Lambda)}$.  This operation can be done efficiently with a Fast Fourier Transform (FFT), but, in practice, we evaluate the expression directly to ensure that the grid of pointings is the same for all frequencies.

\begin{figure}
   \centering \includegraphics[width=0.98\linewidth,keepaspectratio]{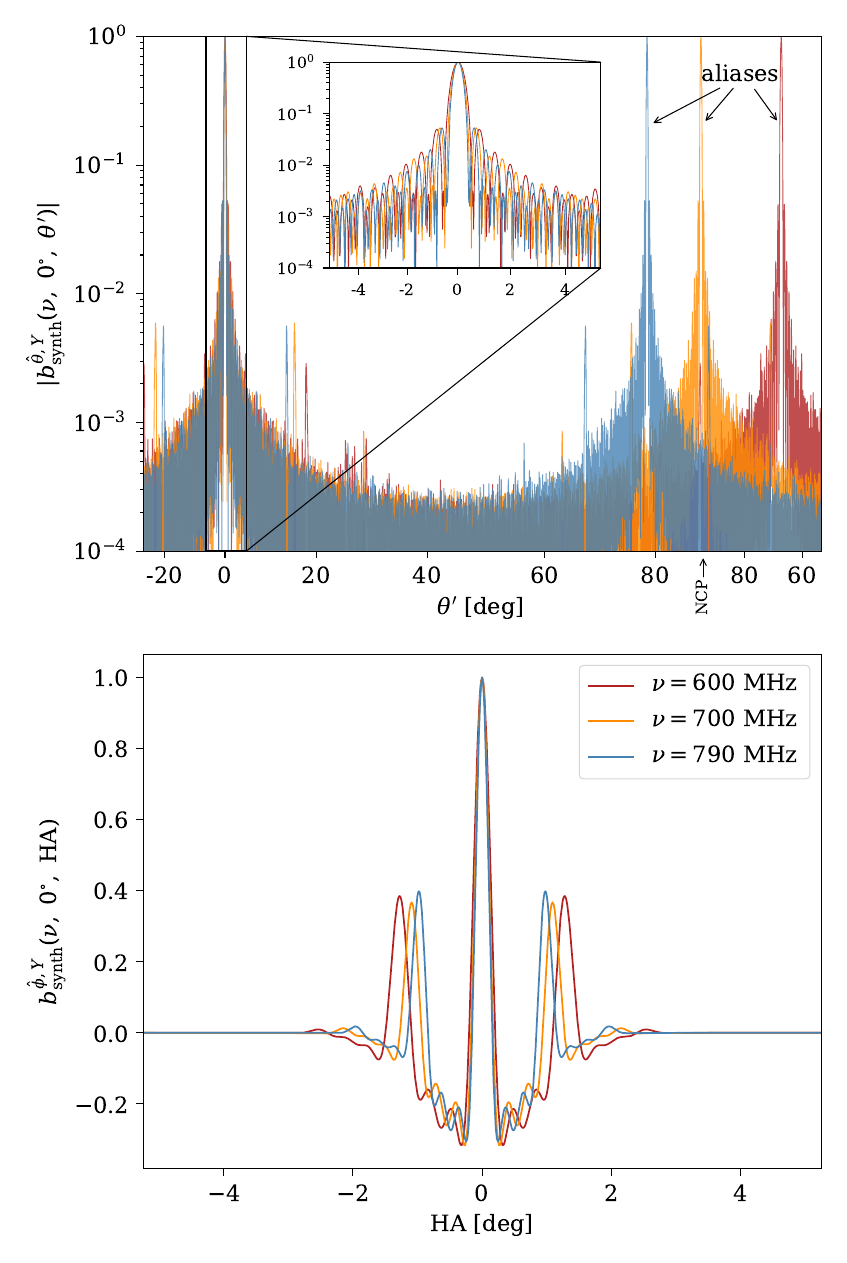}
   \caption{The synthesized beam (i.e., the point spread function of the map) for the $Y$ polarisation array at declination $\theta = \SI{0}{\degree}$.  Each color corresponds to a different frequency as described in the legend in the bottom panel.  The top panel shows the synthesized beam in the $\hat{\theta}$ direction (see \cref{eq:bsynth_theta}).  The x-axis is uniformly spaced in the sine of the zenith angle and spans from horizon to horizon, with the region to the right of the north celestial pole (NCP) annotation corresponding to the antipodal transit at hour angle = \SI{180}{\degree}.  The synthesized beam has sensitivity to both the beamformed declination at $\theta' = \SI{0}{\degree}$ and --- due to aliasing --- a second frequency-dependent declination at $\theta^{\prime}_{\rm alias}(\nu)$ (see \cref{eq:alias_dec}).  The inset panel zooms in on $\pm \SI{5}{\degree}$ from the beamformed declination.  The bottom panel shows the synthesized beam in the $\hat{\phi}$ direction (see \cref{eq:bsynth_phi}).  The exclusion of intra-cylinder baselines in the map making procedure results in negative shoulders on either side of the main lobe.  Aliasing in the $\hat{\phi}$ direction results in two \emph{grating lobes} with amplitudes that are \SI{40}{\percent} of the amplitude of the main lobe.  As the regularization parameter $\eta \rightarrow 0$ (see \cref{eq:ap_deconvolved}), the primary beam is perfectly deconvolved (assuming an accurate primary beam model) and the grating lobes disappear.  For this analysis we have chosen a relatively large value of $\eta$, which results in better point-source sensitivity, but larger grating lobes.}
    \label{fig:synthesized_beam}
\end{figure}

We will refer to $V_{x}^{p}(\nu, \phi, \theta)$ as the hybrid beamformed visibility, since the north-south component of the baseline has been beamformed to a specific declination, but there is still fringing associated with the east-west component of the baseline.  Combining \cref{eq:sky_vis}, \cref{eq:phase_linear}, and \cref{eq:beamform_ns} we obtain the following theoretical expression for the hybrid beamformed visibililties
\begin{align}
    \mathcal{V}_{x}^{p}(\nu, \phi, \theta) = \int & \synth^{\hat{\theta}, p}(\nu, \theta, \theta', \phi) \ B^{p}_{x}(\nu, \theta', \phi - \phi') \nonumber \\
    & S(\nu, \theta', \phi') \ \cos{\theta'} d\theta' d\phi'
    \label{eq:calVpx}
\end{align}
where
\begin{align}
    \label{eq:beam_transfer}
    B^{p}_{x}&(\nu, \theta', \phi - \phi') = \nonumber \\
    & |A^{p}(\nu, \theta', \phi - \phi')|^{2} \ e^{j 2 \pi \nu x d_{x} \left[\sin{(\theta' - \Lambda) \psi - \cos{\theta'} \ (\phi - \phi')}\right] / c}
\end{align}
will be referred to as the beam transfer function, and
\begin{align}
    \label{eq:bsynth_theta}
    b_{{\rm synth},{x}}^{\hat{\theta}, p}&(\nu, \theta, \theta', \phi) = \nonumber \\
    &  \sum_{y} W_{xy}^{p}(\nu, \phi) \ e^{j 2 \pi \nu y d_{y} \left[\sin{(\theta' - \Lambda)} - \sin{(\theta - \Lambda)} \right]/ c}
\end{align}
is the synthesized beam in the $\hat{\theta}$ direction.  The top panel of \cref{fig:synthesized_beam} shows an example of $b_{{\rm synth}}^{\hat{\theta}}$ for the weighting scheme used in this analysis.

The absolute weights in \cref{eq:beamform_ns_weights} are set to the inverse variance of the corresponding visibility, i.e.,
\begin{align}
     w_{xy}^{p}(\nu, \phi) & = \left[\var(V_{xy}^{p}(\nu, \phi))\right]^{-1} \ ,
\end{align}
which will maximize the point source sensitivity since the amplitude of a true point source is the same for all baselines.  We describe how the variance of the visibilities is estimated in \secref{sec:weights}.  The inverse variance weights scale approximately as the number of redundant baselines that are averaged together by the real-time pipeline to produce $V_{xy}^{p}(\nu, \phi)$, which scales with the north-south baseline distance as $(256 - |y|)$.  As a result, the inverse variance weights produce an approximately triangular window function in $y$.  This yields a synthesized beam $\synth^{\hat{\theta}}$ that has a FWHM ranging from \SI{0.35}{\degree} at \SI{585}{\mega\hertz} to \SI{0.25}{\degree} at \SI{800}{\mega\hertz}, and sidelobes that range from $0.05$ to $10^{-4}$ of the peak.  Note that, instead of inverse variance weights, we could set the weights to any window function that further suppresses the sidelobes at the expense of point source sensitivity.

In principle, the synthesized beam $\synth^{\hat{\theta}}$ depends on both the east-west baseline distance $x$ and the local Earth rotation angle $\phi$, because the inverse variance weights change with these parameters.  However, the weights that are used in this analysis yield a synthesized beam that is quite stable with $\phi$ and similar across $x$.  Indeed, the standard deviation of the synthesized beam over $\phi$ is at most \SI{0.1}{\percent} (relative to the peak) over all polarisations, frequencies, and declinations, and the standard deviation over $x$ is at most \SI{2}{\percent}.  In order to simplify the derivation that follows, we will drop the dependence of the synthesized beam on both $\phi$ and $x$.  This assumption can be enforced directly -- while maintaining roughly the same sensitivity -- by explicitly using the triangular window function, or in other words, by setting $w_{xy}(\nu, \phi) = 256 - |y|$.  Doing so, we find no appreciable change in either the signal or noise in the stacks on the eBOSS catalogs.

The regularly gridded baselines do not Nyquist sample the visibility of the sky in the $\hat{y}$ direction for frequencies $\nu \geq \frac{c}{2 d_{y}} \approx \SI{492}{\mega\hertz}$, which includes all frequencies considered in this analysis.  As a result, the hybrid beamformed visibilities will suffer from aliasing.  In this derivation, the effects of aliasing are encoded in the synthesized beam $\synth^{\hat{\theta}}$.  Let $\beta(\nu) \equiv \frac{c}{\nu d_{y}} - 1$.  If $\sinza < -\beta(\nu)$ or $\sinza > \beta(\nu)$, then the synthesized beam will have two main lobes, one centered on the desired declination $\theta^{\prime} = \theta$ and a duplicate centered on the frequency-dependent aliased declination $\theta^{\prime}_{\rm alias}(\nu)$, given by the equation
\begin{multline}
    \label{eq:alias_dec}
    \sinzap = \\
     \begin{cases}
                    \sinza + \frac{c}{\nu d_{y}} & \mbox{if  } -1 \leq \sinza \leq -\beta(\nu) \\
                    \sinza - \frac{c}{\nu d_{y}} & \mbox{if   } \beta(\nu) \leq \sinza \leq 1
           \end{cases} \ .
\end{multline}
Hence, the hybrid beamformed visibility $V_{x}^{p}(\nu, \phi, \theta)$ will contain equal contributions from the sky (modulated by the beam transfer function) at $\theta^{\prime}$ and $\theta_{\rm alias}^{\prime}(\nu)$.  This is illustrated in the top panel of \cref{fig:synthesized_beam}.  At the upper edge of the band, $\beta(\SI{800}{\mega\hertz}) = 0.23$, which implies there is a stripe of the sky centered on zenith (specifically $\theta \in [\SI{36.0}{\degree}, \ \SI{62.6}{\degree}]$) that is free from aliases at all CHIME frequencies.  Outside of this stripe, the aliased sky is heavily attenuated in the inter-cylinder baselines by utilizing the fact that it will fringe at a different rate than the true sky.  This is discussed further below.

The first sidelobe of the synthesized beam in the $\hat{\theta}$ direction has an amplitude that is \SI{5}{\percent} of the amplitude of the main lobe, the next sidelobe is \SIrange[range-units = single, range-phrase=$-$]{1}{2}{\percent}, and beyond roughly \SI{2}{\degree} separation all sidelobes are below \SI{0.5}{\percent}.  We therefore assume that a hybrid beamformed visibility is dominated by the sky at a narrow range of declinations centered on $\theta$.  This assumption will start to break down at right ascensions that coincide with bright foregrounds.  This problem is mitigated to a certain extent by subtracting the four brightest point sources directly from the sidereal visibilities, as explained in the \secref{sec:point_source_subtraction}, and using only inter-cylinder baselines that resolve out the bright, diffuse Galactic emission, which will be explained below.

\begin{figure*}
   \centering \includegraphics[width=0.98\linewidth,keepaspectratio]{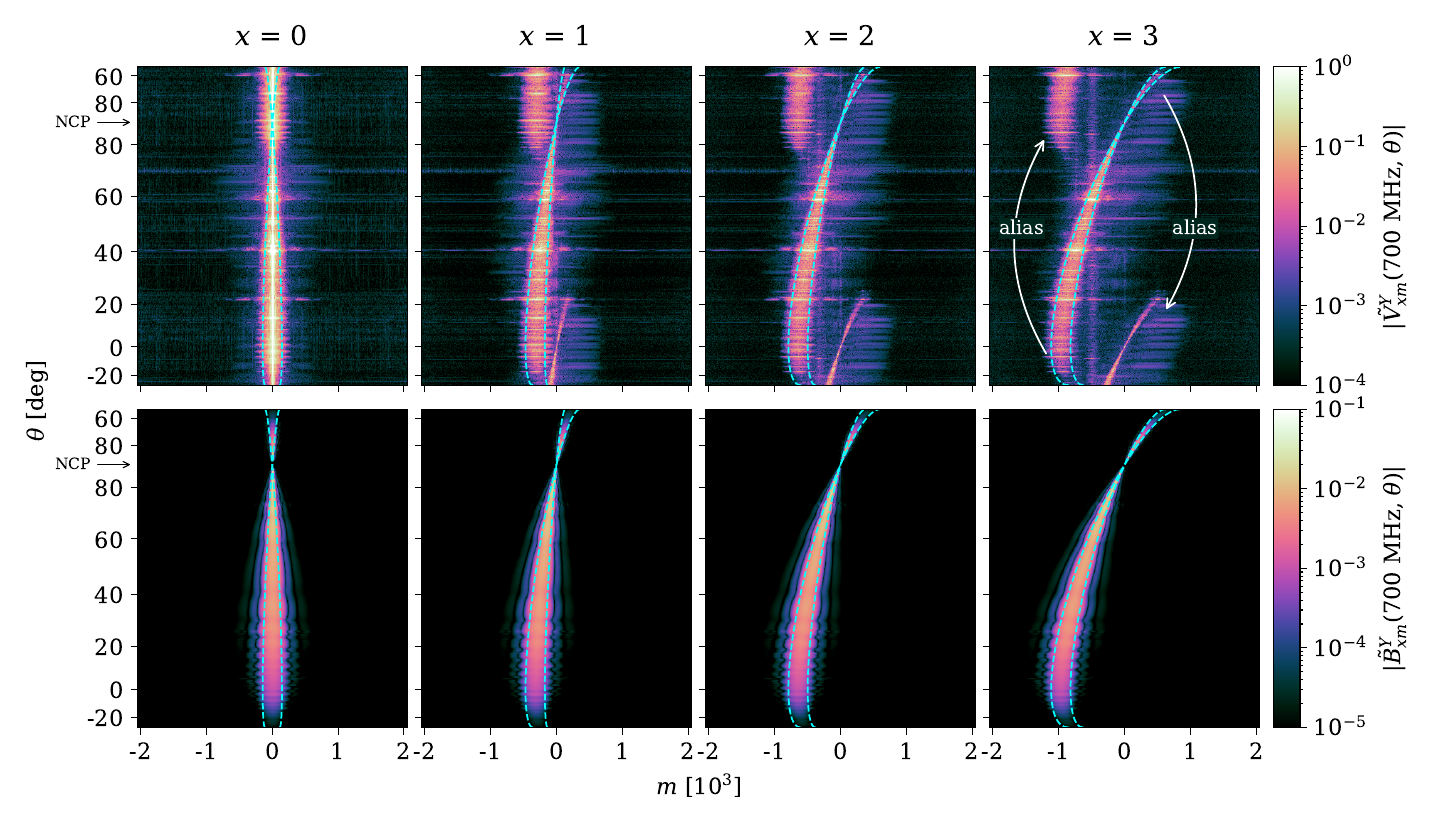}
    \caption{The $m$-mode transform of the hybrid beamformed visibilities (top row) and beam transfer function (bottom row) for the $Y$ polarisation array at frequency $\nu = \SI{700.78125}{\mega\hertz}$.  Each column shows a different east-west baseline separation. The dashed, cyan lines are given by $m_{{\rm center}, x}(\theta, \nu) \pm \frac{1}{2} m_{\rm width}(\theta, \nu)$ (see \cref{eq:m_center} and \cref{eq:m_width}) and enclose the range of $m$ where that baseline has the bulk of its sensitivity to the sky.  Due to our chosen conventions, the positive inter-cylinder baselines (x > 0) measure a negative fringe rate (negative $m$) for the sky south of the north celestial pole (NCP) and a positive fringe-rate (positive $m$) for the sky north of the NCP.  In the top row, the aliased sky is annotated in the $x = 3$ column, and is also clearly visibile in the $x = 2$ and $x = 1$ column, but overlaps entirely with true sky for the intra-cylinder baselines at $x = 0$.  In general, the aliased and true sky are well separated in $m$-space for the inter-cylinder baselines.  The bright features at approximately \SI{20}{\degree},  \SI{40}{\degree}, and \SI{60}{\degree} declination correspond to residual signal from Taurus A, Cygnus A, and Cassiopeia A, respectively.  The leakage from Cassiopeia A to other declinations is clearly visibile in the $x = 3$ panel.  In the bottom row, there is ringing outside of the dashed cyan lines because our model for the primary beam pattern has been truncated so that it only includes the main lobe (see Appendix~\ref{sec:beam_calibration_ptsrc}).}
    \label{fig:mmode}
\end{figure*}

\subsubsection{Primary Beam Deconvolution}

Since the beam transfer function does not change appreciably on scales less than the FWHM of the synthesized beam, it can be brought outside of the integral over $\theta'$ in \cref{eq:calVpx}, resulting in the following equation
\begin{align}
    \label{eq:hybrid_vis_int_phi}
    \mathcal{V}_{x}^{p}(\nu, \phi, \theta) = \int & B^{p}_{x}(\nu, \theta, \phi - \phi') \ \cos{\theta} \ d\phi' \\ \nonumber
    & \int \synth^{\hat{\theta}, p}(\nu, \theta, \theta') \ S(\nu, \theta', \phi') \  d\theta' \ .
\end{align}
Given a model for the primary beam $A$, the beam transfer function is computed using \cref{eq:beam_transfer} and then deconvolved from the hybrid beamformed visibilities at each declination to recover the flux density of the sky $S$ convolved with $\synth^{\hat{\theta}}$.  The construction of the primary beam model will be described in \secref{sec:beams}.  The fast Fourier transform of the hybrid beamformed visibility is taken along the $\phi$ axis,
\begin{align}
    \tilde{V}_{xm}^{p}(\nu, \theta) & = \sum_{n} V_{x}^{p}(\nu, \phi_{n}, \theta) e^{-j m \phi_{n}} \ .
\end{align}
Here the sum runs over the uniformly sampled grid in local Earth-rotation angle and the $m$-modes range over $[-2047, 2048]$.  We will refer to this operation as the $m$-mode transform going forward.  The same operation is performed on the beam transfer function.  \Cref{fig:mmode} provides an example of the $m$-mode transform of both the hybrid beamformed visibility and the beam transfer function.

The beam transfer function is then deconvolved from the data in $m$-space using a Tikhonov regularization scheme
\begin{align}
    \label{eq:mdeconvolve}
    \tilde{M}^{p}_{m}(\nu, \theta) & = \frac{\sum_{x} W^{p}_{x}(\nu) \  \tilde{B}^{p \ *}_{xm}(\nu, \theta) \ \tilde{V}^{p}_{xm}(\nu, \theta)}{\eta + \sum_{x} W^{p}_{x}(\nu) \ |\tilde{B}^{p}_{xm}(\nu, \theta) |^{2}} \,\, ,
\end{align}
where
\begin{align}
    W^{p}_{x}(\nu) & = w^{p}_{x}(\nu) / \sum_{x} w^{p}_{x}(\nu)
\end{align}
denotes the relative weight given to each east-west baseline and $\eta$ is a regularization parameter.  The different east-west baselines measure a largely disjoint set of $m$-modes, with each baseline primarily sensitive to the range of $m$'s centered on
\begin{align}
    \label{eq:m_center}
    m_{{\rm center}, x}(\nu, \theta) = - 2 \pi \nu \cos{\left(\theta\right)} x d_{x} / c \,\,  ,
\end{align}
with width
\begin{align}
    \label{eq:m_width}
    m_{\rm width}(\nu, \theta) = 2 \pi \nu \cos{\left(\theta\right)} w / c \,\, ,
\end{align}
where $w = \SI{20}{\meter}$ is the width of the cylinder.  However, there is some mild overlap that is dependent on the aperture illumination and accounted for by the $m$-mode transform of the primary beam pattern.  \Cref{eq:mdeconvolve} first performs a weighted average of the measurements made by the different east-west baselines, and then deconvolves the primary beam by effectively dividing by the corresponding weighted average of the $m$-mode transform of the beam transfer function.  The regularization parameter $\eta$ is the assumed inverse signal-to-noise.  It defines which $m$-modes are signal dominated, and hence should be divided by the beam, and conversely which $m$-modes are noise dominated, and should not be amplified further by dividing by the beam.

We set
\begin{align}
    w^{p}_{x}(\nu) & = \begin{cases}
                        0 & x = 0 \\
                        \left[\sigma^{p}_{x}(\nu)\right]^{-2} & |x| > 0
                      \end{cases} \ ,
\end{align}
where
\begin{align}
\label{eq:sigma_mmode}
\left[\sigma^{p}_{x}(\nu)\right]^{2}  & = \sum_{n} \left[\sigma_{x}^{p}(\nu, \phi_{n})\right]^{2}
\end{align}
is the variance of the noise in the $m$-mode transform of the hybrid beamformed visibility, and
\begin{align}
    \label{eq:sigma_hybridvis}
    \left[\sigma_{x}^{p}(\nu, \phi)\right]^{2} & = \left(\sum_{y} w_{xy}^{p}(\nu, \phi)\right)^{-1}
\end{align}
is the variance of the noise in the hybrid beamformed visibility.  This weighting scheme masks all intra-cylinder baselines and propagate the inverse variance weights through the beamforming and $m$-mode transform for inter-cylinder baselines.  The redundancy of the array results in $W^{p}_{x}(\nu) \approx \left[0.0, \ 0.5, \ 0.333, \ 0.166\right]$ for $|x| = [0, 1, 2, 3]$ corresponding to the intra-cylinder auto-correlation that is removed, the three-fold redundancy in the one-cylinder separation, two-fold redundancy in the two-cylinder separation, and single appearance of the three-cylinder separation.

The intra-cylinder baselines are masked for this analysis because they contain two sources of contamination that are significantly reduced in the inter-cylinder baselines: (1) large-scale diffuse Galactic emission and (2) noise crosstalk (see \secref{sec:sidereal_avg}).  Since the noise crosstalk changes slowly with time, it contaminates only low $m$, which is where the signal from the sky resides for intra-cylinder baselines.  Note that the signal from the sky at declinations near the north celestial pole (NCP) will also appear at low $m$, even for inter-cylinder baselines.  However, the maximum declination of sources in the eBOSS NGC field is \SI{60}{\degree}, which is far enough from the NCP that the crosstalk contamination in the inter-cylinder measurements is negligible.

The beam transfer function of the inter-cylinder baselines is largely insensitive to the range of $m$-modes occupied by the aliased sky for the declinations considered in this analysis.  This can be shown in a rough way using \cref{eq:alias_dec}, \cref{eq:m_center}, and \cref{eq:m_width}.  The NGC field spans declinations from \SIrange{13}{60}{\degree}, and over this range there is zero overlap between $m_{{\rm center}, x}(\nu, \theta) \pm \frac{1}{2} m_{\rm width}(\nu, \theta)$ and $m_{{\rm center}, x}(\nu, \theta_{\rm alias}(\nu)) \pm \frac{1}{2} m_{\rm width}(\nu, \theta_{\rm alias}(\nu))$ for all $\nu$ and for all $x > 0$.  However, examining \cref{fig:mmode} it is clear that for $\theta \lesssim \SI{20}{\degree}$ our actual beam model does have some sensitivity to the aliased sky for the $x = 1$ baseline.  Therefore, although the deconvolution procedure will heavily attenuate the aliased sky, it is still expected to introduce some non-negligible contamination.

The regularization parameter is set to $\eta = 10^{-4}$.  This value was chosen by constructing a map for several different values of $\eta$ between $10^{-6}$ and $10^{-3}$ and choosing the value that maximizes the point-source sensitivity.  Note that smaller values of the regularization parameter result in better deconvolution of the primary beam in the $\hat{\phi}$ direction, but also higher noise, and were thus disfavored for the analysis presented in this work.

Finally, the deconvolved map is obtained by taking the inverse $m$-mode transform
\begin{align}
    \label{eq:inverse_mmode}
    M^{p}(\nu, \theta, \phi) &= \frac{1}{N_{m}} \sum_{m} \tilde{M}_{m}^{p}(\nu, \theta) e^{j m \phi} \ ,
\end{align}
where $N_{m} = 4096$.

\subsubsection{Map Normalization}

In order to determine the correct normalization for the map, we consider a radio sky that contains a single point source with unit flux density at declination $\theta$ and local Earth rotation angle $\phi'$.  The $m$-mode transform of the hybrid beamformed visibilities at that declination is given by
\begin{align}
    \tilde{\mathcal{V}}^{p}_{xm}(\nu, \theta) & = \tilde{B}^{p}_{xm}(\nu, \theta) \ e^{-j m \phi'} \ .
\end{align}
The source profile along the $\phi$ axis of the resulting map is therefore
\begin{align}
    a^{p}&(\nu, \theta, \phi-\phi') = \nonumber \\
    & \frac{1}{N_{m}} \sum_{m} \frac{\sum_{x} W^{p}_{x}(\nu) \ |\tilde{B}^{p}_{xm}(\nu, \theta) |^{2}}{\eta + \sum_{x} W^{p}_{x}(\nu) \ |\tilde{B}^{p}_{xm}(\nu, \theta) |^{2}} e^{j m (\phi - \phi')} \ ,
    \label{eq:ap_deconvolved}
\end{align}
and the peak flux density of the source is $a^{p}(\nu, \theta, 0)$, which in general is not equal to unity.  Therefore, in order to preserve the point source flux through the map making process, the map is normalized as
\begin{align}
    \label{eq:map_norm}
    M^{p}(\nu, \theta, \phi) & \rightarrow \frac{M^{p}(\nu, \theta, \phi)}{a^{p}(\nu, \theta, 0)} 
\end{align}
and the synthesized beam in the $\hat{\phi}$ direction is given by
\begin{align}
    \label{eq:bsynth_phi}
    \synth^{\hat{\phi}, p}(\nu, \theta, \phi-\phi') & = \frac{a^{p}(\nu, \theta, \phi - \phi')}{ a^{p}(\nu, \theta, 0)} \ .
\end{align}
The bottom panel of \cref{fig:synthesized_beam} shows an example of $\synth^{\hat{\phi}}$ for the weighting scheme, regularization parameter, and beam model employed in this analysis.

The resulting map is modelled as
\begin{align}
    M^{p}(\nu, \theta, \phi) & = \mathcal{M}^{p}(\nu, \theta, \phi) + n^{p}(\nu, \theta, \phi) \ .
\end{align}
Here $\mathcal{M}$ is related to the flux density of the sky through the relation
\begin{align}
    \mathcal{M}^{p}(\nu, \theta,
 \phi) = \int & \synth^{\hat{\theta}, p}(\nu, \theta, \theta') \ \synth^{\hat{\phi}, p}(\nu, \theta, \phi - \phi') \nonumber \\
    & S(\nu, \theta', \phi') \cos{\theta} d\theta' d\phi'
    \label{eq:finalringmap}
\end{align}
where the synthesized beams in the $\hat{\theta}$ and $\hat{\phi}$ directions can be calculated directly from \cref{eq:bsynth_theta} and \cref{eq:bsynth_phi}, respectively.  The quantity $n^{p}(\nu, \theta, \phi)$ represents the noise in the map.

\subsubsection{Variance Estimation}

The variance of the noise in the map is estimated as
\begin{align}
    \label{eq:map_noise_variance}
    \left[\sigma^{p}_{\rm map}(\nu, \theta, \phi)\right]^{2} & = F^{p}(\nu, \phi) \ \left[\sigma^{p}_{\rm map}(\nu, \theta)\right]^{2} \ ,
\end{align}
where $\left[\sigma^{p}_{\rm map}(\nu, \theta)\right]^{2}$ is obtained by propagating the variance given by \Cref{eq:sigma_mmode} through the map making (\cref{eq:mdeconvolve}), inverse $m$-mode transform (\cref{eq:inverse_mmode}), and normalization (\cref{eq:map_norm}) procedure.  The integration time in the sidereal stack is non-uniform, primarily due to seasonal changes in the length of the day and the likelihood of rainfall.  As a result, the variance of the noise depends on the local Earth rotation angle.  This dependence is lost when propagating the variance through the forward and inverse $m$-mode transform.  The factor $F^{p}(\nu, \phi)$ approximately recovers this dependence and is given by the weighted average over east-west baselines of the fractional change in the variance of the noise in the hybrid beamformed visibilities, i.e.,
\begin{align}
    F^{p}(\nu, \phi) & = \sum_{x} W^{p}_{x}(\nu) \frac{\left[\sigma^{p}_{x}(\nu, \phi)\right]^{2}}{\frac{1}{N_{\phi}}\sum_{n} \left[\sigma_{x}^{p}(\nu, \phi_{n})\right]^{2}} \ ,
\end{align}
where $\sigma_{x}^{p}(\nu, \phi)$ is given by \cref{eq:sigma_hybridvis} and the sum runs over the $N_{\phi} = 4096$ samples on the grid.

This procedure for propagating the variance through the map making has been validated as follows.  We generate visibilities that have been randomly drawn from a circularly symmetric, complex Gaussian distribution with mean 0 and variance equal to the expected variance of the radiometric noise,
\begin{multline}
    \label{eq:draw_noise}
    V^{p}_{xy}(\nu, \phi) \sim \\
    \mathcal{N}\left(0, \frac{1}{2} \sigma^{p}_{xy}(\nu, \phi)^{2} \right) + j
    \mathcal{N}\left(0, \frac{1}{2} \sigma^{p}_{xy}(\nu, \phi)^{2} \right) \ ,
\end{multline}
where $\mathcal{N}(\mu, \sigma^{2})$ denotes a Gaussian distribution with mean $\mu$ and variance $\sigma^{2}$.  Estimation of the expected radiometric variance $\sigma^{p}_{xy}(\nu, \phi)^{2} = w^{p}_{xy}(\nu, \phi)^{-1}$ is described in \secref{sec:weights}.  The map making procedure is then applied to this \emph{Gaussian noise realization} in an identical manner as to the data.  The sample variance of the map pixels is calculated in a 2D rolling window in $(\theta, \phi)$ and compared to the estimate given by \cref{eq:map_noise_variance}.  In general we find good agreement ($\lesssim \SI{5}{\percent}$) between the two.  This technique of processing a Gaussian noise realization using the same pipeline that is applied to the data will be used in other comparisons below.

\subsection{Beam Calibration}
\label{sec:beams}

\begin{figure*}
   \centering \includegraphics[width=0.98\linewidth,keepaspectratio]{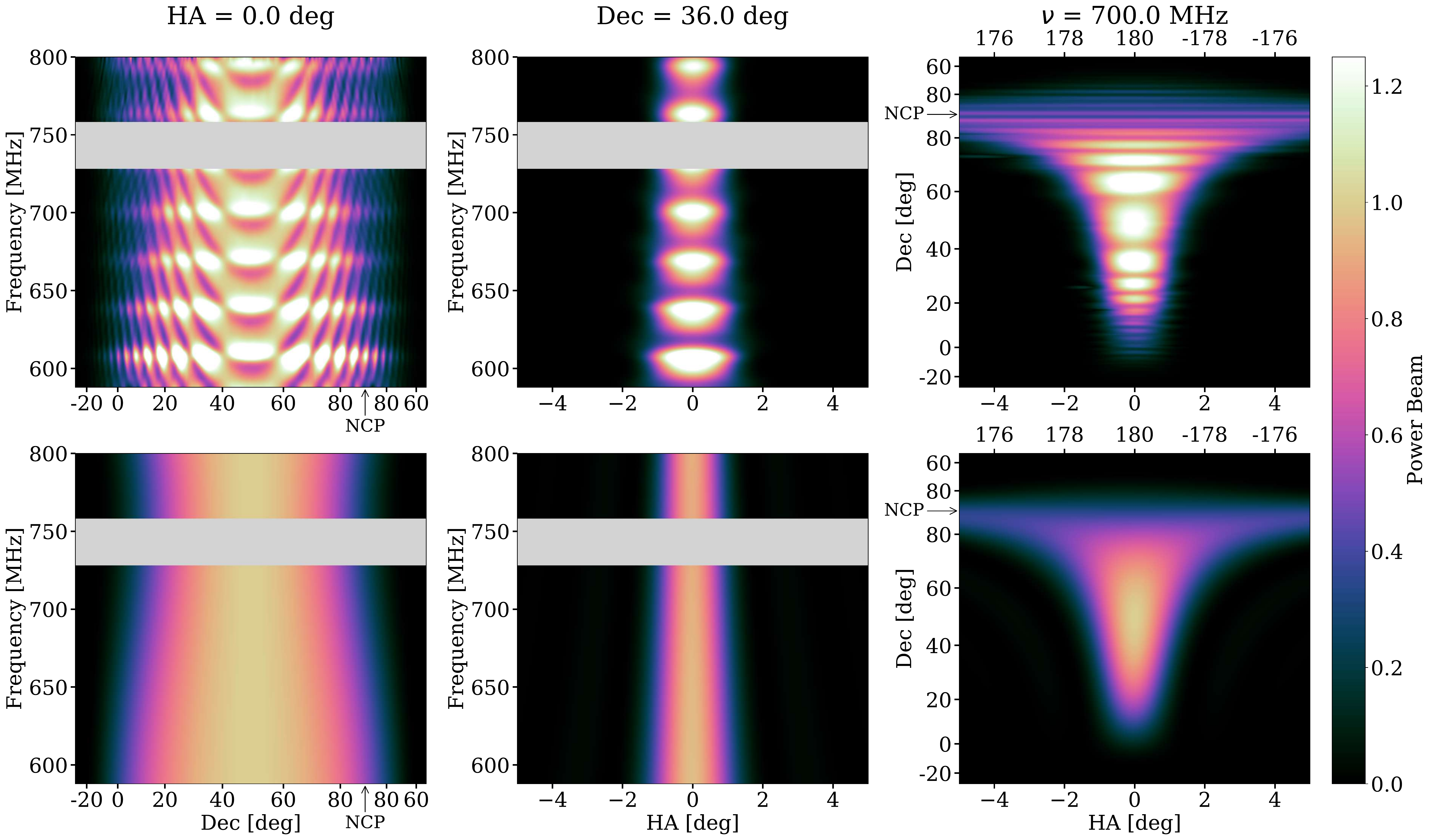}
    \caption{2D slices through the 3D primary beam models. We show the power beam for the Y polarisation array. The top row is the default beam model, obtained by deconvolving a model for the radio emission from extragalactic point sources from the visibilities measured with long east-west baselines.  The bottom row is the control beam model, which has similar global properties as the default, but without the small-scale spectral structure.  \emph{Left:} beam model as a function of declination and frequency on the meridian (hour angle = \SI{0.0}{\degree}).  The declination axis is uniformly spaced in the sine of the zenith angle.  The region to the right of the north celestial pole (NCP) annotation corresponds to the antipodal transit at hour angle = \SI{180}{\degree}.  \emph{Middle:} beam model as a function of hour angle and frequency at a declination of \SI{36.0}{\degree}.  \emph{Right:} beam model as a function of hour angle and declination at \SI{700}{\mega\hertz}.  The region above the NCP annotation corresponds to the antipodal transit at hour angles given by the upper x-axis. The beam has been normalized to 1.0 on meridian at the declination of Cygnus A (\SI{40.73392}{\degree}) at each frequency in order to match how the data are normalized by the calibration procedure.  The gray band denotes frequencies where we do not have a valid model for the beam due to persistent RFI in mobile LTE bands.}
    \label{fig:beam_model}
\end{figure*}

Our primary beam model is obtained by deconvolving a model for the radio sky that consists only of extragalactic point sources from the visibilities measured with baselines that have a large east-west component.  The long baselines resolve out the diffuse Galactic emission, making a point-source-only sky model a reasonable description of the data.  There are several high-resolution, large-area sky surveys that can be interpolated to the \SIrange[range-units = single, range-phrase=$-$]{400}{800}{\mega\hertz} CHIME band to construct this sky model.  At lower frequencies we rely on the VLA Low-frequency Sky Survey (VLSS) \citep{cohen2007} at \SI{74}{\mega\hertz} and the Westerbork Northern Sky Survey (WENSS) \citep{rengelink1997} at \SI{326}{\mega\hertz}.  At higher frequencies we rely on the NRAO VLA Sky Survey (NVSS) \citep{condon1998} at \SI{1400}{\mega\hertz} and the Green Bank survey (GB6) \citep{gregory1996} at \SI{4850}{\mega\hertz}.  The method used to deconvolve the sky model from the data is similar to the method used to construct a map, which was described in the previous section.  Whereas the map maker deconvolves a model for the primary beam from the hybrid beamformed visibilities to estimate the intensity of the sky, the beam calibration deconvolves a model for the sky intensity from the same hybrid beamformed visibilities in order to estimate the primary beam.  Appendix~\ref{sec:beam_calibration_ptsrc} describes this method in detail.

The resulting power beam, $|A^{\rm Y}(\nu, \theta, \phi)|^{2}$, for the Y polarisation array is shown in the top row of \cref{fig:beam_model}.  We briefly describe the main features of the CHIME primary beam pattern, referring the reader to \cite{overview-paper} for a more in-depth discussion.  The large ($~\SI{50}{\percent}$) ripples that are evident in the frequency and declination direction are the result of multi-path interference.  Radiation from the sky can be absorbed and then re-radiated by feeds or reflected off the ground plane.  It then reflects off the cylinder and interferes with the primary path from the sky.  The period of the ripple is $\sim \SI{30}{\mega\hertz}$ and is set by the $\SI{5}{\meter}$ focal length of the CHIME cylinders.  Harmonics at $\SI{60}{\mega\hertz}$ and $\SI{90}{\mega\hertz}$, which arise from multiple reflections off the focal line and cylinder, are also significant, although they may not be distinguishable by eye in \cref{fig:beam_model}.
The narrowing of the beam in the hour angle direction as one moves toward higher frequencies is due to diffraction through the \SI{20}{\meter} aperture.  The apparent widening of the beam in the hour angle direction as one approaches the north celestial pole (NCP) is simply due to the fact that a point at declination $\theta$ travels $\cos{\theta}$ degrees on the sky for every degree in hour angle that elapses.  The beam is normalized to 1.0 on meridian at the declination of Cygnus A (\SI{40.73392}{\degree}) at each frequency in order to match how the data are normalized during complex gain calibration.  This imprints the interference pattern at the declination of Cygnus A onto all other declinations.  The power beam for the X polarisation array exhibits the same general features, but is slightly wider in both the hour angle and declination direction and also has a lower response at zenith because the dipole illuminates the cylinder less efficiently.

In order to characterize the effect that the ripples in the beam have on our final stacking result, we repeat our analysis with a ``control'' beam that has the same large-scale properties as the default beam model, but without the small-scale structure in the frequency and declination direction.  The control beam is a modified version of the analytical beam model proposed in \cite{shaw2015} (henceforth, S15) for cylindrical telescopes.  To briefly summarize the S15 model, the beam pattern of the antenna (henceforth, ``base'' beam) is assumed to be that of a horizontal dipole mounted a distance $\lambda / 4$ over a conducting ground plane.  The response in the east-west direction is the result of solving the Fraunhofer diffraction problem for a dipole illuminating an aperture with width equal to the \SI{20}{\meter} cylinder width.  The response in the north-south direction is simply the reflected amplitude of the base beam.  The primary beam of the telescope is then the outer product of these two 1D functions.

In this work, the S15 model for the base beam is modified to more accurately describe existing measurements of the CHIME primary beam.  The assumption that the base beam for the X polarisation is the base beam for Y polarisation rotated by \SI{90}{\degree} is abandoned.  The FWHM of the base beam in the east-west direction is assumed to be polarisation dependent, but frequency independent, and is obtained by performing a fit to holographic observations of several bright sources made in conjuction with the John~A.~Galt \SI{26}{\meter} telescope (see \cite{overview-paper} for a description of these measurements).  The FWHM of the base beam in the north-south direction is assumed to be polarisation and frequency dependent, and is obtained by fitting a flattened Gaussian to the meridian profile of the default beam at each frequency, and then fitting the resulting FWHM as a function of frequency to a third-order polynomial in order to smooth over the small-scale ripples while retaining large-scale variations observed in the width of the meridian beam with frequency.  The rest of the procedure is unchanged: the beam model is given by the outer product of an east-west response obtained by solving the Fraunhofer diffraction problem and a north-south response obtained from the reflected base-beam amplitude.  The resulting beam model is shown in the bottom row of \cref{fig:beam_model}.

\subsection{Foreground filtering}
\label{sec:filter}

The deconvolved map described in \secref{sec:mapmaking} is dominated by emission from extragalactic point sources, which is expected to be a factor of $\sim 10^{3} - 10^{5}$ brighter than the \tcm signal of interest \citep{santos2005}.  This foreground contamination can be separated from the \tcm signal on the basis of spectral scale; the foregrounds are expected to be spectrally smooth, whereas the \tcm signal varies rapidly with frequency \citep{shaver1999, oh2003,liu2011}.  For each pixel in the map, we apply a high-pass filter along the frequency axis to supress the foregrounds while retaining some fraction of the \tcm signal.

Designing an adequate filter is complicated by the fact that -- as discussed in \secref{sec:freq_mask} -- \SI{47.2}{\percent} of the band has been masked in order to remove RFI-like features and other narrowband, instrumental artifacts.
The DAYENU technique \citep{ewall-wice2021} is used to construct a linear filter for the irregularly-sampled map spectra that achieves the required suppression at large spectral scales.  In what follows, we briefly summarize this technique.

Let $\tau$ denote the delay, which is the Fourier transform dual to frequency $\nu$.  The following simple model is assumed for the covariance of the map as a function of $\tau$:
\begin{align}
    \label{eq:tau_tau_cov}
    \tilde{C}(\tau, \tau') & = \begin{cases}
                                    \frac{\epsilon^{-1}}{2 \taucut} \ \delta^{D}(\tau, \tau') & |\tau| \leq \taucut \\
                                    \Delta \nu \ \delta^{D}(\tau, \tau')  & |\tau| > \taucut
                               \end{cases} \ ,
\end{align}
where the region below $\taucut$ is the region of delay space contaminated by bright foregrounds,
$\epsilon$ is a small number that corresponds to the ratio of the radiometric noise to foreground variance, $\Delta \nu = \SI{0.390625}{\mega\hertz}$ is the width of the frequency channel, and $\delta^{D}(\tau, \tau')$ is the Dirac delta function.  This model results in the following analytical formula for the covariance between frequency channel $\nu_{m}$ and $\nu_{n}$:
\begin{align}
    \label{eq:freq_freq_cov}
    C_{mn} & = \epsilon^{-1} \sinc \left[2 \pi \taucut(\nu_{m} - \nu_{n})\right] + \delta_{mn} \ .
\end{align}
where $\sinc(x) \equiv \sin{(x)} / x$ and $\delta_{mn}$ is the Kronecker delta.

To construct the filter, the delay cut $\taucut^{p}(\theta)$ and stop-band rejection $\epsilon$ are specified.  Note that we allow the delay cut to vary as a function of polarisation and declination. \Cref{eq:freq_freq_cov} is then evaluated for each pair of frequency channels in the \SIrange[range-units = single, range-phrase=$-$]{587.5}{800}{\mega\hertz} band.  Rows and columns of the covariance matrix that correspond to masked frequencies are zeroed and the Moore-Penrose pseudo-inverse is calculated
\begin{align}
    \vec{R}^{p}(\theta) & = \left[\vec{m}^{T} \vec{C}^{p}(\theta) \vec{m}\right]^{+} \ ,
\end{align}
where $\vec{m}$ is a vector that is $1$ for valid frequencies and $0$ for masked frequencies.
The filter is then applied to each map pixel independently,
\begin{align}
    M^{p}_{\rm hpf}(\nu_{m}, \theta, \phi) & = \sum_{n} R^{p}_{mn}(\theta) \ M^{p}(\nu_{n}, \theta, \phi) \ .
\end{align}
The weights are also propagated through the filtering operation according to
\begin{align}
    \label{eq:weight_hpf}
    w^{p}_{\rm hpf}(\nu_{m}, \theta, \phi) & = \left[\sum_{n} \left( R_{mn}^{p}(\theta) \right)^{2} \ \left( \sigma^{p}_{\rm map}(\nu_{n}, \theta, \phi) \right)^{2} \right]^{-1} \ .
\end{align}
where $\sigma^{p}_{\rm map}(\nu, \theta, \phi)$ is given by \cref{eq:map_noise_variance}.

In order to find an appropriate delay cut, the delay power spectrum of the map is estimated as
\begin{align}
    P(\tau, \theta) & = \var_{\phi}\left\{ \tilde{M}^{p}(\tau, \theta, \phi) \right\} \ ,
\end{align}
where $\tilde{M}$ denotes the Fourier transform of the map along the frequency axis.  Direct calculation of $\tilde{M}$ through the fast Fourier transform will result in a point-spread-function in delay space that has large sidelobes due to the band-limited and irregularly spaced nature of our map spectra.  This will leak power from the bright foreground to higher delays, thus biasing our determination of $\taucut$.  To address this, the delay power spectrum is estimated using a Gibbs sampling method, which is described in detail in Appendix~\ref{sec:delay_spectrum}.

\begin{figure}
   \centering \includegraphics[width=0.98\linewidth,keepaspectratio]{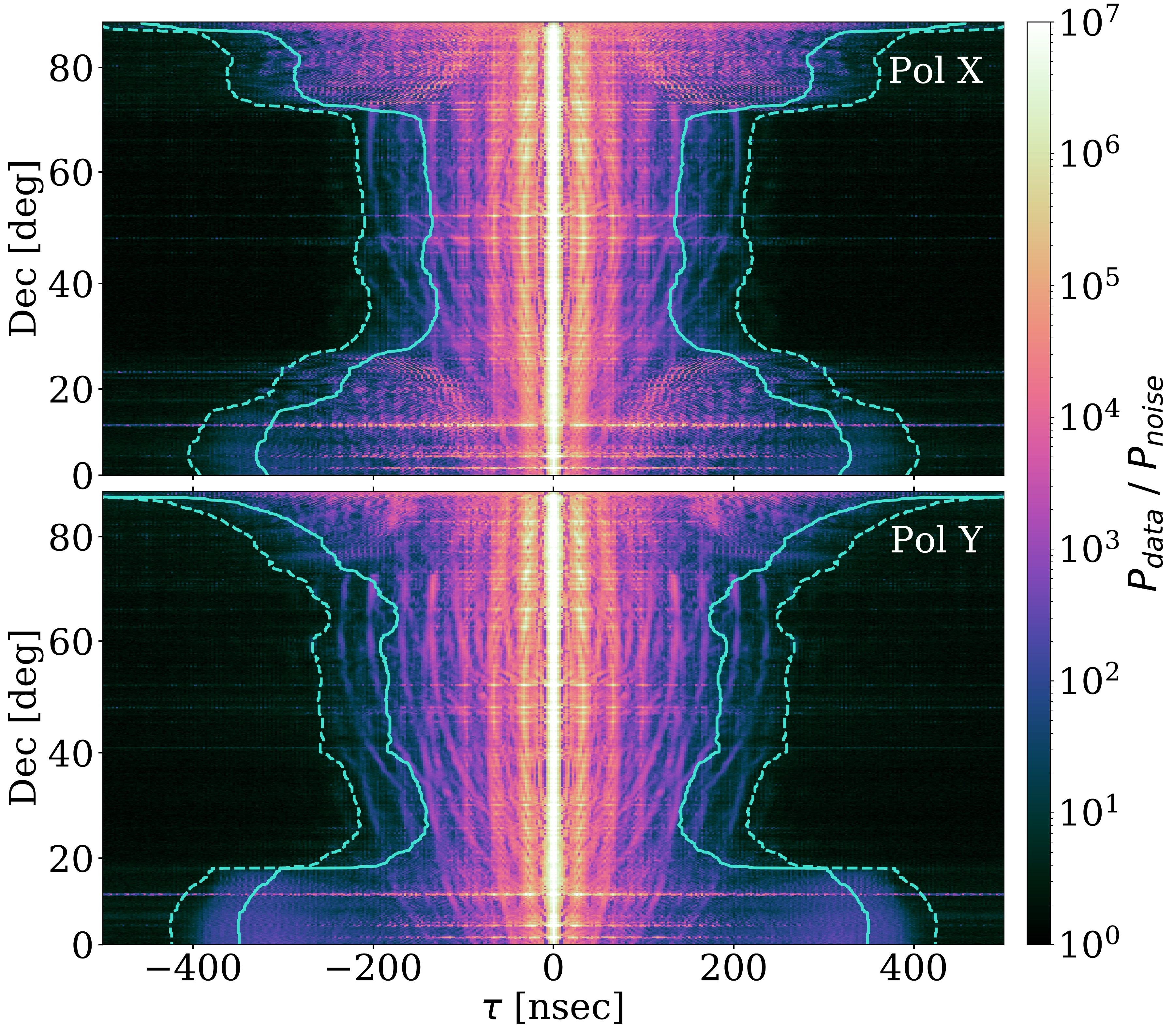}
    \caption{Delay power spectrum of the deconvolved map for the X (top) and Y (bottom) polarisations, normalized by our expectation for the radiometric noise.  The delay power spectrum is obtained by computing the variance of the delay spectrum of the map over $\phi \in [\SI{110}{\degree}, \SI{263}{\degree}]$ (see Appendix~\ref{sec:delay_spectrum} for details).  The dashed line is obtained by finding the minimum delay where $P_{\rm data} / P_{\rm noise} < 3$ at each declination.  The solid line is obtained by subtracting \SI{75}{\nano\second} from the dashed line and corresponds to the delay cut that is used in this analysis.  Note that the x-axis has been restricted in this figure: we are sensitive out to \SI{1250}{\nano\second}, but beyond \SI{500}{\nano\second} the measured spectrum matches our expectation for radiometric noise.}
    \label{fig:delay_spectrum}
\end{figure}

\Cref{fig:delay_spectrum} shows the delay power spectrum of the map as a function of declination for each polarisation.  Note that the variance was calculated over $\phi \in [\SI{110}{\degree}, \SI{263}{\degree}]$, which corresponds to the range of RA covered by the eBOSS NGC field.  The delay power spectrum is normalized by the delay power spectrum of the expected radiometric noise.  This is obtained by applying the map making and delay power spectrum estimation to a Gaussian noise realization randomly drawn according to \cref{eq:draw_noise}.

At high delays, the measured spectrum is in a good agreement with our expectation for the noise, and at low delays we are dominated by foreground emission.  Ideally all foreground power would be contained within the bright peak centered on \SI{0}{\nano\second}.  However, the ripples in the primary beam are imprinted on the foregrounds, leaking power to higher delays.  The three additional peaks observed at integer multiples of $\sim \SI{30}{\nano\second}$ correspond to interference of the primary path through the telescope with secondary paths that have undergone 1, 2, and 3 additional reflections off the focal line and cylinder.  The amplitude of these peaks has been reduced by deconvolving the model for the primary beam; however, they are still significant compared to our expectation for the noise.  We are actively working on improving the accuracy of our beam model and implementing a deconvolution procedure that better addresses the off-axis response (see  \cite{shaw2015} for one example) to further reduce the amplitude of these peaks.  The ``U'' shaped tracks in the delay power spectrum correspond to the brightest point sources moving through the far sidelobes.  In this case, there is a delay associated with the east-west component of the baseline that is not corrected by the map-making procedure because it assumes that the instrument has no sensitivity outside the main lobe of the primary beam.  For each bright source, there are three ``U'' shaped tracks corresponding to the three inter-cylinder, east-west baseline separations which extend out to progressively higher delays.  Finally, at large zenith angle, or low and high declinations, the foreground power extends out to higher delays due to aliasing of the sky in the baselines with one-cylinder east-west separation, as explained in \secref{sec:mapmaking}.

The stop-band rejection is set to $\epsilon = 10^{-12}$, which is much smaller than the inverse of the dynamic range of the delay power spectrum ($\sim 5 \times 10^{-9}$).  This ensures that the brightest foreground features near \SI{0}{\nano\second} delay are attenuated to well below the radiometric noise level.

Our initial delay cut is defined as the minimum delay where the measured power spectrum is less than 3 times the power spectrum of the Gaussian noise realization.  This is indicated by the dashed cyan line in \cref{fig:delay_spectrum} and results in an aggressive filter that yields a map that is dominated by radiometric noise.  However, the dominant contamination at delays just below our aggressive cut originates from a few bright sources in the far sidelobes, which are easily masked.  In an attempt to maximize signal to noise, we examine four different delay cuts that correspond to the aggressive cut minus [\SI{25}{\nano\second}, \SI{50}{\nano\second}, \SI{75}{\nano\second}, \SI{100}{\nano\second}].  For each cut, a foreground filter is constructed and applied to both the data and a simulation of the \tcm signal.  Next, the regions around known bright point sources are masked.  The foreground-filtered data and signal are then pushed through the rest of the analysis pipeline, which is described in the sections that follow.  As the delay cut is reduced, the relative increase in the noise is compared to the relative reduction in the amplitude of the simulated \tcm signal.  The aggressive cut minus \SI{75}{\nano\second} results in the maximum signal-to-noise of the four values tested.  This is indicated by the solid blue line in \cref{fig:delay_spectrum} and will be used as the delay cut $\taucut^{p}(\theta)$ for the rest of the analysis.

\subsection{Additional Masking}
\label{sec:map_mask}

\begin{figure}
    \centering \includegraphics[width=\linewidth,keepaspectratio]{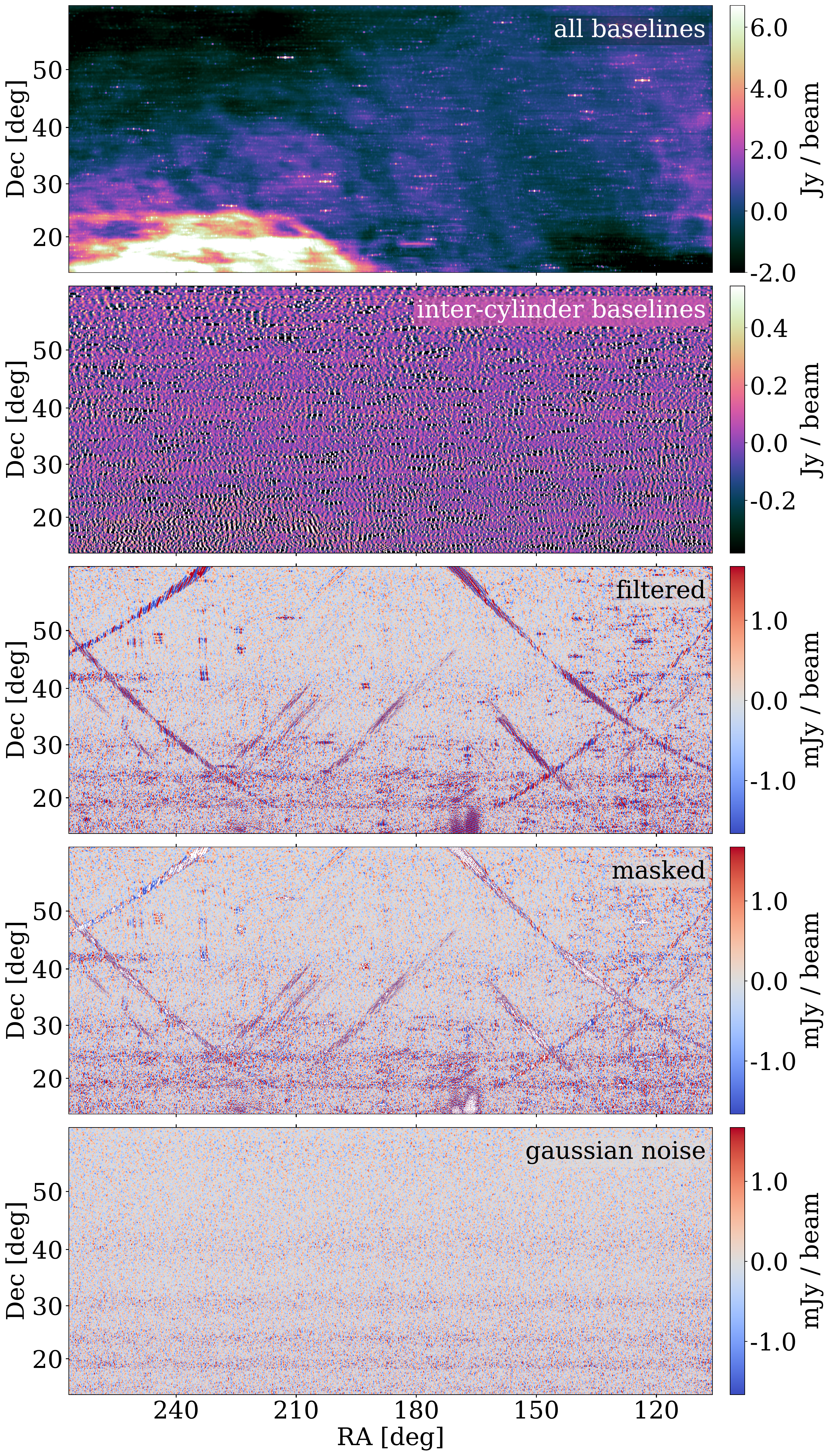}
     \caption{The deconvolved map at \SI{700.78125}{\mega\hertz} at several stages of the processing.  The range of right ascension and declination matches that of the eBOSS NGC field.  The top panel shows the map constructed from all Y polarisation baselines (excluding autocorrelations).  The second panel shows the map constructed from only the inter-cylinder baselines, which resolve out the diffuse Galactic emission, leaving primarily emission from extragalactic point sources.  The range on the color scale has been compressed by a factor of $\sim 15$.  The third panel shows the inter-cylinder map after applying the delay filter.  The range on the color scale has been further compressed by a factor of $\sim 300$.  Residuals associated with very bright sources in the far sidelobes, bright sources in the main lobe, and instrumental artifacts are evident.  However, they are localized and the fourth panel shows the result of masking any pixel that is more than 6 times the standard deviation of the expected radiometric noise (\SI{1.3}{\percent} of the pixels).  This can be compared to the bottom panel, which shows a realization of the radiometric noise generated according to \cref{eq:draw_noise}.  The horizontal features in the bottom two panels are due to the ripples in the primary beam pattern, which are imprinted on the noise during deconvolution.}
     \label{fig:map_stages}
\end{figure}

The foreground filter heavily attenuates the signal at frequencies near the edges of the \SIrange[range-units = single, range-phrase=$-$]{587.5}{800}{\mega\hertz} band and at frequencies neighboring large spans of masked frequencies.   These heavily attenuated frequencies would be improperly upweighted when stacking on external catalogs because the pipeline accounts for the fact that the noise has been attenuated, but does not account for the fact that the signal has also been attenuated.  To address this, at each polarisation and declination the median value of the non-zero diagonal elements of the filter is calculated.  Any frequency where the diagonal element of the filter is less than $\SI{20}{\percent}$ of the median is masked.  This removes approximately \SI{4.2}{\percent} of the \SIrange[range-units = single, range-phrase=$-$]{587.5}{800}{\mega\hertz} band.

The simulations described in \secref{sec:simulations} predict that the RMS of the radiometric noise is more than order of magnitude larger than the RMS of the \tcm signal in the foreground-filtered, deconvolved map.  The distribution of map pixel values is largely set by the radiometric noise and the \tcm signal is a small perturbation that is only evident after averaging over a large number of sources.  The map does contain residual foregrounds, RFI, and instrumental artifacts that are large compared to the propagated fast-cadence estimate for the noise.  However, this excess noise is for the most part restricted to specific frequency bins or localized to regions on the sky.  The subset of the data that exceeds our expectation for the noise is masked using the following procedure.

The foreground-filtered, deconvolved map is standardized by dividing the value of each map pixel by the standard deviation from the fast-cadence estimate.  These standardized maps are examined manually at each frequency channel.  Any channel that contains residuals that are both large compared to the expected noise and corrupt a significant portion of the NGC field are masked.  Note that the delay filter couples frequency channels, so a channel may show significant residuals due to the filter leaking some narrowband artifact from an adjacent channel.  This can be disentangled for the most part by identifying artifacts with a common spatial profile across frequencies and then masking the channel where that artifact has the largest magnitude.  It could also be automated through an iterative procedure of masking and foreground filtering.  In the end, \SI{14.7}{\percent} of the \SIrange[range-units = single, range-phrase=$-$]{587.5}{800}{\mega\hertz} band is discarded in this way.  We note that newer versions of the pipeline with improved RFI excision have reduced this fraction to roughly \SI{5}{\percent}, and most of these frequency channels are believed to be recoverable in future analyses by making additional improvements to the RFI excision algorithm, and by further vetting the time ranges that are included in the sidereal stack.  The frequency mask generated through this procedure is applied to the un-filtered map and the foreground filter is re-applied.

The total fraction of the \SIrange[range-units = single, range-phrase=$-$]{587.5}{800}{\mega\hertz} band that remains after removing persistent RFI bands, frequencies that do not have complete sidereal coverage, frequencies that have low integration time for the NGC field, frequencies that show excess noise, and frequencies near the edges of the large gaps of missing data is \SI{48.6}{\percent}.  Finally, any map pixel whose absolute value is greater than $6 \sigma$ is masked, where $\sigma$ is again obtained from the fast-cadence noise estimate.  This removes \SI{1.4}{\percent} of the remaining map pixels within the NGC field.  The choice of a $6 \sigma$ threshold was informed by signal injection simulations that are described in \secref{sec:linearity}.  The threshold is large enough that the resulting bias in the amplitude of the \tcm signal is small compared to the statistical uncertainty.  \Cref{fig:map_stages} shows the map at several stages of the pipeline processing for a typical frequency channel; the third and fourth panel depict the application of the $6 \sigma$ mask.

\begin{figure}
   \centering \includegraphics[width=0.98\linewidth,keepaspectratio]{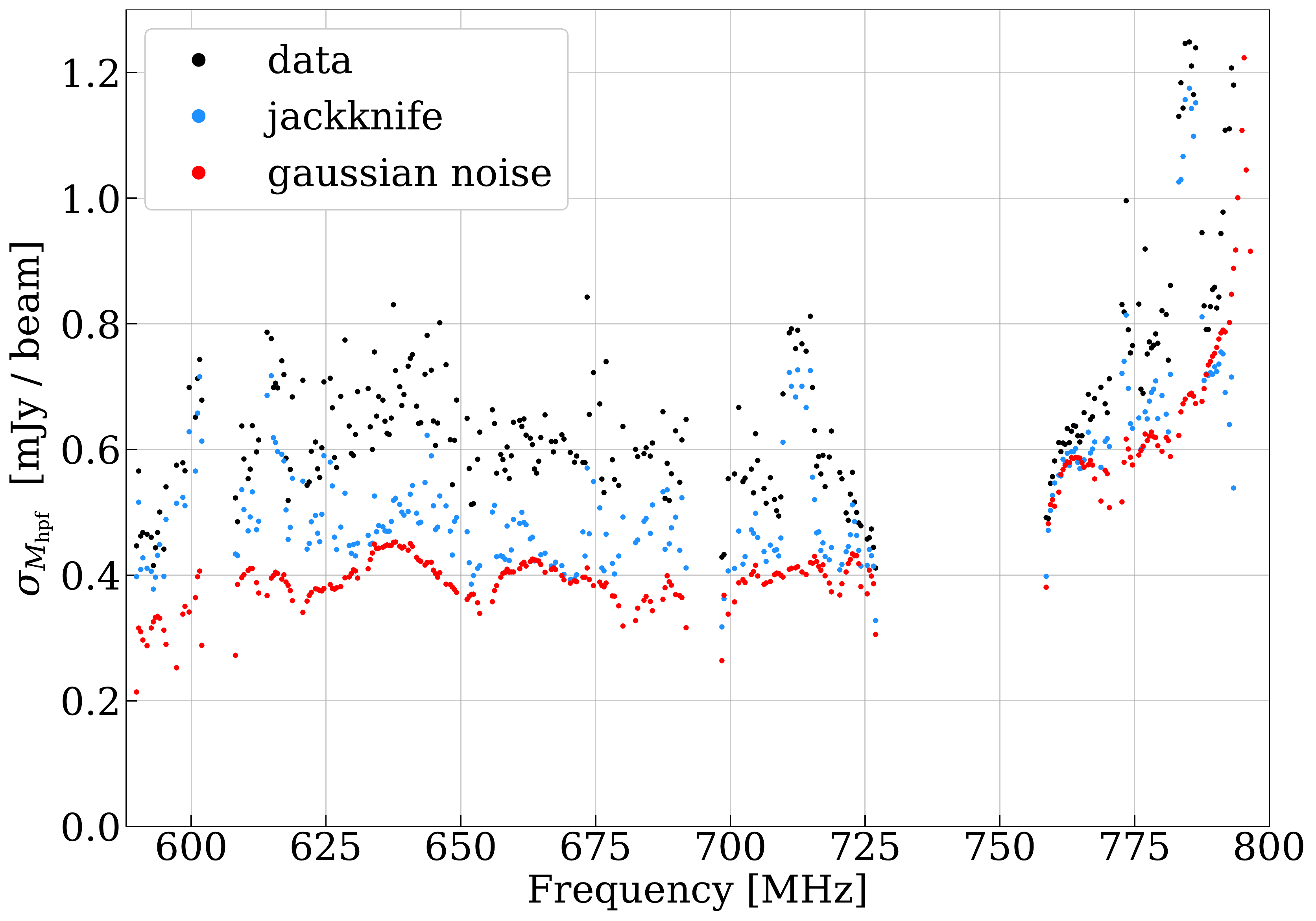}
    \caption{Standard deviation of pixels within the NGC field in the deconvolved, foreground-filtered map.  Black denotes the measured standard deviation.  Red denotes the expected standard deviation due to radiometric noise, which is based on the fast-cadence estimate of the variance.  Blue denotes the measured standard deviation in a jackknife of even and odd days (see \secref{sec:consistent_even_odd}).  The measured noise in a single frequency channel is on average \SI{0.6}{\milli\jansky\per\beam}, with an increase to \SI{1.0}{\milli\jansky\per\beam} in the upper \SI{50}{\mega\hertz} of the band.  This is on average \SI{50}{\percent} greater than the expected radiometric noise, due to residual foregrounds and RFI.  The residual foregrounds are largely the same from day to day, and therefore cancel in the even-odd jackknife, but the RFI does not.  As a result, the jackknife is in better agreement with the radiometric noise outside of certain \SI{6}{\mega\hertz} wide bands that still suffer from unmasked, transient RFI.}
    \label{fig:map_noise}
\end{figure}

\Cref{fig:map_noise} shows in black the standard deviation of the map pixels within the NGC field as a function of frequency after all masking has been applied.  The noise for a single frequency channel is on average \SI{0.6}{\milli\jansky\per\beam}, with an increase to \SI{1.0}{\milli\jansky\per\beam} in the upper \SI{50}{\mega\hertz} of the band.  The increased noise at high frequencies is driven by a reduction in the primary-beam response on meridian when averaged over the declinations spanned by the NGC field, a mild increase in the system temperature, and frequent flagging of the highest frequencies ($\gtrsim \SI{794}{\mega\hertz}$) by the threshold applied in the sidereal regridding stage of the pipeline (see \secref{sec:regrid}).  This last item will be corrected in future revisions of the pipeline.  The noise in the map is on average \SI{50}{\percent} greater than the expected radiometric noise, which is shown in red.  To generate the expected radiometric noise, visibilities are drawn randomly according to \cref{eq:draw_noise} and then propagated through the map making and foreground filtering procedure.  For comparison, we also show in blue the standard deviation of the map pixels in a jackknife of even and odd days.  The procedure for constructing this jackknife will be described in \secref{sec:consistent_even_odd}.  The noise in the jackknife is in better agreement with the expected radiometric noise, except in a few \SI{6}{\mega\hertz} wide bands where there is still unmasked, transient RFI.  This is due to the fact that the residual foregrounds are largely due to instrument chromaticity that is the same from day to day and thus cancels in the jackknife.  Note that a slightly different frequency mask was used for the jackknife because the frequency channels at the upper edge of the band do not have full coverage of the sidereal day in the even or odd split.  This results in the jackknife noise dropping below the expected radiometric noise because the large filter attenuation at the upper edge of the band is pushed to lower frequencies.

Using a more aggressive mask ($3\sigma$) removes \SI{6.6}{\percent} of the map pixels within the NGC field and brings the measured noise to within \SI{22}{\percent} of the expected radiometric noise on average at the expense of introducing a significant non-linearity into the analysis pipeline.  Using a more aggressive mask ($3\sigma$) and more aggressive delay filter with a cutoff that is \SI{75}{\nano\second} larger (indicated by the dashed, cyan line in \cref{fig:delay_spectrum}) brings the measured noise to within \SI{12}{\percent} of the expected radiometric noise on average.  However, this results in a significant reduction in the amplitude of the stacked \tcm signal in simulations, and a better signal to noise ratio is anticipated using the less aggressive delay cut.  Note that with either mask or delay cut, the excess noise from residual foreground, RFI, and instrumental artifacts is comparable to or less than the radiometric noise for the \SI{102}{\day} sidereal stack, assuming that they add in quadrature.

\subsection{Stacking}
\label{sec:stacking}

For each source in a given eBOSS catalog, a spectral cube centered on the source's location is extracted from the deconvolved, foreground-filtered map.  First, the right ascension and declination of the source are converted from ICRS to CIRS coordinates to account for the precession and nutation of the Earth's polar axis.  The redshift of the source is converted to the frequency of the redshifted \tcm emission,
\begin{align}
    \nu_{\rm 21 cm} &= \frac{\SI{1420.406}{\mega\hertz}}{1 + z} \ .
\end{align}
The map pixel and frequency channel closest to these coordinates is found and $\pm 50$ pixels (channels) are extracted in the angular (frequency) directions.  This results in a spectral cube that spans $\pm \SI{3}{\degree}$ in right ascension/declination and $\pm \SI{20}{\mega\hertz}$ in frequency.

The stacked signal is given by the weighted average of the spectral cubes over all sources in the catalog:
\begin{align}
    d^{p}(\Delta \nu, &  \ \Delta \theta, \ \Delta \phi) = \nonumber \\
    \sum_{s} & \ W^{p}_{\rm hpf}(\nu_{s} + \Delta \nu, \ \theta_{s} + \Delta \theta, \ \phi_{s} + \Delta \phi) \nonumber \\
    & \times M^{p}_{\rm hpf}(\nu_{s} + \Delta \nu, \ \theta_{s} + \Delta \theta, \ \phi_{s} + \Delta \phi)
\end{align}
where $(\nu_{s}, \theta_{s}, \phi_{s})$ denote the frequency channel and map pixel closest to the coordinates of source $s$ and
\begin{align}
W^{p}_{\rm hpf}(\nu_{s} & + \Delta \nu, \ \theta_{s} + \Delta \theta, \ \phi_{s} + \Delta \phi) = \\ \nonumber
& \frac{w^{p}_{\rm hpf}(\nu_{s} + \Delta \nu, \ \theta_{s} + \Delta \theta, \ \phi_{s} + \Delta \phi)}{\sum_{s} w^{p}_{\rm hpf}(\nu_{s} + \Delta \nu, \ \theta_{s} + \Delta \theta, \ \phi_{s} + \Delta \phi)}
\end{align}
denotes the relative weight given to source $s$, with the absolute weight $w^{p}_{\rm hpf}$ given by \cref{eq:weight_hpf}.

Note that we make no attempt to interpolate the spectral cubes onto a common grid relative to the coordinates of the source.  Instead we take a forward modeling approach where the stacking procedure is applied to simulations in order to characterize how the pixelization alters a stack of the \tcm signal.  This will result in a small degradation in signal-to-noise because we are not stacking on the true peak, but given the pixelization used we estimate this to be only $\sim \SI{3}{\percent}$ for the NGC field.

For simplicity, all model fitting and parameter estimation uses only the central pixel of the stack as function of frequency offset.  Going forward we will use $d^{p}(\Delta \nu) \equiv d^{p}(\Delta \nu, 0, 0)$ to describe the stack of the pixels closest to the coordinates of the sources.

\subsection{Noise Covariance Estimation}
\label{sec:covariance}

\begin{figure}
    \centering \includegraphics[width=\linewidth,keepaspectratio]{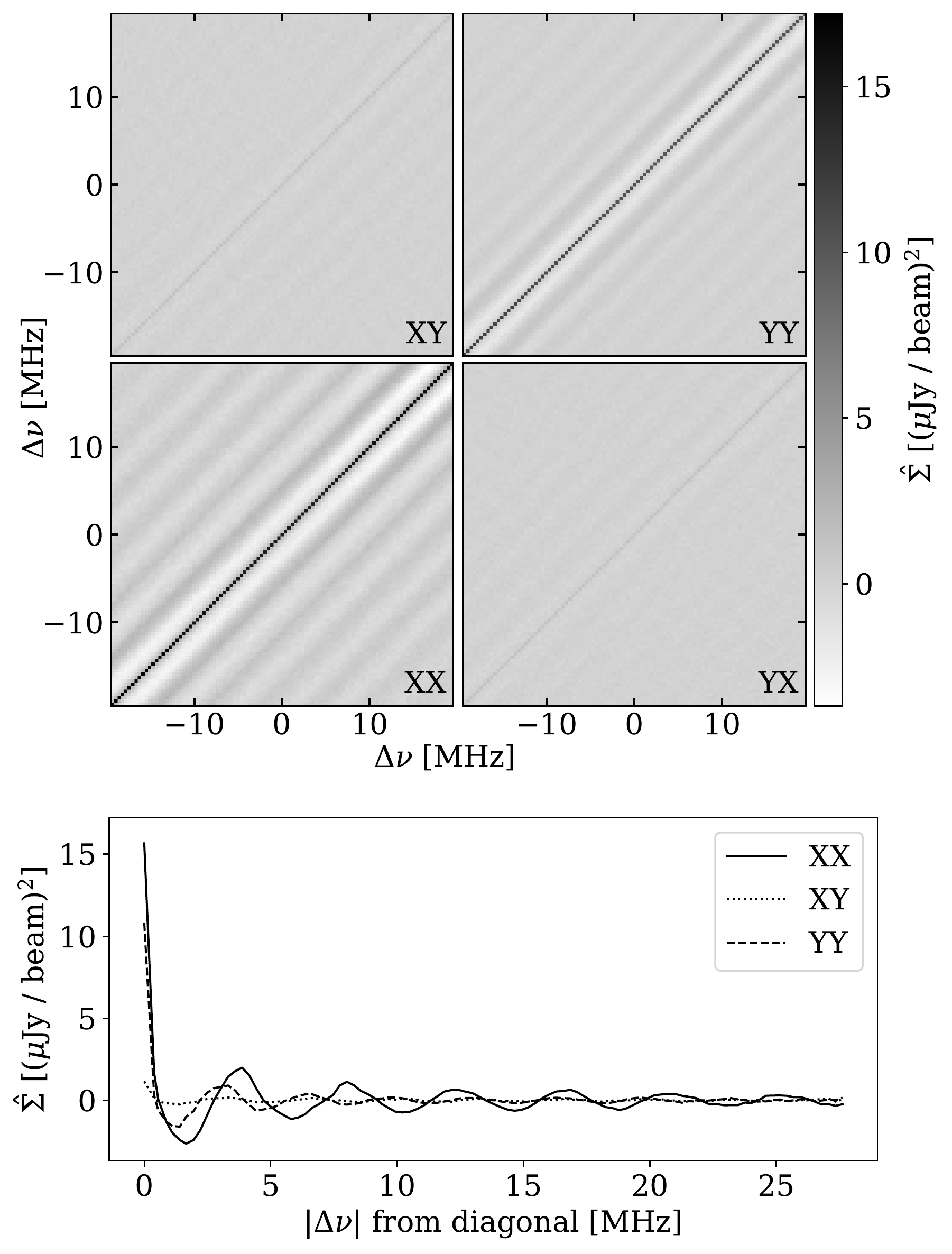}
     \caption{Estimated noise covariance of the NGC QSO stack, obtained by computing the sample covariance of stacks on \SI{10000} random mock catalogs.  Each sub-panel in the upper figure shows the covariance between frequency offsets for a different pair of polarisations.  The bottom figure shows the average value of the covariance as a function of distance from the central diagonal of each sub-panel, with the different polarisation pairs denoted using different linestyles as indicated in the legend.  The noise in the two polarisations is largely independent.  However, there is non-negligible correlation in the noise between frequency offsets within a polarisation.  The $\sinc$ like dependence on frequency offset is introduced by the foreground filter (see \cref{eq:freq_freq_cov}).  The period of the ripple is different for the two polarisations because a different delay cut was used on average.}
     \label{fig:covariance}
\end{figure}

The probability distribution of the noise in the stack must be characterized in order to derive accurate uncertainties on the inferred model parameters.  As discussed in \secref{sec:map_mask}, residual foregrounds and RFI are expected to be sub-dominant but significant contributors to the noise, and both are likely correlated between frequencies.  More generally, the foreground filter couples all frequency channels, ensuring a non-zero correlation between frequency offsets in the stack.  These factors are not accounted for in the propagated fast-cadence noise estimate, which only includes the radiometric contribution to the noise and does not account for the correlation between frequency channels.  To develop an accurate noise model, we stack the data on a large number of random mock catalogs and examine the distribution of values.

Each eBOSS clustering catalog has a corresponding ``random'' catalog that approximates the three-dimensional selection function of the clustering catalog and is more than \SI{40} times as dense \citep{ross2020,raichoor2021}.  We randomly sample the random catalog without replacement to generate a mock catalog that has the same number of sources as the true catalog.  The deconvolved, foreground filtered map is then stacked on the mock catalog following the same procedure described in \secref{sec:stacking}.  This process is repeated $N_{\rm mock} = $~\SI{10000} times.

The noise covariance of the stacked data is estimated using the sample covariance of the mocks
\begin{align}
    \vec{\hat{\Sigma}} & = \frac{1}{N_{\rm mock} - 1} \sum_{m=1}^{N_{\rm mock}} \left(\vec{d}_{m} - \vec{\hat{\mu}} \right)^{T} \left(\vec{d}_{m} - \vec{\hat{\mu}}\right)
\end{align}
where
\begin{align}
    \vec{d}_{m} = \left[d^{p}_{m}(\Delta \nu_{i}, 0, 0) : 0 \leq i \leq 100, \ p \in \{X, Y\}\right]
\end{align}
is a vector containing the stacked signal at the central pixel as a function of frequency offset for both polarisations for the $m$'th mock catalog, and
\begin{align}
    \vec{\hat{\mu}} & = \frac{1}{N_{\rm mock}} \sum_{m=1}^{N_{\rm mock}} \vec{d}_{m}
\end{align}
is the sample mean of the mocks.  An example of the sample covariance is shown in \cref{fig:covariance}.

We find that the sample mean for a given frequency offset and polarisation is non-zero at a level larger than expected given the standard error.  The RMS of the sample mean over all frequency channels and polarisations is $\sigma_{\vec{\hat{\mu}}} = $ \SIrange{0.7}{1.9}{\micro\jansky\per\beam} depending on the tracer, which is roughly \SI{20}{\percent} of the sample standard deviation over mock catalogs and a factor of 20 times larger than the standard error.  This sample mean over mocks is subtracted from the stack on the true catalog to ensure a consistent noise model.

We find that the distribution of values observed in the mocks is consistent with a multivariate Gaussian whose covariance and mean is set to the sample variance and mean as calculated above.

%% file: sections/simulations.tex

\section{Signal Modelling and Simulations}

\label{sec:modelling_simulations}

Interpreting our stacking measurements requires that we are able to predict the cosmological signal within them, and that we understand the performance of our analysis pipeline including any signal loss that has occurred. In this section we discuss the framework to address these: a parameterised model of the cosmological signal, a simulation pipeline producing synthetic time streams and source catalogs, and a scheme for using these simulations to predict the stack signal from the parameters of our model.

\subsection{Cosmological scales being probed}
\label{sec:scales}

\begin{figure}[t]
   \centering \includegraphics[width=0.98\linewidth, keepaspectratio, trim = 10 10 5 0]{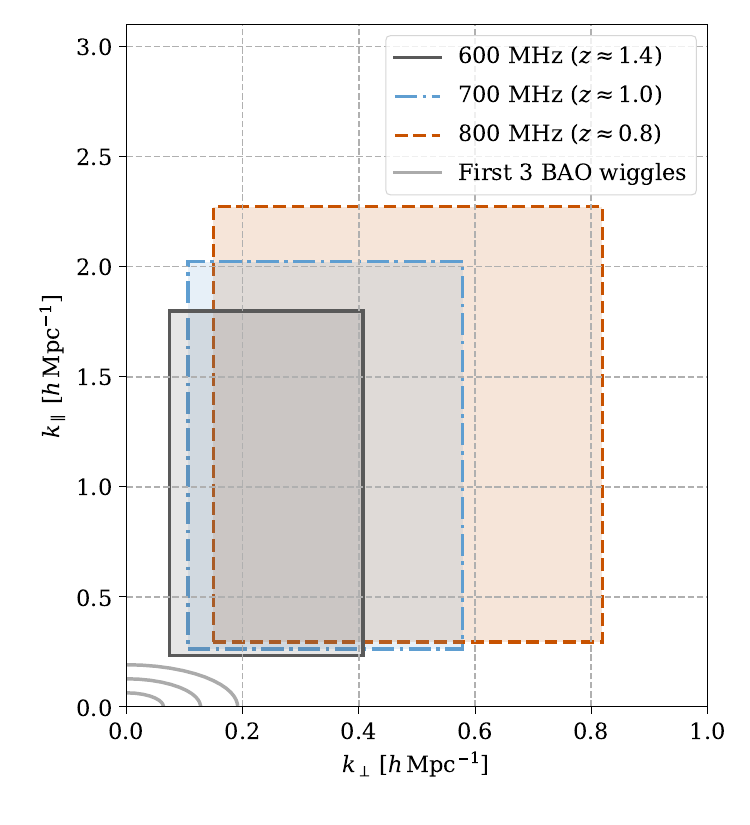}
    \caption{Approximate physical scales probed by the stacking measurements, as comoving wavenumbers along the line of sight ($k_\parallel$) or transverse to it ($k_\perp$). We evaluate the range of scales at three observing frequencies that span the relevant portion of the CHIME band. The accessible values of $k_\parallel$ are determined by the CHIME frequency channel width and delay filtering prescription, while the ranges of $k_\perp$ arise from the synthesized beamwidth and choice to exclude intracylinder baselines; see main text for details. The maxima of the first three BAO wiggles in the matter power spectrum are shown by the grey lines,   making it apparent that the measurements in this work are insensitive to BAO scales, and instead mainly probe the nonlinear regime of structure formation.}
    \label{fig:k_sensitivity}
\end{figure}

To set the stage for the modelling approach described later in this section, in \cref{fig:k_sensitivity} we show the approximate range of physical scales probed by our stacking measurements, represented as comoving wavenumbers $k_\parallel$ (along the line of sight) and $k_\perp$ (transverse to the line of sight). This range depends on observing frequency due to the relationship between frequency and radial distance, and also due to chromaticity of CHIME's beam response, so we show results at three frequencies within the portion of the band used in our analysis.

The foreground filter described in \secref{sec:filter} acts roughly as a high-pass filter in $k_\parallel$, with the minimum accessible $k_\parallel$ determined by the delay cut $\taucut$; for \cref{fig:k_sensitivity}, we use $\taucut = \SI{200}{\nano\second}$, reflective of the typical delay cut within the declinations covered by the eBOSS catalogs. The sensitivity at high $k_\parallel$ is attenuated by the finite width of CHIME's frequency channels, which we approximate as top-hats with width $\SI{390.625}{\kilo\hertz}$.

Similarly, the sensitivity at high $k_\perp$ is determined by the profile of the synthesized beam associated with the maps described in \secref{sec:mapmaking}. For \cref{fig:k_sensitivity}, we use the simplified 1d beamforming result from \cite{masui2017} to obtain NS and EW synthesized beam profiles based on CHIME's feed layout and the analytical (``control") primary beam model discussed in \secref{sec:beams}, take the geometric mean of the full-widths at half maximum in each direction, and translate this into a comoving wavenumber at each plotted frequency. Finally, since intracylinder baselines are excluded from our analysis, we are not sensitive to any angular scales that are only probed by pure NS baselines; these scales are determined by the EW primary beam profile, and we translate the EW full-width at half maximum into a minimum accessible~$k_\perp$.

Note that a more thorough treatment of the scales being probed is possible, in which the stacking measurements can be related to an integral of the galaxy-HI cross-power spectrum multiplied by a transfer function $W(k_\parallel, k_\perp)$ that precisely encodes the sensitivity of our analysis to a given Fourier mode.
Such a treatment in currently under development and will be presented in a forthcoming publication \citep{interpretation-paper}, but preliminary results are in good agreement with the estimates in \cref{fig:k_sensitivity}.

In this figure, we also show the maxima of the first three BAO wiggles in the matter power spectrum, located at multiples of $k_{\rm BAO} = 2\pi/r_{\rm drag} \approx 0.064 h^{-1}{\rm Mpc}$. It is clear that our delay filter and exclusion of intracylinder baselines have effectively filtered out any sensitivity to BAO scales from our stacking measurements. The scales that remain are beyond the reach of analytical perturbative methods for large-scale structure statistics in Fourier space (e.g.~\citealt{damico2020,ivanov2020,chenvlah2021}); while these scales have some overlap with those accessible to hybrid simulation-perturbation theory methods (e.g.~\citealt{kokron2021}), the majority of our signal-to-noise lies at ever smaller scales, implying that we cannot immediately apply those methods in our present analysis.

Halo-based models for HI (e.g.~\citealt{padmanabhan2021}) and galaxy clustering can in principle describe the full range of scales shown in \cref{fig:k_sensitivity}. However, we have found that a simpler model, which makes efficient use of our simulation framework described in \secref{sec:simulations},
is fully capable of describing the observed signal while allowing for marginalization over hard-to-predict properties of nonlinear clustering. We describe this model and its application to our measurements in the following subsections.

\subsection{Signal Model}
\label{sec:skymaps_signal}

Cosmological modelling of the distribution of galaxies\footnote{For brevity, we refer to ELGs, LRGs, and QSOs as ``galaxies" in this section.} and HI typically begins with the matter overdensity $\deltam(\vx, z) \equiv [\rho_{\rm m}(\vx, z) - \bar{\rho}_{\rm m}(z)] / \bar{\rho}_{\rm m}(z)$, where an overbar denotes a spatial average. In our modelling we assume that galaxies and HI are each linearly biased tracers of the total matter density. The overdensity corresponding to galaxy or HI number density, $\deltag$ or $\deltaHI$, can then be written in Fourier space as
\begin{align*}
\delta_X(\vk; z) &=  \left[ b_X(z) + f(z)\mu^2 \right] \tilde{D}_X^{\rm FoG}(k\mu, z) \deltam(\vk, z) \\
&\quad + \epsilon_X(z)
\numberthis
\label{eq:deltaX}
\end{align*}
with $X \in [{\rm g}, {\rm HI}]$. In \cref{eq:deltaX}, $b_X$ is the bias factor (assumed to be scale-independent), and the $f(z)\mu^2$ term encodes the effect of redshift-space distortions at linear order \citep{kaiser1987}, with $f$ as the logarithmic growth rate and $\mu\equiv k_\parallel/k$. We aim to capture the key non-linear contributions to the two-point statistics of the fields: we include real-space non-linear clustering in $\deltam$ itself; the impact of small-scale velocities on redshift-space observations (``Fingers of God"; \citealt{jackson1972}) is modelled with the damping function $\tilde{D}_X^{\rm FoG}$; and finally, we include a term $\epsilon_X$ in \cref{eq:deltaX}, which is uncorrelated with $\deltam$ and represents the contribution of shot noise to $\delta_X$.

In our analysis we will only require the two point statistics of the correlated fields. These are captured entirely by the power spectrum of two fields:
\begin{multline}
    P_{XY}(\vk; z_X, z_Y) = \\
    \bigl[b_X(z_X) + f(z_X) \mu^2 \bigr] \bigl[b_Y(z_Y) + f(z_Y) \mu^2 \bigr] \\
    \times \tilde{D}_X^{\rm FoG}(k\mu, z_X) \tilde{D}_Y^{\rm FoG}(k\mu, z_Y) \, P_{\rm m}(\vk; z_X, z_Y)
    \\ + P^\mathrm{shot}_{XY}(z_X, z_Y) \; .
\end{multline}
The ingredients required to complete our model are functions for the non-linear matter power spectrum $P_{\rm m}$, the linear bias $b_X$, the Fingers of God function $\tilde{D}_X^\mathrm{FoG}$, and the shot noise $P^\mathrm{shot}_{XY}$. We discuss our fiducial choices for these ingredients in the following sections.

\subsubsection{Matter power spectrum}
\label{sec:matterpower}

As input to our simulations, we use the halo model prediction for the nonlinear matter power spectrum from \cite{mead2021}, as implemented in the \texttt{CAMB} code \citep{lewis1999}. We have also considered the Halofit fitting functions from \cite{smith2003} and \cite{takahashi2012}, and have found that these different choices affect the final stacking amplitude in the simulations by at most $\sim 3\%$, with little change in the shape. Thus, the uncertainty arising from the specific choice of nonlinear matter power spectrum is far subdominant to the uncertainty inherent in our assumption of linear, scale-independent bias in \cref{eq:deltaX}.

\subsubsection{Linear bias}
\label{sec:bias}

We assume the following for the linear bias of each eBOSS sample:
\begin{align}
\label{eq:bELG}
b_{\rm ELG}(z) &= 1.5 + 0.7 (z - 0.85)\ , \\
\label{eq:bLRG}
b_{\rm LRG}(z) &= 2.03 + 0.86 (z-0.4) + 0.13(z-0.4)^2\ ,\\
b_{\rm QSO}(z) &=2.38+ 1.4 (z-1.55) + 0.28(z-1.55)^2\ .
\end{align}
The ELG bias uses the redshift evolution of the linear bias predicted by the simulations of \cite{merson2019}, normalized such that \cref{eq:bELG} evaluates to the bias measurement from  \cite{demattia2021} at the mean redshift of the eBOSS ELG sample.
The LRG bias is based on \cite{zhai2017}, who fit a halo occupation distribution model to small-scale clustering of a combined BOSS+eBOSS LRG sample and computed the linear bias from this model. Specifically, \cref{eq:bLRG} is the result of a quadratic fit to the best-performing bias model from Figure~12 of \cite{zhai2017}.
The QSO bias is taken from the fitting function in \cite{laurent2017}, based on measurements of the eBOSS QSO correlation function in four redshift bins.

\begin{figure*}[t]
   \centering \includegraphics[width=0.98\linewidth,keepaspectratio, trim = 0 10 0 0]{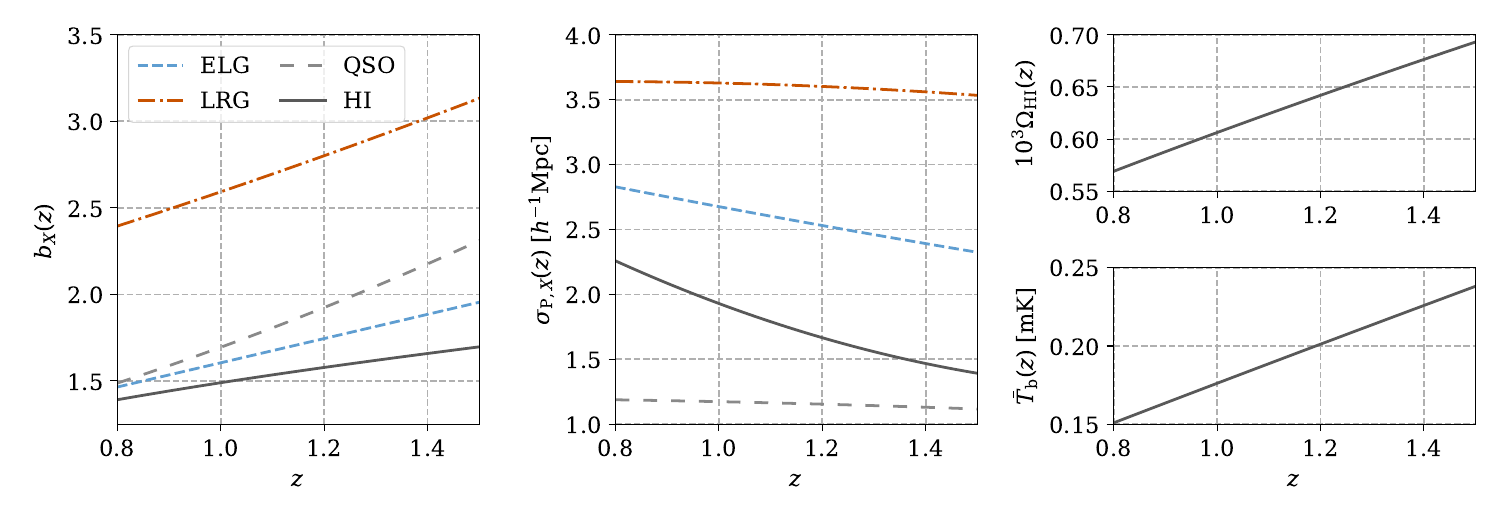}
    \caption{Fiducial models for various redshift-dependent quantities used in our simulated sky maps. See Sections~\ref{sec:bias}-\ref{sec:21Tb} for discussions of how each model was chosen. {\em Left}: Linear bias of each eBOSS sample and HI. {\em Center}: Finger of God damping scale, with same line styles as left panel. {\em Upper right}: HI density, as fraction of the critical density at $z=0$. {\em Lower right}: Mean \tcm brightness temperature.}
    \label{fig:sim_z_models}
\end{figure*}

For the linear bias of HI, we follow \cite{ansari2018} in smoothly interpolating between measurements from the IllustrisTNG simulations \citep{villaescusa-navarro2018} at $z<2$ and the analytical model from \cite{castorina2017} at $z>2$.\footnote{This bias model has been implemented in the \texttt{PUMANoise} code, available from \url{https://github.com/slosar/PUMANoise}.} We show our bias models for HI and each eBOSS sample in the left panel of \cref{fig:sim_z_models}.

\subsubsection{Finger of God models}
\label{sec:fingerofgod}

We model the Finger of God damping in Fourier space as a Lorentzian:
\beq
\tilde{D}_X^{\rm FoG}(k\mu, z) = \frac{1}{1 + k^2\mu^2\sigma_{{\rm P},X}(z)^2 / 2}\ ,
\label{eq:DFoGk}
\eeq
where the damping scale $\sigma_{{\rm P},X}$ can approximately be associated with the pairwise velocity dispersion of galaxies or HI emitters on nonlinear scales. The (constant-redshift) Fourier conjugate of this function is an exponential in comoving distance (e.g.\ \citealt{scoccimarro2004}),
\beq
D_X^{\rm FoG}(x_\parallel, z)
	= \frac{e^{-\lv x_\parallel \rv \sqrt{2} / \sigma_{{\rm P},X}(z)}}{\sqrt{2} \sigma_{{\rm P},X}(z)}\ ,
\eeq
and we implement the Finger of God effect by convolving
our simulated maps with this kernel along the line-of-sight axis. This is equivalent to multiplying the 3d auto-power spectrum of $X$ by $\tilde{D}_X^{\rm FoG}(k\mu, z)^2$, and multiplying the cross-power spectrum of HI and $X$ by $\tilde{D}_\sHI^{\rm FoG}(k\mu, z) \times \tilde{D}_X^{\rm FoG}(k\mu, z)$.

For each eBOSS sample, Fourier-space clustering measurements have been analyzed using Finger of God models similar to what we describe above. For ELGs and LRGs, \cite{demattia2021} and \cite{gilmarin2020} use a squared Lorentzian function multiplied into the 3d galaxy power spectrum, finding best-fit values of $\sigma_{\rm P,ELG} = 2.79 h^{-1}{\rm Mpc}$ at $z_{\rm eff} = 0.85$ and $\sigma_{\rm P,LRG} = 3.64 h^{-1}{\rm Mpc}$ at $z_{\rm eff} = 0.7$ (where we quote the average of separate fits to the NGC and SGC fields). For QSOs, \cite{zarrouk2018} use a Gaussian Finger of God model, and perform fits that isolate the contribution to this model from small-scale velocities (as opposed to QSO redshift errors, which have a similar effect on the observed clustering). Taking the average of their best-fit $\sigma_{\rm P}$ values for the ``3-multipole" and ``3-wedge" analyses yields $\sigma_{\rm P,QSO} = 1.3h^{-1}{\rm Mpc}$ at $z_{\rm eff} = 1.48$. We find that this is roughly equivalent to Lorentzian damping with $\sigma_{\rm P,QSO} = 1.12h^{-1}{\rm Mpc}$.

We use these values to fix the amplitude of our fiducial $\sigma_{\rm P}(z)$ models for each sample. We compute the redshift dependence from a simple model in which $\sigma_{\rm P}(z)$ scales like a weighted average of the velocity dispersion $\sigma_v^2(M,z)$ of a dark matter halo of mass $M$, weighted by the halo mass function $dn/dM$ and the mean satellite occupation in a mass-$M$ halo:
\beq
\sigma_{{\rm P},X}(z) \propto \frac{1+z}{H(z)}
\left[
\frac{\int dM \frac{dn}{dM} N_{{\rm sat},X}(M) \sigma_v^2(M, z)}
{\int dM \frac{dn}{dM} N_{{\rm sat},X}(M)}
\right]^{1/2}\ .
\label{eq:sigmaPzdep}
\eeq
To evaluate \cref{eq:sigmaPzdep}, we use the halo mass function from \cite{tinker2008} and the eBOSS halo occupation distribution models from \cite{alam2020}. The final results for $\sigma_{{\rm P},X}(z)$, incorporating the amplitude constraints described above, are well fit by quadratic functions of redshift, which we present below:
\begin{align}
\frac{\sigma_{\rm P, ELG}(z)}{h^{-1} {\rm Mpc}} &= 2.79 - 0.77 (z - 0.85) + 0.083 (z - 0.85)^2 \ , \\
\frac{\sigma_{\rm P, LRG}(z)}{h^{-1} {\rm Mpc}} &= 3.64 + 0.019 (z-0.7) - 0.19 (z-0.7)^2\ ,\\
\frac{\sigma_{\rm P, QSO}(z)}{h^{-1} {\rm Mpc}} &= 1.12 - 0.14 (z-1.48) - 0.058 (z-1.48)^2\ .
\end{align}

For HI, we choose the damping scale based on simulations from \cite{sarkar2019}, who attempt to account for the motion of HI within galaxies in addition to the contribution from the velocity dispersion within dark matter halos. They assume that the Finger of God damping of the 3d HI power spectrum is given by a Lorentzian, and fit a $\sigma_{\rm P}(z)$ relation to their simulations. We use these results, multiplied by a factor of $2^{-1/2}$ to translate to the damping given by a squared Lorentzian (as implied by our \cref{eq:DFoGk}). The adopted $\sigma_{\rm P}(z)$ model is well fit by a quadratic function of redshift, given by
\beq
\frac{\sigma_{\rm P, \sHI}(z)}{h^{-1} {\rm Mpc}} = 1.93 - 1.48 (z-1) + 0.81 (z-1)^2\ .
\eeq
Over the redshift range of interest, this $\sigma_{\rm P}(z)$ model is within 20\% of the values obtained in \citep{villaescusa-navarro2018} from fits of a squared Lorentzian to measurements from the IllustrisTNG simulations.

We plot our models for eBOSS and HI damping scales in the middle panel of \cref{fig:sim_z_models}.

\subsubsection{\tcm brightness temperature}
\label{sec:21Tb}

We convert simulated maps of $\deltaHI$ into brightness temperature fluctuations by multiplying by the mean \tcm brightness temperature $\Tbar(z)$. Recall that,
after the end of reionization, the spin temperature $T_s$ is high compared to both the background CMB
temperature and $T_\star = h \nuHI / \kB$. In this limit, the \tcm brightness
temperature can be written as (e.g.\ \citealt{bull2015})
\beq
\Tb(\vx, z) = \frac{3 \hbar c^3 A_{10}}{16 \kB \nuHI^2} \frac{(1 + z)^2}{H(z)} \nHI(\vx, z)\ ,
\eeq
where $\nHI$ is the comoving HI number density and $A_{10}$ is the Einstein coefficient for spontaneous emission in the \tcm line.
Using $\nHI(\vx, z) = \nbarHI(z) [ 1 + \deltaHI(\vx, z) ]$, we can write
\begin{align*}
\delta \Tb(\vx, z) &\equiv \Tb(\vx, z) - \Tbar(z) \\
&= \Tbar(z) \deltaHI(\vx, z)\ ,
\numberthis
\end{align*}
which justifies our method of converting maps of $\deltaHI$ into $\Tb$. Using $\nbarHI(z) = \OmegaHI(z) \rho_{\rm c} / (m_{\rm p} + m_{\rm e})$, where $\rho_{\rm c} = 3 H_0^2 / 8 \pi G$ is the critical density today, we can write $\Tbar$ as
\begin{multline}
\Tbar(z) = \left[\frac{9 \hbar c^3 A_{10} H_{100}}{128 \pi k_B \nuHI^2 G (m_{\rm p} + m_{\rm e})} \right]
\\ \times h \frac{H_0}{H(z)} \, \OmegaHI(z) \, (1 + z)^2 \ ,
\label{eq:Tbarlong}
\end{multline}
where $H_{100} = \SI{100}{\kilo\metre \,\second^{-1} \mega\parsec^{-1}}$ and $h = H_0 / 100$.
The prefactor in square brackets is independent of cosmology, consisting only of
fundamental constants and $A_{10}$. Using $A_{10} = \SI{2.8843e-15}{\second^{-1}}$ \citep{Gould1994}, \cref{eq:Tbarlong} can be written more compactly as\footnote{Other versions of \cref{eq:Tbarcompact} in the literature have prefactors that vary significantly from \SIrange{180}{190}{\milli\kelvin}, most of which is accounted for by using values of $A_{10}$ from older calculations. The value quoted in the main text is taken from a recent review of atomic transition properties \citep{Wiese2009}, which takes its $A_{10}$ value for hydrogen from \cite{Gould1994}.}
\beq
\Tbar(z) \approx 191.06 \left[ h \frac{H_0}{H(z)} \, \OmegaHI(z) \, (1 + z)^2 \right] \: \si{\milli\kelvin} \; .
\label{eq:Tbarcompact}
\eeq

For $\OmegaHI(z)$, we use the fitting function from \cite{crighton2015}, which was determined from a compilation of $\OmegaHI$ estimates over $0<z<5$:
\beq
\OmegaHI(z) = 4\times 10^{-4} (1+z)^{0.6}\ .
\label{eq:OmegaHI}
\eeq
We plot Eqs.~\eqref{eq:Tbarcompact} and~\eqref{eq:OmegaHI} in the right panels of \cref{fig:sim_z_models}.

\subsubsection{Shot noise}
\label{sec:shotnoise}

The cross-correlation between maps of HI and the distribution of galaxies in a given sample will be sensitive to the HI content of the galaxies. Specifically, the 3d cross-power spectrum of $\delta \Tb(\vx, z)$ and $\deltag(\vx, z)$ contains a cross shot noise contribution of the form (e.g.\ \citealt{wolz2017})
\beq
P_{T\sg}^{\rm shot}(z_\sHI, z_\sg) = C_\sHI(z_\sHI) \MHIg\ ,
\label{eq:PTgshot}
\eeq
where $\MHIg$ is the mean HI mass per galaxy in the sample, and
\beq
C_\sHI(z) = \frac{3 \hbar c^3 A_{10}}{16 \kB \nuHI^2 (m_{\rm p} + m_{\rm e})} \frac{(1 + z)^2}{H(z)}\ .
\eeq
In principle, $\MHIg$ depends on redshift, but for simplicity, we consider a single value that is averaged over the entire sample. We also write the shot noise contribution as being constant for all $z_\sg$, but note that there is expected to be a gradual, scale-dependent decorrelation as $\lv z_\sHI - z_\sg\rv$ increases, due to relative displacements of sources between different time slices.

\subsubsection{Model Parameters}
\label{sec:freeparameters}

To produce a parameterised model of the \tcm signal, we use the ingredients presented in Sections~\ref{sec:matterpower} to~\ref{sec:shotnoise} as a basis, and introduce a finite number of parameters which will scale their magnitude, but not their redshift dependence. In total, our model contains seven parameters that are used to model the contributions to the cross-power spectrum:
\begin{description}
    \item[$\OmegaHI$] One of the key quantities controlling the stack signal is the total amount of neutral hydrogen in the Universe. Although this quantity is expected to be redshift dependent, in this paper we use the model given in \cref{eq:OmegaHI} as a baseline and use a single redshift-independent parameter $\OmegaHI$ to scale the fiducial model about an effective redshift $\zeff$, which gives
    \begin{equation}
        \OmegaHI(z) = \OmegaHI \biggl[\frac{\OmegaHI^\fid(z)}{\OmegaHI^\fid(\zeff)}\biggr] \; .
    \end{equation}
    \item[$\bHI$, $\bg$] To control the bias of the \tcm field and galaxy density fields which are again expected to be redshift dependent, we scale the models given in \secref{sec:bias}, giving
    \begin{equation}
        \bHI(z) = \bHI \biggl[\frac{\bHI^\fid(z)}{\bHI^\fid(\zeff)}\biggr] \; ,
    \end{equation}
    for the \tcm field and the equivalent definition for the galaxy density,
    \begin{equation}
        \bg(z) = \bg \biggl[\frac{\bg^\fid(z)}{\bg^\fid(\zeff)}\biggr] \; .
    \end{equation}
    \item[$\Mten$] The strength of the shot noise contribution is governed by the mass of neutral hydrogen typically associated with a tracer galaxy $\MHIg$. We control this quantity with the parameter $\Mten$ defined by
    \begin{equation}
        \MHIg = \Mten \times 10^{10}\:\si{\Msolar}\ .
    \end{equation}
    \item[$\alphaNL$] The shape of the high-$k$ real-space cross-power spectrum is uncertain because of non-linear gravitational evolution and baryonic effects. We let this shape vary using a linear mode which interpolates from a linear to a non-linear power spectrum
    \begin{equation}
        P(k) = \alphaNL \, \PNL(k) + (1 - \alphaNL) \, \PL(k) \; .
    \end{equation}
    For $\PNL(k)$ we use the model described in \secref{sec:matterpower}, and for $\PL(k)$ we use a power spectrum with the same parameters but with the Halofit corrections turned off. This parameter is valid for $\alphaNL > 0$ where values above one correspond to increasing the power contributed by non-linear evolution. Although this parameter is not physically motivated, we expect it to capture the effects of non-linearities at the level that can be measured in this work.
    \item[$\alphaFoG{\sHI}$, $\alphaFoG{\sg}$] To account for uncertainties in the Fingers of God smoothing, we allow redshift independent scaling of both the \tcm and tracer velocity dispersion~$\sigma_{\rm P}$:
    \begin{align}
        \sigma_{\rm P,HI}(z) & = \alphaFoG{\sHI} \: \sigma^\fid_{\rm P,\sHI}(z) \; ,\\
        \sigma_{\rm P,\sg}(z) & = \alphaFoG{\sg} \: \sigma^\fid_{\rm P,\sg}(z) \; .
    \end{align}
\end{description}
Put together, these give a model for the cross power spectrum of the \tcm emission and the galactic tracer, controlled by the parameters given above. Written out fully, this gives
\begin{align*}
    &P_{T\sg}(k, \mu; z_1, z_2) \\
    &\quad = \boldsymbol{\OmegaHI} \biggl[\frac{\Tb^\fid(z_1)}{\OmegaHI^\fid(\zeff)}\biggr]
    \, \frac{D^+(z_1)}{D^+(z_{\rm fid})} \frac{D^+(z_2)}{D^+(z_{\rm fid})}  \\
    &\quad\qquad
    \times \left(\boldsymbol{\bHI} \biggl[\frac{\bHI^\fid(z_1)}{\bHI^\fid(\zeff)}\biggr] + f(z_1) \mu^2\right) \\
    &\quad\qquad
    \times \left(\boldsymbol{\bg} \biggl[\frac{\bg^\fid(z_2)}{\bg^\fid(\zeff)}\biggr] + f(z_2) \mu^2\right) \\
    &\quad\qquad
    \times \lp \PL(k, z_{\rm fid}) + \boldsymbol{\alphaNL} \ls \PNL(k, z_{\rm fid}) - \PL(k, z_{\rm fid}) \rs \rp \\
    &\quad\qquad
    \times \DFoG{\sHI}(\boldsymbol{\alphaFoG{\sHI}} \, k \mu; z_1) \: {\DFoG{\sg}}(\boldsymbol{\alphaFoG{\sg}} \, k \mu; z_2) \\
    &\quad\quad
    + \boldsymbol{\Mten} \: \ls C_\sHI(z_1) 10^{10} \: \si{\Msolar} \rs \; ,
    \numberthis
    \label{eq:crossps}
\end{align*}
where we have highlighted the individual parameters in bold. Note that we evaluate the matter power spectrum at a fiducial redshift $z_{\rm fid}$, and apply the linear growth factor $D^+(z)$ to scale it to other redshifts.

We also require one more parameter to describe an apparent frequency or redshift offset between the $\HI$ and galaxies. This will be needed to account for systematic redshift errors in the eBOSS catalogs (see \secref{sec:quasar_redshift_errors}). As it is an observational effect we do not include it in the cross-power spectrum description (where it would manifest itself as a phase rotation).
\begin{description}
    \item[$\Dnu$] This parameter shifts the stack signal away from being centered at zero frequency lag. Positive values of $\Dnu$ move the peak of the signal to higher frequencies, and thus to lower redshifts.
\end{description}

\subsection{Simulations}

\label{sec:simulations}

\begin{figure*}[t]
   \centering \includegraphics[width=1\linewidth,keepaspectratio, trim = 0 10 200 0]{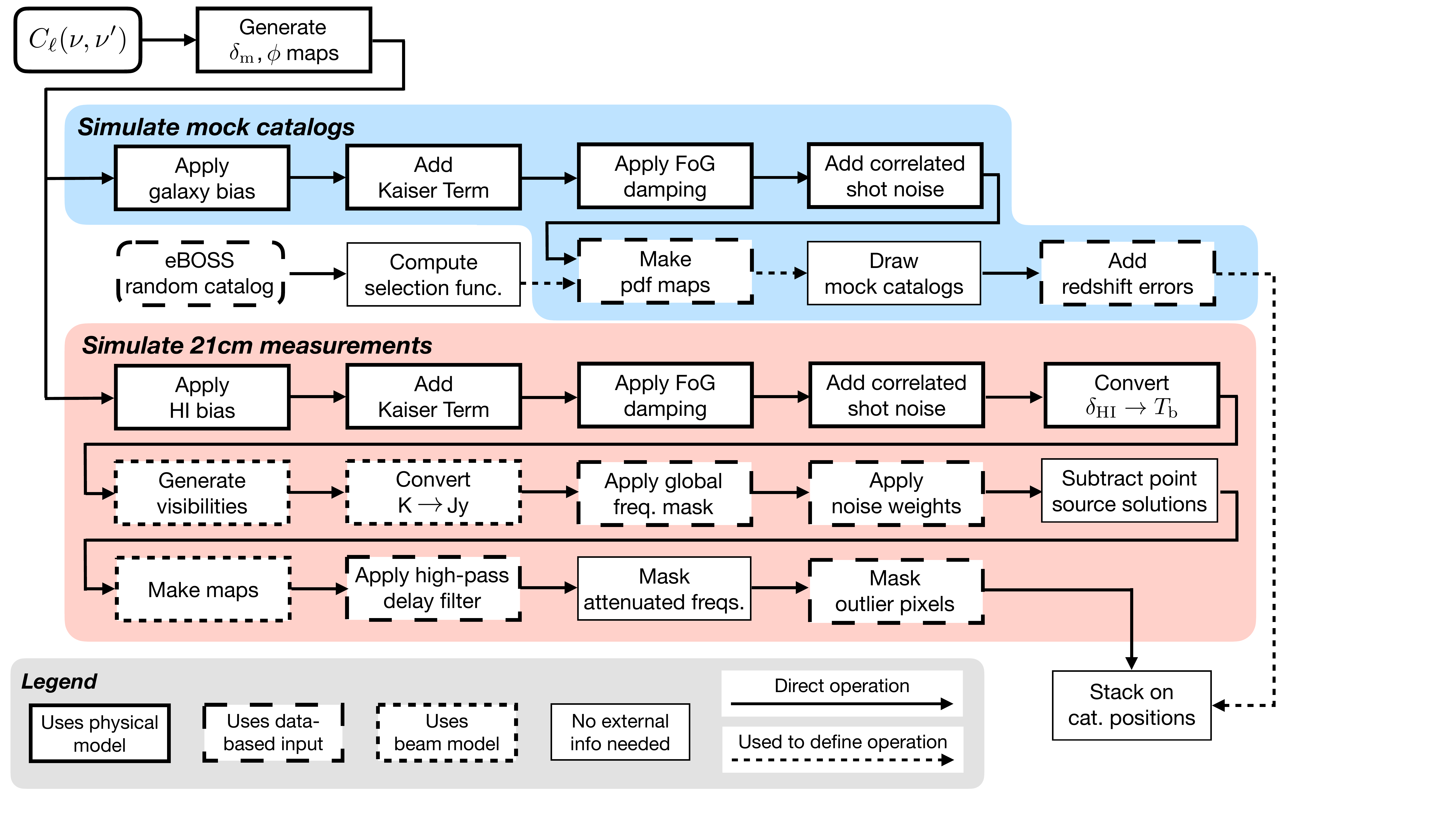}
    \caption{A schematic representation of the simulation pipeline. Starting from the multi-frequency angular power spectrum $C_\ell(\nu,\nu')$ corresponding to an input matter power spectrum, we generate correlated full-sky maps of the matter overdensity $\deltam$ and gravitational potential $\phi$ at redshifts corresponding to each CHIME frequency channel, and transform these into maps of galaxy/quasar overdensity and \tcm brightness temperature using the models described in \secref{sec:skymaps_signal} and the procedures in \secref{sec:skymap_generation}. Mock galaxy/quasar catalogs are constructed from the corresponding maps (\secref{sec:mocks}), while mock CHIME observations are formed from the \tcm maps (\secref{sec:timestreams_etc}), and these observations are then processed with the same stacking pipeline as the data (\secref{sec:mockstacking}). In this diagram, boxes with long dashed outlines are defined using inputs from eBOSS or CHIME observations, rather than simulations (for example, the delay cuts in the delay filter are those from \cref{fig:delay_spectrum}).}
    \label{fig:sim_flowchart}
\end{figure*}

We make extensive use of simulations in this work, both for interpreting our stacking measurements in terms of physical models, determining the signal transfer function, and quantifying the linearity of our analysis pipeline via injection of simulated signals into the data. In this section, we describe our simulation methodology for generating sky maps of \tcm emission and galaxy density (\secref{sec:skymap_generation}), propagating these through to mock galaxy catalogs (\secref{sec:mocks}) and CHIME timestreams (\secref{sec:timestreams_etc}), and finally performing the stacking procedure (\secref{sec:mockstacking}). The associated steps are schematically shown in \cref{fig:sim_flowchart}. Note that we do not attempt to simulate foregrounds, instead relying on several data-based tests to assess the contribution of residual foregrounds to our sky maps and cross-correlation measurements.

\subsubsection{Map Generation}
\label{sec:skymap_generation}

Each simulation produces a pair of correlated $\deltag$ and $\deltaHI$ maps of the sky generated as follows. The input real-space matter power spectrum (\secref{sec:matterpower}), evaluated at $z=1$, is transformed to a 3d correlation function using the \texttt{hankl} Python package \citep{karamanis2021} via the FFTlog method. We additionally employ Richardson extrapolation to repeated computations with increasingly fine $k$ sampling in order to reduce numerical errors. We then transform this to a multi-frequency angular power spectrum $C_\ell(\nu, \nu')$, and perform further frequency integrals over top-hats with width \SI{0.390625}{\mega\hertz} in order to mimic the effect of CHIME's frequency channelization.

We form a set of $N_\nu$ HEALPix maps \citep{HEALPix} from a Gaussian realization of this angular (matter) power spectrum, use the linear growth factor for our fiducial cosmology to scale each map to the redshift corresponding to its frequency, and multiply by the bias $b_X(z)$ (\secref{sec:bias}). In tandem, we generate the same number of maps of the gravitational potential~$\phi$, to which we apply a finite-difference second derivative in the radial direction and appropriate prefactors to generate a velocity field which is added to the biased matter to include linear redshift-space distortions. These maps are then convolved with a frequency kernel designed to reproduce the desired form of Finger-of-God damping in Fourier space (\secref{sec:fingerofgod}).

Finally, the maps corresponding to $\deltaHI$ are multiplied by the mean \tcm brightness temperature $\Tbar(z)$ (\secref{sec:21Tb}), while a lognormal transform is applied to the $\deltag$ maps, to ensure that $\deltag \geq -1$ everywhere; this allows $1+\deltag$ to be used to construct a probability density function from which to draw mock catalogs (see \secref{sec:mocks}). Note that we do not apply a lognormal transform to the $T_{\rm b}$ maps: when Gaussian temperature maps are stacked on mock catalogs generated from lognormal $\deltag$ maps, the two-point statistics are equivalent to the case where both sets of maps are Gaussian (see Appendix~\ref{sec:lognormal_stacking} for details).

Our baseline simulations set the shot noise contribution $\epsilon_X$ to zero, but we require the ability to add shot noise to ascertain its impact on the stacking signal. We incorporate this into our simulations by adding correlated realizations of white noise to each pair of $\deltaHI$ and $\deltag$ maps, such that their cross power spectrum will contain the contribution from \cref{eq:PTgshot} (the auto spectra of these maps are never used). Specifically, for each map voxel, we draw a random number from a Gaussian with $\sigma = [ C_\sHI(z) \MHIg / ( \Tbar(z) V_{\rm vox}) ]^{1/2}$ where $V_{\rm vox}$ is the voxel volume, and add this value to the same voxel in the $\deltaHI$ and $\deltag$ maps.\footnote{This method of adding correlated shot noise adds unphysical contributions to the auto power of the $\deltaHI$ and $\deltag$ maps, which will also affect the variance of the cross power between them, but this effect is completely negligible for our purposes.}

\subsubsection{Mock catalogs}
\label{sec:mocks}

For each galaxy sample we consider, we create mock catalogs of $N$ objects for each pair of simulated $\deltaHI$ and $\deltag$ maps. To do so, we select the pixel indices and frequency channels from a probability density function given by
\beq
\calP(\vx) \propto S(\vx) \ls 1 + \deltag(\vx) \rs\ ,
\eeq
where $S(\vx)$ is a sample-specific selection function.
Once a voxel is selected, galaxies are assigned positions within it according to uniform random distributions, and further displaced by simulated redshift errors as described below.

We obtain approximate galaxy selection functions from the public random catalogs associated with each eBOSS sample. In detail, for each sample, we build a histogram of object positions with $32$ redshift bins from $0.8<z<2.5$ and a HEALPix angular pixelization with $N_{\rm side}=16$ (roughly \SI{3.7}{\degree} resolution). We then form a rank-7 approximation to this distribution by performing a singular value decomposition of the histogram (represented as a $N_z \times N_{\rm pix}$ matrix). Finally, we upsample this to the HEALPix resolution of the input maps, and apply Gaussian smoothing in the angular direction (with width equal to the original pixel size) to apodize any sharp boundaries. Using this as the selection function for generating mocks ensures that we reproduce the large-scale footprint and modulations of each galaxy sample without introducing smaller-scale features of the catalogs into our simulations.

We generate random redshift errors using a separate scheme for each sample, based on estimates of redshift error distributions (represented as line-of-sight velocities) published by the eBOSS team. For LRGs, \cite{ross2020} examined pairs of observations of the same target and found the distribution of redshift differences was well-fit by a Gaussian with $\sigma=\SI{91.8}{\kilo\meter\per\s}$, corresponding to a redshift uncertainty of $\sigma=\SI{65.6}{\kilo\meter\per\s}$ per object. For ELGs, \cite{raichoor2021} quote three redshift error percentiles based on repeated observations; we find that these values are well fit by a Tukey lambda distribution with $\lambda=-0.4$ and $\sigma=\SI{11.88}{\kilo\meter\per\s}$.

For QSOs, \cite{lyke2020} find that, over the entire QSO catalog, the distribution of redshift differences between repeated observations is well fit by a double Gaussian. This implies that the single-observation redshift errors are also described by a double Gaussian, with $\sigma_1=\SI{150}{\kilo\meter\per\s}$, $\sigma_2=\SI{1000}{\kilo\meter\per\s}$, and 18\% of objects having errors drawn from the wider Gaussian%
\footnote{These double-Gaussian parameters are quoted in \cite{lyke2020} as corresponding to the distribution of {\em redshift differences between repeated observations} shown in their Fig.~4, but in our own comparison, we found that the quoted widths of the two Gaussians correspond to the distribution of {\em single-object redshift errors} implied by this figure.}.
Though we use this model for our primary analysis,
there is evidence
that it does not completely capture the distribution of QSO redshift errors. We discuss the discrepancies and the effect on our analysis in \secref{sec:quasar_redshift_errors}.

We do not attempt to simulate catastrophic redshift errors, which the above references estimate to occur in less than 1\% of the LRG and ELG samples and as much as 2\% of the QSO sample. The effect of these errors on our stacking measurements is a simple suppression of the overall amplitude, by an amount equal to the catastrophic error fraction.

\subsubsection{Timestreams}
\label{sec:timestreams_etc}

We make use of the $m$-mode formalism \citep{shaw2015} to translate simulated \tcm maps into visibilities. In this formalism, the spherical harmonic coefficients $a_{\ell m}^P(\nu)$ of sky maps for Stokes parameter $P \in \{ I, Q, U, V \}$ are related to the sidereal-time Fourier transform of the visibility timestream, $\tilde{V}^p_{xy,m}(\nu)$, via multiplication by a beam transfer matrix $B^{p,P}_{xy;\ell m}(\nu)$:
\beq
\tilde{V}^p_{xy,m}(\nu) = \sum_P \sum_\ell B^{p,P}_{xy;\ell m}(\nu) a_{\ell m}^P(\nu) \ .
\eeq
After performing this multiplication, we convert the result to a visibility timestream by inverse Fourier transforming in $m$, applying zero-padding such that the time resolution matches that of the observed sidereal stacks.

\begin{figure*}[t]
   \centering \includegraphics[width=1\linewidth,keepaspectratio, trim = 0 10 0 0]{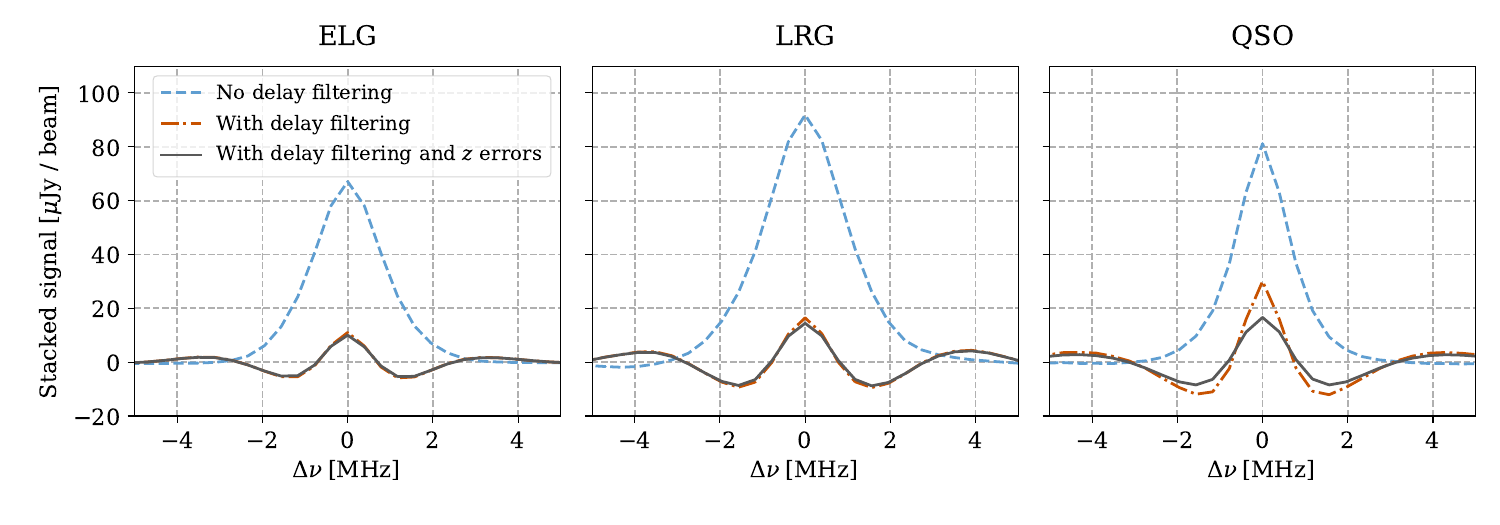}
    \caption{Results of stacking simulated observations containing only \tcm signal on mock galaxy or quasar catalogs correlated with the input signal, generated according to the procedure in \secref{sec:simulations}. The three panels correspond to simulations that use the selection functions and redshift error distributions of the three eBOSS samples we consider. The stacking amplitude in the absence of delay filtering ({\em blue dashed lines}) is heavily suppressed by the delay filter ({\em red dot-dashed lines}), and further suppressed by the inclusion of random redshift errors in the catalogs ({\em black solid lines}).
    }
    \label{fig:sim_filtering_stages}
\end{figure*}

We carry out separate versions of this procedure with beam transfer matrices corresponding to the default or control beam models from \secref{sec:beams}. We compute these matrices using \texttt{driftscan} \citep{driftscan}, with several performance optimizations: precision truncation using the \texttt{bitshuffle} library \citep{2015bitshuffle}, omitting frequencies that fall outside of the mask described in \secref{sec:freq_mask}, and only computing the $P=I$ components (since the \tcm signal is unpolarized).

Up to this point, the simulated data are in temperature units. To transform into spectral flux density units, we first compute the beam solid angle for the assumed beam model:
\beq
\Omega^p(\nu) = \int d\theta\, \cos\theta \int d\phi\, |A^p(\nu,\theta,\phi)|^2\ .
\eeq
We then multiply the visibilities by the standard Rayleigh-Jeans conversion factor and the beam solid angle, normalized by the power beam evaluated at $\ha=0'$ and a reference declination $\theta_{\rm ref}$:
\begin{align*}
&\left. V^p_{xy}(\nu, \phi) \right|_{\rm Jy} \\
&\quad = \frac{2 \times 10^{26} \kB \nu^2}{c^2} \frac{\Omega^p(\nu)}{|A^p(\nu, \theta_{\rm ref}, 0)|^2}
\left. V^p_{xy}(\nu, \phi) \right|_{\rm K}\ .
\numberthis
\end{align*}
With this normalization, a visibility corresponding to a point source that transits at $\theta=\theta_{\rm ref}$ has an amplitude equal to the flux of the source. For consistency with CHIME's beam and complex gain calibration, we set $\theta_{\rm ref}$ to the declination of Cygnus A.

From here, the simulated visibilities are processed in the same way as the real data: a global frequency mask and noise weights described in \secref{sec:freq_mask} and \secref{sec:weights} are applied; the contributions of the four brightest point sources are inferred and subtracted; beam-deconvolved maps are constructed as in \secref{sec:mapmaking}; delay filtering is applied with the declination-dependent delay cuts from \secref{sec:filter}; and the masking operations in \secref{sec:map_mask} are applied. Just as we simulate visibilities for each of the default and control beam models, we also perform two versions of the mapmaking step, assuming either beam model: thus, we obtain four simulated datasets corresponding to each pair of assumed and deconvolved beam, and we compare the results in \secref{sec:beamcalerrors} in order to estimate the systematic uncertainty arising from our choice of beam model.

\subsubsection{Mock Source Stacking}
\label{sec:mockstacking}

Finally, we stack the simulated observations on the associated mock catalogs, following the procedure in \secref{sec:stacking}. \cref{fig:sim_filtering_stages} shows stacking results corresponding to simulations of each eBOSS sample, for a single large-scale structure realization but averaged over \SI{100} mock catalogs of \SI{400000} objects each, in order to suppress shot noise associated with the catalog size. In the absence of delay filtering, the stacking amplitude inferred from these simulations for QSOs is greater than for ELGs and less than for LRGs; the former follows from ELGs having lower bias and higher Finger-of-God suppression than QSOs, while the latter is due to the higher bias of LRGs than QSOs, which wins over the more severe Finger of God effect for LRGs (see \cref{fig:sim_z_models}).

The delay filter significantly suppresses the signal level, reducing the zero-lag amplitude by around 80\% for ELGs and LRGs, and 63\% for QSOs. We attribute the lower suppression for QSOs to their milder Finger-of-God suppression at small scales: the delay filter removes sensitivity to the largest scales (see \secref{sec:scales}), and the remaining smaller-scale contribution is larger for QSOs than for the other tracers due to a smaller amount of suppression. Finally, redshift errors in the simulated catalogs reduce the zero-lag amplitude by no more than 10\% for ELGs and LRGs, but by 40\% for QSOs, thanks to the much wider distribution of QSO redshift errors discussed above.

\subsection{Template Calculation}
\label{sec:template}

To interpret our results, we need to be able to calculate the expected signal from stacking on a given catalog for an underlying set of parameters $\vtheta$. We call this quantity the \emph{template}, denoted by $s(\Delta\nu; \vtheta)$. Though the template is entirely determined by the cross-power spectrum in \cref{eq:crossps}, propagating this through the instrumental transfer function and our analysis procedure is challenging to do both efficiently and accurately, and so will be left to a follow-up paper \citep{interpretation-paper}.

In this work, we instead use our simulation capability to calculate the templates. In brief, we generate large-scale structure realisations
corresponding to several modes, each of which is defined by a specific combination of model parameters;
Monte-Carlo over random mock catalogs to estimate the stack signal for each mode; and calculate the full template for arbitrary parameter values by making linear combinations of the template modes and applying an effective treatment for the Fingers of God. Overall the errors in this approach are $\lesssim 1\%$. We describe this approach in detail in Appendix~\ref{app:template_calculation}.

\begin{figure*}[htbp]
    \centering
    \includegraphics[width=0.94\linewidth]{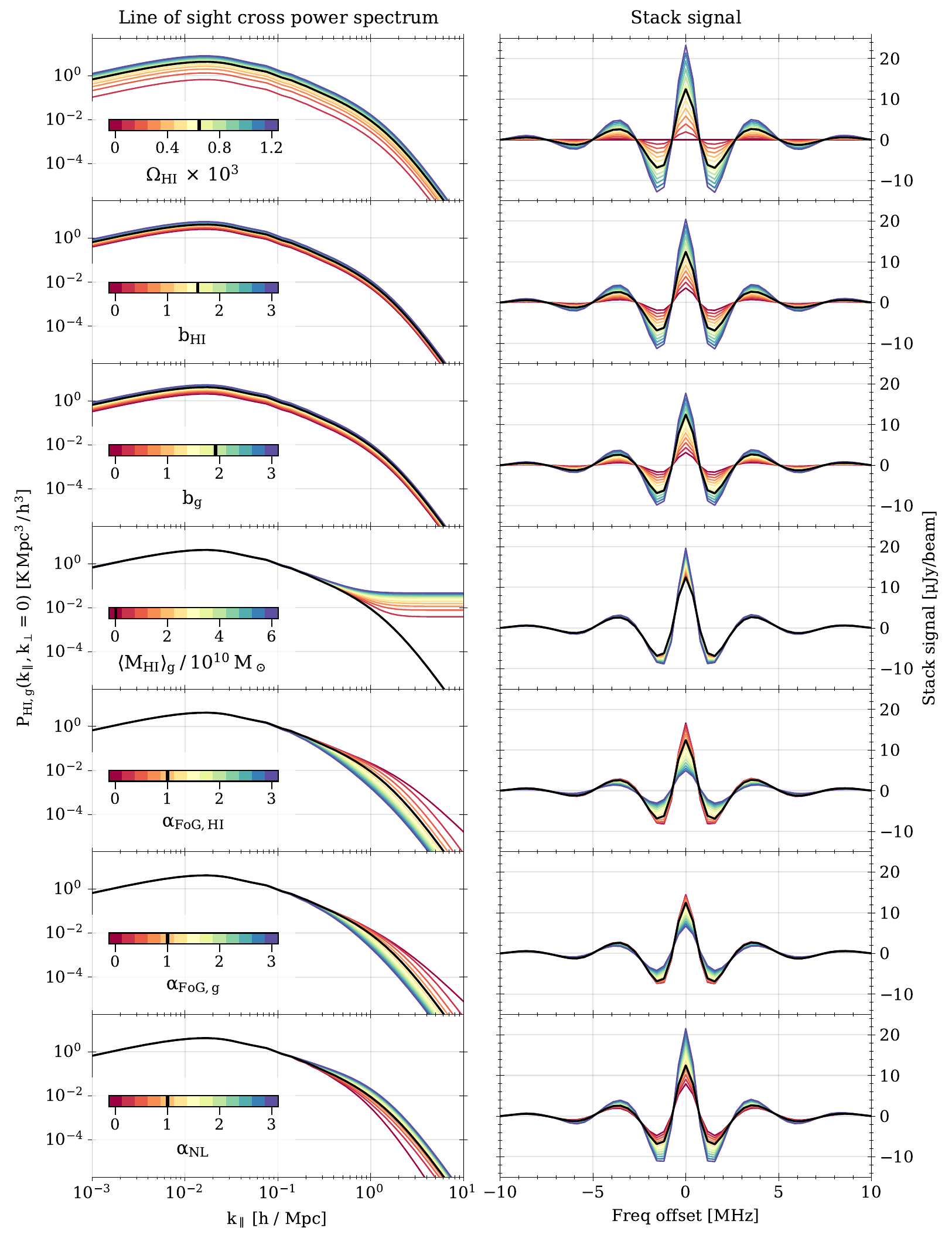}
    \caption{
        The simulated stack signal after processing through the CHIME pipeline. Each row shows the effect of varying a parameter on the theoretical \tcm-QSO cross power spectrum in the left panel and the expected signal observed by CHIME in the right panel. The variation for each parameter is chosen to be over a range consistent with our prior uncertainties. The fiducial model used within our modelling is indicated by the thick black line for each panel, and the location within range of each parameter is indicated by the black line within the color bar.}
    \label{fig:parameters}
\end{figure*}

In \cref{fig:parameters}, we display the change in the HI-tracer cross-power spectrum (left panels) corresponding to variations of each of our 8 model parameters, along with the corresponding change in the predicted stack signal (right panels). Access to the full $k$ range shown in the left panels would allow non-degenerate constraints on several of these parameters, due to their different impacts on the cross power spectrum. However, our filtering choices imply that the stack signal is only sensitive to nonlinear scales ($k\gtrsim 0.3\, h^{-1}{\rm Mpc}$ or so, as shown in \cref{fig:k_sensitivity}), and as a result, we are left with significant parameter degeneracies, which can be inferred from the similar variations in each right-hand panel in \cref{fig:parameters}.

%% file: sections/results.tex

\section{Results}
\label{sec:results}

\subsection{Stacking measurements}
\label{sec:stackmeasurements}

\begin{figure*}
   \centering \includegraphics[width=0.98\linewidth,keepaspectratio]{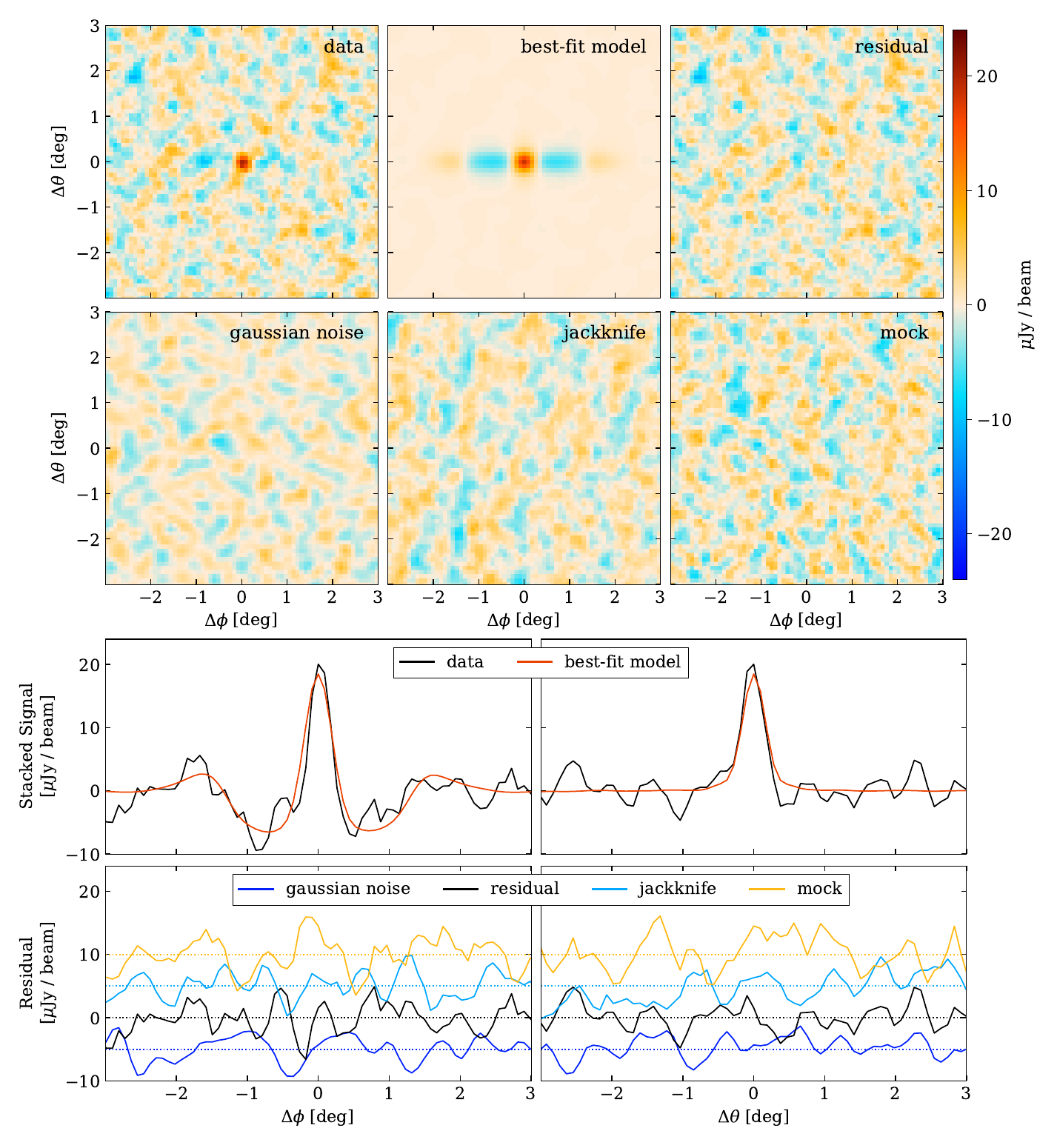}
    \caption{The stacked signal at $\Delta \nu = \SI{0}{\mega\hertz}$ as a function of right ascension offset ($\Delta \phi$) and declination offset ($\Delta \theta$) for the QSO catalog.  The top row shows, from left to right, the data, best-fit model, and residual.  The second row shows, from left to right, the result of stacking the QSO catalog on a Gaussian noise realization, stacking the QSO catalog on a jackknife of even and odd days, and stacking a random mock catalog on the data.  The third row shows a slice of the data in black and best-fit model in red at $\Delta \theta = \SI{0}{\degree}$ on the left and $\Delta \phi = \SI{0}{\degree}$ on the right.  The bottom row shows, for these same slices, the residuals in black compared to the Gaussian noise realization in dark blue, the jackknife in light blue, and the  random mock catalog in orange.  Note that to facilitate the comparison, the slices in the bottom row have been offset by an amount indicated by the dotted line of the same color.}
    \label{fig:stack2d}
\end{figure*}

The top left panel of \cref{fig:stack2d} shows the result of stacking the deconvolved, foreground-filtered maps on the three-dimensional positions in the eBOSS NGC quasar catalog.  It is shown as a function of right ascension offset and declination offset at \SI{0}{\mega\hertz} frequency offset, averaged over the two polarisations, in other words $d(0, \Delta \theta, \Delta \phi)$ in the notation of \secref{sec:stacking}.  Also shown in the top row is our best-fit model for the \tcm emission based on the simulations described in the preceding section and the residuals obtained by subtracting the best-fit model from the data.  The residuals can be compared to the three panels in the second row, which correspond to three different techniques for estimating the noise present in the stack.  The left panel is the result of applying the stacking procedure to a Gaussian noise realization generated according to \cref{eq:draw_noise}.  The middle panel is the result of applying the stacking procedure to a jackknife of even and odd days (see \secref{sec:consistent_even_odd}).  Finally, the right panel is result of stacking the data on a random mock catalog.

The noise in the residuals is consistent with that observed in the random mock catalog.  Both are in excess of the noise in the even-odd jackknife, owing to the fact that residual foregrounds are highly correlated between even and odd days and therefore cancel in the jackknife.  The noise observed in the even-odd jackknife is in excess of that observed in the Gaussian noise realization due to unflagged RFI and variations in the foregrounds from day to day caused by instrument instability.

The third row of \cref{fig:stack2d} shows one-dimensional slices of both data and best-fit model.  The negative shoulders in the right ascension direction that are observed in both the data and model are caused by the exclusion of intra-cylinder baselines from our analysis.  The grating lobes in the right ascension direction, which are shown in \cref{fig:synthesized_beam}, have largely averaged away in the stack because their location varies with frequency and declination.  It is important to note that the angular information displayed in \cref{fig:stack2d} was not used to constrain the model.   For simplicity, the model is only fit to the central pixel of the stack as a function of frequency.  A full three-dimensional fit could further improve the signal-to-noise and help break the degeneracy between the amplitude $\AHI$ and the Fingers-of-God damping, but we leave that for a future analysis.

\begin{figure*}
   \centering \includegraphics[width=0.98\linewidth,keepaspectratio]{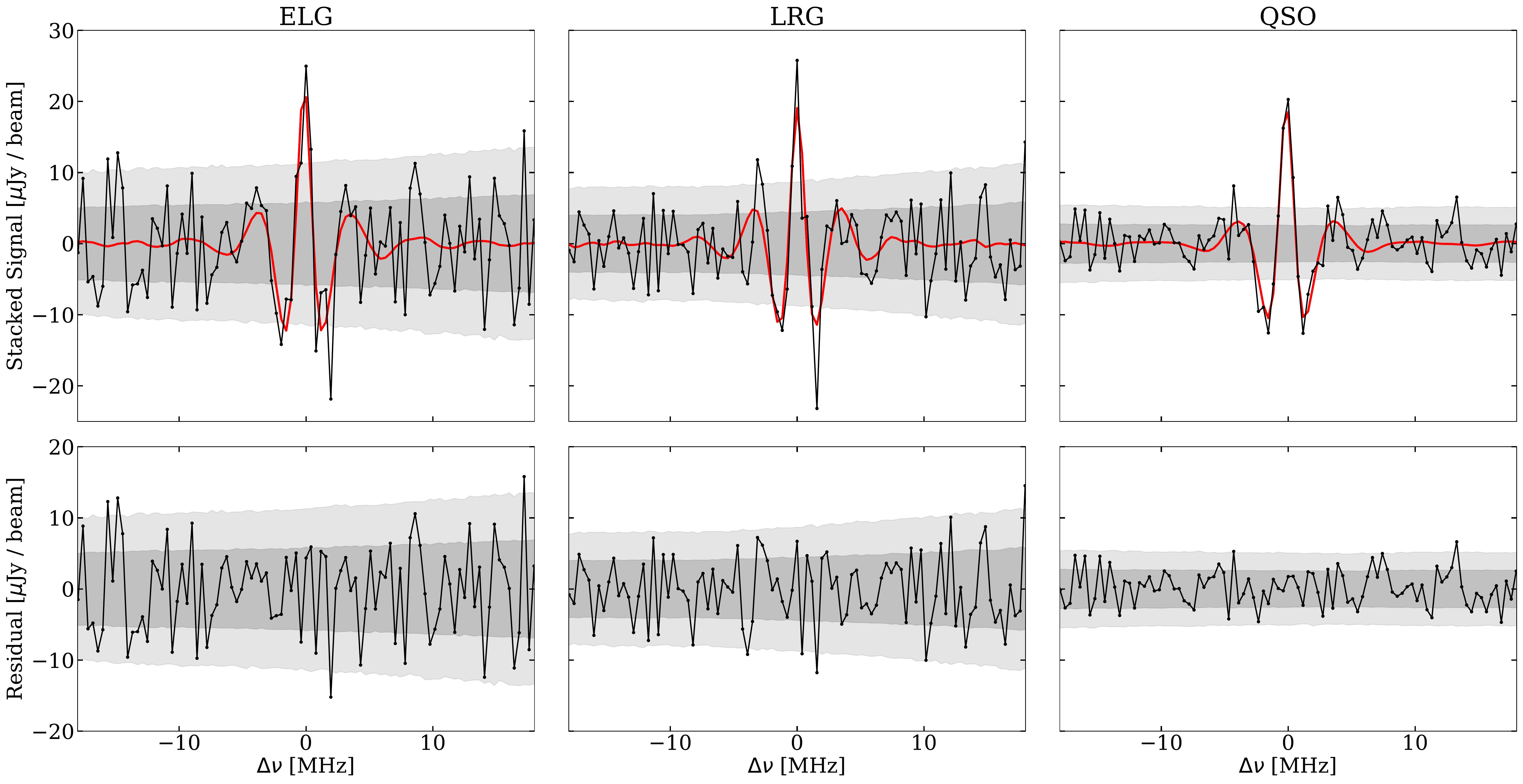}
    \caption{Top:  the stacked signal as a function of frequency offset for the ELG, LRG, and QSO catalogs.  The data are shown in black and the best-fit model is shown in red.  Bottom: the residuals obtained by subtracting the best-fit model from the data.  For both the top and bottom rows, the dark gray and gray bands indicate the central \SI{68}{\percent} and \SI{95}{\percent} of values observed when applying the same stacking procedure to \SI{10000} mock catalogs.}
    \label{fig:stack1d}
\end{figure*}

For all three tracers, the spatial extent of the signal is consistent with the synthesized beam computed directly from \cref{eq:bsynth_phi,eq:bsynth_theta} and averaged over sources, indicating that the \tcm signal is unresolved.  \Cref{fig:stack1d} shows the central pixel of the stack as a function of frequency, i.e., $d(\Delta \nu, 0, 0)$, for the three tracers in black.  The dark gray and light gray contours indicate the central \SI{68}{\percent} and \SI{95}{\percent} of values observed when stacking the maps on \SI{10000}{} random mock catalogs as outlined in \secref{sec:covariance}.  The red line indicates our best-fit model for the signal.  Note that although the two polarisations are fit jointly, to simplify the figure we show only their weighted average, with the weights set to the inverse variance as measured by the random mock catalogs.  Also note that the polarisation and frequency dependent mean value of the noise has been characterized using the random mock catalogs and subtracted from both the stack on the true catalog and the stack on the mock catalogs that are shown in the figure.

The best-fit model shown in both \cref{fig:stack2d} and \cref{fig:stack1d} consists of fixing all non-linear parameters at their fiducial values and allowing the parameters governing the large-scale clustering of HI to vary. This model has been described in \secref{sec:freeparameters}.  The bottom row shows the result of subtracting the best-fit model from the data and compares to the same gray mock catalog contours shown in the top row.  For all tracers, the residuals are consistent with our noise model based on the random mock catalogs.  This is also true for all QSO redshift bins, which are not shown.

\subsection{Model fitting}
\label{sec:modelfitting}

We assume that the noise in the stacked source data is described by a Gaussian, and that the signal is described by the model given in \secref{sec:template}. This means that the likelihood function $\calL(\vd | \vtheta)$ of observing the stacked signal $\vd$ given a template $\vs(\vtheta)$ with model parameters $\vtheta = \ls \Dnu, \OmegaHI, \bHI, \bg, \Mten, \alphaNL, \alphaFoG{\sHI}, \alphaFoG{\sg}\rs$ is described by a multivariate Gaussian
\begin{align}
    \calL(\vtheta) & = \calP(\vd \mid \vtheta) \\
    & =\frac{1}{\lv 2\pi \mSigma \rv^{1/2}} \exp{\lp - \frac{1}{2} \chi^{2} \rp}
\end{align}
with
\begin{align}
    \label{eq:chisq}
    \chi^{2} & = \lp \vd - \vs(\vtheta) - \vmu \rp^T \mSigma^{-1} \lp \vd - \vs(\vtheta) - \vmu \rp \ ,
\end{align}
where $\vec{s}(\vtheta)$ is the model for the \tcm signal, and $\vec{\mu}$ and $\mSigma^{-1}$ are the mean and inverse covariance of the noise, which are estimated using the sample mean and covariance of the mock catalogs as outlined in the \secref{sec:covariance}.

We employ a Markov Chain Monte Carlo (MCMC) to sample from the joint posterior distribution,
\begin{equation}
    \label{eq:posterior}
    \calP(\vtheta \mid \vd) = \frac{1}{\calZ} \, \calL(\vtheta) \, \pi(\vec{\theta}) \ ,
\end{equation}
where $\pi(\vec{\theta})$ is the prior probability distribution over the model parameters, and $\calZ$ is the normalisation constant such that the posterior integrates to unity.

We use non-informative priors for most parameters, ascribing equal prior probability over large ranges. For the non-linear parameters we choose to do this even where there is some external information from either simulations, or more strongly from analysis of the eBOSS data itself (for example on the Fingers of God scale; see \secref{sec:fingerofgod}) as it is difficult to combine the different prescriptions for modelling the non-linear scales. These analyses guide our choice of fiducial model, but we allow a wide range of variation around them when trying to fit the data.

The one exception to this is for the galactic bias $\bg$. As it is a large-scale parameter, it is less susceptible to systematic differences in the modelling, and we instead use a prior informed by modelling of the eBOSS tracers. For the QSOs, our fiducial model is that from \cite{laurent2017}, and to get an uncertainty on this, we fit a shift in the amplitude to the two lowest redshift bins in their analysis (which overlap with that of this paper), which gives an uncertainty of 3\% about the fiducial model. For the LRGs we translate the overall results of \cite{zhai2017} of $b = 2.30 \pm 0.03$ into a 1.3\% uncertainty on the amplitude of the bias model used here. Finally, for the ELGs we symmetrise the measurements of $b_1$ from \cite{tamone2020} to give an uncertainty of 10\% for the ELG linear bias.

For $\OmegaHI$, which gives an overall normalisation to the signal, we use a prior symmetric about zero, despite the fact that physically $\OmegaHI \geq 0$. This is to ensure that our priors do not give an artificial bias towards positive signal and give a more robust estimation of the detection significance. However we do enforce that $\bHI \geq 0$ to exclude an unphysical mode of high probability with both $\OmegaHI < 0$ and $\bHI < 0$.

We summarise our choice of priors in \cref{tab:priors}.

\begin{deluxetable}{Lcp{0.6\linewidth}}[t]
    \tablecaption{The prior placed on each parameter during our analysis.\label{tab:priors}}
    \tablecolumns{3}
    \tablewidth{\linewidth}
    \tablehead{
        \colhead{Parameter} &
        \colhead{Type} &
        \colhead{Description} \\
    }
    \startdata
        \multicolumn{2}{l}{\emph{Standard parameters}} \\[1mm]
        \hspace{4mm} \OmegaHI       & Uniform & Range: $-10^{-2}$ to $10^{-2}$ \\
        \hspace{4mm} \bHI           & Uniform & Range: 0 to 10 \\
        \hspace{4mm} \bg            & Gaussian & Mean: $\bar{b}_\sg = \bg^\fid(\zeff)$ \newline
                                    standard deviation: QSOs 3\%, LRGs 1.3\%, ELGs 10\% \\
        \hspace{4mm} \Dnu           & Uniform & Range: \SIrange{-0.8}{0.8}{\mega\hertz} \\[5mm]
        \multicolumn{3}{l}{\emph{Non-linear parameters}} \\[1mm]
        \hspace{4mm} \Mten          & Uniform & Range: 0 to 20; Fixed: 0 \\
        \hspace{4mm} \alphaNL       & Uniform & Range: 0 to 5; Fixed: 1  \\
        \hspace{4mm} \alphaFoG{\HI} & Uniform & Range: 0 to 5; Fixed: 1 \\
        \hspace{4mm} \alphaFoG{\sg}   & Uniform & Range: 0 to 5; Fixed: 1 \\
    \enddata
    \tablecomments{There are two classes of parameters in our analysis, \emph{standard} parameters that capture the large scale quantities we hope to constrain, and nuisance parameters which model the signal on small, \emph{non-linear}, scales. In our analysis this latter group of parameters will either by marginalised over, or fixed to their fiducial values in order to assess the contribution of modelling uncertainties to our constraints.}
\end{deluxetable}

The affine-invariant ensemble sampler from the \texttt{emcee} package \citep{foremann-mackey2013} is used to sample from the joint posterior distribution.  We run 32 samplers initialized from random locations within the region defined by \cref{tab:priors}. The autocorrelation lengths of the parameter chains are calculated for each sampler, the average is taken over samplers, and the maximum is taken over parameters to obtain a single autocorrelation length, $\zeta$. The first $10 \times \zeta$ samples in each chain are discarded as burn-in.  The chains are then thinned by $\zeta$ and concatenated. The parameter space is high dimensional and has complex degeneracies, which means that the correlation lengths are large, $\zeta \gtrsim 500$ in the full parameter space. We also make extensive use of the GetDist package \citep{GetDist} for analysing the MCMC chains.

\subsection{Parameter Constraints}
\label{sec:constraints}

\begin{figure*}[htbp]
    \centering
    \includegraphics[width=0.98\linewidth]{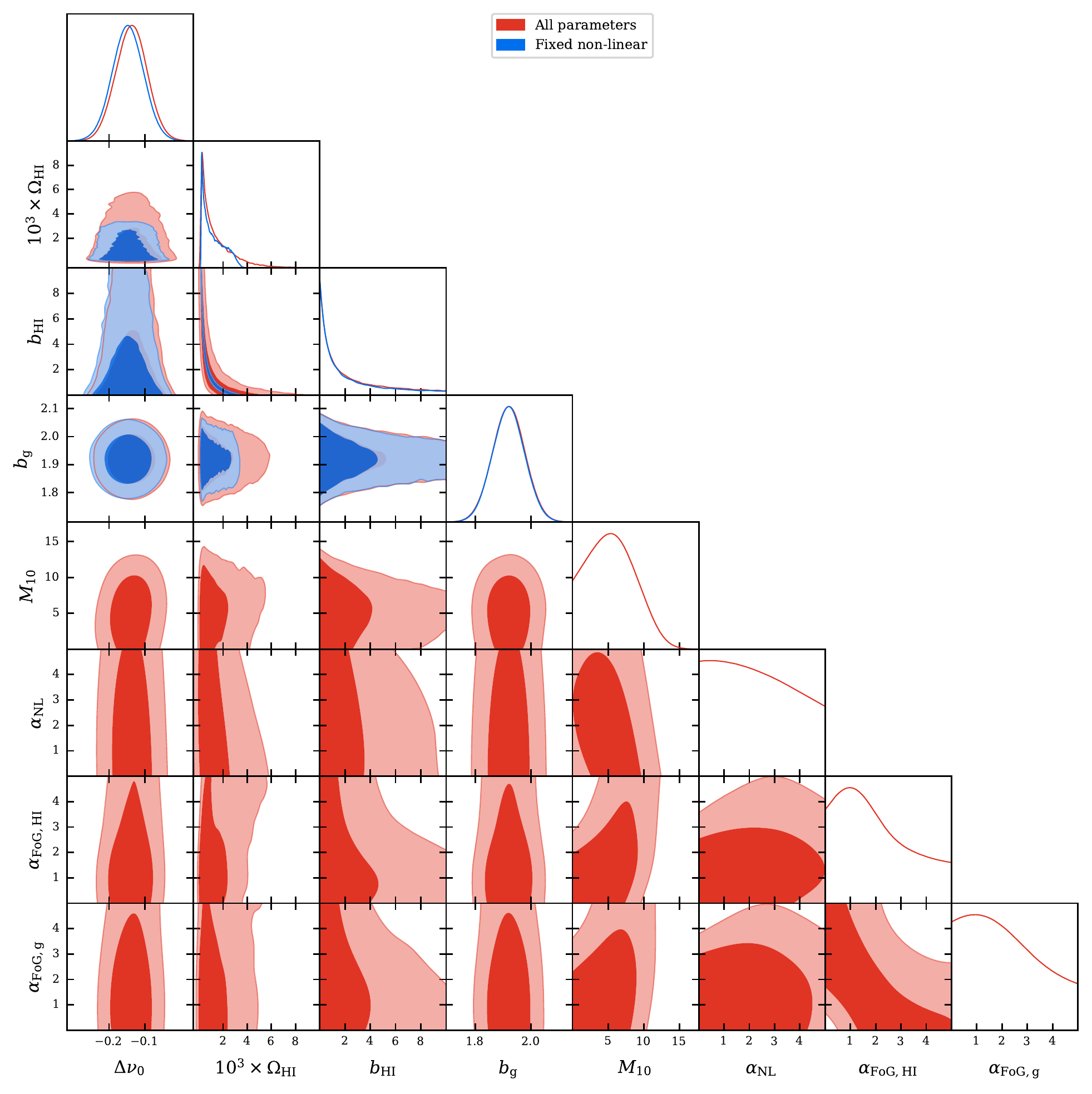}
    \caption{The constraints on the model derived from the cross-correlation of CHIME and the full eBOSS QSO sample. The red contours show the constraints on the full parameter set (described in \secref{sec:freeparameters}), whereas the blue contours show the constraints if we fix the non-linear parameters to their fiducial values and only allow $\Dnu$, $\OmegaHI$, $\bHI$ and $\bg$ to vary. There are significant degeneracies between parameters, notably $\OmegaHI$--$\bHI$, but also within the non-linear parameters such as $\alphaFoG{\sHI}$--$\alphaFoG{\sg}$.}
    \label{fig:qso_orig_params}
\end{figure*}

In \cref{fig:qso_orig_params} we show the constraint on the default model parameters for the QSO catalog. We show constraints for both a model where all parameters are allowed to vary as well as a model where the non-linear parameters are fixed to their fiducial values ($\Mten = 0$, $\alphaNL = 1$, $\alphaFoG{\sHI} = 1$ and $\alphaFoG{\sg} = 1$). The constraints show that certain parameter combinations are highly degenerate, most notably $\OmegaHI$--$\bHI$, but also correlations with the non-linear parameters $\alphaFoG{\sHI}$ and $\alphaFoG{\sg}$. As these degeneracies limit our ability to make a cosmological interpretation of our results, it is worth attempting to understand them.

The most severe degeneracy in our model is between $\OmegaHI$ and $\bHI$, and is clearly apparent in both the full and fixed models. The origin of this can be seen in \cref{eq:crossps}, which, simplified slightly down to the linear terms, has
\begin{equation}
    P_{\HI,\sg}(k, \mu) \propto \OmegaHI (\bHI + f\mu^2) (\bg + f\mu^2) P(k) \; ,
\end{equation}
which contains a multiplicative $\OmegaHI \bHI$ term, responsible for the curved degeneracy seen in the $\OmegaHI$-$\bHI$ panel of \cref{fig:qso_orig_params}. Previous \tcm cross-correlation analyses \citep{masui2013,switzer2013,wolz2021} gave constraints directly on the combination $\OmegaHI \bHI r$, where $r$ is a scale-independent cross-correlation parameter that absorbs modelling uncertainties on non-linear scales; however, this is not sufficient for the analysis here. Although transforming our constraints to be in terms of $\OmegaHI \bHI$ removes the curved degeneracy\footnote{In fact we actually sample within a transformed basis by replacing the parameter $\OmegaHI$ with $\OmegaHI \bHI$. This substantially improves convergence as the remaining linear degeneracy is easily navigated by the affine invariant sampler, where the original curved degeneracy was not. To do this we need to carefully adjust the prior applied in the sampler to ensure that the prior on $\OmegaHI$ remains uniform.} we find that a linear degeneracy against $\OmegaHI$ remains. This can be understood straightforwardly as the effect of the Kaiser redshift-space distortions. As CHIME has higher resolution in the frequency direction versus the angular direction, and we have removed low-$k_\parallel$ modes by foreground filtering, the sensitivity in this analysis is biased towards wavenumbers with higher $\mu$ (which is illustrated in \cref{fig:k_sensitivity}). As both $\bHI$ and $f$ are of order unity, the contribution of the Kaiser term is important and cannot be neglected.

To account for this, we transform to a plane of $(\OmegaHI \bHI)$--$\OmegaHI$ and determine a linear combination of these parameters that minimises their variance. For a single galaxy or quasar sample $g$, the solution for an exactly linear degeneracy can be found by using the MCMC samples to construct the covariance matrix between $\OmegaHI \bHI$ and $\OmegaHI$, which we write as $\mC_{\Omega b,g}$, and then finding the eigenvector with minimal eigenvalue, which gives the linear combination we are searching for. We will use this combination as our primary amplitude parameter
\begin{equation}
  \AHI \equiv 10^3 \, \OmegaHI \left(\bHI + \fmu \right)\ ,
  \label{eq:AHI-definition}
\end{equation}
where we make the interpretation that the coefficient $\fmu$ is the sensitivity-weighted average $f \mu^2$ that this CHIME analysis is probing. We perform this optimisation on the chains with fixed non-linear parameters, as this gives a cleaner separation from other degenerate parameters.

The $\fmu$ coefficient preferred by each tracer differs slightly from $\sim 0.45$ (QSOb00) to $\sim 0.62$ (QSOb2), which we would expect as both $f$ and CHIME's sensitivity change with redshift. As we would like to be able to compare our measurements between tracers we would instead like a single effective $\fmu$. To do this, we minimise the covariance $\mC_{\Omega b,\mathrm{all}}$, defined by
\begin{equation}
    \mC_{\Omega b,\mathrm{all}}^{-1} = \sum_g \mC_{\Omega b,g}^{-1}
\end{equation}
where we sum over the tracers QSOb0, QSOb1, QSOb2, LRG and ELG (we exclude the other QSO tracers to avoid double counting the data). The form of $\mC_{\Omega b,\mathrm{all}}$ is motivated by considering each tracer to be a different measurement in the $(\OmegaHI \bHI)$--$\OmegaHI$ plane: if each distribution was Gaussian, and all were consistent, the covariance on the combined distribution would be given by $\mC_{\Omega b,\mathrm{all}}$. After this procedure we derive an effective $\fmu \approx 0.552$ which we fix for the rest of this analysis. The overall loss of constraining power from fixing a single value is small, with a drop of $\sim 7\%$ for the worst affected tracer (full QSO catalog).

The second degeneracy we focus on is between the Fingers of God parameters. If we examine the cross-power spectrum given by \cref{eq:crossps} and expand the Fingers of God damping factors defined in \cref{eq:DFoGk} assuming $k_\parallel \sigma_P \gg 1$, we find that
\begin{align}
    P_{\sHI,\sg}(k, \mu) & \propto \DFoG{\sHI}(\alphaFoG{\sHI} \mu k) \DFoG{\sg}(\alphaFoG{\sg} \mu k) \\
    & \appropto \frac{4}{k_\parallel^4 \sigma_{P,\sHI}^2 \sigma_{P,\sg}^2} (\alphaFoG{\sHI} \, \alphaFoG{\sg})^{-2} \; . \label{eq:crossps_fog_exp}
\end{align}
For most of the region of CHIME's $k$-space sensitivity (see \cref{fig:k_sensitivity}) we are close to this regime, and so we expect there to be an approximate degeneracy of the form $\alphaFoG{\sHI} \alphaFoG{\sg}$, which can be seen in the $\alphaFoG{\sHI}$--$\alphaFoG{\sg}$ panel of \cref{fig:qso_orig_params}. This motivates us to transform to two new parameters
\begin{align}
    \alphaFoG{+} & = (\alphaFoG{\sHI} \, \alphaFoG{\sg})^{1/2} \\
    \alphaFoG{-} & = \log{(\alphaFoG{\sHI} / \alphaFoG{\sg})}
\end{align}
where in the large $k_\parallel$ limit $\alphaFoG{+}$ controls the amount of damping given by the Fingers of God, and $\alphaFoG{-}$ does not affect the cross-power spectrum. The logarithm in the definition of $\alphaFoG{-}$ is to limit the effect of small $\alphaFoG{\sg}$ values generating extremely large values for this parameter.

\begin{figure*}[htbp]
    \centering
    \includegraphics[width=0.98\linewidth]{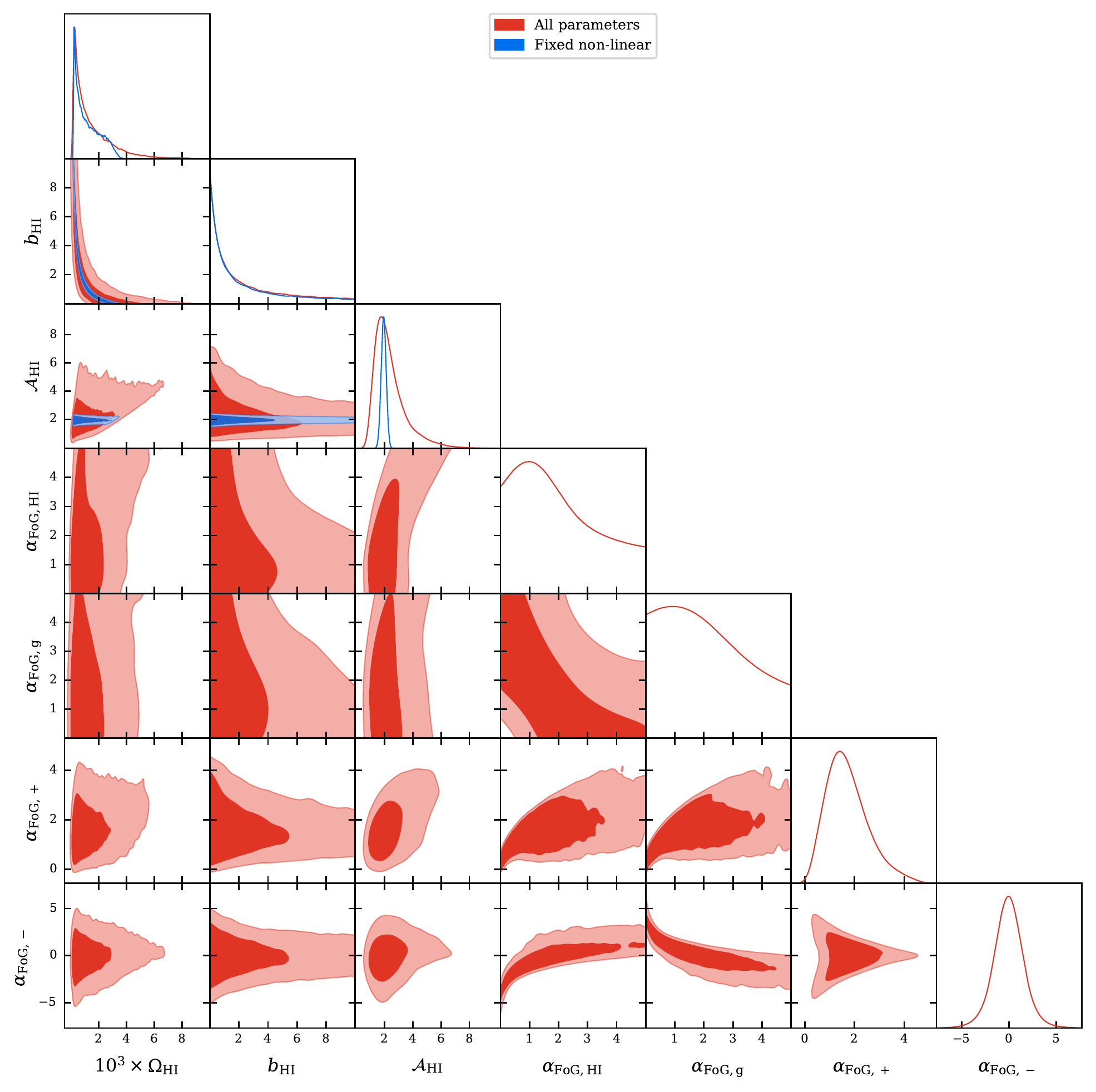}
    \caption{
    The parameter constraints corresponding to the QSO catalog, showing the derived parameters, $\AHI$, $\alphaFoG{+}$, and $\alphaFoG{-}$, and their correlations with the parameters they are derived from ($\OmegaHI$, $\bHI$, $\alphaFoG{\sHI}$, and $\alphaFoG{\sg}$). These new parameters are less degenerate than the original parameters, and show that the key behavior in the constraints can be captured by just two quantities, $\AHI$ and $\alphaFoG{+}$.}
    \label{fig:qso_corr_params}
\end{figure*}

In \cref{fig:qso_corr_params} we show these new parameters and how they are correlated with the parameters they are derived from. The new amplitude-like parameter $\AHI$ clearly flattens the degeneracy, capturing all the information in $\OmegaHI$ and $\bHI$. Similarly, the parameter $\alphaFoG{+}$ correlates with the amplitude parameter whereas the orthogonal combination $\alphaFoG{-}$ does not, although there is interesting behaviour observed at low $\alphaFoG{+}$ where we are even further from the regime where we can make the high-$k_\parallel$ expansion used in \cref{eq:crossps_fog_exp}.

One of the key remaining degeneracies is that between the overall amplitude, $\AHI$, and the combined Fingers of God strength, $\alphaFoG{+}$. This can be understood physically: on the scales that CHIME observes, the Fingers of God damping reduces the stacked signal amplitude, and so an increase in $\alphaFoG{+}$ must be compensated by an increase in the underlying \tcm signal amplitude to remain consistent with the measurements.

\begin{figure*}[htbp]
    \centering
    \includegraphics[width=0.55\linewidth]{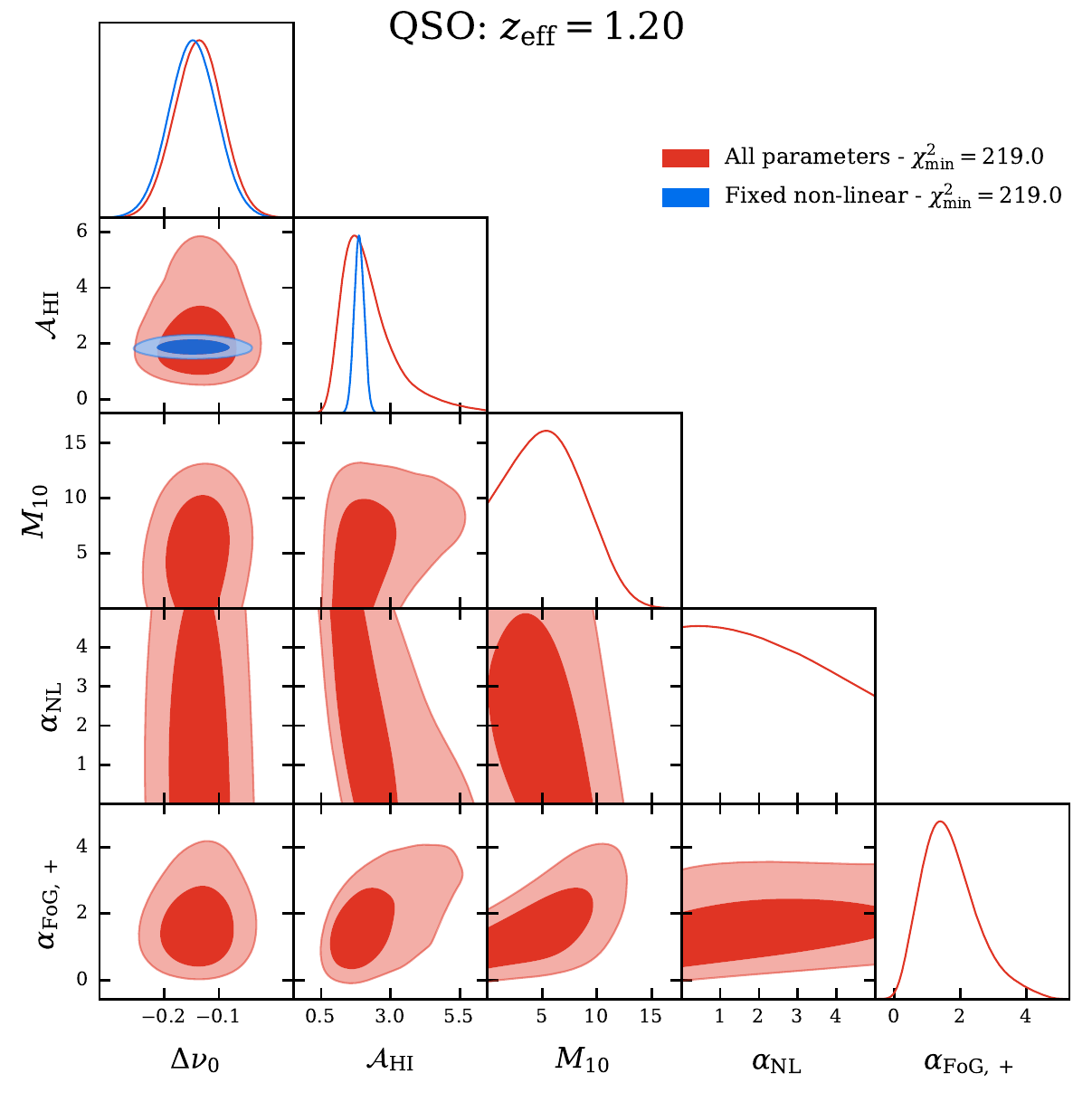}
    \caption{The constraints from stacking the QSO catalog over the full
    frequency band chosen for this analysis. We reduce the original set of eight parameters down to five: the amplitude-like $\AHI$, the frequency offset $\Dnu$, the correlated shot noise $\Mten$, and the two relevant non-linear nuisance parameters $\alphaFoG{+}$ and $\alphaNL$. The fits with all five parameters free (red contours) or with the $\Mten$, $\alphaFoG{+}$, and $\alphaNL$ fixed to their fiducial values (blue contours) both result in an excellent goodness-of-fit, with $\chi^2_\text{min} \approx 219$ for $202$ degrees of freedom. We discuss the physical interpretation of these constraints in \secref{sec:quasar_redshift_errors}, \ref{sec:OmegaHIresults}, and \ref{sec:M10results}.}
    \label{fig:qso_reduced_params}
\end{figure*}

\begin{figure*}[htbp]
    \centering
    \includegraphics[width=0.55\linewidth]{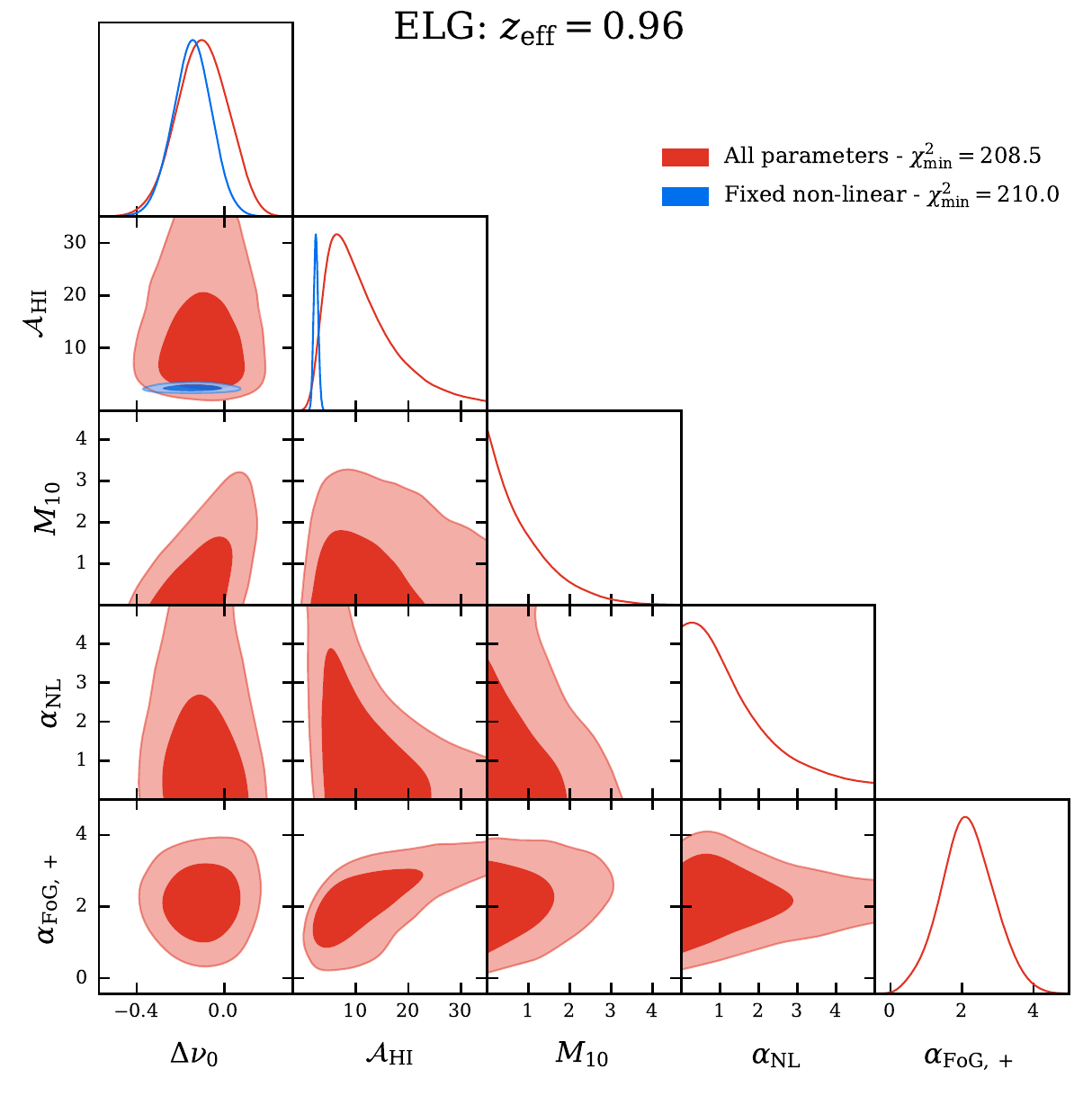}
    \caption{Parameter constraints from stacking the ELG catalog, in the same format as \cref{fig:qso_reduced_params}.}
    \label{fig:elg_reduced_params}
\end{figure*}

\begin{figure*}[htb]
    \centering
    \includegraphics[width=0.55\linewidth]{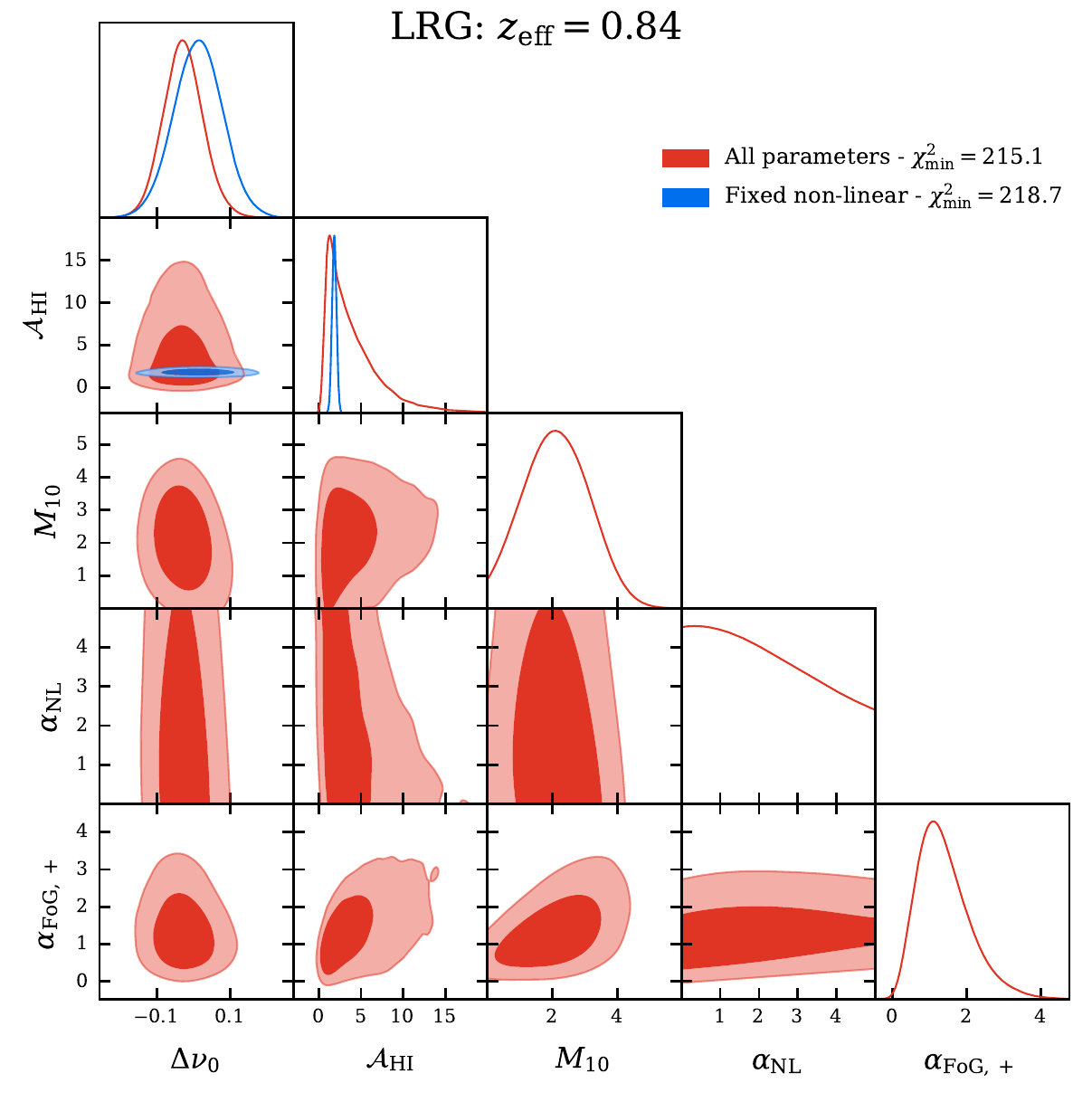}
    \caption{Parameter constraints from stacking the LRG catalog, in the same format as \cref{fig:qso_reduced_params}.}
    \label{fig:lrg_reduced_params}
\end{figure*}

In \cref{fig:qso_reduced_params,fig:elg_reduced_params,fig:lrg_reduced_params}, we show the constraints for the QSO, ELG and LRG tracers stacked over full \SIrange{585}{800}{\mega\hertz} band
for the amplitude parameter, $\AHI$, the frequency offset, $\Dnu$, the shot noise, $\Mten$, and the two non-linear nuisance parameters, $\alphaFoG{+}$ and $\alphaNL$. In all cases we find an excellent goodness-of-fit with $\chi^2_\text{min}$ being close to the 202 degrees of freedom.

For all tracers, the amplitude constraints are significantly weakened by marginalising over the non-linear parameters compared to the case of fixed non-linear parameters. However, the posteriors are non-Gaussian and highly skewed such that, despite the large credible interval, the probability that $\AHI \le 0$ is negligible. Even though the non-linear parameters are degenerate with the amplitude, the amplitude must be non-zero for a signal to be seen.

Interpretation of these constraints is complicated by a volume factor pushing the constraints towards larger FoG smoothing effects. The originally uniform prior on $\pi(\alphaFoG{\HI}, \alphaFoG{\sg})$ transforms to a $\pi(\alphaFoG{+}, \alphaFoG{-}) \propto \alphaFoG{+}$. As the stack signal $\appropto \AHI / \alphaFoG{+}^2$, our broad non-informative priors give an unintentional upward pressure on $\AHI$ as there is more volume at higher FoG damping levels. This can be resolved by placing a flat prior on $\alphaFoG{+}$, but as it is not a physical parameter, this is difficult to justify. Future analysis will need to have data that can break this degeneracy internally, or use better modelling that allows for the prior bounds on $\alphaFoG{\sHI}$ and $\alphaFoG{\sg}$ to be reduced.

\subsection{Detection Significance}
\label{sec:significance}

Assessing the significance of the detection is difficult for two reasons. First, the posterior distributions of the full set of parameters are highly non-Gaussian, which means that a naive ``mean over standard deviation" figure does not accurately represent the significance of a parameter being non-zero. Second, there is not a single amplitude-like parameter that we can use to assess significance. Although we are primarily interested in $\OmegaHI$ or $\AHI$, whose posterior distributions include the projected degeneracies with $\alphaNL$ and $\alphaFoG{x}$, they are also somewhat degenerate with $\Mten$, and this should be captured as any measurement of $\Mten$ should also count towards a detection.

One way of describing the detection significance is by way of a Bayesian model comparison. In this case, we seek to compare two explanations of the data, one in which the signal is represented by the full signal model given above ($\calM_1$), and a null model where the signal is exactly zero and the data are entirely noise ($\calM_0$). To compare these, we need to calculate the marginal likelihood, or \emph{Bayesian Evidence}, $\calZ$ which is the normalisation constant for the posterior distribution shown in \cref{eq:posterior}:
\begin{align}
    \calZ & = \calP(\vd \mid \calM) \\
    & = \int \calP(\vd \mid \vtheta, \calM) \, \calP(\vtheta \mid \calM) \, d^n\theta \\
    & = \int \calL(\vtheta) \,\pi(\vtheta) \, d^n\theta \; .
\end{align}
The evidence allows us to compare the relative probability of two models given the observed data
\begin{align}
    \frac{\calP(\calM_1 \mid \vd)}{\calP(\calM_0 \mid \vd)} & = \frac{\calP(\vd \mid \calM_1)}{\calP(\vd \mid \calM_0)} \times \frac{\calP(\calM_1)}{\calP(\calM_0)} \\
    & = \frac{\calZ_1}{\calZ_0} \times \frac{\calP(\calM_1)}{\calP(\calM_0)} \; ,
\end{align}
where the $\calP(\calM)$ terms give the prior probabilities of the models. We assume that the model prior probabilities are equal from this point on, and focus solely on the Bayesian evidence ratio $\calZ_{10} = \calZ_1 / \calZ_0$ (often termed the Bayes Factor).

Calculating the evidence directly is challenging, as the region of high likelihood is typically much smaller than the prior volume, and so estimates tend to be dominated by sample noise. The standard techniques for evidence calculation are variants on Nested Sampling \citep{Skilling2006} but here we instead use the simpler process of Thermodynamic integration \citep{Gelman1998} as we do not need the extra efficiency of nested sampling-based techniques. To do this, we introduce the quantity
\begin{equation}
    \calZ(\lambda) = \int \calL(\vtheta)^\lambda \,\pi(\vtheta) \, d^n\theta \; .
\end{equation}
Noting that $\calZ(0) = 1$ and $\calZ(1) = \calZ$ we can write the quantity we want to calculate as
\begin{equation}
    \ln{\calZ} = \int_0^1 \frac{\partial \ln{\calZ(\lambda)}}{\partial \lambda} \, d\lambda \; .
\end{equation}
This transformation is useful because we can write the integrand as
\begin{align}
    \frac{\partial \ln{\calZ(\lambda)}}{\partial \lambda} & = \frac{1}{\calZ(\lambda)} \int \ln{\calL(\vtheta)} \: \calL(\vtheta)^\lambda \, \pi(\vtheta)\, d^n\theta \\
    & = \la \ln{\calL(\vtheta)} \ra_\lambda
\end{align}
where the $\la \ldots \ra_\lambda$ denotes an expectation evaluated against a posterior with the likelihood raised to the power $\lambda$. This gives us a straightforward way of calculating $\ln{\calZ}$: first, on a discrete grid in $\lambda$, we use a standard MCMC sampler to draw from the un-normalised distribution $\calL(\vtheta)^\lambda \pi(\vtheta)$, and then estimate $\la \ln{\calL(\vtheta)} \ra_\lambda$ from these samples; second, we numerically integrate over these estimates to calculate $\ln{\calZ}$.

We calculate the evidence for the signal model, $\calM_1$, by multiple sampling runs (as described in \secref{sec:modelfitting}) generated at different $\lambda$. As the bulk of the variation in the integrand is around $\lambda \sim 0$, we use the common choice of a grid regularly spaced in $\lambda^{1/5}$ \citep{Calderhead_Girolami_2009}, and as the integrands are smooth and well behaved, we find a Romberg integration over 33 samples achieves sufficient accuracy. For the evidence calculation, we use shorter chains per $\lambda$ step than for the parameter estimation, with only 15000 samples per chain. After removing the initial samples for burn-in and thinning to the independent samples, this leaves $\sim 700$ samples for each $\lambda$ step. To estimate the error on each evidence calculation, we bootstrap resample the set of points at each $\lambda$ step, integrate over the resampled sets, and estimate the sample variance over bootstrap sets. This gives a typical error in $\ln{\calZ}$ of $\sim 0.1$. In contrast, the null signal model, $\calM_0$, is a zero-parameter model and so its evidence is simply the likelihood of the data evaluated at zero signal. That is,
\begin{equation}
    \ln{\calZ_0} = -\frac{1}{2} \ls \ln{\lv 2 \pi \mSigma \rv} + \chi^2_0 \rs \; ,
\end{equation}
with
\begin{equation}
    \chi^2_0 = (\vd - \vmu)^T \mSigma^{-1} (\vd - \vmu) \; ,
\end{equation}
so that we do not need any MCMC scheme to calculate it.

With $\ln\calZ_0$ and $\ln\calZ_1$, we have both of the ingredients required to give the Bayes factor. To enable a comparison with other significance estimates, we can turn the evidence ratio into an effective ``number of sigma''. Assuming that the only two models that could explain the data are $\calM_0$ and $\calM_1$ and giving them equal prior probabilities, $\calP(\calM_0) = \calP(\calM_1) = 1/2$, we can write the probability of the null model as
\begin{equation}
    \calP(\calM_0 \mid \vd) = \frac{1}{1 + \calZ_1 / \calZ_0} \; .
\end{equation}
We turn this into an effective number of $\sigma$, $N_\calZ$, via
\begin{equation}
    N_\calZ = \Phi^{-1}(1 - \calP(\calM_0 \mid \vd))
\end{equation}
where $\Phi^{-1}(x)$ is the inverse cumulative distribution function of the standard normal distribution.

An alternative, frequentist method of estimating the detection significance is to use a likelihood-ratio test. First we compute the ratio of the maximum likelihood values between a model with no signal and one with the full signal model
\begin{align}
    \lambda & = 2 \ln{\lp\frac{\calL(\hat{\vtheta}_\text{ML})}{\calL_0}\rp} \\
    & = \Delta\chi^2
\end{align}
with $\Delta\chi^2 = \chi^2_0 - \chi^2_\text{min}$. This quantity is asymptotically $\chi^2$ distributed with degrees of freedom equal to the effective number of model parameters. As our model has several notable degeneracies, the effective number of model parameters will be less than the total number of parameters. We use the Bayesian model dimensionality \citep{Handley2019}
\begin{equation}
    d_M = 2 \ls \la (\ln \calL)^2 \ra - \la \ln \calL \ra^2 \rs
\end{equation}
as estimate of the number of parameters, where the expectation $\la \ldots \ra$ is taken over the posterior. Taking an average of this over the set of tracers, we find $d_M \sim 4.4$, and so we use 4 as the effective number of parameters. Using this we can ascribe a detection significance via the probability for a $\chi^2_{\nu=4}$ distribution to exceed $\lambda$. We again turn this into an effective number of sigma using the inverse CDF of a standard normal distribution:
\begin{equation}
\label{NLR}
    N_\mathrm{LR} = \Phi^{-1}\lp 1 - \int_\lambda^\infty \chi^2_4(x) dx \rp \; .
\end{equation}

As a final estimate of the detection significance, we take the best fit (minimum $\chi^2$) template as a fixed single template, and then fit that directly to the data with a varying amplitude, $A$. As the likelihood is Gaussian,  the distribution of $A$ can be computed exactly, and is Gaussian with mean of 1 and variance $(\Delta\chi^2)^{-1}$. This directly gives the number of sigma of detection, $N_A = \sqrt{\Delta\chi^2}$. We expect this quantity to overestimate the detection significance as the template has already been adjusted to fit the data.

\Cref{tab:significance} shows the detection significance for each tracer calculated by each method given above. The Bayes factors, $\ln(\calZ_1 / \calZ_0)$ are $\gtrsim 4.6$ for all tracers which corresponds to \emph{decisive} evidence for a cross correlation detection according to the interpretations of \cite{Jeffreys1961} and \cite{Kass1995}. The number of sigmas for each method are reasonably close, with the Bayesian evidence based number $N_\calZ$ the lowest of the three and the amplitude parameter the highest. The common criticism of evidence calculations is that they are dependent on the prior widths, and, as is the case here, a choice which is intended to be non-informative for the purpose of parameter estimation can significantly lower the evidence compared to a less conservative choice of prior. In our case, parameters like $\Mten$ could be significantly narrower without influencing the parameter estimation, which would boost the Bayes factor. Although we do not attempt it here, one resolution to this for nested comparisons (of which this is one), advocated by \cite{Gordon2007}, is to optimise the prior widths centred on the value implied by the nested model to maximise the Bayes factor.

We also calculate the evidence for the signal model where we fix the non-linear parameters, which we call $\calZ_2$, and give the log Bayes factor comparing it to the full signal model, $\ln{\calZ_{12}}$ in \cref{tab:significance}. In most cases $\ln\calZ_{12}$ is negative, that is, there is not sufficient improvement in the fits to justify the expanded model from statistical arguments alone, and in the remaining cases the evidence is marginal.

Our rationale for varying the non-linear parameters is to explore what our data tells us about the large-scale $\HI$ distribution while including the genuine uncertainties in the modelling. With that in mind, we do not take this as an indication that we should fix these non-linear parameters, but as one that they are not meaningfully constrained as they allow the model to over-fit the data.

\begin{deluxetable}{c DDD DD D}
    \tablecaption{The detection significance for each tracer.\label{tab:significance}}
    \tablecolumns{7}
    \tablehead{
        \colhead{Tracer} &
        \multicolumn{6}{c}{Bayesian} &
        \multicolumn{4}{c}{Likelihood ratio} &
        \multicolumn{2}{c}{Amplitude} \\ \cmidrule(lr){2-7} \cmidrule(lr){8-11} \cmidrule(lr){12-13}
        &
        \multicolumn{2}{c}{$\ln{\calZ_{10}}$} &
        \multicolumn{2}{c}{$\ln{\calZ_{12}}$} &
        \multicolumn{2}{c}{$N_\mathcal{Z}$} &
        \multicolumn{2}{c}{$\Delta\chi^2$} &
        \multicolumn{2}{c}{$N_\mathrm{LR}$} &
        \multicolumn{2}{c}{$N_A$}
    }
    \decimals
    \startdata
   LRG & 18.9 & -1.5 &        5.7 &   60.3 &           7.1 &        7.8 \\
   ELG & 10.8 & -2.4 &        4.1 &   40.8 &           5.7 &        6.4 \\
   QSO & 56.3 & -2.2 &       10.3 &  133.5 &          11.1 &       11.6 \\
   \hline
 QSOb0 & 23.9 & -2.3 &        6.5 &   66.2 &           7.5 &        8.1 \\
 QSOb1 & 19.6 &  0.8 &        5.8 &   53.2 &           6.6 &        7.3 \\
 QSOb2 & 16.9 & -0.9 &        5.3 &   50.0 &           6.4 &        7.1 \\
   \hline
QSOb00 &  7.6 &  1.5 &        3.3 &   27.8 &           4.5 &        5.3 \\
QSOb01 & 14.6 & -1.6 &        4.9 &   46.3 &           6.1 &        6.8 \\
    \enddata
    \tablecomments{We calculate the detection signficance by three different methods, using the Bayesian evidence, a likelihood-ratio test, and the amplitude constraints on the best-fit model. The log Bayes factors, $\ln{\calZ_{10}}$, all exceed the highest threshold $\gtrsim 4.6$ for \emph{decisive} evidence according to the scale of \cite{Jeffreys1961}, and the $\Delta\chi^2$ values all have $p$-values $\lesssim 10^{-5}$. We also convert each to an effective number of sigmas for comparison, giving $N_\calZ$, $N_\mathrm{LR}$ and $N_A$ respectively. The results are similar; however, $N_\calZ$ suffers from the choice of wide priors and the amplitude ratio $N_A$ is overly optimistic as it does not account for the previous fitting of the template. We also give the evidence ratio $\ln{\calZ_{12}}$ comparing the full signal compared to fixing the non-linear parameters. For most catalogs there is no evidence in favor of the full model ($\ln\calZ < 0$), that is the improved fit is not sufficient to support the additional free parameters. Only for the QSOb00 tracer is there moderate evidence to support the full model.
    }
\end{deluxetable}

%% file: sections/validation.tex

\section{Validation}
\label{sec:validation}

In this section we describe several consistency tests that were performed on the analysis and inform the systematic errors that are placed on the result.  These tests consist of evaluating if the measurements made by the two polarisations are consistent, evaluating if the signal is the same from day to day, estimating the uncertainty on the amplitude of the signal due to beam calibration errors, and evaluating the linearity of the analysis pipeline.

\subsection{Consistency Between Polarisations}
\label{sec:consistent_pol}

\input{tables/table_polarisation_comparison}

The following procedure is used to determine if measurements made with the XX and YY baselines are consistent given our model for the noise and \tcm signal.  The two polarisations are jointly fit to a restricted and unrestricted model.  For the restricted model, both polarisations are described by the same set of parameters, $\bm{\theta}$, as outlined in \secref{sec:modelfitting}.  The version of the model that holds the non-linear parameters fixed at their fiducial values is employed for this exercise, since the version that allowed them to vary did not yield a significantly better fit to the data for any tracer or QSO redshift bin.  For the unrestricted model, the polarisations are described by a different set of parameters, $\bm{\theta}_{X}$ and $\bm{\theta}_{Y}$.  The maximum-likelihood estimate of the parameters is obtained for each model using the L-BFGS-B optimization algorithm.  The following test statistic is then calculated
\begin{align}
    \label{eq:delta_chisq}
    \Delta \chi^{2} & = \chi^{2}(\bm{\hat{\theta}}_{\rm res}) - \chi^{2}(\bm{\hat{\theta}}_{\rm unres}) \ ,
\end{align}
where $\chi^{2}$ is given by \cref{eq:chisq}, $\bm{\hat{\theta}}_{\rm res} \equiv \bm{\hat{\theta}}$ denotes the maximum-likelihood parameter estimates for the restricted model, and $\bm{\hat{\theta}}_{\rm unres} \equiv \left[\bm{\hat{\theta}}_{X}, \bm{\hat{\theta}}_{Y} \right]$ denotes the maximum-likelihood parameter estimates for the unrestricted model.  The $\chi^2$ values and the test statistic are quoted in \cref{tab:compare_pol} for all tracers and all QSO redshift bins.

The test statistic will follow a $\chi^{2}$ distribution with $\Delta \nu = \nu_{\rm res} - \nu_{\rm unres}$ degrees of freedom under the null hypothesis that the two polarisations are described by the same model.   Naively we expect $\Delta \nu$ to be equal to the number of model parameters, since the unrestricted model has twice the number of parameters as the restricted model.  However, the model parameters are highly degenerate, so that using the number of parameters would likely overestimate $\Delta \nu$ and bias the test towards accepting the null hypothesis.

To avoid this, the distribution of the test statistic under the null hypothesis is empirically measured using the random mock catalogs.  We generate \SI{10000}{} realizations of our data by adding the best-fit, restricted model and a stack on a random mock catalog.  We then fit each realization to the restricted and unrestricted model and calculate the test statistic.  The probability to observe a value of the test statistic in excess of that observed in the data is then determined from the empirical cumulative distribution function.  The results are presented in the last column of \cref{tab:compare_pol}.  For all tracers and QSO redshift bins, the null hypothesis that the two polarisations are described by the same set of model parameters is accepted with the probability to exceed (PTE) $> 0.05$.  We also note that the empirical distributions are reasonably well described by a $\chi^{2}$ distribution with $\Delta \nu \approx 2.3$ degrees of freedom.

\subsection{Consistency Between Even and Odd Days}
\label{sec:consistent_even_odd}

The \SI{102}{} sidereal days that were used to construct the sidereal stack are split into two subsets by chronologically ordering the days that went into each seasonal stack and then separating the even days into one set and the odd days into the other set (see \secref{sec:sidereal_avg_seasonal}).  The two sets have size \SI{50}{} and \SI{52}{} sidereal days, and a mean date that differs by \SI{53}{\hour}.  Each set is averaged using the procedure outlined in \secref{sec:sidereal_avg}.  This yields two estimates of the visibilities which are then differenced according to
\begin{align}
    \label{eq:even_odd_jackknife}
    \Delta V &= \frac{1}{c} \left( V_{\rm even} -  V_{\rm odd}\right)
\end{align}
with
\begin{align}
    c & = \begin{cases}
            \dfrac{w_{\rm even} + w_{\rm odd}}{\sqrt{w_{\rm even} w_{\rm odd}}} & \mbox{if } (w_{\rm even} > 0) \wedge (w_{\rm odd} > 0) \mbox{;} \\[+2ex]
            0 & \mbox{otherwise.}
            \end{cases}
\end{align}
Here $V \equiv V^{p}_{xy}(\nu, \phi)$ and $w \equiv w^{p}_{xy}(\nu, \phi)$ denote the visibilities and corresponding weights.  The quantity $c$ is a scale factor that will set the variance of the radiometric noise in the difference equal to that in the weighted average.  In the limit that the even and odd splits have equal radiometric noise, and hence equal weight, then $c = 2$.  In reality, the two splits have slightly different weights such that $c = 2.0065 \pm 0.018$ over the baselines, frequency, and right ascensions examined.  The processing described in \secref{sec:sidereal_avg} through \secref{sec:stacking} is applied to the differenced visibility, with the caveat that we use the global frequency mask, delay cut, and primary beam model that were previously derived from the weighted average of the full set of days.

The cosmological \tcm signal is constant as a function of sidereal day and is expected to cancel in the difference.  The radiometric noise, on the other hand, will be independent in the two subsets, and therefore will remain in the difference.  Transient RFI is also expected to be independent in the two subsets and remain in the difference.  Residual foregrounds caused by spectral leakage due to a chromatic instrument transfer function will be the same from day to day and hence cancel in the difference.  Residual foregrounds due to seasonal changes in the instrument transfer function will also cancel.  On the other hand, residual foregrounds due to changes in the instrument transfer function from day to day will remain.

Since a significant portion of the noise in the stack is due to residual foregrounds that will be mitigated by the differencing procedure, the covariance matrix of the even-odd difference is expected to change relative to the covariance matrix of the full data set.  We recalibrate the covariance matrix with mock catalogs as outlined in \secref{sec:covariance}.  We find better agreement between the even-odd difference covariance and the expected radiometric noise, suggesting that the majority of $\sim \SI{50}{\percent}$ excess noise in the full set is primarily due to foregrounds that are static from one day to the next.

\input{tables/table_even_odd_split_null}

Under the null hypothesis that the observed signal is the same on even and odd days, stacking the even-odd difference on the true catalog should be statistically equivalent to stacking on a random mock catalog.  The distribution of the $\chi^{2}$ test statistic for the random mock catalogs is well described by a theoretical $\chi^{2}$ distribution with 202 degrees of freedom.  \Cref{tab:even_odd_split_null} quotes the $\chi^{2}$ value of the stack on each tracer and QSO redshift bin, as well as the fraction of the \SI{10000}{} random mock catalogs that have a $\chi^2$ test statistic in excess of that observed for the true catalog.  We find that all tracers and redshifts bins have a PTE greater than $0.05$, except for the LRG catalog, which has a PTE of 0.025, and the subset of QSO catalog with a redshift between 0.91 and 1.03 (QSOb01), which has a PTE of 0.023.  If the large $\chi^{2}$ values are due to differences in the observed signal on even and odd days, then we would expect to see a copy of the signal in the jackknife.  We recompute the test statistic using only frequencies $|\Delta \nu| < \SI{5}{\mega\hertz}$ where the magnitude of the signal is largest.   We find that the PTE for the LRG catalog increases to 0.12, the QSOb01 catalog decreases to 0.009, and all others tracers and QSO redshift bins have a value greater than $0.05$.  This leads us to conclude that the large $\chi^{2}$ observed when stacking the jackknife on the LRG catalog originates from a rare noise fluctuation rather than differences in the signal on even and odd days.  However, the large $\chi^{2}$ for the QSOb01 catalog warrants additional investigation.

\input{tables/table_even_odd_split_model}

To explore this further, we perform a model-dependent analysis of the even and odd days that is similar to the analysis used to check for consistency between polarisations, which was described in \secref{sec:consistent_pol}.  Each split is processed independently through the pipeline, and then stacked on the true catalog and the random mock catalogs.  We use the same random mock catalogs for both splits to ensure that the covariance matrix captures correlated noise between them.  The two splits are jointly fit to both a restricted and unrestricted model.  The restricted model describes both splits with the same set of parameters.  The unrestricted model describes each split with a different set of parameters.  We employ the version of our model where the non-linear parameters are held fixed at their fiducial values (see \cref{tab:priors}).  We compute $\Delta \chi^{2}$ as given by \cref{eq:delta_chisq} and calibrate its distribution under the null hypothesis using the random mock catalogs.  The results are presented in \cref{tab:even_odd_split_model}.

As anticipated, the LRG catalog passes the test (PTE~$=0.71$) and the QSOb01 catalog fails the test (PTE~$=0.02$). The QSOb0 catalog, which contains all quasars with a redshift between 0.80 and 1.03 and is a superset of QSOb01, also fails the test (PTE~$=0.005$).  The discrepancy appears primarily in the amplitude of the signal, with the even-day split exhibiting an approximately \SI{50}{\percent} larger amplitude than the odd-day split for these two redshift bins.  We perform an MCMC fit of both the restricted and unrestricted models and use the posterior distributions of the amplitude parameter $\AHI$ to characterize the fractional error in $\AHI$ implied by this discrepancy.  This is defined as half the difference in $\AHI$ between the even and odd splits as measured by the unrestricted model fit divided by the most likely value from the restricted model fit, and is quoted in the last column of \cref{tab:even_odd_split_model}.

The discrepancy is suggestive of a \SI{50}{\percent} difference in our calibration between even and odd days at frequencies between \SI{700}{\mega\hertz} and \SI{745}{\mega\hertz}.  However, we have ruled out an error in the relative calibration of this magnitude by examining the spectra of \SI{34}{} bright point sources in the NGC field extracted from the maps prior to foreground filtering.  We find that the difference in spectra between the even and odd days is at most \SI{1}{\percent} over all sources and frequencies.

In order to account for the observed discrepancy, we will assume an additional \SI{25}{\percent} systematic error on the $\AHI$ constraint for the QSOb0 and QSOb01 catalogs.

\subsection{Beam Calibration Errors}
\label{sec:beamcalerrors}

In order to estimate the uncertainty on the default beam model described in \secref{sec:beams}, it is compared to independent measurements of the beam from observations of the Sun and holographic observations of bright point sources made in conjunction with the John~A.~Galt \SI{26}{\meter} telescope \citep{overview-paper}.  Based on these comparisons we estimate that within the main lobe the beam model is accurate to $\lesssim$\SI{5}{\percent} relative to the beam on meridian at the declination of Cygnus A.  Currently our beam calibration technique is unable to constrain the sidelobes of the beam (for details see Appendix~\ref{sec:beam_calibration_ptsrc}).  The solar and holographic data both suggest that the sidelobes are $\lesssim$\SI{1}{\percent} at hour angles $\lesssim \SI{30}{\degree}$ and $\lesssim \SI{0.1}{\percent}$ at hour angles $\gtrsim \SI{30}{\degree}$.  It is estimated that approximately $\SI{10}{\percent}$ of the beam solid angle lies outside the region that we are able to measure with the default beam model.

The solar beam measurements are described in \cite{solar-beam-paper} and span $\SI{-23.5}{\degree} \leq \theta \leq \SI{23.5}{\degree}$, which corresponds to the range of apparent declinations that the Sun travels between winter and summer solstice.  The RMS difference between the solar and default beam model is \SI{4}{\percent} (relative to the beam on meridian at the declination of Cygnus A) in the region $|\ha| \lesssim \SI{3}{\degree}$, $|\theta| < \SI{23.5}{\degree}$, $\SI{587.5}{\mega\hertz} < \nu < \SI{800}{\mega\hertz}$.  However, the inferred amplitude of the \tcm signal is primarily sensitive to the fractional error in the beam on meridian when averaged over the large range of declinations and frequencies covered by the eBOSS catalogs.  In order to estimate the systematic error on the \tcm amplitude due to beam uncertainties, the fractional difference between the default and solar beam model on meridian at the declination and \tcm frequency of each source in each catalog is extracted and then averaged using the same weights that are used in the stacking procedure described in \secref{sec:stacking}.  Only \SI{3}{\percent} of the QSOs and LRGs in the NGC field are at declinations that overlap with the solar data.  However, \SI{42}{\percent} of the ELGs in the NGC field lie at declinations where there are two independent measurements of the beam, and the average fractional difference between these two measurements is \SI{6}{\percent}.

We have also compared the flux density of the brightest radio sources in the deconvolved map to their expected flux density in order to obtain an additional estimate of the systematic uncertainty on the \tcm amplitude.  The expected flux densities are obtained by interpolating recent measurements made by the Very Large Array (VLA) to frequencies in the CHIME band \citep{perley2017}.  There are 14 sources in total used for this purpose with an average declination separation of \SI{5}{\degree}.  These sources do not provide a dense sampling of the declination axis like the solar data, but they do cover the full range of declinations spanned by the eBOSS catalogs.  All 14 sources have interpolated flux densities that are accurate at the sub-percent level and are greater than \SI{10}{\jansky} at \SI{600}{\mega\hertz}.  The RMS of the fractional error in the flux density of these sources in the deconvolved map is \SI{5.0}{\percent}, \SI{6.4}{\percent}, and \SI{7.4}{\percent} for the range of frequencies and declinations spanned by QSO, LRG, and ELG catalogs in the NGC field, respectively.  Taking instead the weighted average of the fractional error in the flux at the declinations and \tcm frequencies nearest to the sources in each catalog yields \SI{0.6}{\percent}, \SI{2}{\percent}, and \SI{0.5}{\percent} for the QSO, LRG, and ELG catalogs in the NGC field.  Note that this is an end-to-end test of our ability to recover the true flux of point sources and is sensitive to a variety of potential sources of systematic error including beam calibration errors, but also complex gain errors and regridding artifacts.

As a final check, we simulate observations of the fiducial \tcm signal using both the default beam and the control beam.  For each of these simulations, we construct a map by deconvolving both the default beam and control beam, and then stack the map on simulated catalogs.  We then examine the fractional difference in the amplitude of the stacked signal for the four different pairs of (simulation beam model, deconvolution beam model) relative to the (default, default) pair that was used in the actual analysis.  For all four pairs and all three tracers the observed difference is less than \SI{6}{\percent}.  This provides an estimate of the systematic error due to uncertainty in the interference pattern that modulates the beam.  Note that this is a very conservative estimate because the uncertainty on the interference pattern is roughly a factor of 10 less than than the amplitude of the interference pattern itself.

Based on the solar comparison, bright point source comparisons, and simulations of different beam models, a conservative \SI{8}{\percent} systematic error on the amplitude of the \tcm signal will be assumed for all fields.

\subsection{Linearity}
\label{sec:linearity}

\begin{figure}
   \centering \includegraphics[width=0.98\linewidth,keepaspectratio]{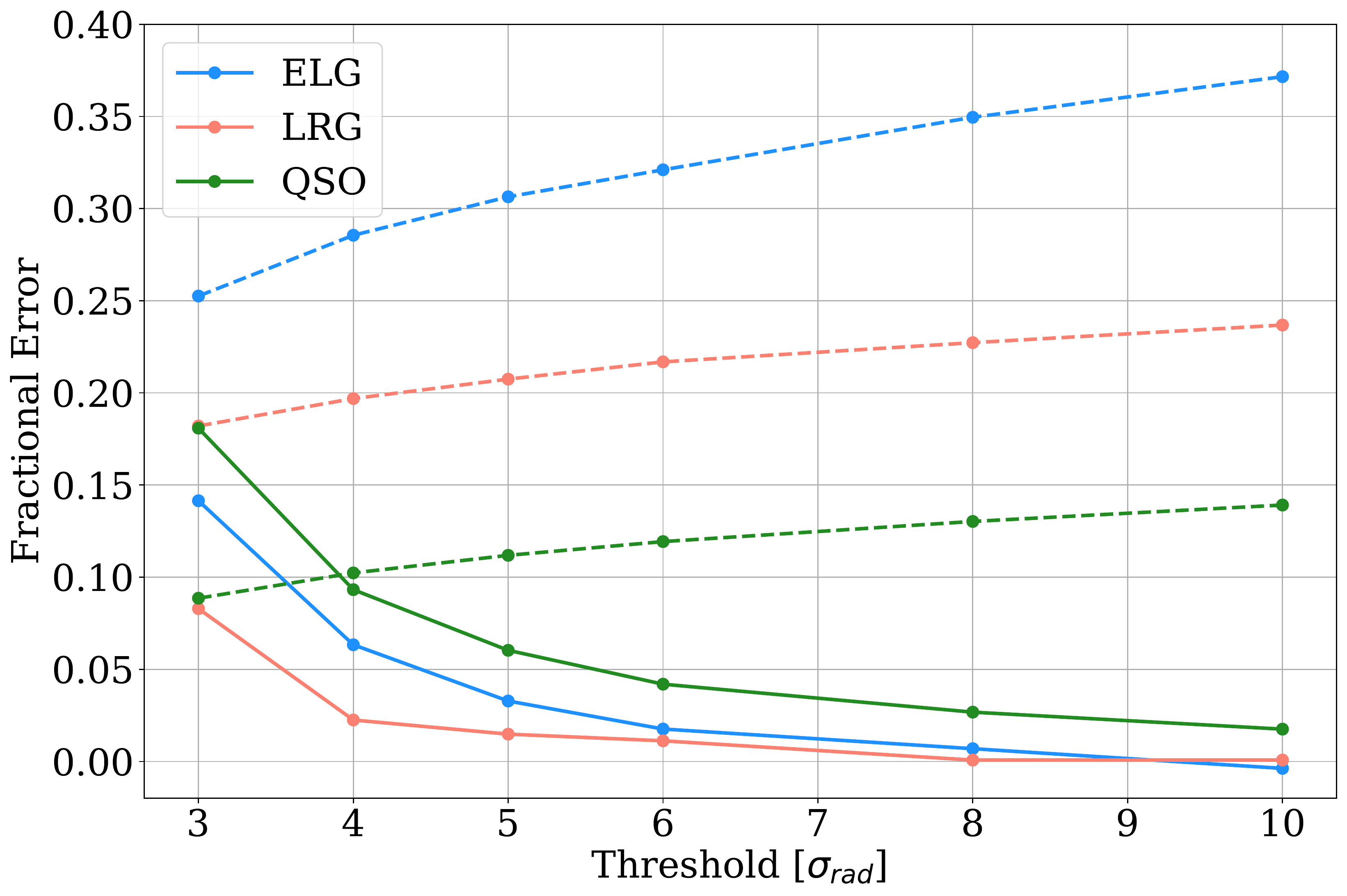}
    \caption{Comparison of the statistical uncertainty on the amplitude of the stacked signal (dashed line) to the bias in the amplitude caused by application of the outlier mask (solid lines) as a function of the threshold used to generate the mask.  The threshold is defined in units of the standard deviation of the radiometric noise.  The different colors correspond to different tracers of large-scale structure.  These measurements were made using the signal injection technique described in the text, wherein we stack on the sum of the data and the fiducial simulation for the \tcm signal.  Increasing the threshold reduces the bias in the recovered \tcm amplitude, but also increases the statistical uncertainty due to inclusion of residual foregrounds.  A threshold of six times the expected radiometric noise was chosen for this analysis, which results in a bias in the amplitude that is $< \SI{4}{\percent}$ and small relative to the statistical error for all tracers.}
    \label{fig:threshold_error}
\end{figure}

Many of the elements in our analysis pipeline, such as the delay filtering, are explicitly linear, meaning that they operate independently on the \tcm signal and foregrounds present in the data. To characterize the linearity of the entire analysis, we inject simulated \tcm signal into the data, process the signal+data combination in the same way as the data, and stack the results on mock eBOSS catalogs that are correlated with the simulated signal. We also separately perform the stacking on mock catalogs using the data without injected signal, and using mock observations containing only the injected signal. In a perfectly linear analysis, the difference of the signal+data and data stacks will be equal to the signal-only stacks, while non-linearities will cause a violation of this equality. This method has previously been used to characterize signal loss in \tcm analyses that rely on strongly nonlinear foreground filtering techniques (e.g.\ \citealt{masui2013,paciga2013}).

In detail, we generate correlated \tcm and galaxy number density sky maps, and propagate them through to simulated timestreams and mock galaxy or quasar catalogs following the procedures in \secref{sec:simulations}. The signal-only timestream is added to the sidereal stack derived from the data prior to subtraction of the brightest point sources (i.e.\ in the first box in \cref{fig:stacking_flowchart}), and this combined timestream is passed through the same analysis pipeline as the data, culminating in the beam-deconvolved, filtered, masked map being stacked on the mock catalogs. Prior to delay filtering, the signal-only map has RMS $\sim$\SI{0.3}{\milli\jansky\per\beam}, compared to $\sim$\SI{3}{\jansky\per\beam} for the data map; therefore, the addition of signal to the data map has negligible effect on the determination of the elevation-dependent delay cut (\secref{sec:filter}), or on which frequencies are identified as outliers (\secref{sec:map_mask}), so these aspects of the analysis are not regenerated for the signal+data combination.

However, the final masking step---which masks map pixels whose absolute value exceeds a chosen threshold, based on the estimated map noise level---is explicitly non-linear, so we recompute this mask to determine the impact of this nonlinearity on the recovered signal. We find that this impact is significant, which can be explained as follows. The distribution of pixel values in the signal-only map is symmetric about zero, but this distribution is skewed positive if one only considers pixels containing an object in a given mock catalog, since these objects are more likely to occupy pixels corresponding to matter overdensities, which are also correlated with \tcm emission. Thus, if a given pixel (containing a catalog object) in the data map is positive and just below the mask threshold, it is more likely to be perturbed above the threshold by the addition of the signal-only map; similarly, a given negative pixel that is just beyond the threshold is more likely to be perturbed within the threshold.

The net effect is that the signal injection increases the number of negative near-threshold unmasked pixels and decreases the number of such positive pixels, resulting in an artificial attenuation of the overall stacking amplitude. A lower threshold will result in a greater number of affected pixels and more severe attenuation, while a higher threshold will mitigate this, but at the expense of decreasing the signal-to-noise ratio due to a greater number of anomalous pixels being included in the stack.

\Cref{fig:threshold_error} quantifies these two effects. For each threshold in the figure, we compute the difference of stacks on signal+data and data-only maps, form a ``prediction" given by a stack on the corresponding signal-only map, and fit the overall amplitude of the prediction to the stack difference, using the data stack covariance matrix described in \secref{sec:covariance}. This amplitude indicates the amount of attenuation (shown as solid lines in  \cref{fig:threshold_error}) induced by the outlier mask, while the fractional uncertainty on this amplitude (shown as dashed lines) indicates the effect of the mask threshold on the statistical significance of the stacking measurement.

Based on these results, we choose a mask threshold of $6\sigma$, where $\sigma$ is the estimated radiometric noise in the maps (see \secref{sec:weights}). For the fiducial \tcm model assumed in our simulations, this results in signal attenuation of less than 4\% for each eBOSS tracer, which is at least a factor of three smaller than the statistical uncertainty. Note that in our actual fits to data, presented in \secref{sec:results}, the stacking amplitudes are factors of $(1.9, 1.5, 1.4)$, for the (ELG, LRG, QSO) stacks, greater than in our simulations. We have re-run the test in \cref{fig:threshold_error}, modifying the amplitude of the injected signal accordingly, and have verified that the attenuation level is unchanged, while the fractional uncertainty decreases by the quoted factors. Even with this change, the attenuation is still less than half of the uncertainty for each tracer, which we deem to be acceptable for this analysis.

%% file: tables/table_polarisation_comparison.tex
\begin{deluxetable}{c c c c c c}
\tablecolumns{6}
\tablecaption{Model-dependent test for consistency between polarisations. \label{tab:compare_pol}}
\tablehead{\colhead{ } & \multicolumn{2}{c}{$\chi^{2}$} & \colhead{} & \colhead{} & \colhead{} \\ \cmidrule{2-3}
\colhead{Tracer} & \colhead{Restricted} & \colhead{Unrestricted} & \colhead{$\Delta \chi^{2}$} & \colhead{$\Delta \nu$} & \colhead{PTE}}
\startdata
LRG & 218.6 & 214.2 & 4.4 & 2.3 & 0.12 \\
ELG & 210.1 & 209.3 & 0.7 & 2.3 & 0.77 \\
QSO & 219.0 & 213.9 & 5.0 & 2.3 & 0.10 \\
\hline
\hspace{2mm}QSOb0 & 210.5 & 208.3 & 2.2 & 2.3 & 0.37 \\
\hspace{2mm}QSOb1 & 202.1 & 200.8 & 1.2 & 2.3 & 0.62 \\
\hspace{2mm}QSOb2 & 220.4 & 214.3 & 6.1 & 2.3 & 0.05 \\
\hline
\hspace{2mm}QSOb00 & 185.5 & 184.8 & 0.6 & 2.2 & 0.78 \\
\hspace{2mm}QSOb01 & 235.8 & 233.4 & 2.3 & 2.2 & 0.35 \\
\enddata
\tablecomments{For each tracer and redshift bin, we report the minimum $\chi^{2}$ obtained when fitting a model in which the two polarisations are described by the same set of parameters (restricted) and a different set of parameters (unrestricted).  The distribution of the difference, $\Delta \chi^{2}$, under the null hypothesis that the polarisations are described by the same set of parameters is calibrated using random mock catalogs and approximately follows a theoretical $\chi^{2}$ distribution with the quoted $\Delta \nu$ degrees of freedom.  The PTE provides the fraction of random mock catalogs that exceed the value observed in the data.}
\end{deluxetable}

%% file: tables/table_even_odd_split_null.tex
\begin{deluxetable}{c c c c c}
\tablecolumns{5}
\tablecaption{Model-independent test for consistency between even and odd days. \label{tab:even_odd_split_null}}
\tablehead{\colhead{ } & \multicolumn{2}{c}{$|\Delta \nu| \leq \SI{20}{\mega\hertz}$ (202 dof)} & \multicolumn{2}{c}{$|\Delta \nu| \leq \SI{5}{\mega\hertz}$ (50 dof)} \\ \cmidrule(l{3pt}r{3pt}){2-3} \cmidrule(l{3pt}r{3pt}){4-5}
\colhead{Tracer} & \colhead{$\chi^{2}$} & \colhead{PTE} & \colhead{$\chi^{2}$} & \colhead{PTE}}
\startdata
LRG & 242.4 & 0.025 & 62.0 & 0.12 \\
ELG & 199.1 & 0.54 & 47.2 & 0.58 \\
QSO & 202.0 & 0.49 & 59.2 & 0.17 \\
\hline
QSOb0 & 233.5 & 0.064 & 66.3 & 0.062 \\
QSOb1 & 177.2 & 0.90 & 35.1 & 0.95 \\
QSOb2 & 190.5 & 0.71 & 57.9 & 0.21 \\
\hline
QSOb00 & 207.5 & 0.38 & 55.9 & 0.26 \\
QSOb01 & 243.5 & 0.023 & 76.3 & 0.009 \\
\enddata
\tablecomments{For each tracer and redshift bin, we report the $\chi^{2}$ test statistic when stacking the catalog on a jackknife of even and odd days.  Under the null hypothesis that the \tcm signal is the same on even and odd days, this will follow a $\chi^{2}$ distribution with the stated degrees of freedom (dof). The PTE provides the fraction of random mock catalogs that exceed the value observed when stacking on the true catalog.}
\end{deluxetable}

%% file: tables/table_even_odd_split_model.tex
\begin{deluxetable}{c c c c c c c}
\tabletypesize{\footnotesize}
\tablecolumns{7}
\tablecaption{Model-dependent test for consistency between even and odd days. \label{tab:even_odd_split_model}}
\tablehead{\colhead{ } & \multicolumn{2}{c}{$\chi^{2}$} & \colhead{} & \colhead{} & \colhead{} & \colhead{} \\ \cmidrule{2-3}
\colhead{Tracer} & \colhead{Restricted} & \colhead{Unrestricted} & \colhead{$\Delta \chi^{2}$} & \colhead{$\Delta \nu$} & \colhead{PTE} & \colhead{$\frac{\Delta \AHI}{2\AHI}$}}
\startdata
LRG & 484.0 & 483.0 & 1.0 & 2.4 & 0.71 & $-0.11^{+0.16}_{-0.14}$ \\ [+1mm]
ELG & 414.7 & 412.7 & 2.1 & 2.3 & 0.42 & $-0.15^{+0.22}_{-0.11}$ \\ [+1mm]
QSO & 425.2 & 422.4 & 2.8 & 2.6 & 0.34 & $-0.06^{+0.11}_{-0.10}$ \\ [+1mm]
\hline
QSOb0 & 447.9 & 436.7 & 11.2 & 2.4 & 0.005 & $0.26^{+0.12}_{-0.12}$ \\ [+1mm]
QSOb1 & 370.6 & 368.3 & 2.4 & 2.5 & 0.41 & $-0.12^{+0.14}_{-0.12}$ \\ [+1mm]
QSOb2 & 452.7 & 446.4 & 6.3 & 2.4 & 0.052 & $-0.16^{+0.16}_{-0.13}$ \\ [+1mm]
\hline
QSOb00 & 378.7 & 378.1 & 0.6 & 2.3 & 0.81 & $0.11^{+0.20}_{-0.17}$ \\ [+1mm]
QSOb01 & 468.6 & 460.1 & 8.5 & 2.4 & 0.020 & $0.25^{+0.13}_{-0.11}$ \\ [+1mm]
\enddata
\tablecomments{For each tracer and redshift bin, we report the minimum $\chi^{2}$ obtained when fitting a model in which the even and odd splits are described by the same set of parameters (restricted) and a different set of parameters (unrestricted).  The distribution of the difference, $\Delta \chi^{2}$, under the null hypothesis that the splits are described by the same set of parameters is calibrated using random mock catalogs and approximately follows a theoretical $\chi^{2}$ distribution with the quoted $\Delta \nu$ degrees of freedom.  The PTE provides the fraction of random mock catalogs that exceed the value observed in the data.  The last column provides the fractional error on the amplitude parameter, $\AHI$, inferred from this comparison.}
\end{deluxetable}

%% file: sections/cosmological.tex

\section{Discussion}
\label{sec:interpretation}

\subsection{Quasar Redshift Errors}

\label{sec:quasar_redshift_errors}

As illustrated in \cref{fig:qso_reduced_params}, there is a statistically significant frequency offset in the QSO stacks of $\Dnu \approx \SI{-0.2}{\mega\hertz}$, equal to roughly half the width of a CHIME frequency channel. As this is only seen in the QSO stacks and not within the overlapping LRG and ELG measurements, it is difficult to explain this as an instrumental issue within CHIME. Instead we interpret this as being a systematic bias in the eBOSS quasar redshifts, stemming from the difficulty of determining a redshift from the complex processes producing a quasar spectrum \cite[see][Section~4.6]{lyke2020}. Quasar emission lines such as \ion{C}{4} are frequently blueshifted from the host galaxy redshift by dynamical and radiative processes within the quasar's accretion disk and outflowing winds \citep{Shen2016,Richards2011}.

\begin{figure}[tbp]
    \centering
    \includegraphics[width=\linewidth]{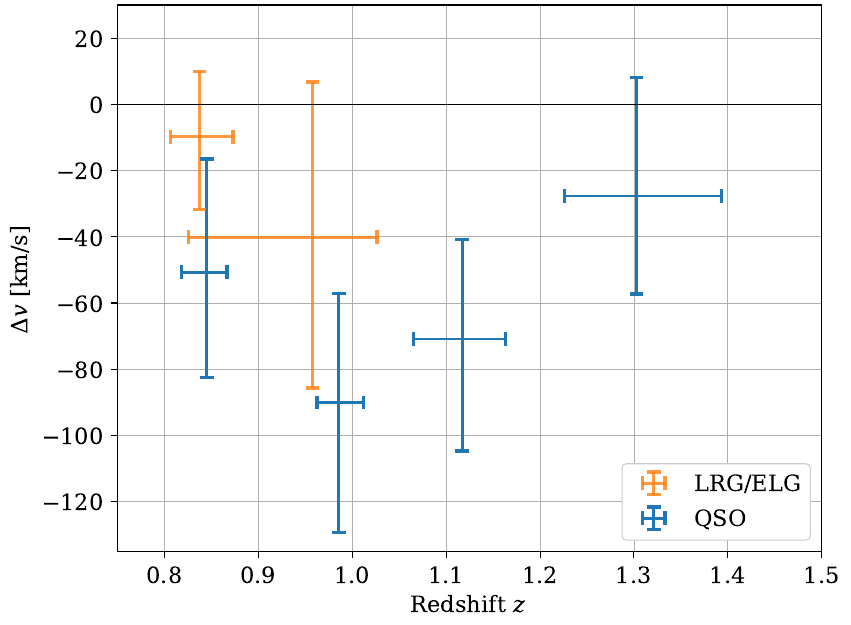}
    \caption{
        The derived redshift bias for each tracer given as a velocity shift $\Delta{v} = c \Delta{z} / (1 + z)$. We derived this from the frequency offset $\Dnu$ measured from each tracer assuming that the source is a systematic bias in the eBOSS catalog redshifts. For both the ELG and LRG catalogs (orange points) there is no discernible bias, but the QSO catalogs (blue points; from left to right, QSOb00, QSOb01, QSOb1, and QSOb2) have a significant bias. This is in agreement with the bias of the $z_\mathrm{PCA}$ redshift estimates shown in Figure 3 of \cite{lyke2020}.}
    \label{fig:redshift_bias}
\end{figure}

Similar to \cite{lyke2020} we express the redshift error as a velocity which can be connected to our measured frequency offset
\begin{align}
    \Delta{v} & = c \frac{\Delta{z}}{1 + z} \\
    & = c (1 + z) \frac{\Dnu}{\nuHI} \; .
\end{align}
In \cref{fig:redshift_bias}, we show the inferred velocity bias for the QSOb00, QSOb01, QSOb1 and QSOb2 stacks, which give non-overlapping measurements in redshift. Overall, we measure $\Delta{v} \sim \SI{-66}{\kilo\metre\per\second}$ at $\sim 3.3 \sigma$, with individual bins ranging from $0.8 \sigma$ (QSOb2) to $2.5\sigma$ (QSOb01). Our analysis does not account for the Doppler shift from the Earth's motion around the solar system barycentre; however, while the shift on any individual source may be up to $\sim \SI{30}{\kilo\metre\per\second}$, on average this effect is small. Taking a weighted mean of the Doppler shift towards each source for each night of observation, we find an average Doppler correction of \SI{-3.1}{\kilo\metre\per\second}.

Overall the results in \cref{fig:redshift_bias} are consistent with those of \citet[Fig.~3]{lyke2020}, who estimated the systematic bias in the $z_{\rm PCA}$ redshift estimates (which we used for stacking) as compared to redshifts of quasar host galaxies measured using stellar absorption lines. We anticipate that future quasar cross-correlation analyses with higher source numbers and improved processing of the CHIME data will be able to provide useful measurements of this bias across a broad range of redshifts.

The eBOSS quasar redshifts are significantly noisier than those of LRG and ELG samples due to the broader emission lines, with significant long tails of poor redshift estimates \citep{lyke2020}. This has a noticeable effect on the  stack signal (recall \cref{fig:sim_filtering_stages}) as the convolutional effect of the redshift errors broadens and suppresses the peak of the stack signal. Uncertainties in the quasar redshift error model can therefore give sizable changes in the constraints on the signal amplitude.

In our primary analysis we use the ``double Gaussian'' model of \citet[Eq.~A2]{lyke2020} to describe the QSO redshift uncertainties. However, the model as presented does not seem to match their measurements of the redshift errors in ways which are significant for our analysis. There are two clear differences: first, that the fraction of observations which have errors drawn from the wider Gaussian component appears to be smaller in the data \cite[Fig.~4]{lyke2020} than the $\sim 18\%$ quoted in the model; second, there appears to be a significant reduction in the errors at low redshift compared to the rest of the sample \cite[Fig.~9]{lyke2020}, which is expected from the presence of \ion{O}{3} and H$\beta$ in the wavelength range of the spectrograph at redshifts $z \lesssim 1$ (\'{E}tienne Burtin, private communication).

To assess the importance of this, we modify the $z_\mathrm{PCA}$ redshift error model to capture these effects. This change is intended to give a plausible alternative consistent with the data presented in \citet{lyke2020}, though we do not claim it is more realistic. Producing an improved model would require repeating the analysis of \citealt{lyke2020} and is beyond the scope of this paper. Our model is a straightforward modification of the published ``double Gaussian'' where we allow the coefficients to be redshift dependent. The redshift error on a single observation of a quasar, as given by a velocity error $\delta{v}$, is drawn from a redshift probability distribution
\begin{multline}
    \calP(\delta{v} \mid z) = \frac{1}{1 + f^{-1}(z)} \\ \times \ls \calG(\delta{v}, \sigma_1(z)^2) + f^{-1}(z) \calG(\delta{v}, \sigma_2(z)^2) \rs
\end{multline}
where $\calG$ is the standard normalised Gaussian with
\begin{equation}
    \calG(x, \sigma^2) = \frac{1}{\sqrt{2 \pi \sigma^2}} e^{- \frac{x^2}{2\sigma^2}} \; ,
\end{equation}
and $\sigma_1(z)$ and $f(z)$ are smooth step functions centered at $z = 1.0$ with $\sigma_1$ transitioning from \SIrange{90}{150}{\kilo\metre\per\second}
\begin{equation}
    \sigma_1(z) = \ls 90 + 30 \lp 1 + \tanh\lp \frac{z - 1.0}{0.05} \rp \rp \rs \: \si{\kilo\metre\per\second}
\end{equation}
and $f^{-1}(z)$ changing from zero at low redshift (that is, no errors in the broad distribution) to $\sim 0.03$ (a value which approximately matches the data of \citealt[Fig.~4]{lyke2020})
\begin{equation}
    f^{-1}(z) = \frac{1}{35} \lp 1 + \tanh\lp \frac{z - 1.0}{0.05} \rp \rp \; .
\end{equation}
It is necessary to change both $\sigma_1$ and $f$ as no single change is able to reproduce the observed low redshift uncertainties. However, we do leave $\sigma_2$ unchanged with a redshift independent $\sigma_2(z) = \SI{1000}{\kilo\metre\per\second}$.

\begin{figure}[tbp]
    \centering
    \includegraphics[width=\linewidth]{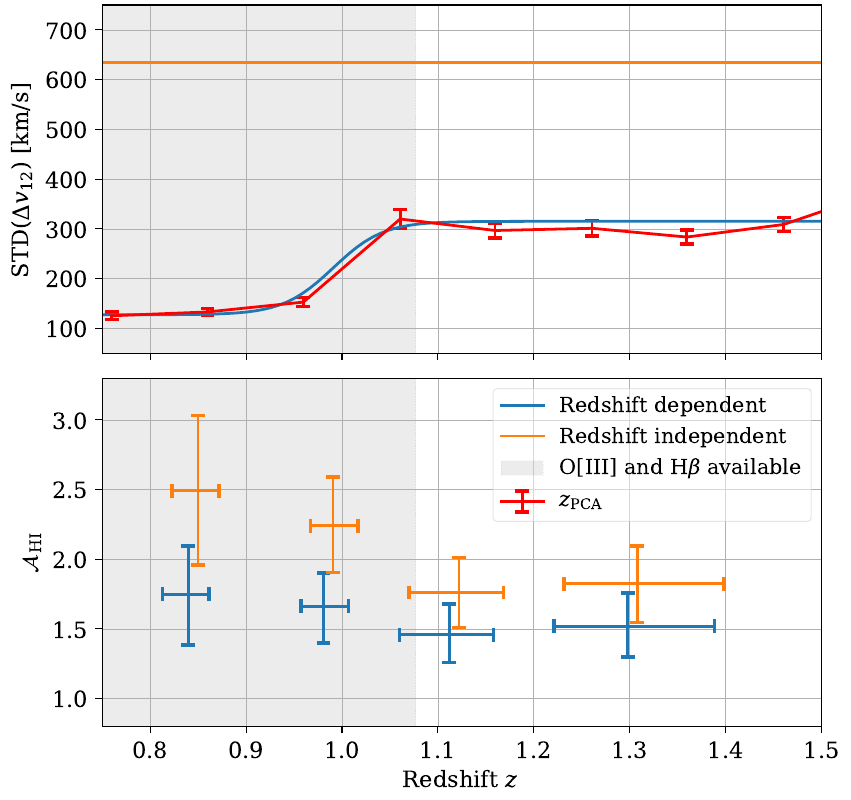}
    \caption{The QSO amplitude constraints are strongly dependent on the model for the QSO redshift errors. In the top panel we show the statistical error on the QSO $z_\mathrm{PCA}$ redshift estimates given by differences between repeated measurements of the same source. These estimates improve at $z\lesssim 1.1$ (grey shaded region) due to the availability of the O[III] and H$\beta$ lines. The measured distribution (red line) is taken from \citet[Fig.~9]{lyke2020}, the orange line shows the standard deviation of the published redshift independent error model \citep[Eq.~A1]{lyke2020}, and the blue line is a redshift dependent model described in the text, that gives a plausible fit to the $z_\mathrm{PCA}$ errors. In the lower panel we show the amplitude constraints (with fixed non-linear parameters) for assuming each of these redshift error models. The shifts are significant and are redshift dependent. This suggests that the modelling of the QSO redshift errors is a larger source of uncertainty than the statistical error in our measurements.}
    \label{fig:redshift_error}
\end{figure}

In \cref{fig:redshift_error} we show the change in our constraints that occur if we switch to this modified model. The top panel compares the redshift dependence of our new model to the \cite{lyke2020} model and the measured uncertainties in their Fig.~9. The lower panel shows the change in the inferred amplitude, $\AHI$, between the two models. We have fixed the non-linear parameters in these constraints which gives an indication of the statistical error on our constraints. At all redshifts the difference between the published model and our alternative is larger than the statistical uncertainty on $\AHI$, and suggests that redshift error distribution is significant source of systematic uncertainty in our analysis. Future analyses will need to resolve the questions in this modelling to make precision constraints on $\AHI$.

We use the differences observed in \cref{fig:redshift_error} to estimate a systematic uncertainty from the redshift error modelling of $\lv \AHI^\text{alt} - \AHI^\text{norm} \rv / \sqrt{2}$, where the $\sqrt{2}$ comes from an argument that the models considered are samples from some distribution of plausible models.

\subsection{$\AHI$ constraints and sources of error}
\label{sec:AHI_errors}

We are interested in learning about the amplitude of fluctuations in the HI distribution, which is probed most effectively by the parameter $\AHI$ (see Eq.~\ref{eq:AHI-definition}) in our analysis. In \secref{sec:constraints} we discuss the constraints on $\AHI$ in the case where we allow the full set of parameters to vary, and where we pin the non-linear parameters to their fiducial values.

The uncertainty in the case with fixed non-linear parameters is dominated by the statistical uncertainty in the data, and from the prior on the galactic bias. We assume that this statistical contribution is the same in the case where we allow the non-linear parameters to vary, with the weaker constraints coming from modelling uncertainties. In this case, and assuming the modelling errors are multiplicative within the degenerate regions of parameter space, we can roughly separate the uncertainty in the full parameter constraints into statistical and modelling contributions.

There are many potential additional sources of systematic errors in our analysis that have been discussed beyond the modelling uncertainty. These are listed, along with the statistical and modelling uncertainty breakdown, in \cref{tab:systematics}. This error budget is dominated by the modelling uncertainties; however, both the systematic error added to cover unexpected validation failures (labelled \emph{Consistency} in \cref{tab:systematics} and discussed in \secref{sec:consistent_even_odd}), and the error from uncertainties in the quasar redshift error model (\secref{sec:quasar_redshift_errors}) are larger than the statistical error, and thus could be the limiting sources if the modelling of non-linear scales can be improved.

\begin{deluxetable*}{c C CCC C CCCCCCC}[htb]
    \tablecaption{Sources of uncertainty.\label{tab:systematics}}
    \tablecolumns{10}
    \tablewidth{0.8\linewidth}
    \tablehead{
        \colhead{Tracer} &
        \multicolumn{9}{c}{Fractional errors [\%]}
        \\
        \cmidrule(lr){2-10}
        &
        \colhead{Statistical} &
        \colhead{Modelling} &
        \colhead{Flux} &
        \colhead{Template} &
        \colhead{Consistency} &
        \colhead{Beam} &
        \colhead{Linearity} &
        \colhead{Redshift errors} &
        \colhead{Total}
    }

    \startdata
   LRG &      14 &        150 &     4 &    1 &     0 &     8 &          1 &       0 &  151 \\
   ELG &      18 &         93 &     4 &    0 &     0 &     8 &          2 &       0 &   95 \\
   QSO &      10 &         49 &     4 &    0 &     0 &     8 &          4 &      14 &   52 \\
    \hline
 QSOb0 &      13 &         73 &     4 &    0 &    25 &     8 &          2 &      24 &   82 \\
 QSOb1 &      14 &         76 &     4 &    0 &     0 &     8 &          4 &      13 &   79 \\
 QSOb2 &      15 &         73 &     4 &    0 &     0 &     8 &          5 &      13 &   76 \\
    \hline
QSOb00 &      21 &        191 &     4 &    0 &     0 &     8 &          1 &      25 &  194 \\
QSOb01 &      15 &         76 &     4 &    0 &    25 &     8 &          2 &      21 &   85 \\
    \enddata
    \tablecomments{In this table we quantify the sources of error in our measurement. From left to right the sources are:
    \emph{Statistical}, inferred from the constraints with fixed non-linear parameters;
    \emph{Modelling} is the symmetrised error from the constraints varying all parameters, after removing the statistical contribution;
    \emph{Flux} is from uncertainty in the absolute flux scale (\secref{sec:real_time});
    \emph{Template} is from errors in the template calculation (\secref{sec:template} and Appendix~\ref{app:template_calculation});
    \emph{Consistency} gives systematic errors inferred from issues observed in data validation (\secref{sec:consistent_even_odd});
    \emph{Beam} lists the uncertainties from an imperfect beam model (\secref{sec:beamcalerrors});
    \emph{Linearity} gives a systematic error to incorporate the effect of signal loss during our analysis that is not fully captured by our template calculation (\secref{sec:linearity}); and
    \emph{Redshift errors} adds a systematic error to account for the difference in inferred amplitudes across plausible alternatives to the quasar redshift error model (\secref{sec:quasar_redshift_errors}). The final column, \emph{Total}, combines the extra sources of systematic error with those from the full parameter constraints to give an estimate of the symmetrized fractional error.}
\end{deluxetable*}

In \cref{tab:constraints}, we summarise the constraints on $\AHI$ for all the tracer catalogs. We show the case with both the non-linear parameters fixed, and when the full set is allowed to vary, illustrating again the substantial increase in the uncertainties from these parameters. We also give a final case including the total error budget from all the systematic contributions above (we have assumed they are all multiplicative effects). As the modelling uncertainties are dominant, including these extra sources of error gives only marginal increases to the total uncertainty. The most severely affected catalogs are QSOb0 and QSOb01, due to the systematic error contributions from both issues in the quasar redshift error model (which is worse at low redshifts) and from the consistency test failures.

\begin{deluxetable}{c C CCCC}[htb]
    \tablecaption{Parameter constraints for each tracer.\label{tab:constraints}}
    \tablecolumns{6}
    \tablehead{
        \colhead{Tracer} &
        \colhead{$\zeff$} &
        \multicolumn{4}{c}{$\AHI$}
        \\
        \cmidrule(lr){3-6}
        &
        &
        \colhead{Fiducial} &
        \colhead{Fixed NL} &
        \colhead{Full NL} &
        \colhead{Full + systematics}
    }
    \startdata
    LRG &   0.84 &   1.13 &  1.82_{-0.25}^{+0.26} &  1.51_{-0.96}^{+3.60} &  1.51_{-0.97}^{+3.60} \\
    ELG &   0.96 &   1.21 &  2.35_{-0.42}^{+0.43} &  6.76_{-3.74}^{+9.01} &  6.76_{-3.79}^{+9.04} \\
    QSO &   1.20 &   1.37 &  1.86_{-0.17}^{+0.18} &  1.68_{-0.60}^{+1.06} &  1.68_{-0.67}^{+1.10} \\
    \hline
  QSOb0 &   0.97 &   1.22 &  2.27_{-0.28}^{+0.31} &  2.04_{-0.94}^{+2.09} &  2.04_{-1.19}^{+2.21} \\
  QSOb1 &   1.12 &   1.31 &  1.75_{-0.25}^{+0.25} &  2.89_{-1.36}^{+3.13} &  2.89_{-1.44}^{+3.17} \\
  QSOb2 &   1.30 &   1.43 &  1.81_{-0.28}^{+0.27} &  1.63_{-0.86}^{+1.55} &  1.63_{-0.90}^{+1.57} \\
    \hline
 QSOb00 &   0.84 &   1.14 &  2.49_{-0.54}^{+0.52} &  1.49_{-1.65}^{+4.06} &  1.49_{-1.69}^{+4.08} \\
 QSOb01 &   0.99 &   1.23 &  2.23_{-0.34}^{+0.35} &  3.23_{-1.56}^{+3.47} &  3.23_{-1.91}^{+3.64} \\
    \enddata

    \tablecomments{After reparameterisation to avoid degeneracies, the physically interesting parameter is the \tcm amplitude $\AHI$. We show the highest-posterior-density 68\% credible intervals for these parameters for both a prior with the non-linear parameters \emph{fixed} and for the \emph{full} parameter space. Comparing the $\AHI$ constraints for the cases of fixed and varying non-linear parameters, we can see that there is a substantial increase in the uncertainty from modelling the small-scale structure. We also show estimates for $\AHI$ including the effects of the systematic errors listed in \cref{tab:systematics}. As the modelling errors are large, the additional uncertainty from this is small.}
\end{deluxetable}

\subsection{$\OmegaHI$ comparisons}
\label{sec:OmegaHIresults}

To be able to compare our results directly to measurements of $\OmegaHI$ from other probes, we need to be able to break the degeneracy between $\OmegaHI$ and $\bHI$. Although our measurements are unable to do this internally, and there are no external measurements of $\bHI$, we can use simulations as a guide.

As an indicator of the uncertainty on the bias, we use the bias measured at $z = 1$ from various simulations. \cite{villaescusa-navarro2018} use the IllustrisTNG hydrodynamic simulation and find that $\bHI(z=1) \approx 1.49$, and \cite{ando2019} use another hydrodynamic simulation, the Osaka simulation, to find that $\bHI(z=1) \approx 1.26$ (from their $b_0$ measurements). Another approach uses semi-analytic prescriptions on top of dark-matter-only simulations, such as \cite{spinelli2020} who find $\bHI(z=1) \approx 1.22$ or $1.31$ (depending if the Millennium I or II simulation is used), or \cite{wang2021} who use an empirically calibrated star formation model to find $\bHI(z=1) \approx 1.27$. Collectively these prescriptions have a mean of $\approx 1.3$ and a standard deviation of $\approx 0.1$. With this in mind, we place a simulation derived Gaussian prior on the bias with a conservative width of 20\%, i.e $\sigma_{\bHI} / \bHI^\fid = 0.2$.

\begin{figure*}[htbp]
    \centering
    \includegraphics[width=0.75\linewidth]{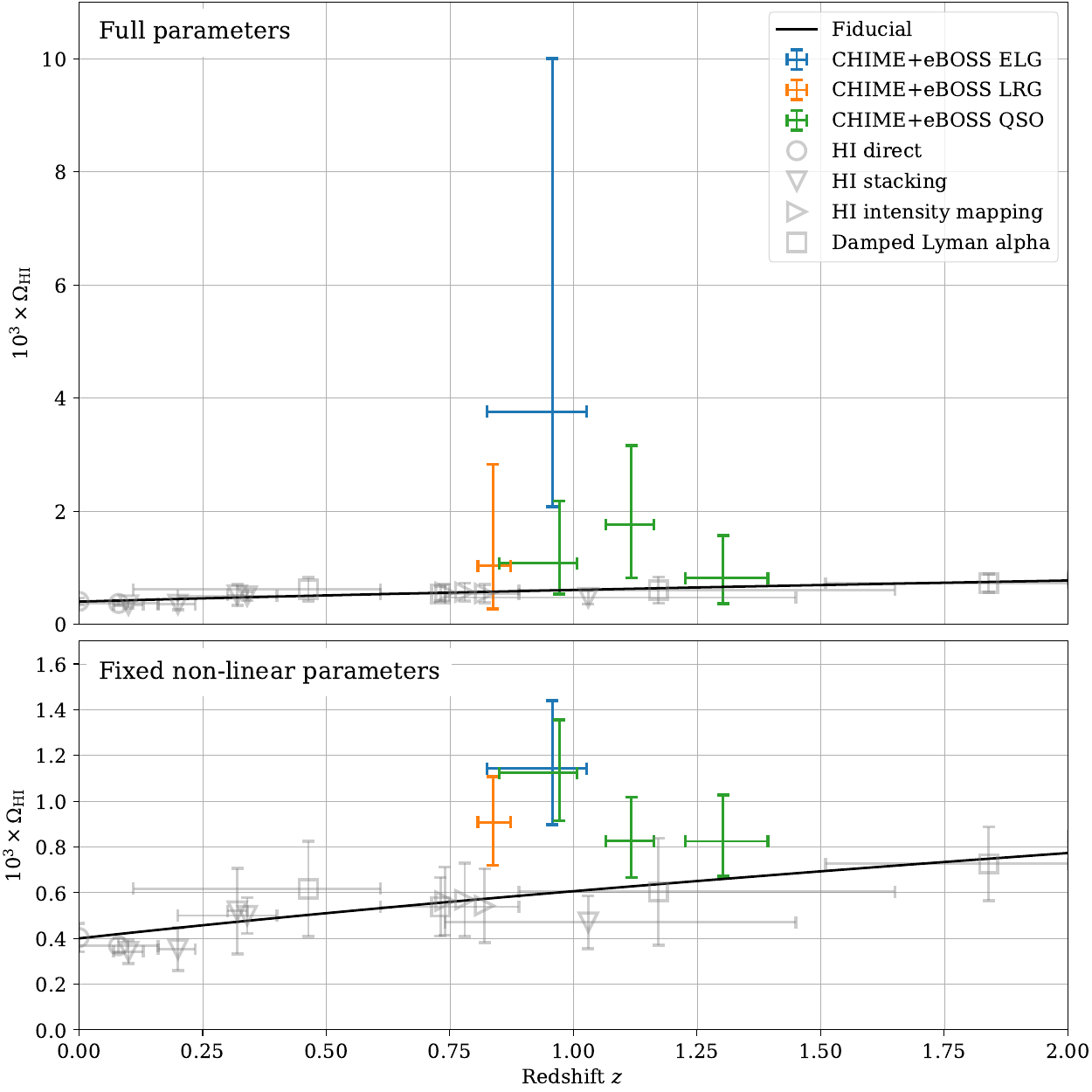}
    \caption{
        Constraints on $\OmegaHI$ from this analysis compared to other experiments. In the top panel we show the constraints in this work when varying the full set of modelling parameters, in the lower panel we fix the non-linear parameters which considerably reduced the uncertainties at the expense of hidden systematic errors. We have selected a representative sample of measurements using independent datasets to place in this figure. The datasets are of four types: at the lowest redshift there are direct \tcm observations, such as those from ALFALFA \citep{jones2018} and the Arecibo Ultra Deep Survey \citep{xi2020the}; at intermediate redshifts source stacking of individual galaxies such as \cite{rhee2013} who use Westerbork data and low redshift galaxies observed with CFHT-MOS, and three studies combining GMRT radio data with different optical catalogs, VVDS optical data taken at VIMOS \citep{rhee2018}, DEEP2 and DEEP3 at low redshift \citep{bera2019}, and DEEP2 at high redshifts \citep{chowdhury2020}; also at intermediate redshifts are HI intensity mapping cross correlations like \cite{wolz2021} who cross correlate GBT intensity mapping data against eBOSS and WiggleZ catalogs; at the highest redshifts, measurements are from surveys of Damped Lyman-alpha systems such using HST ACS and Galex data \citep{rao2017} and using ESO UVES \citep{zafar2013}.
    }
    \label{fig:omegaHI_comparison}
\end{figure*}

We re-weight the MCMC chains from our analysis to apply the updated prior on $\bHI$ and marginalise over all the other parameters to derive constraints on $\OmegaHI$. We give our measurements as the highest posterior density credible interval about the mode of the distribution. In \cref{fig:omegaHI_comparison} we show the measurements for the LRG and ELG samples, as well as the QSOs split across three redshift bins, compared to measurements from other experiments.

There are four main methods for measuring $\OmegaHI$ that we include in \cref{fig:omegaHI_comparison} for comparison:
\begin{description}
    \item[Direct HI surveys] At the lowest redshifts, blind surveys of the \tcm line can measure the HI mass function directly which can be integrated to obtain estimates of $\OmegaHI$.
    \item[HI stacking] At intermediate redshifts, it is difficult to detect individual galaxies in their \tcm emission; to get around this, high-resolution radio data can be stacked on the positions of galaxies found in optical catalogs to get an estimate of the average amount of HI per galaxy in the sample. This can then be combined with an optical luminosity function for the sample, and corrected for completeness to give an estimate of $\OmegaHI$.
    \item[HI intensity mapping] Another method is to cross correlate HI intensity mapping data with optical catalogs. These are distinct from the HI stacking measurements described above in that they do not resolve the emission in the (average of) individual galaxies, but instead are sensitive to the correlated HI mass in the vicinity of the galaxy. Our results are an example of this technique.
    \item[Damped Ly$\alpha$] At this highest redshifts Damped Ly$\alpha$ systems are detected in optical and UV quasar spectra, and the distribution of their observed column densities can be integrated to find $\OmegaHI$.
\end{description}
For all measurements, we convert into the Planck 2018 cosmology used in this paper. In each case, the measurements are effectively a flux like quantity, that is multiplied by an area to give an HI mass, divided by a survey volume to give a density and then divided by the critical density to give $\OmegaHI$, though some of these steps are implicit (this is still true for the Damped Ly$\alpha$ analysis, though the ``area'' in the mass is cancelled with the one implicit in the volume). If we approximate the observations as coming from a narrow band in redshift, then the cosmology dependence is
\begin{equation}
    \OmegaHI(z) \propto \frac{1}{H_0^2} \frac{dz}{d\chi} \; .
\end{equation}
For the \cite{wolz2021} intensity mapping points we convert their $\OmegaHI \bHI r$ measurements into $\OmegaHI$ constraints with the fiducial bias model we use in this paper to allow a consistent comparison.

In \cref{fig:omegaHI_comparison} for the constraints both when varying the non-linear parameters (top panel) and when fixing them (lower panel) our results are in broad agreement with other $\OmegaHI$ constraints. As we would expect the uncertainties are much larger when allowing the non-linear parameters to vary though the distributions are non-Gaussian and the probability of $\OmegaHI \le 0$ is still negligible. We note that while we expect all points to be pushed towards higher values of $\OmegaHI$ by the prior volume effect in the FoG parameters discussed in \secref{sec:constraints}, the CHIME+eBOSS ELG point is noticeably discrepant when the non-linear parameters are varied. We believe this is a chance fluctuation where the region further along the $\AHI$-$\alphaFoG{+}$ degeneracy is preferred, and leads to a $\sim 2 \sigma$ shift from the $\OmegaHI$ values preferred by the other tracers. This can also be seen clearly in \cref{fig:elg_reduced_params} where the preferred range of $\alphaFoG{+}$ values is higher than in the QSO and LRG cases (\cref{fig:qso_reduced_params} and \cref{fig:lrg_reduced_params} respectively). When fixing the non-linear parameters the ELG constraints are much more consistent with both the other CHIME tracers and the external datasets.

As the constraints with fixed non-linear parameters do not include the full modelling uncertainties they show the internal consistency and significance of our measurements, but are not good indicators of the plausible range of $\OmegaHI$ determined from our data. In all cases we are showing constraints derived with the fiducial eBOSS quasar error model. As discussed in \secref{sec:quasar_redshift_errors} we believe that this model may bias the quasar constraints (particularly the lowest redshift bin) to higher values of $\OmegaHI$.

\subsection{Atomic hydrogen content of galaxies and quasars}
\label{sec:M10results}

As mentioned previously, our stacking analysis probes not only the correlated clustering of eBOSS catalog objects and HI, but is also sensitive to the HI associated with the objects themselves, which sets the value of our $\Mten$ parameter for each sample (recall that $\Mten$ is defined as the mean HI mass per catalog object, in units of $10^{10}\:\si{\Msolar}$). \Cref{fig:qso_reduced_params,fig:elg_reduced_params,fig:lrg_reduced_params} show that the posteriors for $\Mten$ peak at nonzero values for the QSO and LRG stacks, while for the ELG stack the posterior peaks at $\Mten=0$. In each case, however, the model where $\Mten$ and the other nonlinear parameters ($\alphaFoG{+}$ and $\alphaNL$) are allowed to vary is not strongly preferred over the case where these parameters are fixed to their fiducial values (see \secref{sec:significance}); thus, we cannot interpret the posteriors of $\Mten$ as providing definitive information about the HI content of the objects in each catalog.

Nevertheless, the finite width of these posteriors indicates that future analyses may hold the promise of interesting constraints. In particular, for the ELG stack, the highest-posterior-density 68\% credible interval is $\Mten < 1.04$. This is consistent with the simulations of \cite{wolz2021}, which were based on the \texttt{DARK SAGE} semi-analytical galaxy evolution model \citep{stevens2016} and predicted a shot noise contribution to the HI-ELG cross-power spectrum equivalent to $\Mten \approx 0.8$ (as inferred from their Fig.~12). It is also consistent with the analysis of \cite{chowdhury2020}, who stacked GMRT \tcm observations on star-forming galaxies from the DEEP2 survey and found $\Mten = 1.19 \pm 0.26$ at an effective redshift $z_{\rm eff} = 1.03$. A cross-correlation analysis with greater power to break the parameter degeneracies in our model would likely improve the constraint on $\Mten$ to a level where it could fruitfully be compared with these other values.

Empirical information on the HI content of LRGs at $z \sim 1$ is scarce: direct stacking analogous to \cite{chowdhury2020} has only been carried out at lower redshifts for such red galaxies (e.g.\  \citealt{rhee2018}). Thus, constraints on $\Mten$ for LRGs (and QSOs) would provide valuable information about the evolution and environments of these objects. On the other hand, inclusion of an external prior on $\Mten$, obtainable from, for example, stacking GMRT observations on a subset of objects from each eBOSS catalog, would help to break the degeneracies in our model (or other, more detailed models of HI-galaxy cross-correlations), and we see this as a promising avenue for future investigation.

%% file: sections/conclusions.tex

\section{Conclusions}
\label{sec:conclusions}

In this paper, we have presented the first detection of cosmological \tcm emission with the CHIME telescope. This detection is the result of constructing sky maps from CHIME data, filtering and cleaning these maps in various ways, and performing a cross-correlation analysis with catalogs of galaxy and quasar positions from the eBOSS survey. We have described several aspects of CHIME data processing that have not previously appeared in the literature: these include our procedures for combining multiple sidereal days of observations (\secref{sec:sidereal_avg}), forming beam-deconvolved sky maps from measured visibilities (\secref{sec:mapmaking}), measuring delay power spectra using Gibbs sampling (Appendix~\ref{sec:delay_spectrum}),
and inferring the primary beam pattern based on external measurements of many radio point sources (Appendix~\ref{sec:beam_calibration_ptsrc}).

We have filtered bright foregrounds  out of the measurements with a high-pass delay filter using the approach of \cite{ewall-wice2021}, with a declination-dependent delay cutoff that selects the regime where the fluctuations in the data are close to the expected noise level. This filtering has the effect of removing any sensitivity to linear cosmological scales related to baryonic acoustic oscillations, such that the signal-to-noise is concentrated at nonlinear scales ($0.3h\,{\rm Mpc}^{-1} \lesssim k \lesssim {\rm few}\, h\,{\rm Mpc}^{-1}$; see \cref{fig:k_sensitivity}).

We perform the cross-correlation by separately stacking CHIME sky maps at the angular and spectral locations of the objects in eBOSS catalogs of ELGs, LRGs, and QSOs. In each case, the spatial extent of the signal is consistent with an unresolved point source (\cref{fig:stack2d}), so we present our main results as one-dimensional stacking profiles as a function of frequency offset from the locations of the catalog objects (\cref{fig:stack1d}). We achieve significant detections for each catalog, as indicated by  Bayes factors $\mathcal{Z}_1 / \mathcal{Z}_0$  of $\ln{(\mathcal{Z}_1 / \mathcal{Z}_0)} \approx 18.8$ (LRGs), $10.8$ (ELGs), and $56.3$ (QSOs), computed by comparing our signal model with a noise-only model; alternatively, a frequentist likelihood ratio test gives signal-to-noise ratios of $7.1$ (LRGs), $5.7$ (ELGs), and $11.1$ (QSOs).

We interpret these measurements using a simulation-based framework (\secref{sec:simulations} and~\ref{sec:template}), within a model that considers HI and galaxies to be linearly biased tracers of the underlying matter distribution, including the leading effects of redshift-space distortions and a correlated shot noise contribution related to the mean HI mass of the objects in each catalog (\secref{sec:skymaps_signal}). We are able to constrain an effective HI clustering amplitude $\AHI \equiv 10^3 \, \OmegaHI (\bHI + \langle f\mu^2 \rangle)$, where $\OmegaHI$ is the cosmic abundance of HI, $\bHI$ is the linear bias of HI, and $\langle f\mu^2 \rangle$ (equal to $0.552$ in this analysis) is an average over the linear growth rate $f$ and an angular factor $\mu^2$ related to the line-of-sight components of the Fourier modes probed in the stacks (\secref{sec:modelfitting}). We constrain this amplitude separately for each eBOSS catalog, marginalizing over parameters controlling the scale dependence of non-linear clustering, obtaining $\AHI=1.51_{-0.97}^{+3.60}$ (LRGs), $\AHI=6.76_{-3.79}^{+9.04}$ (ELGs), and $\AHI=1.68_{-0.67}^{+1.10}$ (QSOs). (See Table~\ref{tab:constraints}.) Previous cross-correlations between GBT \tcm maps and galaxy catalogs have measured $\OmegaHI \bHI r$ (where $r$ is a phenomenological cross-correlation parameter) with 15\% to 25\% precision \citep{chang2010,masui2013,wolz2021}; our constraints on $\AHI$ are weaker than this, but only due to our more detailed modelling of small-scale clustering, which requires marginalization over several parameters.

We also constrain an overall frequency offset $\Delta\nu$ of the stacking profile. This offset is consistent with zero for ELGs and LRGs, while for QSOs we find $\Delta\nu \approx -\SI{0.2}{\mega\hertz}$. We interpret this as a systematic bias in the measured redshifts of the QSOs, corresponding to $\Delta{v} \approx \SI{-66}{\kilo\metre\per\second}$ in velocity units. As discussed in \secref{sec:quasar_redshift_errors}, this is consistent with what was found by the eBOSS team in \cite{lyke2020}.

Our results point to several interesting directions for future investigation. Our present analysis only considered CHIME frequencies above \SI{585}{\mega\hertz}, corresponding to redshifts less than \SI{1.42}, but the eBOSS QSO catalog contains a significant number of QSOs at higher redshift (see \cref{fig:zdist}), and it would be worthwhile to repeat the stacking procedure using these objects, after additional effort to remove transient RFI in CHIME data at the relevant frequencies. Also, similar future analyses have the potential to constrain the mean HI mass per catalog object. This would provide opportunities for coordination with stacking analyses from higher-resolution interferometers like GMRT (e.g.\ \citealt{chowdhury2020}), which could help to disentangle the contributions from large-scale structure and correlated shot noise, and also provide new information about the evolution and properties of galaxy and quasar samples. In parallel, future cross-correlation analyses could be used to obtain more detailed information about systematic errors in spectroscopic redshifts obtained from optical instruments.

More broadly, many of the methods developed for this analysis are not specific to CHIME, but could also be applied to other low-redshift interferometric \tcm surveys, such as CHORD \citep{vanderlinde2019}, Tianlai \citep{li2020-tianlaicyl,wu2021-tianlaidish}, HIRAX \citep{crichton2021}, uGMRT \citep{chakraborty2021}, and the Ooty Wide Field Array \citep{subrahmanya2017}, as well as
higher-redshift surveys like HERA \citep{deboer2017} and
potential future projects \citep{ansari2018}.

Finally, we note that this paper has made use of only a small fraction of the total amount of data collected by CHIME in the last three years. Future improvements in data processing will be focused not only on enabling much more detailed cross-correlation measurements, but also on the ultimate goal of measuring baryon acoustic oscillations in the auto-power spectrum of \tcm emission, providing important clues as to the nature of dark energy and the  properties of the low-redshift universe.

%% file: sections/acknowledgment.tex

\acknowledgments

We thank \'{E}tienne Burtin for useful discussions.


We thank the Dominion Radio Astrophysical Observatory, operated by the National Research Council Canada, for gracious hospitality and expertise. The DRAO is situated  on the traditional, ancestral, and unceded territory of the Syilx Okanagan people.   We are fortunate to live and work on these lands.
%

CHIME is funded by  grants from the Canada Foundation for Innovation (CFI) 2012 Leading Edge Fund (Project 31170), the CFI 2015 Innovation Fund (Project 33213), and by contributions from the provinces of British Columbia, Qu\'ebec, and Ontario. Long-term data storage and computational support for analysis is provided by WestGrid\footnote{\url{https://www.westgrid.ca}}, SciNet\footnote{\url{https://www.scinethpc.ca/}} and Compute Canada\footnote{\url{https://www.computecanada.ca}}, and we thank their staff for flexibility and technical expertise that has been essential to this work, particularly Martin Siegert, Lixin Liu, and Lance Couture.

Additional support was provided by the University  of British Columbia, McGill University, and the University of Toronto. CHIME also benefits from NSERC Discovery Grants to several researchers,  funding from the Canadian Institute for Advanced Research (CIFAR), and from the Dunlap Institute for Astronomy and Astrophysics at the University of Toronto, which is funded through an endowment established by the David Dunlap family.
This material is partly based on work supported by the  NSF through   grants (2008031)  (2006911) and  (2006548) and by the Perimeter Institute for Theoretical Physics, which in turn is supported by the Government of Canada through Industry Canada and by the Province of Ontario through the Ministry of Research and Innovation.

We thank the Sloan Digital Sky Survey and eBOSS collaborations for publicly releasing the galaxy and quasar catalogs and supporting mock catalogs used in this work.
Funding for the Sloan Digital Sky  Survey IV  has been provided by the  Alfred P. Sloan Foundation, the U.S. Department of Energy Office of  Science, and the Participating  Institutions. SDSS-IV acknowledges support and resources from the Center for High Performance Computing  at the  University of Utah. The SDSS website is www.sdss.org.

\software{
bitshuffle \citep{2015bitshuffle},
CAMB \citep{lewis1999},
caput \citep{caput},
ch\_pipeline \citep{ch_pipeline},
cora \citep{cora},
Cython \citep{Cython},
draco \citep{draco},
driftscan \citep{driftscan},
emcee \citep{foremann-mackey2013},
GetDist \citep{GetDist},
hankl \citep{karamanis2021},
h5py \citep{h5py},
HDF5 \citep{HDF5},
HEALPix \citep{HEALPix},
healpy \citep{healpy},
Matplotlib \citep{Matplotlib},
mpi4py \citep{mpi4py},
networkx \citep{NetworkX},
NumPy \citep{NumPy},
OpenMPI \citep{OpenMPI},
pandas \citep{pandas,pandas_paper},
peewee \citep{Peewee},
SciPy \citep{SciPy},
Skyfield \citep{Skyfield},
}


%% file: appendix/delay_spectrum.tex

\section{Delay Power Spectrum Estimation via Gibbs Sampling}
\label{sec:delay_spectrum}

Delay power spectra%
\footnote{For clarity, we will use \emph{delay spectrum} to refer only to the direct Fourier transform of a frequency \emph{intensity} or \emph{flux} spectrum. The \emph{delay power spectrum} will refer only to the variance of this quantity. Though the intensity and flux are both second-order statistics of the electric field and thus are power-like quantities in a physical sense, we do not think this is ambiguous anywhere in this text.}
measure the power at different time lags observed within a frequency spectrum, and are an extremely powerful tool for investigating instrumental effects as well as the frequency structure of radio emission from the sky (see \cref{fig:delay_spectrum} for an example). Superficially estimating a delay power spectrum involves taking a Fourier transform of a frequency spectrum and estimating the resulting power in the time domain. However, in the presence of interference that causes certain frequencies to be masked out, and a large dynamic range between the power at different delays, significant care must be taken to avoid mixing of power between different delays. There are several existing strategies for dealing with this such as using a CLEAN-like algorithm in delay space \citep{Parsons2014}, and Least-squares Spectral Analysis \citep[LSSA, see][]{Vanicek1969,Trott2016}.

To understand the challenges involved, consider a noisy observation $\vf$ of a frequency spectrum with length $N_\mathrm{f}$. This is related to an underlying delay spectrum $\vd$ by
\begin{equation}
    \label{eq:def_fi}
    \vf = \mF \vd + \vn
\end{equation}
with noise $\vn$ and where the delay spectrum is assumed to be drawn from the input delay power spectrum $D[\tau_a]$:
\begin{equation}
D[\tau_a] = \left\langle \big| d_a \big|^2 \right\rangle\ .
\end{equation}
The noise is described by covariance $\mN$, and for the moment we treat the noise as being uniform except for entirely missing frequencies which we give infinite noise. We write this as $\mN^{-1} = \sigma^{-2} \mM$ where $\mM$ is a diagonal masking matrix with ones for included frequencies and zeros for missing frequencies. The matrix $\mF$, with  $F_{ab} = e^{-2 \pi j \tau_a \nu_b} / N_\mathrm{f}^{1/2}$, is unitary and performs a discrete Fourier transform from the time (delay) domain to the conjugate frequency domain.

A first attempt to estimate the delay power spectrum might start by simply applying the mask $\mM$ to the observed frequency spectrum and performing an inverse Fourier transform,
\begin{equation}
    \hat{\vd}_\text{inv} = \mF^\hconj \mM \vf \ .
\end{equation}
With this estimate of the delay spectrum, we can then infer the delay power spectrum $D[\tau_a]$ by using a variance over $N_\mathrm{obs}$ observations, indexed by $i$:
\begin{equation}
    \label{eq:D_samplevar}
    \hat{D}[\tau_a] = \frac{1}{N_\mathrm{obs}} \sum_{i=1}^{N_\mathrm{obs}} \lv \hat{d}^i_a \rv^2 \ ,
\end{equation}
with $\hat{d}^i_a$ being any estimator for $d_a$ such as $\hat{\vd}_\text{inv}$ defined above (later we will introduce additional estimators). However, this procedure generates significant leakage between delay channels, with a delay spread function given by the matrix $\mF^\hconj \mM \mF$. In the case of random masking of $N_\text{masked}$ single frequencies, it can be shown that this gives leakage at the level of $\sim N_\text{masked} \sum_a D[\tau_a] / N_\mathrm{f}^2 $ uniformly across delays, and we are not able to see any structure in the delay power spectrum below this level.

To improve this, we could modify the delay spectrum estimate by deconvolving the delay spread function by its pseudo-inverse, or equivalently use a maximum-likelihood estimator\footnote{We note that this is similar to LSSA, though LSSA considers more general cases such as irregular sampling, and typically restricts the range of delays being solved for to minimise correlations and leakage.}
\begin{equation}
    \hat{\vd}_\text{ml} = (\mF^\hconj \mN^{-1} \mF)^+ \mF^\hconj \mN^{-1} \, \vf \ .
\end{equation}
However, as $\mF$ is unitary, the pseudo-inverse $(\mF^\hconj \mN^{-1} \mF)^+$ is equal to $\mF^\hconj (\mN^{-1})^+ \mF$, and as the noise matrix is diagonal with zeros where samples are masked, $(\mN^{-1})^+ \mN^{-1} = \mM$; together, these imply that $\hat{\vd}_\text{ml} = \hat{\vd}_\text{inv}$. In words, the maximum-likelihood estimator is exactly equivalent to inverse Fourier transforming the masked frequency spectra.

Another option is using a Wiener filter instead of a maximum-likelihood type filter:
\begin{equation}
    \label{eq:vd_wiener}
   \hat{\vd}_\text{w} = (\mD^{-1} + \mF^\hconj \mN^{-1} \mF)^{-1} \mF^\hconj \mN^{-1} \vf \ ,
\end{equation}
where $D_{ab} = D[\tau_a] \delta_{ab}$ is the covariance matrix of the delay spectrum signal. By providing information about the distribution of power at various delays, the filter can distinguish delays related to true signal in the masked frequency spectra, resulting in delay spectra with significantly lower leakage and hence cleaner power spectra. However, constructing this requires that we already know the delay power spectrum $D[\tau_a]$, which is the quantity that we are trying to estimate. A close enough guess may minimise the leakage enough to produce accurate delay power spectrum estimates, but there is no knowing in advance if this is the case.

A resolution to this is to jointly solve for both the delay spectrum and the delay power spectrum, a problem which is tractable by Gibbs sampling \citep{Geman1984}, a Markov Chain Monte Carlo technique for drawing samples from a joint distribution where the conditional distributions are easily sampled. In particular, we draw inspiration from techniques used for power spectrum estimation of the Cosmic Microwave Background (e.g.\ \citealt{eriksen2004,wandelt2004}).

We want to infer both the delay spectrum $\vd$ and the delay power spectrum (equivalent to the diagonal matrix $\mD$) by drawing samples from the joint probability distribution $\calP(\vd, \mD \mid \vf)$. Gibbs sampling allows us to do that by alternately drawing from the conditional distributions $\calP(\vd \mid \mD, \vf)$ and $\calP(\mD \mid \vd, \vf)$; the ensuing set of samples will eventually converge to the joint distribution, and we can take the mean over $\mD$ samples as an estimate of the delay power spectrum. We now describe how to sample from each conditional distribution.

Starting with $\calP(\vd \mid \mD, \vf)$, we can use Bayes' theorem to write
\begin{equation}
    \calP(\vd \mid \mD, \vf) \propto \calP(\vf \mid \vd, \mD) \, \calP(\vd \mid \mD) \ .
\end{equation}
The first term on the right hand side is the likelihood function for the frequency spectrum, which for Gaussian noise can be written as $\calP(\vf \mid \vd, \mD) = \calG_C(\vf - \mF \vd, \mN)$, where $\calG_C$ is a circularly symmetric complex Gaussian distribution:
\begin{equation}
    \calG_C(\vz, \mC) = \frac{1}{\lv \pi \mC \rv} e^{-\vz^\hconj \mC^{-1} \vz} \ .
\end{equation}
We will also model the conditional prior distribution for the delay spectrum as Gaussian, with $\calP(\vd \mid \mD) = \calG_C(\vd, \mD)$. Combining these together and grouping the terms in $\vd$, we find that the conditional distribution is
\begin{equation}
    \calP(\vd \mid \mD, \vf) = \calG_C(\vd - \hat{\vd}_\text{w}, \mC)\ ,
\end{equation}
where $\mC^{-1} = \mD^{-1} + \mF^\hconj \mN^{-1} \mF$. Thus, the mean of the conditional distribution is just the Wiener filter of \cref{eq:vd_wiener}, with the standard covariance. Although drawing from this can be done by solving for the mean, followed by inversion and factorization of $\mC^{-1}$ to add a random fluctuation, it is more efficiently done by constructing
\begin{equation}
    \mC^{-1} \, \vd = \mF^\hconj \mN^{-1} \vf + \mD^{-1/2} \vw_1 + \mF^\hconj \mN^{-1/2} \vw_2
    \label{eq:specsample}
\end{equation}
where $\vw_1$ and $\vw_2$ are standard Gaussian random samples, and then solving for $\vd$ \citep{Jewell2004}.

The conditional distribution for the delay power spectrum is more straightforward. We wish to calculate the conditional distribution $\calP(\mD \mid \vd, \vf)$, which is independent of $\vf$ as all the information about $\mD$ is contained within $\vd$. Using a flat prior on the elements of $\mD$, and the prior $\calP(\vd \mid \mD)$ we find that $\calP(\mD \mid \vd, \vf) \propto \calG_C(\vd, \mD)$.
Assuming that $\mD$ is diagonal we can rewrite this in terms of the sample variance estimates $\hat{D}[\tau_a]$ for each delay $\tau_a$ which are sufficient statistics for the diagonal elements of $\mD$ itself, $D[\tau_a]$. The sample variance $\hat{D}[\tau_a]$ has a chi-squared distribution,
\begin{equation}
N_\mathrm{obs} \frac{\hat{D}[\tau_a]}{D[\tau_a]} \sim \chi^2(N_\mathrm{obs}) \ ,
\label{eq:powspecchi2}
\end{equation}
and so we can draw samples from $\calP(\mD \mid \vd) \propto \calP(\hat{D}[\tau_a] \mid D[\tau_a])$ by drawing a standard chi-squared deviate for each delay $x_a \sim \chi^2(N_\mathrm{obs})$ and setting the new sample for $D[\tau_a]$ to be $N_\mathrm{obs} \hat{D}[\tau_a] / x_a$.

Our practical implementation of this algorithm is as follows:
\begin{enumerate}
\item Pick a set of data whose delay spectra are expected to be similar enough that we can average over them. For computing delay spectra from visibilities, this might consist of all RA samples for individual baselines (after stacking over redundant copies). For the map delay spectrum described in \secref{sec:filter}, we choose the set of RA samples at each polarization and declination. As above, we use $N_\mathrm{obs}$ to denote the size of this set.
\item Apply an apodization window to each frequency spectrum, if desired. A Nuttall window is used \secref{sec:filter}.
\item Choose an initial guess $D_0[\tau_a]$ for the delay power spectrum. We use a white spectrum with amplitude \SI{10}{\jansky\per\beam} in this work.
\item Loop over the following steps until convergence has been achieved:
\begin{enumerate}
\item For each element $i$ of the set of $N_\mathrm{obs}$ spectra, draw the $n$th delay spectrum sample $\vd_n^i \leftarrow \calP(\vd \mid \vf, \mD_{n})$ using \cref{eq:specsample}. Note that each $\vd_n^i$ is a delay spectrum with $N_\mathrm{f}$ elements.
\item Draw the $(n+1)$th delay power spectrum sample $D_{n+1}[\tau_a] \leftarrow \calP(D[\tau_a] \mid \{\vd_n^i\}_{i=1}^{N_\mathrm{obs}})$, using the $N_\mathrm{obs}$ delay spectra drawn at step $n$ to compute $\hat{D}[\tau_a]$ in \cref{eq:powspecchi2}.
\end{enumerate}
\item Take the average of the converged samples, after removing burn-in and performing any necessary thinning. In \secref{sec:filter}, we halt after 100 samples, and take the median over the final 50 samples as an estimate of the delay power spectrum.
\end{enumerate}

In summary, the Gibbs sampling approach is a statistically well-motivated technique that iteratively deconvolves the delay spectra, uses them to update a delay power spectrum, and uses this to improve the next deconvolution round.

\begin{figure}[htbp]
    \centering
    \includegraphics[width=0.7\textwidth]{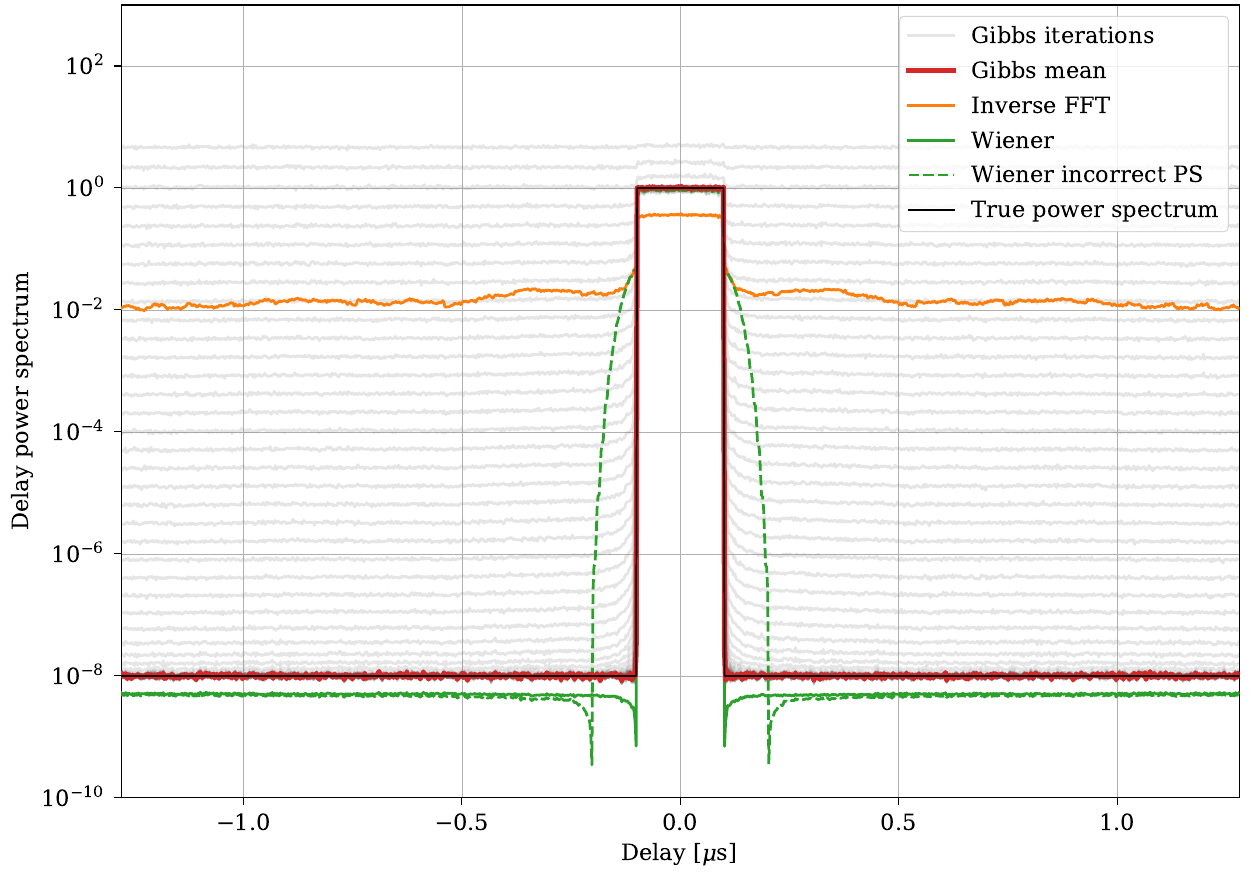}

    \caption{To test the performance of the delay power spectrum estimation techniques discussed in Appendix~\ref{sec:delay_spectrum}, we generate a set of random delay spectra with a true delay power spectrum (black line) consisting of very high power at low delays, and a plateau of low power outside this region. We Fourier transform these into frequency spectra and apply the CHIME RFI mask used in this analysis (\secref{sec:rfi}). In orange we show the direct inverse estimate, which has significant leakage at the $\sim 10^{-2}$ level. The Wiener filter estimate (green solid) produces a much closer estimate, correcting most of the leakage effects, at the expense of needing a good starting guess. If generated with a poor initial delay power spectrum (high power over twice the range of delays as the true power spectrum) the estimate is significantly worse (green dashed). The Gibbs sampler produces an estimate that is much closer to the true power spectrum (red). Like any MCMC scheme, attention must be paid to the convergence of the chain. We used 100 samples and derived our estimate from the median of the last half. The full chain is shown in gray and can be seen converging from a poor starting guess of a flat delay power spectrum to the true power spectrum over $\sim 30$ samples.}
    \label{fig:gibbs_example}
\end{figure}

In \cref{fig:gibbs_example} we apply the various estimators discussed above  to a synthetic dataset with high dynamic range in delay space and a realistic frequency mask. We clearly see that the Gibbs sampling based estimator is able to accurately recover the input spectrum, while the na\"{i}ve inverse-Fourier and Wiener estimators show various degrees of discrepancy.

%% file: appendix/beam_calibration.tex

\section{Estimating the Primary Beam by Deconvolving a Model for the Point Source Sky}
\label{sec:beam_calibration_ptsrc}

In this appendix, we describe the algorithm that is used to directly reconstruct the average primary beam pattern of the CHIME antennas.  First, a model for the radio emission from extragalactic point sources is constructed from measurements made by other telescopes.  The \emph{specfind v2} table \citep{vollmer2010} in the Vizier database is queried for flux measurements of all known sources between declinations \SIrange{-40}{85}{\degree}.  For each source, all available measurements of the flux are fit to a power-law with frequency
\begin{align}
    s(\nu) & = a \left(\frac{\nu}{\SI{600}{\mega\hertz}}\right)^{\gamma} \ ,
\end{align}
where the amplitude $a$ and exponent $\gamma$ are allowed to float.  The fit is done by performing a weighted linear regression of the logarithm of the flux to the logarithm of the frequency.  The uncertainties provided in the \emph{specfind v2} table are used to construct inverse variance weights.  These uncertainties are \SI{20}{\percent} of the measured flux \citep{vollmer2005}, and the power-law model is in general a good fit given these large uncertainties.  Only sources with $s(\SI{600}{\mega\hertz}) > \SI{15}{\milli\jansky}$ that have at least one measurement on either side of the CHIME band are included in the sky model.  There are \SI{97941} sources in total that meet these criteria.  All of the sources have at least 3 flux measurements, with 6 flux measurements on average.

Our model for the visibility measured by baseline $\vec{b}$ at frequency $\nu$ and local Earth rotation angle $\phi$ is then given by
\begin{align}
    \label{eq:sky_model}
    S(\vec{b}, \nu, \phi) & = \sum_{i} s_{i}(\nu) \ e^{j 2 \pi \nu \vec{b} \cdot \vec{\hat{n}}(\theta_{i}, \phi - \phi_{i}) / c} \ \delta\left(\phi, \phi_{i}\right)
\end{align}
where $s_{i}(\nu)$ is a power-law model for the flux of the $i$'th source, $\vec{\hat{n}'}(\theta_{i}, \phi - \phi_{i})$ is the unit vector pointing in the direction of the $i$'th source and is given by \Cref{eq:sky_unit_vector} with $\phi_{i}$ and $\theta_{i}$ denoting the source's right ascension and declination in CIRS coordinates, and
\begin{align}
    \delta\left(\phi, \phi_{i}\right) & = \begin{cases}
                                            1 & |\phi - \phi_{i}| < \frac{1}{2} \Delta \phi \\
                                            0 & \mbox{otherwise} \\
                                           \end{cases}
\end{align}
with $\Delta \phi = \SI{0.0879}{\degree}$ denoting the sample spacing of the data in local sidereal angle.  The sum in \Cref{eq:sky_model} runs over all sources.

The following identical operations are then performed on the sidereal visibilities $V$ and sky model $S$.  First we arrange the baselines onto a 2D grid and then beamform in the $\vec{\hat{y}}$ direction using \Cref{eq:beamform_ns}.  The weights used in the beamformer are given by
\begin{align}
    \label{eq:weights_ns_beamcal}
    w_{xy}^{p}(\nu, \phi) & = W_{\alpha}\left(\frac{1}{2} \left[1 + \frac{\nu y}{\nu_{\rm min} y_{\rm max}}\right]\right)
\end{align}
where $W_{\alpha}$ denotes the Dolph-Chebyshev window, $\alpha = \SI{60}{\decibel}$ is the peak-to-sidelobe ratio, $\nu_{\rm min} = \SI{587.5}{\mega\hertz}$ is the minimum frequency examined, and $y_{\rm max} = 255$ corresponds to the maximum baseline distance in the $\vec{\hat{y}}$ direction.

The window function in \Cref{eq:weights_ns_beamcal} will result in a frequency-independent synthesized beam in the $\hat{\theta}$ direction that has a $\mbox{FWHM} = \SI{0.385}{\degree}$ and sidelobes that are $\lesssim 10^{-3}$ of the peak amplitude.  The Dolph-Chebyshev window minimizes the main lobe width for a given number of baselines and equiripple peak-to-sidelobe ratio.  It will degrade the point-source sensitivity relative to the inverse variance weighting scheme discussed in \S\ref{sec:mapmaking}, however the loss of sensitivity is not problematic for beam calibration because it relies on a foreground signal that is $\gtrsim 500$ times brighter than the noise.  The low equiripple sidelobes help to ensure that each formed beam is sensitive to the primary beam at a narrow range of declinations.

The argument of the window function is scaled with frequency so that the synthesized beam in the $\hat{\theta}$ direction is frequency independent.  Essentially the resolution at every frequency is degraded to the resolution at the lowest frequency.  This ensures that all frequencies are sensitive to the same declinations, so that any errors in our sky model are not further modulated by a frequency-dependent synthesized beam pattern.

\begin{figure*}[t]
   \centering \includegraphics[width=1\linewidth,keepaspectratio, trim = 0 430 0 0]{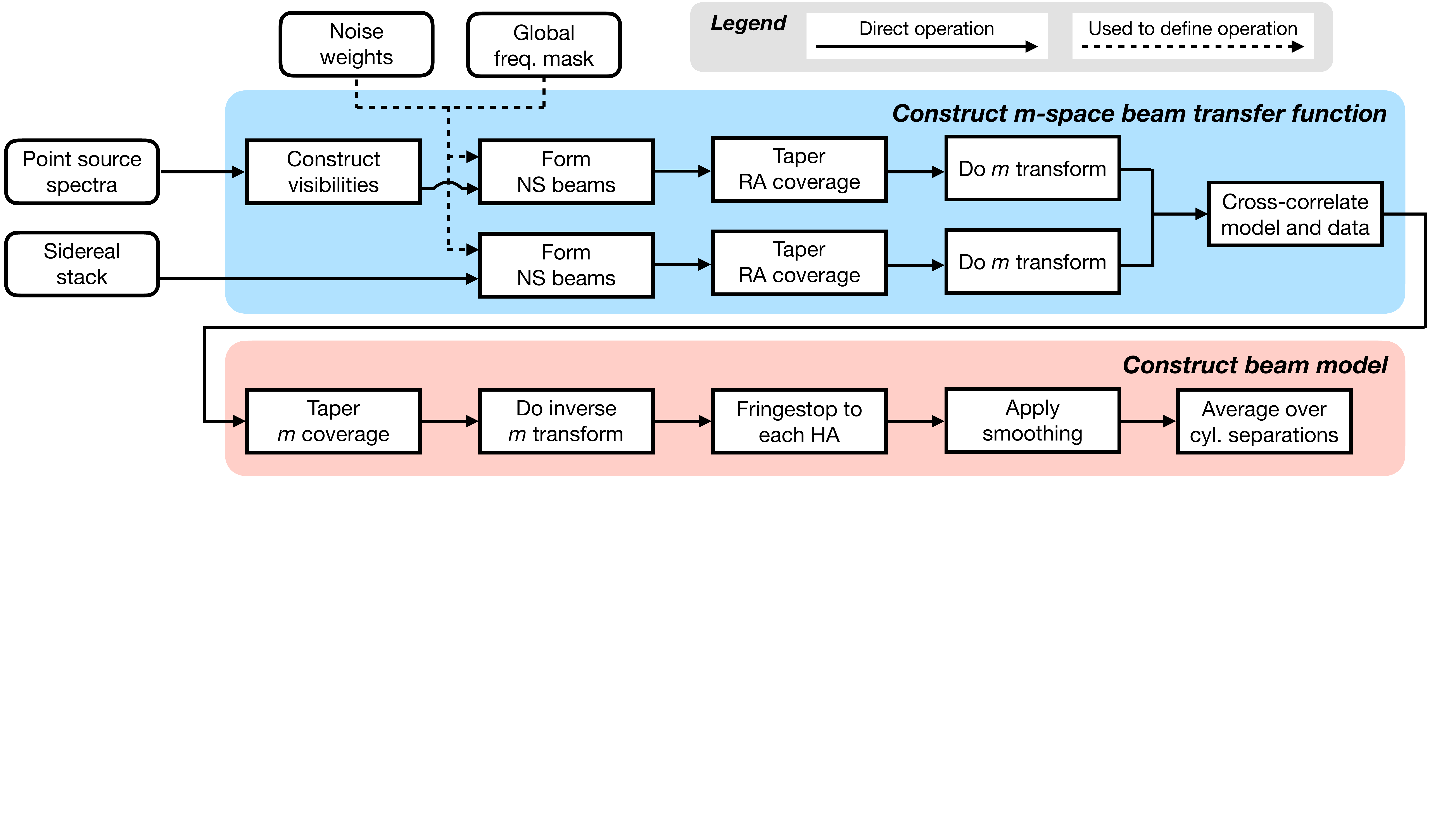}
    \caption{A schematic representation of the construction of the primary beam model used in our stacking analysis. Externally-measured spectra of \SI{97941} radio point sources are propagated into mock visibilities, which, after several transformations, are cross-correlated with CHIME observations to construct a beam transfer function that assumes that the sky is solely composed of these sources. Further transformations are applied to minimize sensitivity to this assumption and remove artifacts.}
    \label{fig:psbeam_flowchart}
\end{figure*}

Next we multiply the hybrid beamformed visibilities by a cosine-tapered window that is unity for $\SI{125}{\degree} < \phi < \SI{255}{\degree}$ and transitions to zero over a span of \SI{15}{\degree}.  This restricts our attention to a relatively quiet portion of the radio sky, avoiding sharp features in the Galactic emission that are present in the data but not in our model, and also avoiding regions of the sky contaminated by Cygnus A and Casseopia A in the sidelobes, which this technique is unable to account for properly.  This range of $\phi$ also coincides with the range covered by the eBOSS NGC field.  The $m$-mode transform is then taken.

The model for the primary beam is obtained by cross-correlating the sky model and the visibilities in $m$-mode space,
\begin{align}
    \tilde{B}^{p}_{xm}(\nu, \theta) & = \frac{\tilde{S}^{*}_{xm}(\nu, \theta) \ \tilde{V}^{p}_{xm}(\nu, \theta)}{|\tilde{S}_{xm}(\nu, \theta)|^{2} + \sigma^{p}_{x}(\nu)^{2}} \ ,
\end{align}
where $\tilde{S}$, $\tilde{V}$, and $\tilde{B}$ denote the $m$-mode transform of the hybrid beamformed visibilities for the sky model, data, and beam transfer function, respectively, and $\sigma$ is an estimate of the noise in $\tilde{V}$.

The $m$-mode transform of the beam transfer function, $\tilde{B}$, is multiplied by a cosine-tapered mask to remove any $m$-modes that cannot originate from the sky near meridian.  This mask is unity for $m_{{\rm center}, x} \pm 0.75 m_{\rm width}$ and then smoothly transitions to zero by $m_{{\rm center}, x} \pm m_{\rm width}$ (see \Cref{eq:m_center} and \Cref{eq:m_width}).  The inverse $m$-mode transform is then calculated to obtain our estimate of the beam transfer function $B$ for each east-west baseline separation $x$.  The beam transfer function is then ``fringestopped'', or in other words, is multiplied by the complex conjugate of the exponential term in \Cref{eq:beam_transfer}, to recover $|A^{p}(\nu, \theta, \phi)|^{2}$, which we will refer to as the power beam.  Note that this method yields 4 distinct estimates of the power beam, one for each east-west baseline separation.

We find that the resulting estimate of the power beam exhibits small scale variations along the declination axis that are highly correlated as a function of frequency and hour angle.  We suspect that these variations are due to errors in the flux of the sources in the sky model, and remove them as follows.  At each declination, the logarithm of the power beam at $\phi = \SI{0}{\degree}$ is fit to a fourth order polynomial in frequency.  This logarithmic polynomial model is then high-pass filtered along the $\hat{\theta}$ direction so that only variations on scales $\lesssim \SI{3}{\degree}$ are preserved.  The power beam at each declination is then divided by the exponential of the high-pass filtered, logarithmic polynomial model.

The uncertainty in the power beam is estimated at each frequency and declination by examining the variance at large hour angle ($0.087 \leq |\cos{\theta} \sin{\phi}| \leq 0.42$).  This uncertainty varies significantly as a function of declination based on the brightness of the sources at that declination.  We apply a 2D Savitzky-Golay filter in ($\nu$, $\theta$) space to low-pass filter the beam model.  For each ($\nu$, $\theta$, $\phi$) a 4th-order Chebyshev polynomial in both $\nu$ and $\theta$ is fit to a small window centered on that location.  The best-fit polynomial model is evaluated at that location to obtain the low-pass filtered version of the beam model.  The variance in the beam model at large hour angle is used to estimate the weights in the fit and properly account for the declination dependent uncertainties.  The size of the window changes between three distinct values based on the declination and frequency in order to retain features in the beam at progressively smaller scales as one moves to lower declinations.  In addition to smoothing the beam, the low-pass filter interpolates the beam to the majority of the frequencies that have been masked because of missing data or RFI.

Even after applying the 2D smoothing operation, there are still sharp features in the beam along the frequency axis that we believe originate from unflagged RFI present in the sidereal visibilities.  These sharp features will leak foreground power to small-spectral scales when the beam model is deconvolved from the data.  To address this, at each $(\theta, \phi)$ we apply an 8th order low-pass Butterworth filter along the frequency axis.  The cutoff used  for the low-pass filter is declination dependent in order to retain what we suspect are actual features of the beam.  The cutoff ranges from \SIrange{125}{200}{\nano\second}.

The final estimate of power beam is obtained from a weighted average of the estimate from baselines with a \SI{44}{\meter} and \SI{66}{\meter} east-west component.  The baselines with \SI{0}{\meter} east-west component are contaminated by diffuse Galactic emission, which is not present in our sky model, and also coupled noise that varies slowly as a function of Earth-rotation angle and thus appears at low-$m$'s that overlap with the range of $m$ at which the meridian sky fringes. The baselines with \SI{22}{\meter} east-west components are also contaminated by coupled noise, albeit to a lesser extent.

This technique is currently unable to measure the primary beam accurately at hour angles greater than $\approx \SI{2.0}{\degree}$, where the first-order approximation for the geometric phase given in \Cref{eq:phase_linear} begins to break down and an additional term that depends on the north-south baseline distance, declination, and hour angle becomes relevant.  The phase due to this term will be equal to the first-order phase at a new ``effective'' declination given by
\begin{eqnarray}
    \theta_{\rm eff}' & = \arcsin{\left( \cos{\Lambda} \sin{\theta'} - \sin{\Lambda} \cos{\theta'} \cos{\left(\phi - \phi'\right)}\right)} + \Lambda \ .
\end{eqnarray}
As a result, bright sources will exhibit a ``U'' shape track in the hybrid beamformed visibilities as they move out of the meridian beam centered on their true declination $\theta'$ and into meridian beams centered on more northern declinations at $\theta_{\rm eff}'$.  The recovered primary beam will be attenuated at large hour angles by a factor $\synth^{\hat{\theta}, p}(\nu, \theta, \theta_{\rm eff}', \phi) \ / \ \synth^{\hat{\theta}, p}(\nu, \theta, \theta', \phi)$.  In the main lobe of the primary beam the attenuation is less than \SI{6}{\percent} for the polarisations, frequencies, and declinations considered in this work, but in the sidelobes it quickly becomes significant.  We are actively exploring extensions to this algorithm that are capable of recovering the side lobes as well.

%% file: appendix/lognormal_stacking.tex

\section{Stacking on Lognormal Galaxy Density Realizations}
\label{sec:lognormal_stacking}

In \secref{sec:skymap_generation}, we made the following statement: if simulated galaxy catalogs are drawn from lognormal realizations of the galaxy density $\deltag$, and correlated Gaussian-distributed HI maps are stacked on the resulting galaxy positions, the measured stacking signal is the same as it would be if the galaxy catalogs were drawn from Gaussian realizations of the galaxy density. In this appendix, we justify this statement. We will make use of the following Gaussian integrals: if $\vec{\delta}$, $\vec{\alpha}$ are $n$-component vectors and $\mC$ is a symmetric, positive-definite $n\times n$ matrix, we can write
\begin{align}
\label{eq:gaussianint1}
\int d^n\vec{\delta}\, e^{-\frac{1}{2} \vec{\delta}^T \mC^{-1} \vec{\delta} + \vec{\alpha} \cdot \vec{\delta}}
	&= \sqrt{ (2\pi)^{n} \det \mC }\, e^{ \frac{1}{2} \vec{\alpha}^T \mC \vec{\alpha} }\ , \\
\label{eq:gaussianint2}
\int d^n\vec{\delta}\, \delta_i\, e^{-\frac{1}{2} \vec{\delta}^T \mC^{-1} \vec{\delta} + \vec{\alpha} \cdot \vec{\delta}}
	&= \sqrt{ (2\pi)^{n} \det \mC }\, e^{ \frac{1}{2} \vec{\alpha}^T \mC \vec{\alpha} }
	\sum_j C_{ij} \alpha_j\ .
\end{align}

Consider an idealized version of the stacking analysis, in which we average the HI overdensity $\deltaHI$ at a 3d separation $\vr$ from the location of each of $N$ galaxies in a catalog:
\beq
S(\vr) = \frac{1}{N} \sum_{i=1}^N \deltaHI(\vx_i+\vr)\ .
\label{eq:idealstack}
\eeq
Suppose that $\deltaHI$ has Gaussian statistics, while the galaxy overdensity $\deltag$, from which the galaxy positions are drawn, is lognormal, related to a Gaussian field $\deltaG$ by
\beq
1+\deltag(\vx) = e^{\deltaG(\vx) + \mu(\vx)}\ .
\label{eq:deltag-lognormal}
\eeq
Also, let $C_{ab}(\vx,\vx') \equiv \langle \delta_a(\vx) \delta_b(\vx') \rangle$, where $a, b \in \{ {\rm G}, {\rm HI} \}$. Following the standard procedure for lognormal fields, we fix $\mu(\vx)$ in \cref{eq:deltag-lognormal} such that $\langle 1+\deltag(\vx) \rangle = 1$. We can compute the relevant ensemble average by defining $\vec{\delta}$ to be $\deltag$ evaluated at a finite number of points, and writing
\begin{align*}
\left\langle 1+\deltag(\vx) \right\rangle
	&= \int d^n\vec{\delta}\, \left[ 1+\deltag(\vx) \right] \calP(\vec{\delta}) \\
&= \int d^n\vec{\delta}\, e^{\deltaG(\vx) + \mu(\vx)}
	\frac{1}{\sqrt{(2\pi)^n \det \mC_{\rm GG} }}
	e^{ -\frac{1}{2} \vec{\delta}^T \mC_{\rm GG}^{-1} \vec{\delta} } \\
&= e^{\mu(\vx) + \frac{1}{2} C_{\rm GG}(\vx, \vx) }\ ,
\numberthis
\end{align*}
where in the second equality, we substituted \cref{eq:deltag-lognormal} and the pdf for a Gaussian random field, and in the final equality we used \cref{eq:gaussianint1}. Setting this to unity implies that $\mu(\vx) = -(1/2) C_{\rm GG}(\vx, \vx)$.

We wish to show that stacked HI overdensity in \cref{eq:idealstack} approaches the same result whether the galaxy positions are drawn from the lognormal field in \cref{eq:deltag-lognormal} or from the Gaussian field $\deltaG$ itself. To do so, we first consider an ensemble average over galaxy positions in the catalog, keeping the underlying fields ($\deltag$ and $\deltaHI$) fixed:
\beq
\left\langle S(\vr) \right\rangle_{\rm cat}
	= \frac{1}{N} \sum_{i=1}^N  \left\langle \deltaHI(\vx_i+\vr) \right\rangle_{\rm cat}
	= \frac{1}{N} \sum_{i=1}^N \int_V d^3\vx_i\, \deltaHI(\vx_i + \vr) \, \calP(\vx_i | \deltag)\ ,
	\label{eq:cataverage1}
\eeq
with the integral evaluated over the survey volume $V$.
The probability distribution function of galaxy positions, $\calP(\vx | \deltag)$, is given by $\calP(\vx | \deltag) = V^{-1} [1+\deltag(\vx)]$. By combining Eqs.~\eqref{eq:deltag-lognormal}-\eqref{eq:cataverage1} and the expressions for $\calP(\vx | \deltag)$ and $\mu(\vx)$, we obtain
\beq
\left\langle S(\vr) \right\rangle_{\rm cat} = \frac{1}{V} \int_V d^3\vx\,
	\deltaHI(\vx+\vr) e^{\deltaG(\vx) - \frac{1}{2} C_{\rm GG}(\vx, \vx)}\ .
	\label{eq:cataverage2}
\eeq

We now take the ensemble average of \cref{eq:cataverage2} over the density fields as well as the catalog positions. Defining $\vec{\delta}$ to contain both $\deltaHI$ and $\deltaG$, and letting $\mC$ represent the joint covariance, we have
\begin{align*}
\left\langle \deltaHI(\vx+\vr) e^{\deltaG(\vx)} \right\rangle_{\rm fields}
	&= \int d^n\vec{\delta}\, \deltaHI(\vx+\vr) e^{\deltaG(\vx)}
	\frac{1}{\sqrt{(2\pi)^n \det \mC }}
	e^{ -\frac{1}{2} \vec{\delta}^T \mC^{-1} \vec{\delta} } \\
&= e^{ \frac{1}{2} C_{\rm GG}(\vx, \vx) } C_{\rm HI, G}(\vx+\vr, \vx)\ ,
\numberthis
\end{align*}
where we used \cref{eq:gaussianint2} in the second equality. Combining this with \cref{eq:cataverage2} and assuming that the fields have translation-invariant statistics, we arrive at
\beq
\left\langle S(\vr) \right\rangle_\text{cat, fields} = C_{\rm HI, G}(\vr, \vec{0})\ ,
\eeq
which is also what we would obtain if the galaxy positions were drawn directly from $\deltaG$ itself.

%% file: appendix/template_calculation.tex

\section{Template Calculation}
\label{app:template_calculation}

In this appendix, we discuss the challenge of calculating the signal templates for arbitrary parameter combinations and the approach we take in this work. Other than the frequency bias parameter $\Dnu$, the parameters described in \secref{sec:freeparameters} affect the properties of the underlying large-scale structure, or the \tcm or tracer density fields. That suggests that one way of calculating the template is to produce a realisation of the \tcm field and a correlated tracer catalog given a set of parameters, and then simulate a CHIME timestream from the \tcm field using a model for the instrumental transfer function, repeat the analysis procedure done to the actual data (flagging, filtering and map making), and finally stacking the output on the mock catalog. By repeating this procedure and averaging the results, we can estimate the expected signal.

Unfortunately, a full Monte-Carlo of this procedure is challenging, as even a single iteration requires around 900 core-hours of compute time (dominated by the timestream generation from input sky maps).
Instead we utilize the ergodic principle. We have an overlapping volume of $\gtrsim \SI{10}{\giga\parsec^3\per\h^3}$ (covered by the eBOSS quasar sample), but the stacking is probing scales $\sim \SI{10}{\mega\parsec\per\h}$. This gives many quasi-independent regions of that size within the volume, and so on those scales we expect the volume average to approach the ensemble average, or equivalently, that averaging over independent mock source catalogs drawn from a single large-scale structure realisation should give the same as averaging over completely independent large-scale structure realisations. Though this naive picture will break down on larger scales where the cosmic variance contribution is significant, we find that the cosmic variation in the stack signal is small:
around 0.8\% for the LRG sample, 0.2\% for the ELGs and 0.3\% for QSOs (estimated by comparing the zero-lag amplitude for stacks drawn from distinct LSS realisations).
This is no more than the variation between single catalogs drawn from the same LSS realisation
($\sim 0.7\%$ for LRGs and QSOs, $\sim 1.2\%$ for ELGs)
although we average over a sufficiently large number of catalogs to reduce this contribution to well below the cosmic variance level.

While this gives us a tractable method of computing the template for a given set of parameters, it still requires a costly timestream simulation for each set of values. To avoid this, we note that as both the process of observation and analysis are linear (other than data-derived RFI and bright pixel masking), if we can isolate individual terms in the cross-power spectrum description, they map to distinct contributions to the stacked signal.

For the moment we will fix the Fingers of God parameters $\alphaFoG{\sHI}$ and $\alphaFoG{\sg}$ as well as the non-linear power spectrum parameter $\alphaNL$, and to make the notation more compact we will define scaling parameters about the fiducial model: $\alphaOmega = \OmegaHI / \OmegaHI^\fid(\zeff)$, $\alphah = \bHI / \bHI^\fid(\zeff)$ and $\alphag = \bg / \bg^\fid(\zeff)$. With this we can rewrite \cref{eq:crossps} as
\begin{equation}
    \label{eq:crossps_sep}
    P_{\sHI,\sg}(\alphaOmega, \alphah, \alphag, \Mten) =
    \alphaOmega (\alphah \alphag \, P_{hg} + \alphah \, P_{hv} + \alphag \, P_{vg} + P_{vv}) + \Mten \, P_\mathrm{sn}
\end{equation}
where we have left the $k$, $\mu$ and $z$ dependence implicit. The power spectrum terms on the right hand side are
\begin{align}
    P_{hg} & = \Tb^\fid(z_1) \bHI^\fid(z_1) \bg^\fid(z_2) P \\
    P_{hv} & = \Tb^\fid(z_1) \bHI^\fid(z_1) f(z_2) \mu^2 P \\
    P_{vg} & = \Tb^\fid(z_1) \bg^\fid(z_2) f(z_1) \mu^2 P \\
    P_{vv} & = \Tb^\fid(z_1) f(z_1) f(z_2) \mu^4 P \\
    P_\mathrm{sn} & = C_\sHI(z_1) 10^{10} \: \si{\Msolar}
\end{align}
with
\begin{equation}
    \label{eq:P_sep}
    P = \ls \PL(k) + \alphaNL \lp \PNL(k) - \PL(k) \rp \rs \DFoG{\sHI}(\alphaFoG{\sHI} \, k \mu; z_1) \: \DFoG{\sg}(\alphaFoG{\sg} \, k \mu; z_2) \; .
\end{equation}
The linearity of the simulation and analysis procedure means that the stack signal should be separable into distinct terms like \cref{eq:crossps_sep}. If we write the template generated by the given parameters as $s(\alphaOmega, \alphah, \alphag, \Mten)$, where we make the dependence on $\Delta\nu$ implicit, we find
\begin{equation}
    \label{eq:template_sep}
    s(\alphaOmega, \alphah, \alphag, \Mten) = \alphaOmega (\alphah \alphag \, s_{hg} + \alphah \, s_{hv} + \alphag \, s_{vg} + s_{vv}) + \Mten \, s_\mathrm{sn}
\end{equation}
where each $s_{xy}$ term is the stack signal corresponding to cross-power spectrum term $P_{xy}$ as defined above. However, as we cannot directly propagate a cross-power spectrum into a stack signal, we must determine these terms indirectly. This can be done by running simulations through with specific $\alpha$ parameters that generate known linear combinations of the $s_{xy}$. By choosing these simulated parameters judiciously we can easily invert these combinations to generate the individual $s_{xy}$ terms. One such choice is
\begin{align}
    s_{hg} & = s(1, 1, 1, 0) - s(1, 1, 0, 0) - s(1, 0, 1, 0) + s(1, 0, 0, 0) \\
    s_{hv} & = s(1, 1, 0, 0) - s(1, 0, 0, 0) \\
    s_{vg} & = s(1, 0, 1, 0) - s(1, 0, 0, 0) \\
    s_{vv} & = s(1, 0, 0, 0) \\
    s_\mathrm{sn} & = s(0, 0, 0, 1) \; .
\end{align}
Each of the five unique combinations of parameters passed to $s(\alphaOmega, \alphah \alphag, \Mten)$ in the equation above requires a separate simulation to determine, but after that, we can use these modes and \cref{eq:template_sep} to determine the stacked template for any combination of parameters.

This scheme allows us to exactly treat the effects of the three linear parameters and the shot-noise contribution on the template. Incorporating the effect of the non-linear power spectrum shape is straightforward as the parameterisation used for $\alphaNL$ means that the output stack signal is a simple linear mixing of two terms, as it is in the cross power spectrum \cref{eq:P_sep}. This means that four more simulations can be used to generate templates at any $\alphaNL$. However, the template is a non-linear function of the Fingers of God parameters and so cannot be exactly generated in a finite number of modes.

To account for this we start by noting that the effect of the Fingers of God treatment we use (see \secref{sec:fingerofgod}), at constant time and in comoving distance, is a convolution of the underlying fields along the line of sight. In a narrow enough interval in redshift, such that we could ignore evolutionary effects and the constant frequency spacing of our measurements maps to a constant separation in comoving distance, this effect commutes with the stacking and we could apply it via convolving a post-simulation $s_{xy}$ template to the desired $\alphaFoG{x}$ value, rather than needing to incorporate it into the simulations directly. However, as our sources are located over wide redshift intervals, evolution of the cosmological fields as well as the pairwise velocity dispersion $\sigma_P$ cannot be neglected, and in addition the RFI masking, redshift-dependent source number density, and sensitivity further break the stationarity of the radial axis.

However, even if there is no exact mapping from the Fingers of God effects into a convolution on the template modes, we can still attempt to find an effective one that is accurate in the vicinity of the fiducial Fingers of God parameters $\alphaFoG{x} = 1$. To do this we use use a transfer function in delay, $\tau$, the Fourier conjugate of the frequency separation $\Delta\nu$, of the form
\begin{equation}
    D^\mathrm{eff}_{x}(\tau, \alphaFoG{x}) = \frac{2 + \tau^2 \sigma_\mathrm{eff}^2}{2 + \alphaFoG{x} \tau^2 \sigma_\mathrm{eff}^2} \; ,
\end{equation}
that will be applied to the templates with the fiducial Fingers of God strength.
To motivate this choice, we note that within a short redshift interval,
\begin{equation}
   \tau \approx - \frac{1}{2 \pi \nuHI} \frac{c}{H(z)} (1 + z)^2 k_\parallel
\end{equation}
and so if we set
\begin{equation}
    \sigma_\mathrm{eff} = 2 \pi \nu_0 \frac{H(z)}{c (1 + z)^2} \sigma_{\rm P}\
\end{equation}
and then apply the transfer function above to the template, the numerator would effectively undo the Lorentzian Fingers of God model with the fiducial $\alphaFoG{x} = 1$ and the denominator would reapply it with the desired $\alphaFoG{x}$. Thus we would have transformed the template mode from the fiducial to the the desired $\alphaFoG{x}$ parameter. To take into account the wide redshift range and non-stationarity, we estimate an effective smoothing width $\sigma_\mathrm{eff}$ by finding the value which minimises the template error at a higher $\alphaFoG{x} = 1.2$ compared to an exact simulation at the same value. This effective convolution approach is applicable over a wide range of values of $\alphaFoG{x}$, with errors at $\alphaFoG{x} = 0$ or $3$ of $\lesssim 0.5 \%$ and much smaller around the pivot $\alphaFoG{x} = 1$. Computationally this requires an additional eight simulations, one for each of the four $(\alphah$, $\alphag)$ combinations with a perturbed value of $\alphaFoG{\sHI} = 1.2$, and an additional four with $\alphaFoG{\sg} = 1.2$

The final effect we need to apply is the frequency shift $\Dnu$. This is performed in Fourier space by phase rotating the delay transform of the template.